\documentclass[11pt]{book}

\usepackage{graphicx,siunitx,array,makecell,hhline,multirow,url,fancyhdr,lastpage,listings,color,textcomp,nicefrac,booktabs,mathdots,bm,mathtools,titlesec,rotating,setspace,amssymb,diagbox,float,xcolor,standalone,import,nameref,pdfpages}
\usepackage[margin=2.5cm]{geometry}
\usepackage[T1]{fontenc}
\usepackage[labelformat=simple,labelsep=colon]{subcaption}
\usepackage[labelfont=small,labelfont=bf]{caption}
\usepackage[none]{hyphenat}
\usepackage[tt=false]{libertine}
\usepackage[useregional=text,calc,en-GB]{datetime2}
\usepackage[bottom]{footmisc}
\usepackage[ %
  colorlinks=true,
  breaklinks=true,
  linktocpage=true
]{hyperref}

\hypersetup{ %
	pdftitle={Traceable thermal imaging in harsh environments},
	pdfauthor={Jamie McMillan},
	pdfdisplaydoctitle=true,
	pdfpagelayout=TwoColumnLeft
}
\DeclareSIUnit\px{px}
\DeclareSIUnit\frame{frame}
\DeclareSIUnit\digitallevel{DL}
\DTMlangsetup[en-GB]{ord=omit}
\DeclarePairedDelimiter\abs{\lvert}{\rvert}
\newcolumntype{L}[1]{>{\sloppy\let\newline\\\arraybackslash\hspace{0pt}}p{#1}}
\newcolumntype{M}[1]{>{\centering\arraybackslash}m{#1}}
\makeatletter
\renewcommand\p@subfigure{}
\makeatother

\graphicspath{{./Images/}}

\fancypagestyle{fancyStyle}{%
\fancyhf{}%
\fancyhead[R]{\nouppercase{\leftmark}}
\fancyhead[L]{Traceable thermal imaging in harsh environments}
\fancyfoot[L]{Jamie McMillan}
\fancyfoot[R]{{\thepage} of \pageref*{LastPage}}
\fancyfoot[C]{Version 3.1 - \DTMtoday}

}

\fancypagestyle{plain}{}

\fancypagestyle{frontmatterStyle}{%
\fancyhf{}%
\fancyhead[L]{Traceable thermal imaging in harsh environments}
\fancyfoot[C]{\thepage}
}

\newcommand{\markedchapter}[2]{\chapter[#2]{#2%
\chaptermark{#1}}%
\chaptermark{#1}}

\makeatletter
\def\cleardoublepage{\clearpage\if@twoside \ifodd\c@page\else
\begingroup
\mbox{}
\thispagestyle{empty}
\newpage
\if@twocolumn\mbox{}\newpage\fi
\endgroup\fi\fi}
\makeatother

\titleformat{\paragraph}
{\normalfont\normalsize\bfseries}{\theparagraph}{1em}{}
\titlespacing*{\paragraph}
{0pt}{3.25ex plus 1ex minus 0.2ex}{1.5ex plus 0.2ex}

\title{\vspace{-2em} Traceable thermal imaging in harsh environments \vspace{-1em}}
\author{J McMillan \vspace{-2em}}
\date{\DTMtoday \vspace{-2em}}

\begin{document}
\includepdf{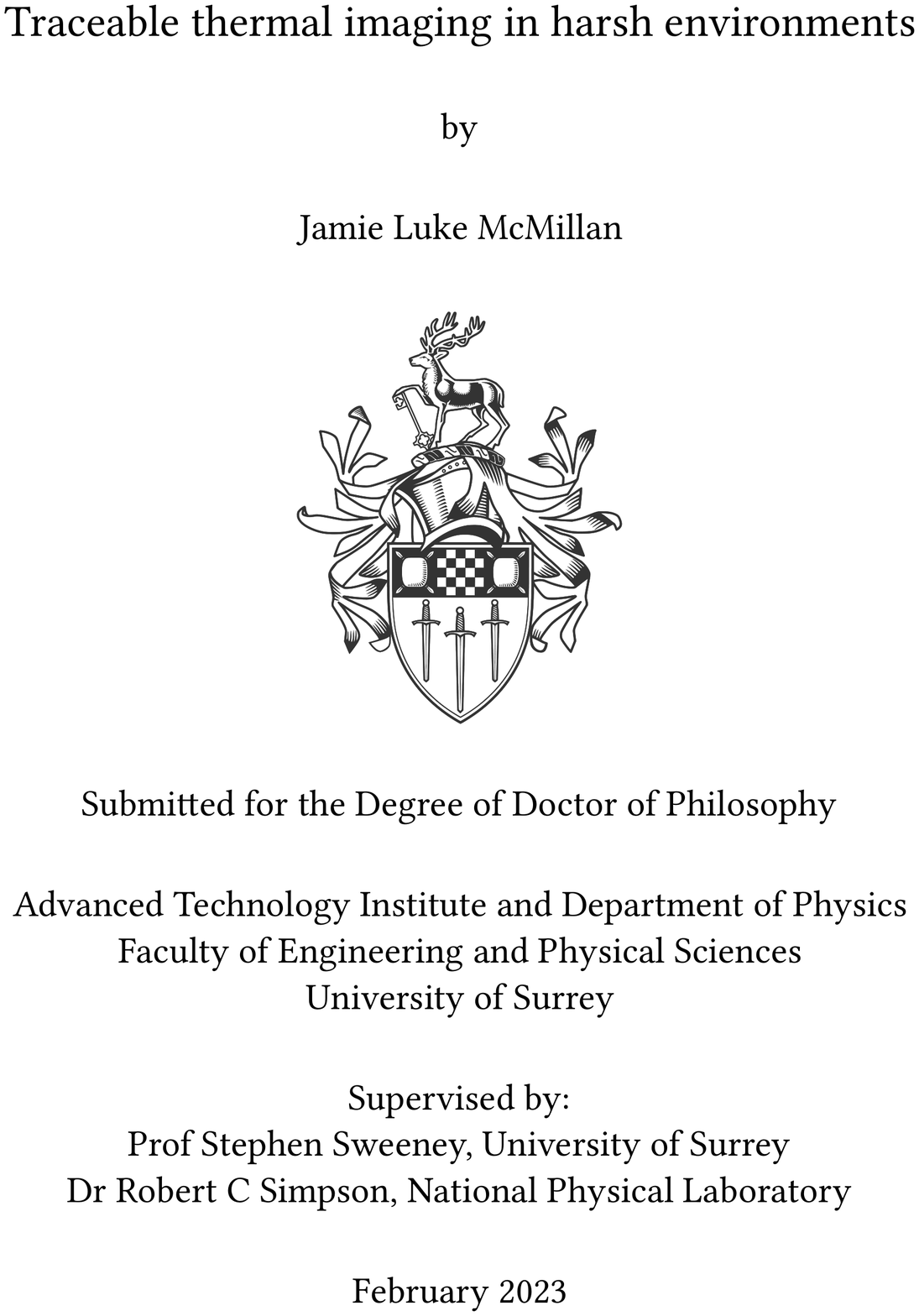}

\frontmatter
{\let\newpage\relax\maketitle}
\thispagestyle{empty}
\pagestyle{frontmatterStyle}
\tableofcontents

\clearpage
\newpage
\chapter*{Abstract}\label{chap:abstract}
Despite being regarded as a well-established field, temperature measurement continues to pose significant challenges for many professionals in the metrology industry. Thermal imagers enable fast, non-contact and a full field measurement, however there is a lack of metrological development to support their use. Here, thermal imagers have been examined for the monitoring of special nuclear material containers; the surface temperature is an important parameter for store management decisions. Throughout this research: a selection of thermal imagers were calibrated and made traceable to the International Temperature Scale of 1990; laboratory observations of a proxy steel plate were made; initial measurement of nuclear material storage containers were made; then a deployment to an inactive store was demonstrated. For this technique to be feasible, uncertainties less than \SI{10}{\celsius} would be required.

During the laboratory calibration of an uncooled and cooled thermal imager against blackbody reference sources, across the measured temperature range of \SIrange{10}{100}{\celsius} the uncertainties were less than \SI{3.20}{\celsius} (\(k=2\)) and \SI{0.50}{\celsius} (\(k=2\)) respectively. Here \(k\) is the uncertainty coverage factor. When these calibrations were applied to the plate, regions of steel and higher emissivity coating were evaluated. These uncoated regions were measured with a thermal imager to demonstrate temperature differences compared to surface mounted thermocouples of \SI{8.3}{\celsius} and uncertainties up to \SI{30.1}{\celsius} (\(k=2\)). For the coated regions this temperature difference was reduced to \SI{1.8}{\celsius} with uncertainties up to \SI{6.8}{\celsius} (\(k=2\)).

Extending this approach to store containers -- each instrumented with internal heaters and thermocouples -- yielded poor comparability between thermocouples and an external thermal imager. Using a revised container instrumentation, deployment of two uncooled thermal imagers to an inactive store to observe the container was completed. From this measurement campaign the surface temperature determined using a thermal imager for a container ranged from \SI{4.9}{\celsius} (\(k=2\)) to \SI{20.5}{\celsius} (\(k=2\)). These results demonstrate the feasibility of thermal imagers being deployed to nuclear material stores for the assessment of radioactive material behaviour in storage containers.

\newpage

\section*{Declaration of Originality}
This thesis and the work to which it refers are the results of my own efforts. Any ideas, data, images or text resulting from the work of others (whether published or unpublished) are fully identified as such within the work and attributed to their originator in the text, bibliography or in footnotes. This thesis has not been submitted in whole or in part for any other academic degree or professional qualification. I agree that the University has the right to submit my work to the plagiarism detection service TurnitinUK for originality checks. Whether or not drafts have been so-assessed, the university reserves the right to require an electronic version of the final document (as submitted) for assessment as above.

\vspace{2em}
\begin{table}[h]
\renewcommand{\arraystretch}{2.5}
\begin{tabular}{ l l }
  Signature	&	\includegraphics[width=0.2\textwidth, keepaspectratio]{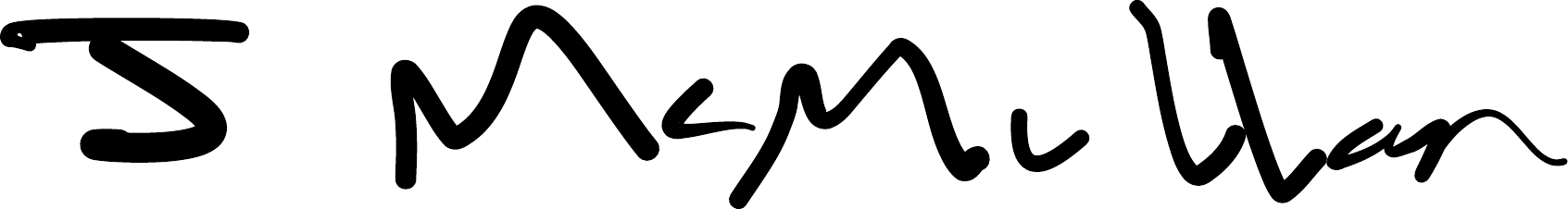}	\\
  Date		&	\DTMtoday		\\
\end{tabular}
\label{tab:statementOfOriginality}
\end{table}

\newpage
\section*{Acknowledgements}
The tangible output from this doctorate pales when compared to the joy and value from the journey. The countless catch ups and brainstorming with the people I care about are the most important results from my research.

\newpage
\section*{Publications}\label{sec:publications}
The following instances of external recognitions are detailed in Table~\ref{tab:publications}, conference presentations shown in Table~\ref{tab:conferences} and posters submitted to conferences in Table~\ref{tab:posters}.

\begin{table}[ht]
		\renewcommand{\arraystretch}{1.5}
		\centering 
		\caption{Publications submitted through this research project.}
		\begin{tabular}{ L{1cm}L{12.0cm}L{2.0cm}}
		\toprule
		Index & Publication & Status \\
		\cmidrule(lr){1-3}
		1 & {\bf McMillan J L}, Hayes M, Hornby R, Korniliou S, Jones C, O'Connor D, Simpson R, Machin G, Bernard R and Gallagher C. Thermal and dimensional evaluation of a test plate for assessing the measurement capability of a thermal imager within nuclear decommissioning storage. {\em Measurement}, 202:111903, 2022. & Accepted \\
		2 & {\bf McMillan J L}, Hayes M, Simpson R and Sweeney S. Radiance correction methods and a cross-comparison of applicability. {\em Quantitative InfraRed Thermography Journal}. & Draft \\
		3 & {\bf McMillan J L}, Hayes M, Korniliou S, Simpson R, Machin G, Bernard R and Gallagher C. Evaluating nuclear material storage containers using thermal imagers. {\em Nuclear Engineering and Design}. & Draft \\
		\bottomrule
		\end{tabular} 
		\label{tab:publications}
\end{table}

\begin{table}[ht]
		\renewcommand{\arraystretch}{1.5}
		\centering 
		\caption{Conferences attended through this research project.}
		\begin{tabular}{ L{6.5cm}L{2.0cm}L{6.5cm} }
		\toprule
		Conference & Date & Presentation Title \\
		\cmidrule(lr){1-3}
		\href{https://www.npl.co.uk/pgi/conference}{PostGraduate Institute Conference 2021}, Teddington & October 2021 & So you think you can measure temperature? \\
		\bottomrule
		\end{tabular} 
		\label{tab:conferences}
\end{table}

\begin{table}[ht]
		\renewcommand{\arraystretch}{1.5}
		\centering 
		\caption{Posters submitted to conferences through this research project.}
		\begin{tabular}{ L{6.5cm}L{2.0cm}L{6.5cm} }
		\toprule
		Conference & Date & Poster Title \\
		\cmidrule(lr){1-3}
		\href{https://www.npl.co.uk/pgi/conference-2020}{PostGraduate Institute Conference 2020}, Teddington & October 2020 & Towards traceable thermal imaging of nuclear waste containers \\
		\bottomrule
		\end{tabular} 
		\label{tab:posters}
\end{table}

\clearpage
\newpage
\pagestyle{fancyStyle}
\mainmatter
\chapter{Introduction}\label{chap:introduction}
Temperature measurement has a storied past with many scientists defining their own realisation of a temperature scale, but with the turn of the 20\textsuperscript{th} century these were unified through an International Temperature Scale. The most recent of which, the International Temperature Scale of 1990 is the practical realisation of temperature achievable using contact thermometers (metal in glass, thermocouples or resistance thermometers) as well as radiation thermometers.

Thermal imaging technology development has been growing since the 1940s with roots in observation and detection. As such the principle requirements of hardware included temperature sensitivity, linearity, responsivity and low size, weight and power. Increasing commercial demand has made the technology more accessible both through cost, instrument footprint and critically through the omission of liquid nitrogen cooling; this has led to deployment in a diverse range of applications, more of which now use the absolute temperature determination capability of the equipment.

In comparison to radiation thermometry, there are many unique characteristics of thermal imagers that do not facilitate straightforward surface temperature determination (e.g. focal plane uniformity, pixel cross-talk). On top of this there are other phenomena they share with radiation thermometers explored in Chapter~\ref{chap:radiance} which challenge low uncertainty temperature measurement such as influence on the measurement due to surface emissivity, size-of-source effect, detector temperature sensitivity.

Existing application of thermal imagers for surface temperature determination has been demonstrated in each: healthcare \cite{ref:medicalImagingDevice}, ground based satellite testing \cite{ref:3d_thermal_imaging} and nuclear material storage container assessment \cite{ref:ilw_paper_2018}.

The research presented within this thesis focuses on temperature metrology and outlines a robust surface temperature measurement methodology using thermal imagers. Experimental detail in the application to both controlled laboratory and semi-controlled in-situ measurements is presented, with computational extrapolation to a wider set of thermal environments. These should provide an understanding of the current state-of-the-art in temperature metrology and insight into the largest sources of experimental uncertainty that require consideration.

The primary narrative through this research comprises the progression from the surface temperature measurement in a laboratory of a proxy stainless steel plate in Chapter~\ref{chap:laboratoryMeasurement}, to a nuclear material container and finally to this container in an inactive store environment in Chapter~\ref{chap:inSituMeasurement}. This is with the objective to evaluate the suitability of using a thermal imager for surface temperature determination in a harsh environment. To demonstrate the feasibility of thermal imaging in this application, both agreement with a surface mounted thermometer within their respective uncertainties and measurement uncertainties below \SI{10}{\celsius} would be necessary.

Apparent radiance temperature as measured by thermal imagers can be considered using the Planck distribution law. Through both this relationship between object temperature, wavelength and spectral radiance, as well as the understanding of radiation heat transfer, the contribution from surface emitted radiance and any number of reflected components can be described. Emissivity of a surface describes the ability of a surface to absorb and emit electromagnetic radiation, which governs the photo-thermal behaviour of a surface with regard to its environment. A good understanding of the material properties and its environment are necessary in order to reliably use thermal imaging to determine surface temperature. Emissivity corrections are explored in this work using a range of assumptions and under more general cases. This experimental detail in a laboratory and extrapolation to a simulated environment are explored in Chapter~\ref{chap:laboratoryMeasurement}.

Essential to low uncertainty surface temperature measurement using thermal imagers is a low uncertainty apparent radiance temperature traceable calibration to an international temperature scale. Through comprehensive experience with dissemination of temperature to radiation thermometers, a range of calibration processes have been explored and discussed for thermal imaging systems; this includes both uncooled detector packages and cooled proprietary software systems. In addition to a comparison to an international temperature scale, further system characteristics including the calibration fit, detector temperature susceptibility, imager non-uniformity and size-of-source effect are investigated.

Deployment of the radiance temperature assessment and instrumentation characterisation techniques to the temperature determination of nuclear material storage containers has been demonstrated in Chapter~\ref{chap:inSituMeasurement}. A transition from controlled laboratory experiments using a representative stainless steel plate to an inactive storage facility permitted a detailed investigation into the sources of error encountered in surface temperature measurement in an uncontrolled environment.

\clearpage
\newpage
\pagestyle{fancyStyle}
\mainmatter
\chapter{Thermal Imaging Metrology}\label{chap:radiance}
\section{Introduction}\label{sec:introduction}

Measured temperature is a regular occurrence for most people either through local thermometer measurements or regional meteorology; accurate temperature measurement requires careful consideration with respect to the measurement application. All measurements are required to be traceable in order to be measuring an internationally recognised value and this can be achieved through traceable instrument calibration to internationally recognised standards; this enables both lateral comparison between different instruments as well temporal comparison using the same instrument over a long period of time. This level of comparability is not possible for uncalibrated instruments. 

Through rigorous examination of the calibration parameters for various instruments, a robust methodology to instrument characterisation can be implemented and the results deplyed to real-world applications where confidence in temperature measurement is critical.

For thermometers this traceability is demonstrated through calibration to the International Temperature Scale of 1990 (ITS-90) \cite{ref:npl_its90}. Rather than the thermodynamic temperature explored in literature and realised in a laboratory (e.g. through acoustic thermometry \cite{ref:acousticThermometry} or Johnson noise thermometry \cite{ref:johnsonNoiseThermometry}), thermometry using contact thermometers (e.g. thermocouples and resistance thermometers) or non-contact radiation thermometers is made traceable to a practical temperature scale comprising a set of fixed temperature points (analogous to a set of rungs on a ladder). Fixed-points refer to the melting or freezing temperatures at which a material passes through a phase transition; at these temperatures the energy heating or cooling the material is used to break or form structural bonds and so demonstrate a repeatable known temperature. These fixed points up to the freezing point of silver (\SI{961.78}{\celsius}) define ITS-90 and beyond this the scale is defined by the radiation emitted by an object using the Planck distribution law to relate this radiance and temperature.

This chapter will explore the calibration of radiation thermometers and thermal imagers. Each of these will be substantiated through a case study of an instrument calibration then comparisons and discrepancies between the two will be detailed. Textbook literature to support the physics described here is presented in Section~\ref{sec:radiation_heat_transfer} and Section~\ref{sec:photothermal_properties} where each radiation heat transfer and the description of emissivity are introduced. The concepts of emissivity are less pertinent during calibration due to the high emissivity nature of the reference sources used. However during the subsequent chapters where in-situ applications are investigated, emissivity plays a greater role.

\newpage
\section{Radiation thermometer calibration}\label{sec:thermometer_calibration}
Radiation thermometers capture infrared radiance to measure the apparent radiance temperature of an object. Beyond its dependence on the temperature of the surface and its environment, apparent radiance temperature is also a function of the surface emissivity defined (detailed in Section~\ref{sec:photothermal_properties}), it comprises a proportion of radiance from the observed surface and that from reflected contributions (discussed in Section~\ref{sec:radianceCorrection}). Calibration of a radiation thermometer is carried out using high emissivity (typically greater than 0.999) temperature reference blackbody sources and so throughout this chapter the reflected component will be omitted. The design of these sources and their subsequent use for the calibration of a Device Under Test (DUT) is described in this section.

\subsection{Calibration facilities}\label{subsec:thermometer_facilities}
For the calibration of an optical instrument that measures thermal radiation, the use of a blackbody source can be used. It should have the following properties (in addition to the characteristics outlined in Section \ref{subsec:radiation_definition_blackbody}) \cite{ref:blackbody_and_cal_sources,ref:npl_reference_sources}:

\begin{enumerate}
  \item High hemispherical total emissivity over the appropriate spectral range and emission direction
  \item Sufficiently large aperture size for the DUT optical characteristics
  \item Low temperature non-uniformity radially; and longitudinally for a cavity
  \item Have demonstrable traceability to the International Temperature Scale of 1990
\end{enumerate}

\begin{figure}[H]
  \centering
  \includegraphics[width=0.7\textwidth,keepaspectratio]{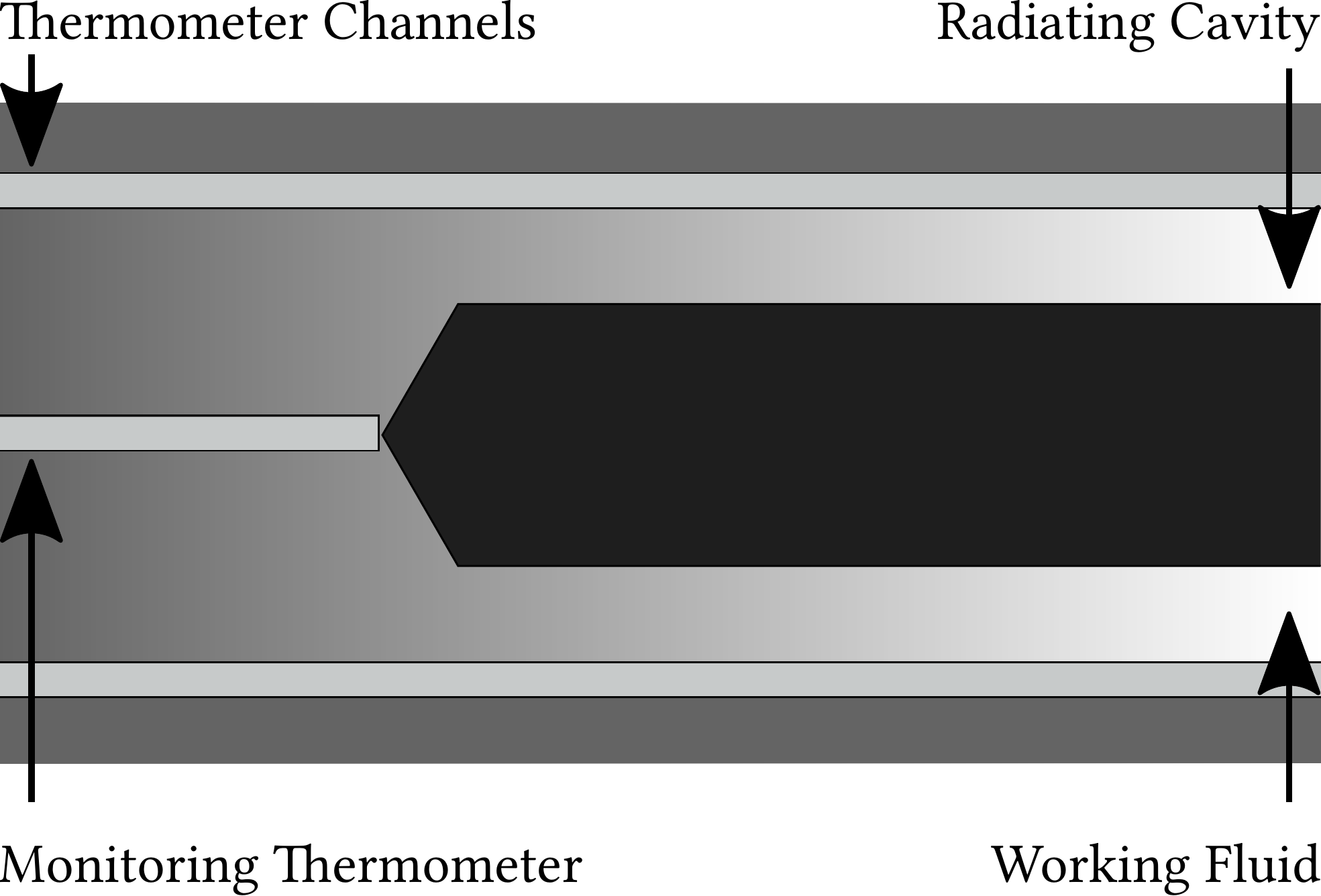}
  \caption{Typical construction of a heat-pipe cavity for radiance temperature reference standards. This cavity would be placed in a uniform thermal environment and the passive heat transfer from the working fluid would ensure an isothermal radiating cavity.}
  \label{fig:heatPipeConstruction}
\end{figure}

These criteria can be achieved for example by using variable temperature heat-pipe blackbody cavities in the temperature range from \SIrange{-60}{1000}{\celsius}; a typical design is shown in Figure~\ref{fig:heatPipeConstruction}. To accommodate this wide temperature range a series of working fluids are used and these fluids surround the radiating cavity \cite{ref:heatPipeIntroduction}; these fluids each operate in a specific operating temperature range: ammonia from \SIrange{-60}{50}{\celsius}; water from \SIrange{50}{270}{\celsius}; caesium from \SIrange{250}{650}{\celsius}; and sodium from \SIrange{500}{1000}{\celsius} \cite{ref:blackbody_and_cal_sources}. The high emissivity value of the cavity is achieved through: high cavity surface emissivity (e.g. 0.9), high temperature uniformity along the cavity length (below \SI{10}{\celsius} \cite{ref:euromet658}) and increased effective cavity emissivity using geometrical enhancement. Typical cavity surfaces are high emissivity paints where suitable or oxidised surfaces at higher temperatures. The longitudinal and radial temperature uniformity along the cavity is managed through the behaviour of the heat-pipe where it maintains high thermal conductivity along the condenser length. Finally, cavity emissivity can be enhanced through the use of an elongation of cavity length with respect to opening aperture diameter (refer to Eq.~\ref{eq:cavity_emissivity} \cite{ref:emissivityOfACavity}) or more often through complex cavity structures: pyramids at the cavity rear, ridges along the cavity length, asymmetric conical cavities, narrower opening apertures than cavity diameter or spherical cavity recesses between pyramid bases at the rear wall \cite{ref:large_area_blackbody_in_flight_calibration},

\begin{equation}
  \varepsilon_{cavity} = 1 - \frac{1 - \varepsilon_{wall}}{\varepsilon_{wall}} \frac{1}{1 + \nicefrac{l^2}{r^2}} \mathrm{.}
  \label{eq:cavity_emissivity}
\end{equation}

Here the geometry of a simple cavity -- of length \(l\) and radius \(r\) -- is enhanced from \(\varepsilon_{wall}\) to \(\varepsilon_{cavity}\) solely through the object geometry \cite{ref:blackbody_and_cal_sources}. The radiance temperature may be traceable to ITS-90 either through direct radiance comparison with a calibrated radiation thermometer or through a calibrated thermometer in close proximity to the back wall of the radiating cavity (as depicted in Figure~\ref{fig:heatPipeConstruction}).

Through each of these design considerations, a blackbody characterisation source with a measurement uncertainty from \SIrange{0.02}{1.5}{\celsius} (\(k = 2\)) can be achieved \cite{ref:comparison_cavity_ear_thermometer,ref:rbcf_detectors_radiators_ptb,ref:comparison_radiation_ptb_npl,ref:euramet_radiance_temperature,ref:development_bath_vertical_cavity} and the calibration of radiation thermometers can be demonstrated to achieve uncertainties from \SIrange{0.05}{0.3}{\celsius} (\(k = 2\)) in the temperature range from \SIrange{-60}{1000}{\celsius} \cite{ref:npl_radiation_thermometers_cmc}. 

\subsection{ITS-90 calibration}\label{subsec:thermometer_its90}
Using the facilities detailed in Section \ref{subsec:thermometer_facilities}, the ITS-90 below the silver point can be transferred either from an ITS-90 calibrated contact thermometer or by comparison using an ITS-90 calibrated radiation thermometer. For the DUT, the instrument is aligned to the reference source at a prescribed distance, with a temperature-regulated aperture plate placed in front of the cavity. This aperture plate reduces the reflected environmental radiance, the radiance from objects in the background (e.g. neighbouring furnace enclosure) and often to provide a cold aperture radiance signal compared to the hotter furnace radiance from within the aperture. At each calibration temperature setpoint, the blackbody temperature and DUT temperature reading are recorded for an identical duration, either through direct comparison or subsequent measurement of the radiating cavity.

If the DUT offers an analogue output capability then this response can also be calibrated against ITS-90. This requires the sensitivity \(\delta\) (temperature response per volt) to either be linear across its operational temperature range, or to be fully characterised in this range. The sensitivity values can be calculated from the \(i^{th}\) measurement by calculating the localised sensitivity within nominally \SI{1}{\celsius}, this is shown in Eq.~\ref{eq:radiance_sensitivity},

\begin{equation}
  \delta_i = \abs*{ \frac{ T_i - T_{i+1} }{ V_i - V_{i+1} } } \mathrm{.}
  \label{eq:radiance_sensitivity}
\end{equation}

Where the measured voltage \(V_i\) is from the instrument analogue output, \(T_i\) is the equivalent ITS-90 measured temperature from the reference source and \(i+1\) corresponds to the next higher temperature setpoint.

\subsection{Size-of-source effect}\label{subsec:thermometer_sse}
The Size-of-Source Effect (SSE) arises because the indicated reading of an instrument is affected by radiation from outside the nominal field of view and leads to a distorting effect on the measurement \cite{ref:sse}. The define measurement region within the field of view of the thermometer is likely to measure radiation outside of this area, conversely radiation from within the field of view may not reach the detector. This effect arises from scattering in the optical path and for a given optical system when the surface area of an observed source decreases, the contribution due to this effect increases.

This should be evaluated, for all such instruments being calibrated as an ongoing monitoring and assessment of the system to identify any gradual changes to the optical performance. The SSE measurement procedure is often performed at the highest temperature used in the calibration and becomes more sensitive to variation as the apparent temperature approaches ambient temperatures; due to the radiance contributions becoming comparable. The SSE, \(\sigma\), is described by Eq.~\ref{eq:sse} which is determined by the ratio of the radiances (Eq.~\ref{eq:planck_distribution_law}) at two aperture diameters,

\begin{equation}
  \sigma = \frac{ e^{\frac{c_2}{\lambda T_{\varphi_i}}} - 1 }{ e^{\frac{c_2}{\lambda T_{\varphi_{max}}}} - 1 } \mathrm{.}
  \label{eq:sse}
\end{equation}

Here \(T_{\varphi_i}\) is the measured temperature in kelvin at the \(i^{th}\) aperture and \(T_{\varphi_{max}}\) is the equivalent temperature at the greatest aperture diameter. \(\lambda\) is the instrument effective wavelength and \(c_2\) is the second radiation constant. Evaluation of the SSE results is carried out by determining the smallest aperture with which the DUT provides an acceptable measurement. An indication of a satisfactory SSE value for a particular aperture is to be within \SI{2}{\percent} of the maximum aperture value (e.g. above 0.98 for temperatures greater than ambient and below 1.02 for temperatures below ambient) \cite{ref:sse}.

An SSE profile measured at a particular temperature may not be representative of the profile at other temperatures if the signal from the blackbody is not much greater than that from the surrounding field of view. It is possible to infer a temperature invariant instrument SSE through the method described in \cite{ref:sse_correction_corrupted_background}. Through the use of this SSE measurement it is possible to correct for a measurement at one target size to another; for example if a calibration is performed at one aperture size, this can be applied to all target sizes within the tested range of aperture sizes.

\subsection{Case study: radiation thermometer}\label{subsec:case_study_thermometer}
A Heitronics TRT IV.82 radiation thermometer (S/N: 4182) was calibrated over the temperature range from \SIrange{-30}{1000}{\celsius} using National Physical Laboratory (NPL) standard blackbody sources. The temperatures of the blackbodies were determined through the use of ITS-90 calibrated thermometers (as shown in Figure~\ref{fig:heatPipeConstruction}). The aperture diameter of each blackbody was set to nominally \SI{40}{\milli\metre} using blackened apertures. For the calibration points greater than \SI{50}{\celsius} the apertures were water-cooled. The distance from the front of the thermometer lens to the aperture was nominally \SI{504}{\milli\metre}. The thermometer was aligned to the centre of the aperture using the sighting optics.

The calibrated digital voltmeter was connected to the output terminals of the thermometer. The calibration was performed by recording at each temperature the output signal of the test thermometer and the temperature of the blackbody source. The mean of at least 60 readings was used to determine each calibration point.

\begin{figure}[H]
  \centering
  \input{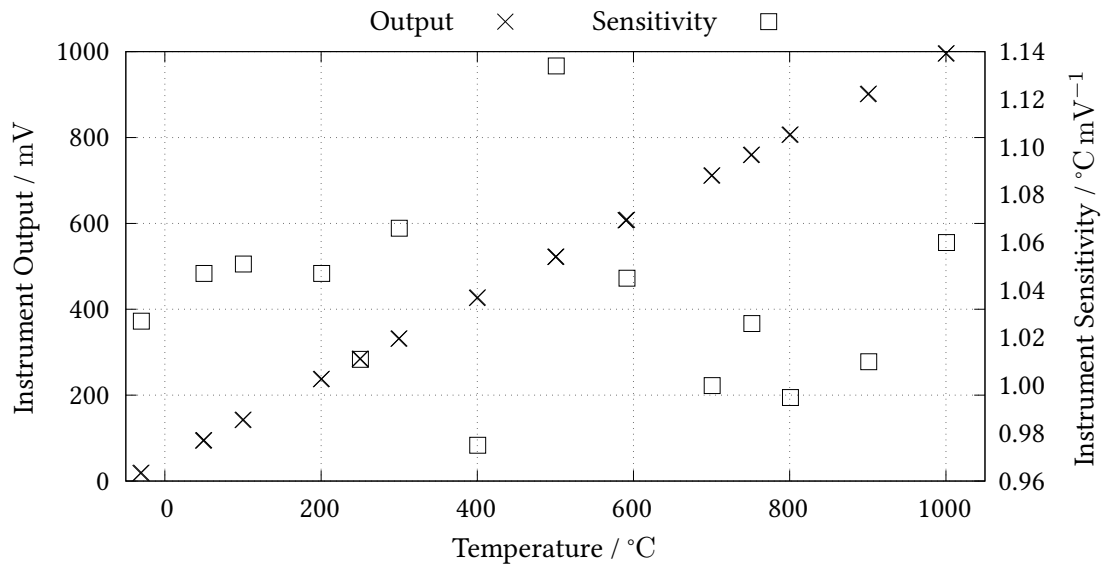}
  \caption{Measurement data from the radiation thermometer calibration presented in Table~\ref{tab:radiation_thermometer_calibration}. Output is presented on the left y-axis and the sensitivity on the right y-axis.}
  \label{fig:radition_thermometer_calibration}
\end{figure}

\begin{figure}[H]
  \centering
  \input{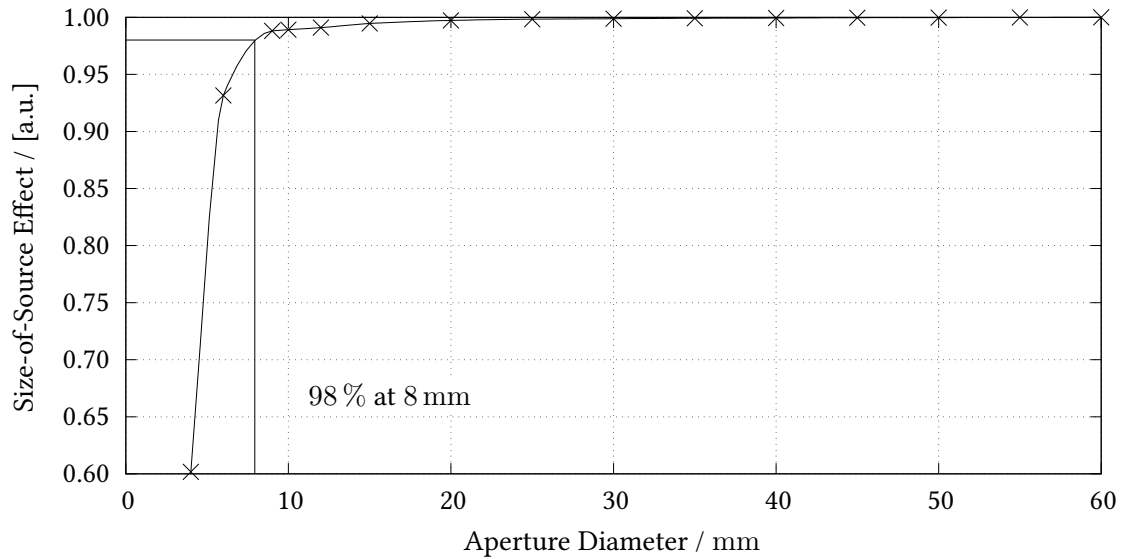}
  \caption{Size-of-source effect data from the radiation thermometer calibration using an aperture set ranging from \SIrange{4}{60}{\milli\metre} diameter. The SSE was measured using a source at \SI{1000}{\celsius}.}
  \label{fig:radition_thermometer_sse}
\end{figure}

Results of both an ITS-90 calibration and SSE assessment of a Heitronics TRT IV.82 radiation thermometer are described here. These results are shown in Figure~\ref{fig:radition_thermometer_calibration}, the associated SSE results are displayed in Figure~\ref{fig:radition_thermometer_sse}.

The calibration uncertainties are given in Table~\ref{tab:radiation_thermometer_uncertainties}. The reported expanded uncertainties are based on standard uncertainties multiplied by the coverage factor \(k\) given in Table~\ref{tab:radiation_thermometer_uncertainties}, providing a coverage probability of approximately \SI{95}{\percent}. The expanded uncertainty comprises the uncertainty associated with the NPL standards and the uncertainty associated with the instrument under test. Instrument uncertainty includes components for the resolution and stability of the test instrument, the alignment on the calibration source and the residual from the transfer function from the calibration.

\begin{figure}[H]
  \centering
  \input{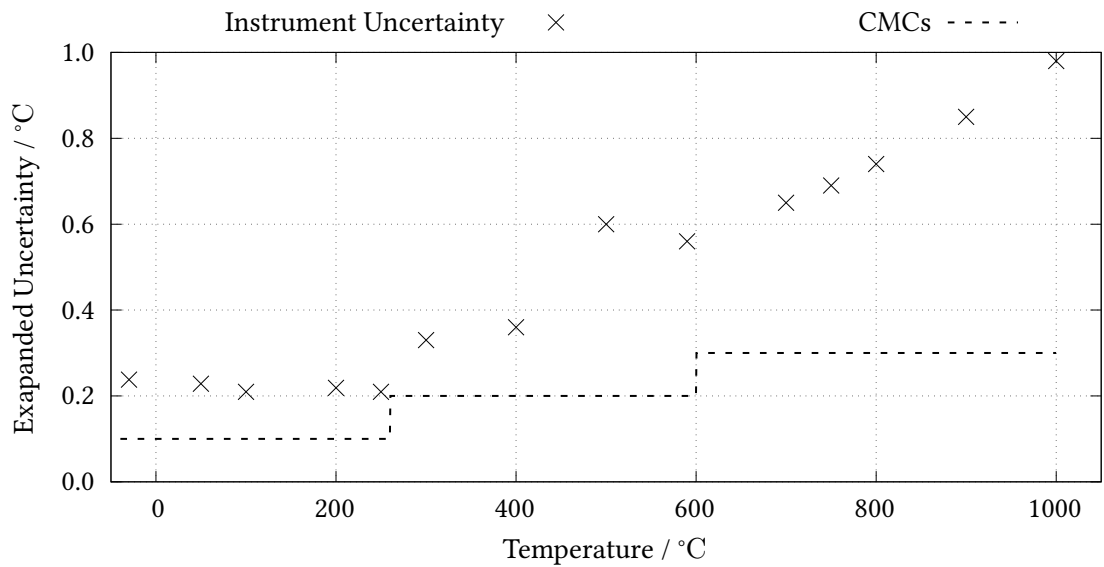}
  \caption{Measurement uncertainties for the Heitronics TRT IV.82 across the calibration temperature range, normalised to a coverage factor of \num{2.0}. The CMCs are representative of the lowest uncertainties that can be issued by NPL for this type of measurement \cite{ref:npl_radiation_thermometers_cmc}.}
  \label{fig:thermometer_case_study_uncertainty_comparison}
\end{figure}

Calibration and Measurement Capability (CMC) describes the best achievable uncertainty result demonstrated through an independent audit that has been accredited by an organisation such as the United Kingdom Accreditation Service (UKAS) \cite{ref:ukas}. Specifically with respect to an NPL calibration of radiation thermometers these are shown in Figure~\ref{fig:thermometer_case_study_uncertainty_comparison} against the measured uncertainties. These CMCs have been determined through extensive rigour and demonstrated by international comparisons between accredited laboratories \cite{ref:thermometry_cross_comparison}.

The results shown provide context for the achievable apparent radiance temperature measurement uncertainty for the instrument, here the instrument approaches these boundaries but does exceed.

\clearpage
\section{Thermal imager calibration}\label{sec:thermal_imager_calibration}
Calibration of radiation thermometers is a mature field within which metrological instrumentation is used to measure and disseminate the ITS-90. Similarities between radiation thermometers and thermal imagers enable their behaviour to be characterised using identical reference sources. Extending the calibration from the comparatively narrow field of view of the single spot infrared radiation thermometer to the extended field of view for the thermal imager introduces particular limitations to the calibration of a thermal imager and these will be detailed here, alongside unique developments to calibrate this instrumentation.

This section will explore the history of thermal imagers, nuances to calibration facilities beyond those used for radiation thermometers, and the characterisation of each: signal transfer function, non-uniformity, size-of-source effect, effect from distance and housing temperature effect. These are explored within a case study for a cooled and uncooled thermal imager and evaluated with respect to a radiation thermometer.

\subsection{Introduction to thermal imagers}\label{subsec:thermal_imager_introduction}
Thermal imaging was first developed as a military technology in the early twentieth century and in the 1950s it was declassified and broader research began. Initial instruments comprised a mirror mounted on a motor to scan across a scene and build a thermal image. During the 1960s indium antimonide based detectors were developed, these were cooled by liquid nitrogen manually decanted into the detector reservoir \cite{ref:thermalHumanBody}.

Two infrared radiation detection methods employed within thermal imagers are photon-based and thermal-based. Photon detection is observed when a photon interacts with an electron in a material, if the photon energy exceeds the ionisation energy of the electron an hole-electron pair is generated and this can be detected by the read-out integrated circuit. Thermal detection refers to any mechanism through which a change in temperature leads to a measurable material property change. One example of this is the resistive bolometer, here the electrical resistance of a film changes when absorbing infrared radiation. This change in material behaviour is measured by the read-out integrated circuit \cite{ref:uncooled_thermal_imaging}.

\subsection{Calibration facilities}\label{subsec:thermal_imager_facilities}
Similar to radiation thermometer calibration, the use of a blackbody source (with identical properties as shown in Section~\ref{subsec:thermometer_facilities}) is typical for the calibration of a thermal imager; the second requisite property, a {\em sufficiently large aperture size for the DUT field of view} introduces a limitation to these sources for thermal imager calibration. Due to design constraints, it is challenging to manufacture a sufficiently large heat-pipe cavity that demonstrates a high thermal stability and rapid temperature changes (due to its large mass).

Attempts to design large aperture cavity reference have been shown to be successful at temperatures close to ambient, where heat transfer losses are less restrictive. These take the form of blackbody cavities submerged within stirred fluid baths; the greatest source of uncertainty for these reference sources are the greater impact from convective losses and condensation at temperatures below the dew point \cite{ref:large_aperture_blackbody_bath}.

\begin{figure}[H]
  \centering
  \includegraphics[width=0.5\textwidth,keepaspectratio]{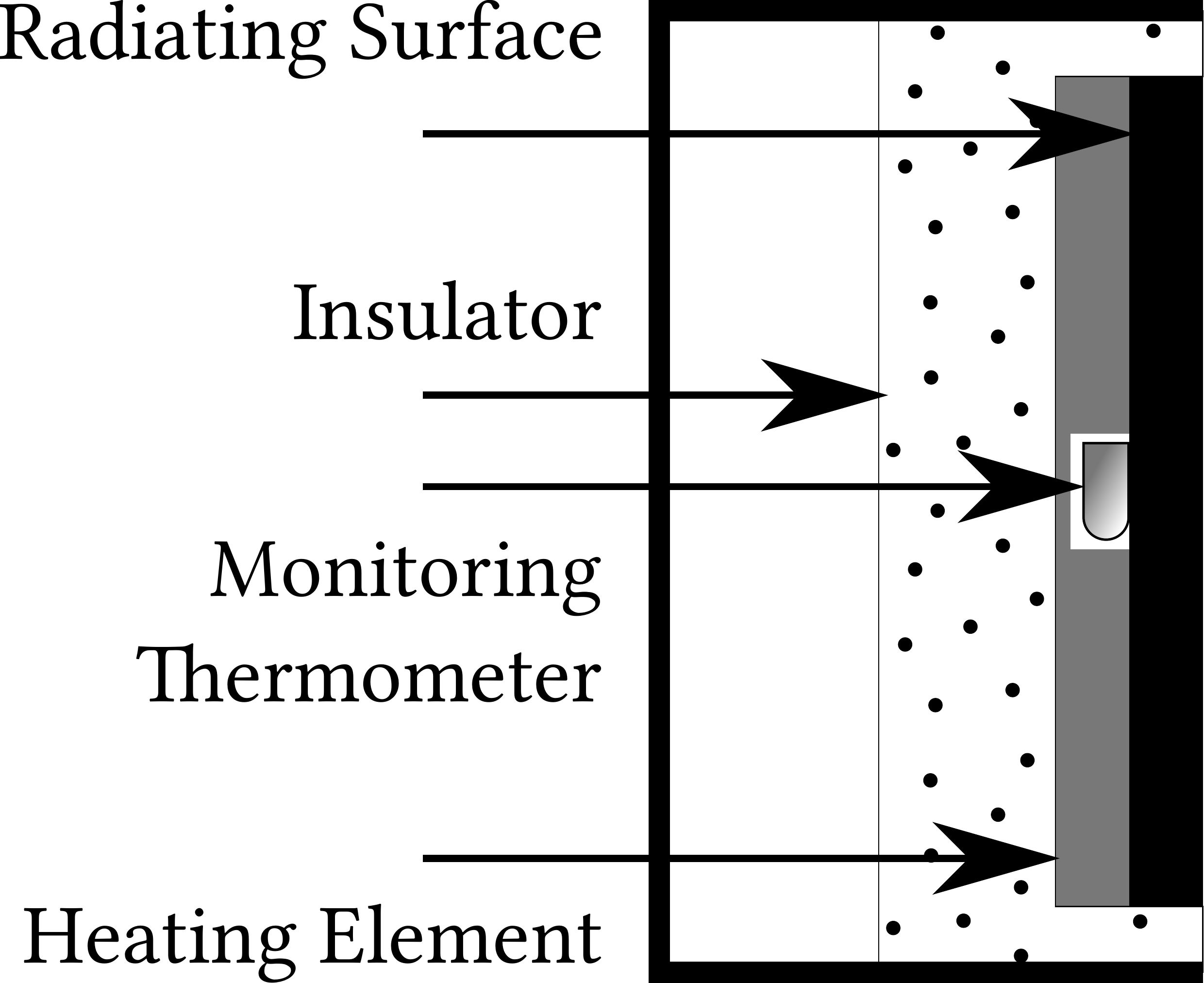}
  \caption{Typical construction of a flat-plate radiator for radiance temperature reference standards. This radiator comprise a thermal insulated and heated radiating surface with a monitoring thermometer in close proximity.}
  \label{fig:flatPlateConstruction}
\end{figure}

\begin{figure}[H]
  \centering
  \includegraphics[width=0.9\textwidth,keepaspectratio]{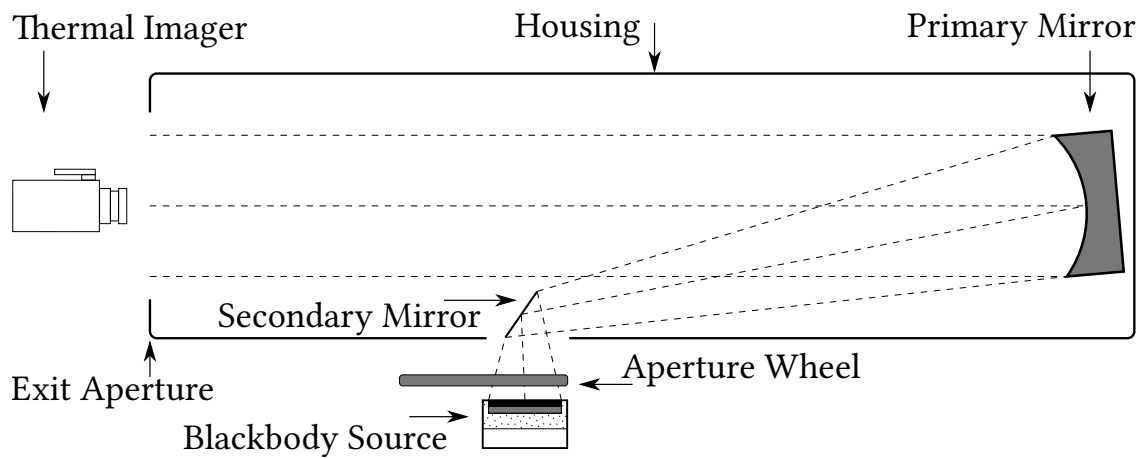}
  \caption{Example construction of a collimator source. The radiation from a source blackbody (here a flat-plate radiator is shown) is projected onto a primary collimating mirror to produce a collimated beam of radiation for calibration.}
  \label{fig:collimatorConstruction}
\end{figure}

Alternative radiation sources include: flat-plate radiators and collimating projection systems \cite{ref:analysis_ref_source_cal_infrared}. Commercially available flat-plate sources comprise a thermally-regulated metal plate covered with a high emissivity coating \cite{ref:infrared_calibrator_fluke,ref:infrared_calibrator_flir}; these systems offer a larger radiating surface in turn for lower source emissivities, poorer thermal stability and greater radial non-uniformity. A diagram of a typical flat-plate radiator is shown in Figure~\ref{fig:flatPlateConstruction}. Achieving greater radiating source sizes than typical flat-plate calibrators and offering a collimated beam profile, an infrared collimator can be used for the calibration of a thermal imager and many other characterisation tests (e.g. Noise Equivalent Temperature Difference (NETD), Minimum Resolvable Temperature Difference (MRTD)). A collimator system can comprise a radiating source (often a flat-plate calibrator) and through the use of a primary collimating mirror, project a collimated beam of a greater size (displayed in Figure~\ref{fig:collimatorConstruction}) \cite{ref:electro_optical_performance,ref:testing_thermal_imagers_guidebook,ref:cal_inv_camera_blackbody}.

Detailed in Section~\ref{subsec:thermometer_facilities}, using state-of-the-art reference sources thermometers are able to demonstrate uncertainties as low as \SI{0.3}{\celsius}. Given equivalent sources, thermal imaging systems are able to demonstrate uncertainties between \SI{0.45}{\celsius} and \SI{2.1}{\celsius} (\(k=2\)) in the limited temperature range from \SIrange{10}{120}{\celsius} \cite{ref:thermalImagerCertificate2020030001,ref:thermalImagerCertificate2021040168_1,ref:thermalImagerCertificate2021040168_2}. This discrepancy between what is achievable between radiation thermometers and thermal imagers is large; there are approaches that can be taken to reduce this variation.

\subsection{Signal transfer function characterisation}\label{subsec:thermal_imager_signal_characterisation}
In the same manner that the response of the radiation thermometer against ITS-90 was described and demonstrated in Section~\ref{subsec:thermometer_its90} and Section~\ref{subsec:case_study_thermometer}, the response signal from a thermal imager may be correlated against temperature. Consideration should be given to the specific instrument and its internal processing architecture -- if the detector response is interrogated -- that intermediary proprietary processing steps are reduced. This access to a raw signal will vary between instruments but is critical to reduce detrimental unknown influences on the signal transfer function \cite{ref:uncooled_thermal_imaging}.

Two specific examples are explored here, the use of an uncooled microbolometer and a cooled photon detector based thermal imager. The microbolometer instrument response is interrogated from an early stage of the response processing pipeline. The cooled thermal imager has been calibrated using the manufacturer proprietary software and calibration process. Both were calibrated against heat-pipe cavity reference sources.

\subsubsection{Uncooled thermal imagers}\label{subsubsec:signal_transfer_uncooled}
The DUT can be calibrated by positioning against the blackbody cavity reference sources, then recording both the DUT and ITS-90 measurement at a number of temperature setpoints.

Consideration of the DUT software and system features enabled to ensure measurement traceability is maintained from the laboratory environment to deployment. Specifically for uncooled thermal imagers, the flat-field correction implementation must be determined. This may be that the shutter is operated once, a fixed time period prior to measurements or that the shutter is automatically operated when the system requires this (often from either an internal temperature drift or regular time period parameter). This shutter operation has been observed to affect temperature measurement and would be recommended to be avoided during a calibration measurement setpoint \cite{ref:analysisThermalImagers}.

The temperatures of the blackbodies should be determined in terms of ITS-90 and the measurements carried out in accordance with ISO 17025 \cite{ref:bsiCalibration}.

The measurements can be used to generate a DUT transfer function presented in Eq.~\ref{eq:transfer_function}, where digital level \(S_{DUT} [\mathrm{DL}]\) and measured Focal Plane Array (FPA) temperature \(T_{FPA}\) are used to calculate \(t_{90}\) ITS-90 temperatures alongside the calibration coefficients \(a\) through \(f\),

\begin{equation}
\label{eq:transfer_function}
\begin{split}
	t_{90}\left(S_{DUT}, T_{FPA}\right) & = \\
	& \hspace{1.25em} a\cdot S_{DUT}^2 + b\cdot S_{DUT} \\
	& + c\cdot S_{DUT} \cdot T_{FPA} \\
	& + d\cdot T_{FPA}^2 + e\cdot T_{FPA} \\
	& + f \mathrm{.}
\end{split}
\end{equation}

This formula is an empiracal fitting function used to described two co-dependent second order polynomials.

\subsubsection{Cooled thermal imagers}\label{subsubsec:signal_transfer_cooled}
\begin{figure}[h]
  \centering
  \includegraphics[width=1.0\textwidth,keepaspectratio]{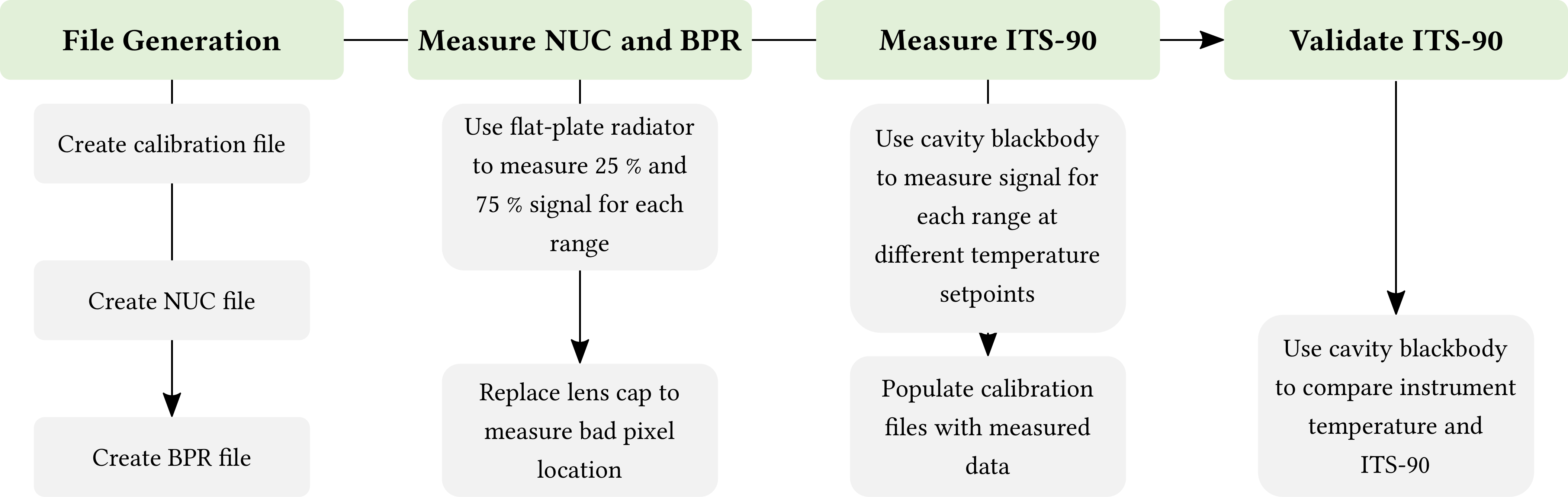}
  \caption{The calibration procedure for the thermal imager. First the initial calibration files are prepared and then the non-uniformity and bad pixel replacement are characterised. These characterisations are then utilised during the comparison of digital level to ITS-90. Following the primary calibration stage, the measured ITS-90 and digital level values are input to the calibration files. These updated calibration files (utilising the measured calibration) are then validated to ensure suitability.}
  \label{fig:thermal_imager_cooled_calibration}
\end{figure}

An example of a thermal imager calibration using the manufacturer guidance \cite{ref:infratec_proprietary_calibration} and heat-pipe cavity blackbodies described in Section \ref{sec:thermometer_calibration} is described here. The suggested stages of instrument calibration from the manufacturer are depicted in Figure~\ref{fig:thermal_imager_cooled_calibration}.

The first three phases are described in the manufacturer guidance material, however the dissemination of the International Temperature Scale of 1990 and final validation are an extension to the prescribed procedure.

The file generation phase generates the \texttt{cal} calibration file that describes integration mode, integration time, temperature range, framerate, relevant Non-Uniformity Correction (NUC) and Bad Pixel Replacement (BPR) files and calibration housing temperature. Integration mode describes the detector readout method, either integrate while read or integrate then read, the former was used during this calibration. Integration time defines the period the detector will be exposed for and is typically defined in microseconds. The temperature range is limited by the saturation temperatures and noise floor for the specific integration time. The NUC and BPR files specified within this file are those measured at the relevant integration time. The housing temperature of the imager during the calibration is measured from the internal thermometer and is defined here to ensure the imager in the application environment is within calibration.

The \texttt{coe} NUC file is created by filling the imager field of view with a blackbody source and following the factory NUC method (refer to Section~\ref{subsec:nuc}). The \texttt{pix} BPR file was created by replacing the lens cap for each integration time and using the auto BPR function.

To populate the calibrated temperature and digital level response for each of the integration times, the thermal imager should be aligned to the respective heat-pipe cavity blackbody at a prescribed distance. At each temperature setpoint the internal imager temperature was recorded as well as the digital level response and blackbody temperature, over a period of \SI{2}{\minute}. Table~\ref{tab:thermal_imager_cooled_calibration_ranges} depicts the temperature setpoints measured using each of the four integration times. The saturation limits were specified to be below \SI{18500}{\digitallevel} and above \SI{28500}{\digitallevel}.

\begin{table}[H]
  \renewcommand{\arraystretch}{0.75}
  \centering
  \caption{The temperature setpoints measured using each of the four integration times for the cooled thermal imager.}
  \vspace*{\floatsep}
  \begin{tabular}{ M{5.0cm}M{2.0cm}M{2.0cm}M{2.0cm}M{2.0cm} }
    \toprule
    Temperature Setpoint / \SI{}{\celsius} & \SI{500}{\micro\second} & \SI{175}{\micro\second} & \SI{75}{\micro\second} & \SI{50}{\micro\second} \\
    \cmidrule(lr){1-5}
	5	&	\(\checkmark\)	&	\(\checkmark\)	&					&					\\
    20	&	\(\checkmark\)	&	\(\checkmark\)	&					&					\\
	\cmidrule(lr){2-3}
    40	&	\(\checkmark\)	&	\(\checkmark\)	&	\(\checkmark\)	&					\\
	\cmidrule(lr){2-4}
    60	&	\(\checkmark\)	&	\(\checkmark\)	&	\(\checkmark\)	&	\(\checkmark\)	\\
    80	&	\(\checkmark\)	&	\(\checkmark\)	&	\(\checkmark\)	&	\(\checkmark\)	\\
    100	&	\(\checkmark\)	&	\(\checkmark\)	&	\(\checkmark\)	&	\(\checkmark\)	\\
	\cmidrule(lr){3-5}
    125	&					&	\(\checkmark\)	&	\(\checkmark\)	&	\(\checkmark\)	\\
    150	&					&	\(\checkmark\)	&	\(\checkmark\)	&	\(\checkmark\)	\\
	\cmidrule(lr){4-5}
    175	&					&					&	\(\checkmark\)	&	\(\checkmark\)	\\
    200	&					&					&	\(\checkmark\)	&	\(\checkmark\)	\\
	\cmidrule(lr){5-5}
    225	&					&					&					&	\(\checkmark\)	\\
    \bottomrule
  \end{tabular}
  \label{tab:thermal_imager_cooled_calibration_ranges}
\end{table}

\begin{figure}[h]
  \centering
  \begin{subfigure}[t]{0.8\textwidth}
	\centering
	\includegraphics[width=0.7\textwidth,keepaspectratio]{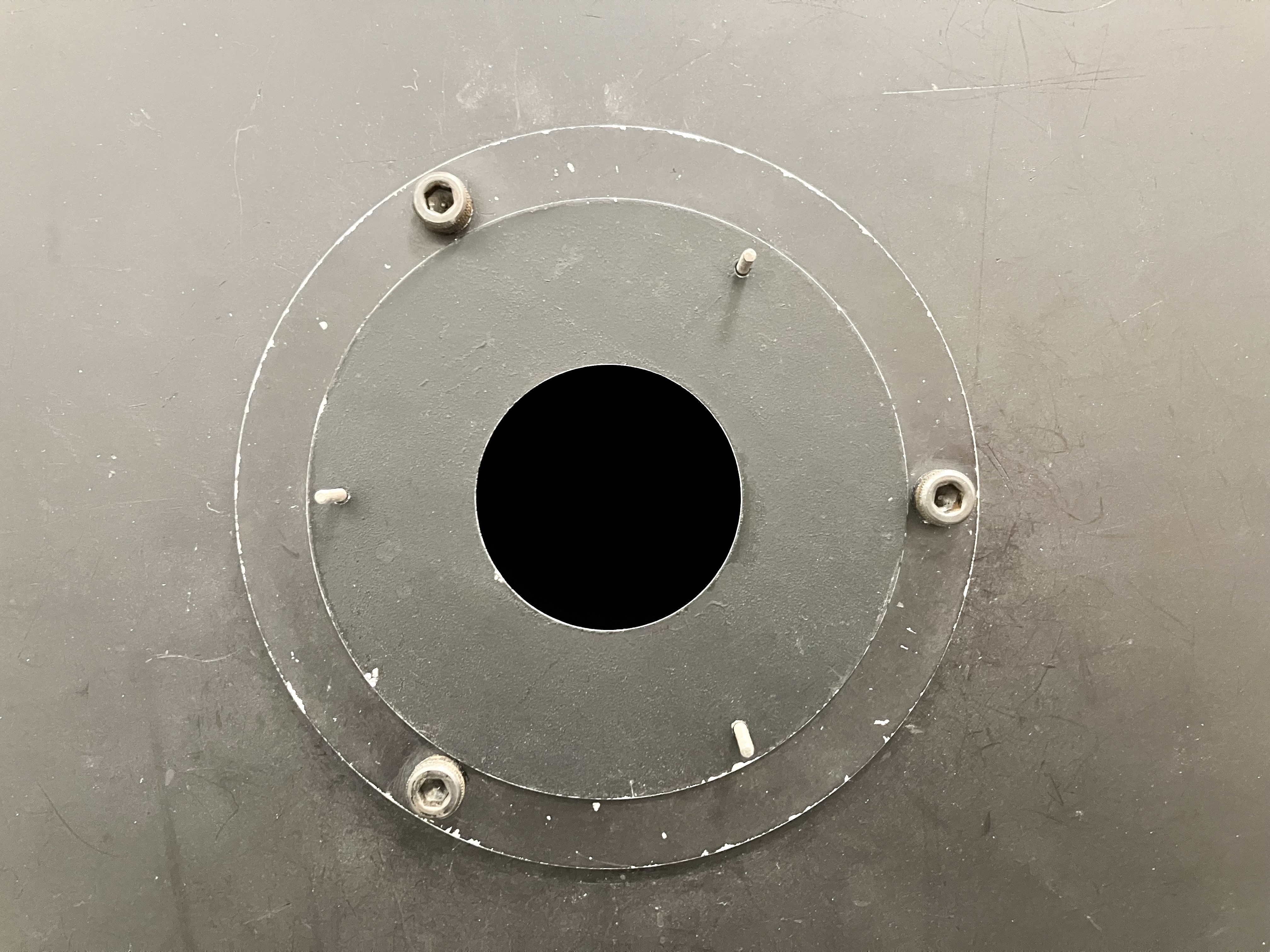}
	\caption{A visible image of the aperture for an ammonia heat-pipe reference source.}
  	\label{fig:thermal_imager_roi_heat_pipe_thermal}
  \end{subfigure}
  \vspace{\floatsep}
  \begin{subfigure}[t]{0.8\textwidth}
	\centering
  	\input{./Images/thermalImagerRoiHeatPipe}
	\caption{A region of interest evaluated for blackbody cavity measurements is highlighted by the dashed square.}
  	\label{fig:thermal_imager_roi_heat_pipe_thermal}
  \end{subfigure}
  \caption{This measurement is of a \SI{40}{\milli\metre} aperture for the ammonia heat-pipe blackbody cavity at \SI{40}{\celsius}.}
  \label{fig:thermal_imager_roi_heat_pipe}
\end{figure}

For each blackbody cavity measurement the imager was aligned to be central to the cavity aperture and perpendicular to its surface and positioned so the cavity aperture was central to the field of view. The region of interest used to evaluate the thermal imager measurement was maintained through both the calibration and validation measurements; this ensured that the same region of the detector was observed in order to minimise non-uniformity effects. An example Region Of Interest (ROI) is overlaid a measurement of a cavity in Figure~\ref{fig:thermal_imager_roi_heat_pipe}.

\clearpage

\subsection{Non-uniformity correction}\label{subsec:nuc}
Image non-uniformity of the apparent radiance temperature measured can comprise a static offset from the on board flat-field pattern. This component is intrinsic to the data pipeline and often can not be isolated. If this component should be isolated then a uniform scene must be projected onto the imager field of view; the measured radiance will now consist of both this on board static offset and a dynamic scene offset.

A decision needs to be taken at the beginning of a calibration process whether to implement a new non-uniformity correction file or not. This will affect the output of the calibration process and reduce the ease with which a historical comparison to previous calibration can be completed.

The literature details a range of methods to identify this non-uniformity both through laboratory evaluation \cite{ref:responsivityNonUniformity,ref:largeAreaMicroCavities} and in-situ scene-based approaches \cite{ref:sceneBasedNUCGated,ref:adaptiveNUC} as well as the sources for these non-uniformity contributions \cite{ref:nonUniformitySourceModeling}. Three methods to evaluate the on board instrument non-uniformity are detailed here.

\subsubsection{Factory method analysis}\label{subsubsec:nuc_factory}
A two-point NUC allows the reduction in fixed pattern noise and non-uniformity by calculating a gain and offset value for each pixel. This works on the assumption that the response of each pixel can be modelled as having a multiplicative gain and additive offset.

The following describes the procedure that can be implemented within the manufacturer proprietary software.

The \texttt{coe} NUC file is created by filling the imager field of view with a blackbody source and exposing the detector to a low and high temperature source. This two-point NUC method requires both a large aperture source and two disparate temperatures, due to these constraints it is preferable to use a flat-plate calibrator as opposed to the heat-pipe cavity blackbodies. The imager may be focused at its calibration distance and then positioned closer to the calibrator surface to fill the field of view (often this is out of focus). The dynamic range of the imager will be limited by its noise floor and saturation point in order to maximise the analogue to digital conversion range for the readout integrated circuit range. The optimal levels to perform the two-point NUC is at \SI{25}{\percent} and \SI{75}{\percent} of the dynamic range of the detector. For each integration time used the two-point NUC can be carried out \cite{ref:infratec_proprietary_calibration}.

\subsubsection{Flooded field of view method analysis}\label{subsubsec:nuc_flooded_fov}
Another method to measure the effect from non-uniformity is to carry out a single point non-uniformity measurement. This can be completed using a flat-plate calibrator and flooding the field of view (as per the factory method) and directly measuring the response. This requires a sufficiently large radiating source and is limited by the uncertainty of the radiating source.

\begin{figure}[H]
\centering
\input{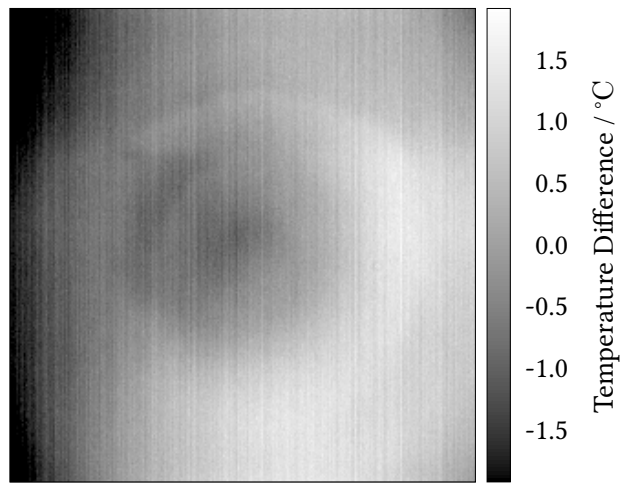}
\caption{The temperature difference measured across a reference source at \SI{80}{\celsius}.}
\label{fig:bosonCalibrationUniformity}
\end{figure}

An example uniformity of a DUT is shown in Figure \ref{fig:bosonCalibrationUniformity}. The spatial standard deviation across this set of measurements is presented in Table \ref{tab:bosonCalibrationUniformity}. It can be shown that this non-uniformity is minimised towards the centre of the calibration range where the digital level is furthest from its saturation and noise floor.

\begin{table}[H]
\centering
\setlength\extrarowheight{-3pt}
\caption{Results from a non-uniformity assessment using a single point measurement. The spatial standard deviation across the region of interest is reported.}
\begin{tabular}{ M{3.00cm}M{6.00cm}M{6.00cm}}
\toprule
{\em t}\textsubscript{90} / \SI{}{\celsius} & DUT Digital Level Measurement / [a.u.] & DUT Standard Deviation / \SI{}{\celsius} \\
    \cmidrule(lr){1-3}
	9.8 & 21275 & 0.99 \\ 
	19.8 & 22379 & 0.74 \\ 
	29.9 & 23643 & 0.80 \\ 
	39.9 & 24913 & 0.57 \\ 
	49.9 & 26224 & 0.64 \\ 
	59.9 & 28115 & 0.75 \\ 
	69.9 & 29475 & 0.90 \\ 
	79.9 & 31550 & 0.96 \\ 
\bottomrule
\end{tabular}
\label{tab:bosonCalibrationUniformity}
\end{table}

An approach often observed is to translate a narrow reference source across the field of view of a thermal imager and collating these together \cite{ref:large_aperture_blackbody_bath}. Whilst this is possible according to the isotropic requirements outlined in Section~\ref{subsec:radiation_definition_blackbody} in practice this would introduce large error from the radiance emitted at angles of observation away from the cavity normal. Although if a collimated projection was used in this approach these errors would not be observed.

\subsubsection{Translation-correction method analysis}\label{subsubsec:nuc_translation_correction}

The following describes the procedure for the independent user measurement assessment of NUC, for benchmarking and monitoring.

\begin{figure}[!htbp]
  \centering
  \includegraphics[width=0.6\textwidth,keepaspectratio]{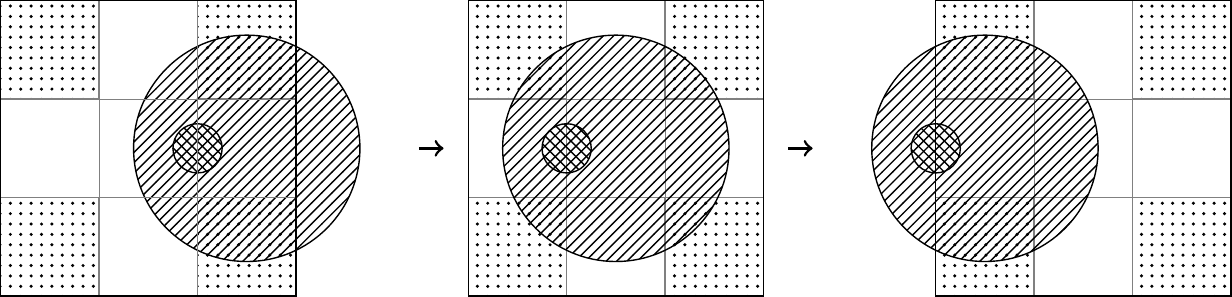}
  \caption{The translation approach employed to assess non-uniformity as described in \cite{ref:towards_small_scale}. The square matrix represents a non-uniform time-invariant surface and the circle represents the non-uniform imager detector as it passes across the surface.}
  \label{fig:nuc_translation_method}
\end{figure}

In order to assess the uniformity of the thermal imager focal plane array, the method outlined in \cite{ref:towards_small_scale} was employed. This technique exploits the statistical independence of two fixed patterns of noise observed in a typical thermal image by translating one across the other (Figure~\ref{fig:nuc_translation_method}).

\begin{figure}[!htbp]
  \centering
  \input{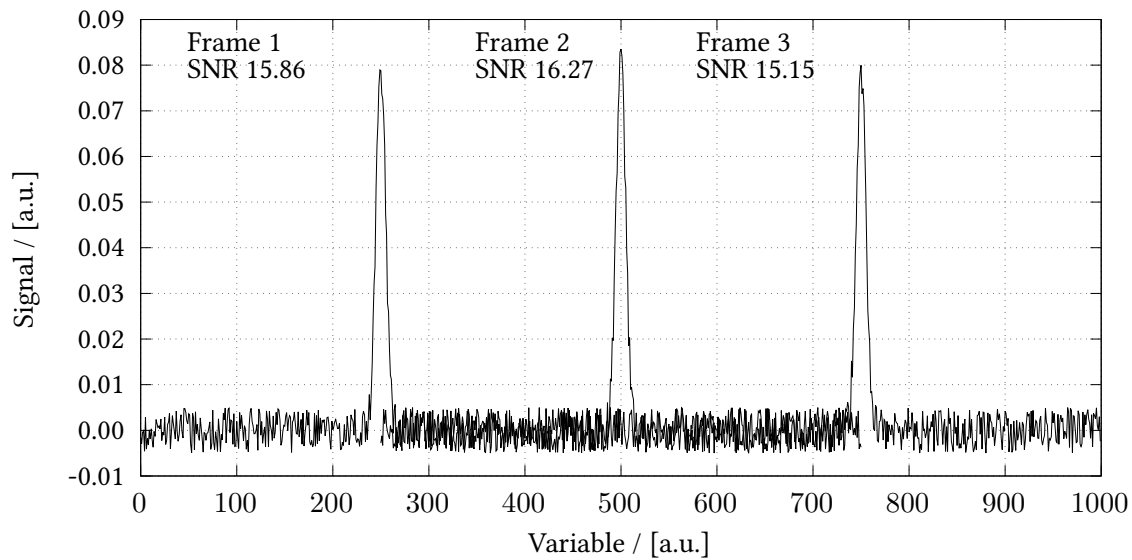}
  \caption{Three arbitrarily valued signals, each with a random background noise. Each of these signals represents a frame from Figure~\ref{fig:nuc_translation_method} where the peak is indicative of the imager non-uniformity and the noise depicts the blackbody uniformity.}
  \label{fig:nuc_snr_summed}
\end{figure}

For example, representing the three frames depicted in Figure~\ref{fig:nuc_translation_method} as Gaussian responses within a background of random noise (refer to Figure~\ref{fig:nuc_snr_summed}), the method can be demonstrated. The Gaussian peaks represent the imager non-uniform response and the background noise depicts the blackbody non-uniformity; given three images of the blackbody as the imager is translated across the surface, the misaligned imager non-uniformity can be pictured.

\begin{figure}[!htbp]
  \centering
  \input{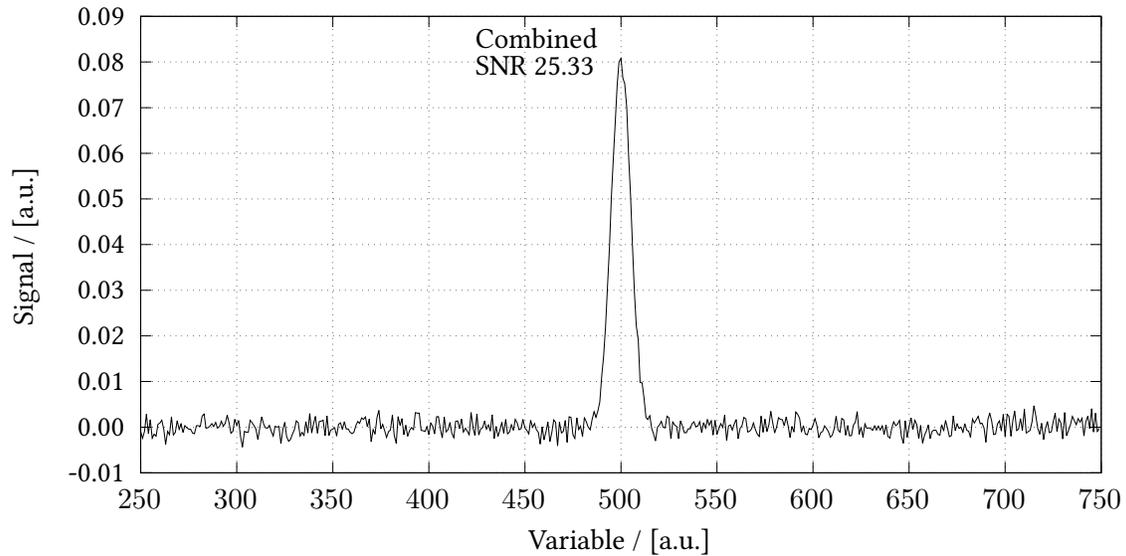}
  \caption{Each of the three frames from Figure~\ref{fig:nuc_snr_summed} has been aligned, summed and normalised. This has increased the signal-to-noise ratio of the Gaussian response.}
  \label{fig:nuc_snr_aligned}
\end{figure}

By aligning frames one and three onto frame two, summing them and dividing by three, the signal-to-noise ratio of the imager non-uniformity can be increased. This is visually demonstrated in Figure~\ref{fig:nuc_snr_aligned}. In this particular example, the signal-to-noise ratio of the Gaussian peak increased by over \SI{60}{\percent}.

The process outlined above has been translated to two-dimensional image arrays. Below is a description of the mathematical process, and a visualisation using image data from the InfraTec ImageIR 8300.

\begin{figure}[!htbp]
  \centering
  \input{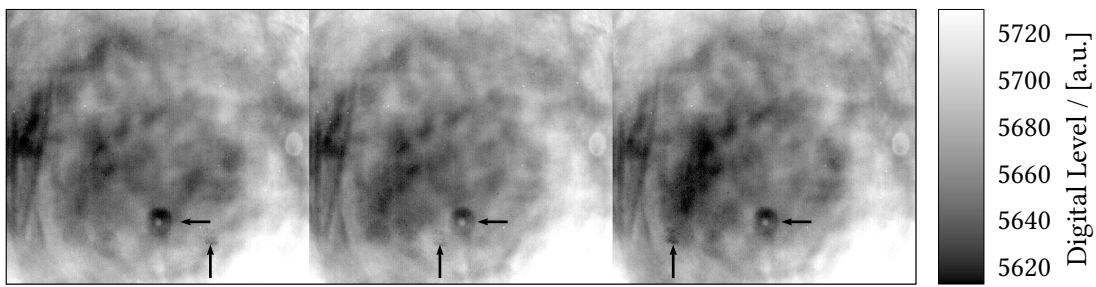}
  \caption{Thermal image of a large area blackbody source at \SI{150}{\celsius} at three positions during a translation data set. From left to right, the imager is being moved from left to right across the blackbody.  In particular, the circular artefact (horizontal arrow) towards the centre of the frame is a source of non-uniformity from the imager. The amorphous region (vertical arrow) located towards the top of the image can be attributed to the radiating surface.}
  \label{fig:nuc_example_translation}
\end{figure}

A set of thermal images observing a flat plate blackbody source (Fluke 4181 Precision IR Calibrator) is shown in Figure~\ref{fig:nuc_example_translation}. This is representative of the data measured within the non-uniformity assessment and is the measured visualisation of Figures~\ref{fig:nuc_snr_summed} and \ref{fig:nuc_snr_aligned}. From left to right the imager is being translated from left to right across the radiating blackbody. The circular artefact (horizontal arrow) located towards the centre of the frame can be attributed to the imager non-uniformity and its position remains fixed through the data set. The dark amorphous region (vertical arrow) towards the top of the frame is a component of the blackbody and is shown to move across the frame throughout the series of images.

As detailed in \cite{ref:towards_small_scale}, the number of pixels the blackbody artefact was displaced between each frame of the (nominally thirty image) sequence is described by \(\delta\) [\SI{}{\px\per\frame}]. This component can be estimated to lie within a particular range through preliminary assessment of the optical setup.

The spatial resolution of the array \(\varphi\) [\SI{}{\metre\per\px}] in a given configuration can be described by Eq.~\ref{eq:spatial_resolution},

\begin{equation}
  \varphi = \frac{p \cdot d}{ f } \mathrm{.}
  \label{eq:spatial_resolution}
\end{equation}

Where \(p\) is the detector pitch, \(d\) is the distance from the imager to the surface of measurement and \(f\) is the lens focal length. The distance is defined as the distance from the detector itself, however this can vary between imager models and even between identical models for different lens architectures. Due to the true distance being non-trivial to determine, the distance from the front of the lens will be used throughout this work and the sensitivity of the algorithm will be assessed to account for this. The number of pixels translated through by a particular displacement \(x\) [\SI{}{\metre\per\frame}] can be described by Eq.~\ref{eq:displacement_variable},

\begin{equation}
  \delta = \frac{ x }{ \varphi } \mathrm{.}
  \label{eq:displacement_variable}
\end{equation}

Given a sequence of images of a blackbody as the imager translates across its surface, each frame can be aligned onto the central frame of the sequence by applying an array translation of \(\pm k \cdot \delta\) columns, where \(k\) is the integer number of frames from the (central) frame to be aligned to. This summed array can then be normalised column-by-column, accounting for the particular number of frames that contributed to each summed column.

\begin{figure}[!htbp]
	\centering
	\begin{subfigure}[t]{0.475\textwidth}
		\centering
		\input{./Images/nucBlackbodyUniformity}
		\caption{The blackbody non-uniformity calculated by applying the translation-correction technique described above.}
		\label{fig:nuc_blackbody_uniformity}
	\end{subfigure}
	\hfill
	\begin{subfigure}[t]{0.475\textwidth}
		\centering
		\input{./Images/nucImagerUniformity}
		\caption{The imager non-uniformity calculated by applying the translation-correction technique described above.}
		\label{fig:nuc_imager_uniformity}
	\end{subfigure}
	\caption{Non-uniformity measured using the translation-correction method.}
	\label{fig:nuc_uniformity}
\end{figure}

This normalised array then depicts the blackbody non-uniformity independent of the detector non-uniformity and is shown in \ref{fig:nuc_blackbody_uniformity}. The translation-correction approach has statistically increased the signal-to-noise ratio of the imager components, where the fluctuations of signal from the detector is considered as background noise.

By subtracting the blackbody non-uniformity (\ref{fig:nuc_blackbody_uniformity}) from the central frame of the data sequence, the imager non-uniformity can be calculated. This imager non-uniformity is shown in \ref{fig:nuc_imager_uniformity}.

\begin{figure}[H]
  \centering
  \input{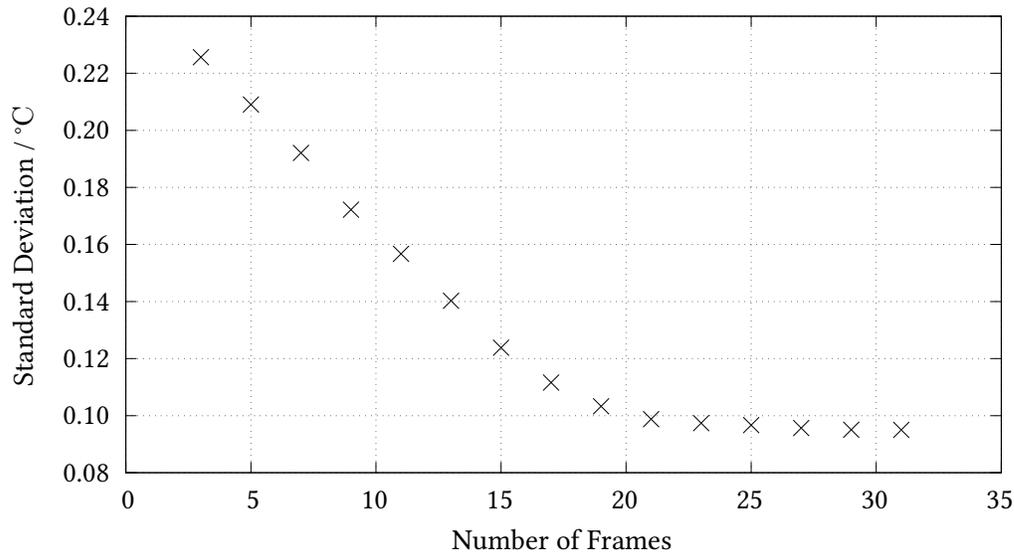}
  \caption{The standard deviation across the entire blackbody non-uniformity evaluated with reducing numbers of sequence sizes. This employed a different thermal imager but the same blackbody source. Nominally twenty frames are required to minimise the statistical fluctuations of the technique.}
  \label{fig:nuc_reduced_count_variable}
\end{figure}

To identify the minimum number of frames required to build a correction, the standard deviation of the entire imager non-uniformity matrix was calculated from three frames up to the complete sequence set. This data is presented in Figure~\ref{fig:nuc_reduced_count_variable}. This data shows that as the number of frames used increases, the standard deviation of the imager non-uniformity decreases. This plateaus at nominally twenty frames. Therefore, a minimum of twenty measured frames is recommended to reduce the statistical anomalies incurred by the technique.

\subsection{Size-of-source and distance effect}\label{subsec:thermal_imager_sse_distance}

Continuing from Section~\ref{subsec:thermometer_sse}, thermal imagers demonstrate similar effects due to imperfect optical behaviour. In comparison to infrared radiation thermometers that are defined as a single point detector, thermal imagers comprise an array of individual detectors; here the redefinition of SSE requires precise definition.

\begin{figure}[H]
  \centering
  \input{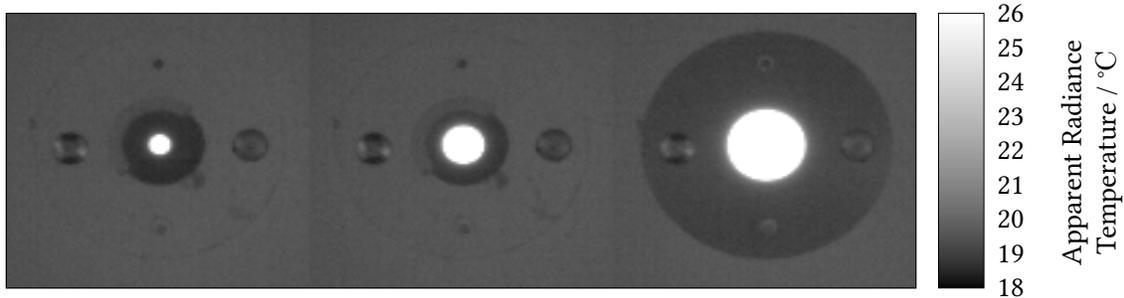}
  \caption{An assessment of the SSE using a thermal imager. The aperture from left to right are \SI{9}{\milli\metre}, \SI{20}{\milli\metre} and \SI{40}{\milli\metre}.}
  \label{fig:thermalImagerSSE}
\end{figure}

For imaging systems, SSE is equivalent to point spread function definitions, which have a wealth of literature describing how to evaluate and compensate for this \cite{ref:enhancementThroughDeconvolution,ref:SSEImageProcessing}. Complications arising from the different optical configuration are that the position in the focal plane array that SSE is evaluated is a single assessment and does not describe the behaviour at other locations or how they should be compensated \cite{ref:saunders_TI_SSE}. The definition of SSE does not detail when evaluating images of increasing aperture diameters whether the region of interest size should be fixed or scale accordingly with the aperture size. This is presented in Figure~\ref{fig:thermalImagerSSE}, the blackbody temperature is at \SI{65}{\celsius} and the aperture diameter is increasing from \SIrange{9}{40}{\milli\metre}. As a figure of merit, for a square \(3\times\)\SI{3}{\px} region of interest, the object under interrogation should occupy a \(10\times\)\SI{10}{\px} region \cite{ref:eval_sse_thermal_imagers}.

For the instruments investigated within the case study, they were evaluated as an extension to a radiation thermometer where the central set of pixels are defined as a fixed ROI throughout the SSE measurements.

As a corollary to SSE the subtended coverage of the thermal imager detector also varies with distance from the object in the scene. It is often challenging to distinguish between the distance effect and the size-of-source effect from the measurement data alone and either requires {\em a priori} information or measurements of the surrounding scene; refer to Figure~\ref{fig:thermalImagerSSE} as an example, by measuring only the cavity it would not be possible to determine which effect is taking place. As the distance from an object to thermal imager increases, the spectral radiance decreases; this sensitivity to distance may be greater depending on the optical configuration or the transmission media.

In typical laboratory conditions it is not likely the transmission media contributes a measurable effect but in applications where the object is greater than \SI{2}{\metre} from the observer this should be considered and assessed with respect to the transmissivity. 

\subsection{Housing temperature variation}\label{subsec:housing_temperature_variation}

Both detector types are susceptible to variation in output response with respect to the internal hardware temperature; this behaviour introduces a particular challenge when deploying the instrument in an uncontrolled environment. There are common and unique sources to this effect for each thermal imager type, each are impacted by the change in lens refractive index with temperature and the incident radiance from the internal housing on the detector. Both detector types are affected by their responsivity change with temperature; microbolometers can be particularly sensitive to thermal perturbations from the shutter during a flat-field correction routine and this effect varies with the shutter temperature.

This response can be characterised through calibration by modifying the environment temperature within the expected excursion range, often this is achieved through using an environmental chamber local to the instrument to induce this variation \cite{ref:effect_environmental_ti,ref:correcting_fpa_dependence_bolometer}.

Accounting for this effect can be carried out through either the proprietary manufacturer calibration software (described in Figure~\ref{fig:thermal_imager_cooled_calibration}) or by accounting for this internal temperature variation in the signal transfer function (as shown in Eq.~\ref{eq:transfer_function} by the \(T_{FPA}\) components). 

\subsection{Additional contributors}\label{subsec:additional_contributors}
Further to the sources of error discussed, additional contributors are: warm-up stabilisation, data pipeline processing, factory default parameters, pixel gain and bias, read-out integrated circuit, integration time interpolation and NUC look-up tables.

\begin{figure}[H]
  \centering
  \input{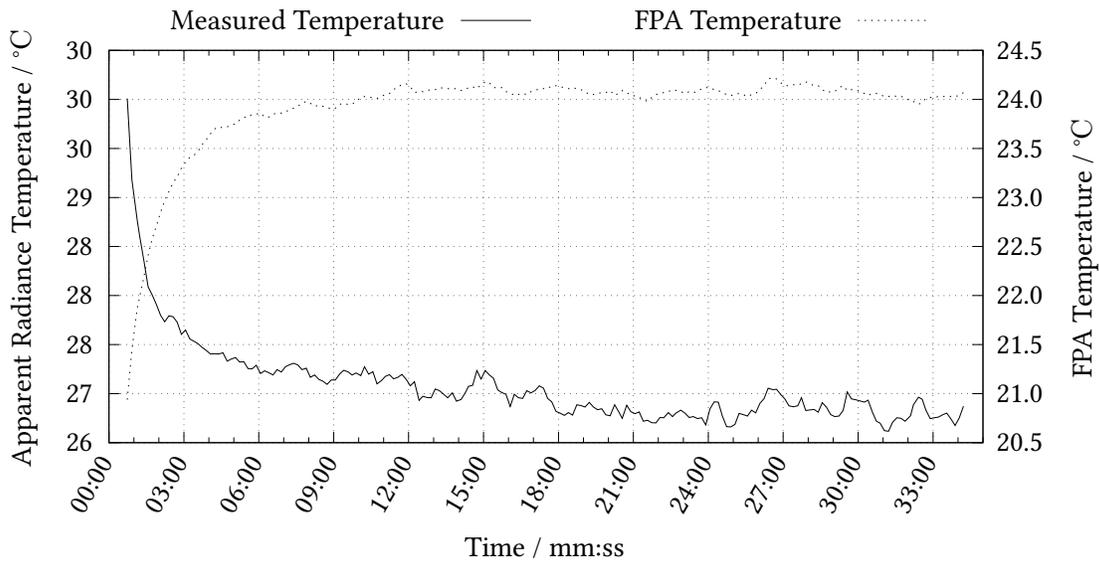}
  \caption{Warm-up profile for an uncooled microbolometer thermal imager. This FLIR Lepton was observing a \SI{30}{\celsius} target and the recommended \SI{15}{\minute} warm-up period recommended by literature is supported by this data \cite{ref:FlirOne}.}
  \label{fig:thermalImagerWarmUp}
\end{figure}

It is recommended that a measurement instrument such as a thermal imager should be powered up for an appropriate time period prior to measurement to ensure the local enclosure reaches a thermal equilibrium (for example due to stray radiation from the internal housing). This is shown in Figure~\ref{fig:thermalImagerWarmUp} where a microbolometer thermal imager was observing a \SI{30}{\celsius} reference target; this data was collected during a collaborative research project with Ben Kluwe, University of South Wales \cite{ref:thesisKluwe}. Literature supports this measurement by recommending a \SI{15}{\minute} initial stabilisation period for this instrument \cite{ref:FlirOne}. In a laboratory environment this is often trivial to achieve, however during many applications this is not feasible. When an application cannot achieve this, the initial stabilisation period should be characterised as it is likely to be unique to that instrument and environment \cite{ref:medicalImagingDevice}.

For low uncertainty and repeatable temperature measurements with a thermal imager, it is prudent to minimise non-systematic effects on measurement traceability. Operating software control parameters that are used should be considered with respect to the repeatability and effect on measurements.

For cooled thermal imagers the signal transfer function is determined for fixed integration times and then interpolated. The uncertainty contribution from this interpolation should be considered where possible. For uncooled thermal imagers a look-up table of digital values is implemented for temperature measurement and this may comprise a step-wise function, the uncertainty contribution local to these discontinuities may be large and understanding of where these step changes occur is critical.

\subsection{Case study: thermal imager}\label{subsec:case_study_thermal_imager}
To contrast the evaluation techniques used for radiation thermometers, a case study for thermal imager calibration is presented for both detector types. Measurement results from the previously outlined sections are detailed.

\subsubsection{Signal transfer calibration}\label{subsubsec:caseStudySignalTransfer}
{\em Uncooled}

\begin{figure}[H]
  \centering
  \includegraphics[width=\textwidth,keepaspectratio]{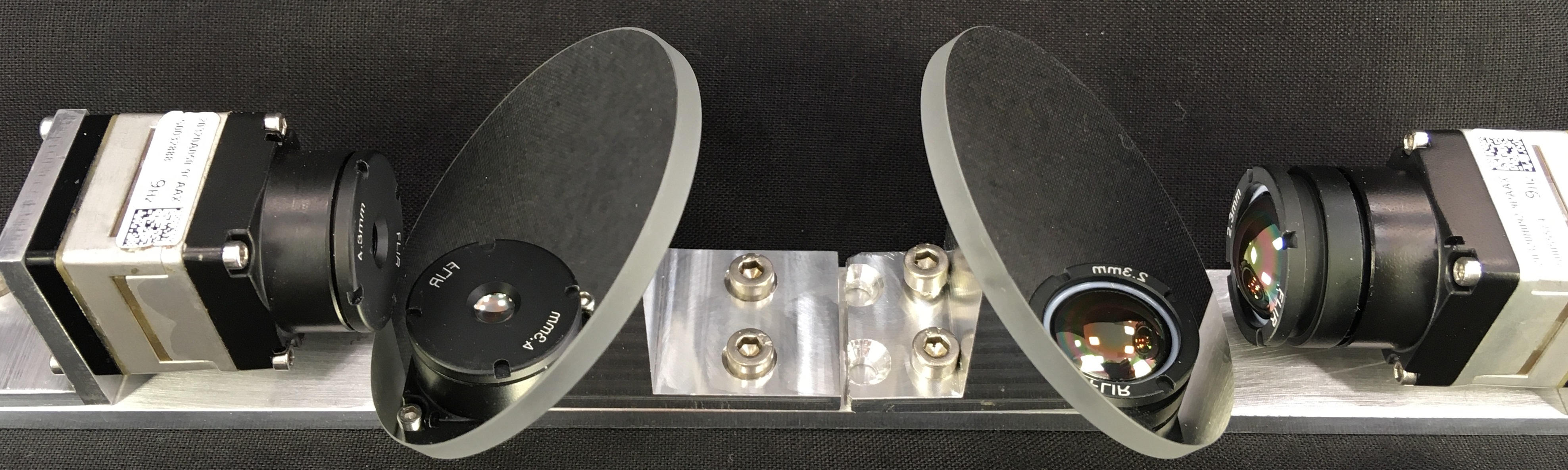}
  \caption{An optical carriage to mount a pair of uncooled thermal imagers during an end use application. The imagers were calibrated using this assembly by positioning the blackbody cavity perpendicular to the axis of the carriage.}
  \label{fig:imagerOpticalCarriage}
\end{figure}

The FLIR Boson 20320A050-9CAAX was controlled using the FLIR Boson Application (version 1.4.4). The DUT was calibrated by: recording the digital level measured by the detector at a framerate of \SI{9}{\hertz} as a \texttt{tiff} file, using high gain mode, manual flat-field correction, without a supplementary flat-field correction, averager disabled, spatial pattern and spatial column noise reduction disabled. A flat-field correction was manually triggered \SI{30}{\second} prior to each measurement setpoint. For the points at \SI{80}{\celsius} and below the aperture size of the blackbody was set to nominally \SI{160}{\milli\metre} diameter. For the points at \SI{90}{\celsius} and above the aperture size of the blackbody was set to nominally \SI{40}{\milli\metre} diameter using water-cooled blackened aperture plates; no correction for size-of-source effect was made. The measured distance from the aperture of the blackbody cavity to the mirror was nominally \SI{75}{\milli\metre}, the distance from the mirror to the lens housing of the DUT was nominally \SI{15}{\milli\metre}. This mirror assembly is shown in Figure~\ref{fig:imagerOpticalCarriage} and was used to support the end use case for this instrument and optimise footprint along the axis of the carriage in its narrow deployment location. The measurement region was taken as the \(\num{71}\times\SI{71}{\px}\) region at the centrally-aligned coordinates \((162,125)\); subsequent coordinates were also centrally-aligned. The DUT was aligned so that the centre of the field of view was positioned at the centre of the blackbody cavity and the DUT focused on the aperture plate.

The device under test was calibrated over the temperature range from nominally \SIrange{10}{100}{\celsius} using blackbody sources with calculated emissivities greater than 0.999. The temperatures of the blackbodies were determined in terms of ITS-90 and the measurements were carried out in accordance with ISO 17025 \cite{ref:bsiCalibration}.

The calibration was performed by recording at each radiance temperature the mean digital level of the ROI over the measurement period. The average of twelve measurements over a period of \SI{2}{\minute} was used for each radiance temperature setpoint and the temperature of the blackbody source during the measurements was also recorded. The results of the calibration are given in Table \ref{tab:caseStudySignalTransferUncooled}. Digital level \(S_{DUT}\) and measured FPA temperature \(T_{FPA}\) refer to the measurements from the DUT at the respective temperature setpoint. These measurements were then used to determine the calibration coefficients (from Eq.~\ref{eq:transfer_function}) and these are presented in Table~\ref{tab:caseStudySignalTransferUncooledCoefficients}.

\begin{table}[H]
\centering
\setlength\extrarowheight{-3pt}
\caption{The results from the calibration of an uncooled thermal imager as described in Section~\ref{subsubsec:signal_transfer_uncooled}. \textsuperscript{\dag} denotes repeat measurements.}
\begin{tabular}{ M{3.00cm}M{6.00cm}M{6.00cm}}
\toprule
{\em t}\textsubscript{90} / \SI{}{\celsius} & DUT Digital Level Measurement / [a.u.] & FPA Temperature / \SI{}{\celsius} \\
    \cmidrule(lr){1-3}
	9.8 & 21200 & 32.1 \\ 
	19.8 & 22300 & 32.4 \\ 
	29.9 & 23500 & 32.9 \\ 
	39.9 & 24800 & 33.2 \\ 
	49.9 & 26400 & 33.6 \\ 
	59.9 & 28000 & 33.8 \\ 
	69.9 & 29400 & 33.8 \\ 
	79.9 & 31500 & 34.5 \\ 
	90.1 & 33900 & 33.6 \\ 
	99.7 & 36300 & 32.6 \\ 
	\cmidrule(lr){1-3}
	9.8 \textsuperscript{\dag} & 21200 & 32.2 \\ 
	9.8 \textsuperscript{\dag} & 21300 & 31.7 \\ 
	19.9 \textsuperscript{\dag} & 22400 & 31.7 \\ 
	29.9 \textsuperscript{\dag} & 23600 & 31.7 \\ 
	49.9 \textsuperscript{\dag} & 26100 & 33.3 \\ 
	99.9 \textsuperscript{\dag} & 36200 & 33.1 \\ 
\bottomrule
\end{tabular}
\label{tab:caseStudySignalTransferUncooled}
\end{table}

\begin{table}[H]
\centering
\setlength\extrarowheight{-3pt}
\caption{The calibration coefficient results for the signal transfer measurement of an uncooled thermal imager using the data in Table~\ref{tab:caseStudySignalTransferUncooled}. For use with Eq.~\ref{eq:transfer_function}.}
\begin{tabular}{ M{4.00cm}rl}
\toprule
Coefficient Variable	&	\multicolumn{2}{c}{Coefficient Value} \\
    \cmidrule(lr){1-3}
	\(a\)	&	\(-\)	&	\num{1.874e-7}	\\
	\(b\)	&			&	\num{1.832e-2}	\\
	\(c\)	&	\(-\)	&	\num{4.906e-5}	\\
	\(d\)	&	\(-\)	&	\num{2.479e-1}	\\
	\(e\)	&			&	\num{1.767e1}	\\
	\(f\)	&	\(-\)	&	\num{5.727e2}	\\
\bottomrule
\end{tabular}
\label{tab:caseStudySignalTransferUncooledCoefficients}
\end{table}

{\em Cooled}

The InfraTec ImageIR 8300 was controlled using IRBIS 3.1 Professional (version 3.1.90). The DUT was calibrated by recording the digital level measured by the detector at a framerate of \SI{60}{\hertz}. The calibration tables: \SI{500}{\micro\second}, \SI{175}{\micro\second}, \SI{75}{\micro\second} and \SI{50}{\micro\second}, were used with {\em DV} units selected in the software as detailed in Table~\ref{tab:caseStudySignalTransferCooled}. The aperture size of each blackbody was set to nominally \SI{40}{\milli\metre} diameter using blackened aperture plates. The measured distance from the aperture of the blackbody cavity to the front of the lens of the DUT was nominally \SI{450}{\milli\metre}. The measurement region was taken to be the central \(\num{92}\times\SI{92}{\px}\) region at the centre of the field-of-view. The DUT was aligned so that the centre of the field-of-view was positioned at the centre of the blackbody cavity and the DUT focused on the aperture plate. The calibration was performed using a combination of calibration tables as detailed in the result tables. 

The calibration was performed by recording, at each radiance temperature: the mean digital level of the measurement region over the measurement period. The average of twelve measurements over a period of \SI{2}{\minute} was used for each radiance temperature setpoint.

\begin{table}[H]
\centering
\setlength\extrarowheight{-3pt}
\caption{The results from calibration using each configuration for the cooled thermal imager, with the Integration Time (IT) defined in each column. \textsuperscript{\dag} denotes repeat measurements.}
\begin{tabular}{ M{3.00cm}M{3.00cm}M{3.00cm}M{3.00cm}M{3.00cm}}
\toprule
& \multicolumn{4}{c}{DUT Measurement / Digital Level} \\
\cmidrule(lr){1-5}
Range / \SI{}{\celsius} & 5 \ldots 100 & 5 \ldots 150 & 40 \ldots 200 & 60 \ldots 225 \\
\cmidrule(lr){1-5}
\diagbox{{\em t}\textsubscript{90} / \SI{}{\celsius}}{IT / \SI{}{\micro\second}} & 500 & 175 & 75 & 50 \\
\cmidrule(lr){1-5}
   5.0 & 19293 & 18627 & - & - \\ 
   5.0 \textsuperscript{\dag} & 19294 & - & - & - \\ 
   5.1 \textsuperscript{\dag} & - & 18637 & - & - \\ 
   20.0 & 19766 & 18787 & - & - \\ 
   40.1 & 20758 & 19123 & - & - \\ 
     \cmidrule(lr){2-2}
   59.8 \textsuperscript{\dag} & 22320 & 19686 & 18842 & 18659 \\ 
   60.0 & 22283 & 19653 & 18840 & 18657 \\ 
   79.8 & 24774 & 20526 & - & - \\ 
   79.9 & - & - & 19211 & 18904 \\ 
   99.1 & 28174 & 21716 & 19722 & 19242 \\ 
     \cmidrule(lr){2-3}
   124.4 & - & 24059 & - & - \\ 
   124.5 & - & - & 20729 & 19913 \\ 
   150.0 & - & 27536 & 22218 & 20906 \\ 
   150.0 \textsuperscript{\dag} & - & 27500 & 22202 & 20896 \\ 
     \cmidrule(lr){3-4}
   174.6 & - & - & 24197 & 22229 \\ 
   199.5 & - & - & 26926 & 24050 \\ 
   199.6 \textsuperscript{\dag} & - & - & 26892 & 24025 \\ 
     \cmidrule(lr){5-5}
   225.3 & - & - & - & 26470 \\ 
\bottomrule
\end{tabular}
\label{tab:caseStudySignalTransferCooled}
\end{table}

\begin{figure}[H]
  \centering
  \input{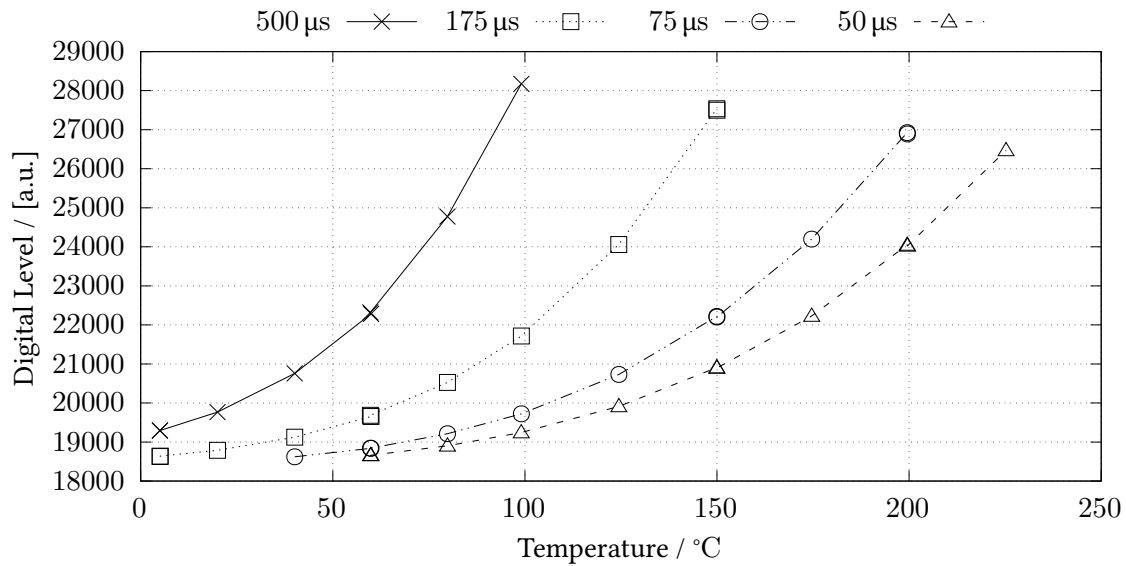}
  \caption{Measurements from the calibration of the cooled thermal imager. The digital level measured using the prepared calibration files at increasing temperatures of ITS-90.}
  \label{fig:caseStudySignalTransferCooled}
\end{figure}

\begin{figure}[H]
  \centering
  \input{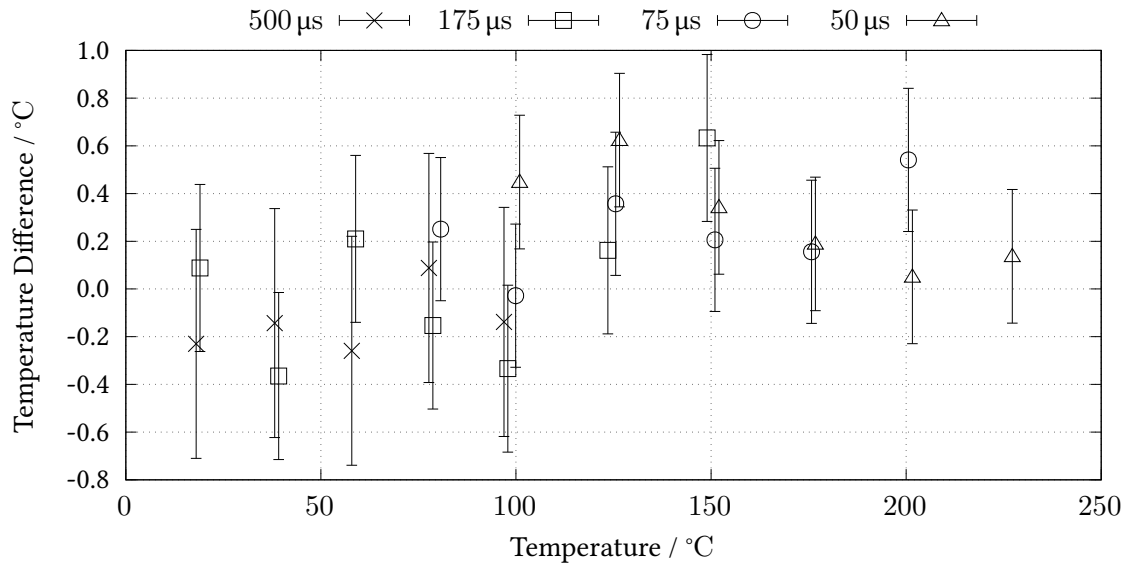}
  \caption{The validation measurements of the calibrated thermal imager. These measurements show the temperature difference between ITS-90 and the apparent radiance temperature from the thermal imager. The lowest temperature setpoint for each integration time has been omitted due to a poor calibration fit. The error bars indicate the expanded uncertainty. The markers have been artificially offset to improve error bar discernability.}
  \label{fig:thermalImagerCooledValidation}
\end{figure}

The measurements depicted in Figure~\ref{fig:caseStudySignalTransferCooled} were used to populate the \texttt{cal} calibration files of the thermal imager and a validation of this calibration was then carried out. At each of the temperature setpoints from the calibration measurements, the instrument apparent radiance temperature was measured against ITS-90 (presented in Table~\ref{tab:caseStudyValidationCooled}). The difference between the ITS-90 and instrument temperature is shown in Figure~\ref{fig:thermalImagerCooledValidation}. The error bars indicate the standard uncertainty multiplied by a coverage factor, providing a coverage probability of approximately \SI{95}{\percent}.

\begin{table}[H]
\centering
\setlength\extrarowheight{-3pt}
\caption{Results from the validation using each configuration for the cooled thermal imager, with the integration time defined in each column.}
\begin{tabular}{ M{3.00cm}M{3.00cm}M{3.00cm}M{3.00cm}M{3.00cm}}
	\toprule
	& \multicolumn{4}{c}{DUT Measurement / \SI{}{\celsius}} \\
	\cmidrule(lr){1-5}
Range / \SI{}{\celsius} & 5 \ldots 100 & 5 \ldots 150 & 40 \ldots 200 & 60 \ldots 225 \\
	\cmidrule(lr){1-5}
\diagbox{{\em t}\textsubscript{90} / \SI{}{\celsius}}{IT / \SI{}{\micro\second}} & 500 & 175 & 75 & 50 \\
\cmidrule(lr){1-5}
	5.0 & 7.7 & 9.5 & - & - \\ 
	20.0 & 20.2 & 19.9 & - & - \\ 
    \cmidrule(lr){2-3}
	40.2 & 40.3 & 40.5 & 43.8 & - \\ 
	59.9 & 60.2 & 59.7 & 58.0 & 61.6 \\ 
	79.7 & 79.6 & 79.9 & 79.5 & 78.7 \\ 
	98.9 & 99.1 & 99.3 & 98.9 & 98.5 \\ 
    \cmidrule(lr){3-4}
	124.6 & - & 124.4 & 124.2 & 123.9 \\ 
	150.0 & - & 149.4 & 149.8 & 149.7 \\ 
    \cmidrule(lr){4-5}
	174.8 & - & - & 174.6 & 174.6 \\ 
	199.6 & - & - & 199.1 & 199.5 \\ 
    \cmidrule(lr){5-5}
	225.2 & - & - & - & 225.1 \\ 
\bottomrule
\end{tabular}
\label{tab:caseStudyValidationCooled}
\end{table}

The validation measurements were carried out at each temperature setpoint the calibration was carried out at. Major offsets up to \SI{-4.5}{\celsius} at the lowest temperature setpoint for each integration time were observed, and large offsets for the second lowest for both the \SI{50}{\micro\second} and \SI{75}{\micro\second} were also observed. This is likely due to the lower saturation point of \SI{18500}{\digitallevel} being too low and causing an inadequate fit from digital level to ITS-90; this noise floor was increased to \SI{19300}{\digitallevel}. Because of this the temperature ranges of each integration time for later measurements were changed to reflect this. For the majority of the integration times across their complete temperature range the temperature difference is centred about \SI{0}{\celsius} with respect to the uncertainties.

\subsubsection{Distance results}\label{subsubsec:caseStudyDistance}
{\em Uncooled}

A measurement was carried out at five distances between the DUT and blackbody cavity at a temperature of nominally \SI{99.9}{\celsius}. These measurements are presented in Table \ref{tab:caseStudyDistanceUncooled}.

\begin{table}[H]
\centering
\setlength\extrarowheight{-3pt}
\caption{The results from distance effect assessment for the uncooled thermal imager. The DUT temperature was determined using Eq.~\ref{eq:transfer_function} with the measured coefficients (Table~\ref{tab:caseStudySignalTransferUncooledCoefficients}).}
\begin{tabular}{ M{3.00cm}M{4.00cm}M{3.00cm}M{3.00cm}}
\toprule
Distance / \SI{}{\milli\metre} & DUT Digital Level Measurement / [a.u.] & FPA Temperature / \SI{}{\celsius} & DUT Measurement / \SI{}{\celsius} \\
\cmidrule(lr){1-4}
75 & 36800 & 31.4 & 101.5 \\ 
85 & 36800 & 31.4 & 101.4 \\ 
95 & 36800 & 31.0 & 101.3 \\ 
105 & 36800 & 31.1 & 101.2 \\ 
115 & 36700 & 31.2 & 100.9 \\ 
\bottomrule
\end{tabular}
\label{tab:caseStudyDistanceUncooled}
\end{table}

The effect of distance variation is nominally \SI{0.6}{\celsius} across this range of distances.

{\em Cooled}

A measurement was carried out at four distances between the DUT and blackbody cavity at a temperature of \SI{170}{\celsius} using the two appropriate integration times. These measurements are presented in Table \ref{tab:caseStudyDistanceCooled}.

\begin{table}[H]
\centering
\setlength\extrarowheight{-3pt}
\caption{The results from distance effect assessment for the cooled thermal imager.}
\begin{tabular}{ M{4.00cm}M{4.00cm}M{4.00cm}}
\toprule
\multirow{2}{*}{Distance / \SI{}{\milli\metre}} & \multicolumn{2}{c}{DUT Measurement / \SI{}{\celsius}} \\
 & \SI{50}{\micro\second} & \SI{75}{\micro\second} \\
\cmidrule(lr){1-3}
450	&	169.40	&	169.46	\\
470	&	169.32	&	169.39	\\
500	&	169.20	&	169.27	\\
520	&	169.11	&	169.19	\\
\bottomrule
\end{tabular}
\label{tab:caseStudyDistanceCooled}
\end{table}

The results demonstrate good agreement between the two integration times and that the effect of distance variation is less than \SI{0.29}{\celsius}.

\subsubsection{SSE results}\label{subsubsec:caseStudySSE}

{\em Uncooled}

An assessment of the size-of-source effect was carried out at \SI{100}{\celsius} using apertures from \SI{9}{\milli\metre} to \SI{40}{\milli\metre}. The measurement region was taken to be the \(\num{13}\times\SI{13}{\px}\) region at the centrally-aligned coordinates \((161,130)\). These measurements are presented in Table~\ref{tab:caseStudySSEUncooled}. The size-of-source effect measurements in Table~\ref{tab:caseStudySSEUncooled} are the mean of three independent measurement sets and the size-of-source effect value is calculated as the ratio of indicated radiance measurement compared to the radiance measurement at the largest aperture.

\begin{table}[H]
\centering
\setlength\extrarowheight{-3pt}
\caption{The results from the size-of-source effect assessment for the uncooled thermal imager. The DUT Temperature was determined using Eq.~\ref{eq:transfer_function}.}
\begin{tabular}{ M{4.00cm}M{4.00cm}M{4.00cm}}
\toprule
Aperture / \SI{}{\milli\metre} & DUT Measurement / \SI{}{\celsius} & Size-of-Source Effect \\
\cmidrule(lr){1-3}
	9 & 97.8 & 0.9788 \\ 
	12 & 98.5 & 0.9855 \\ 
	15 & 99.0 & 0.9903 \\ 
	20 & 99.5 & 0.9953 \\ 
	25 & 99.7 & 0.9980 \\ 
	30 & 99.9 & 0.9996 \\ 
	40 & 99.9 & 1.0000 \\ 
\bottomrule
\end{tabular}
\label{tab:caseStudySSEUncooled}
\end{table}

{\em Cooled}

An assessment of the size-of-source effect was carried out at \SI{170}{\celsius} using apertures from \SI{9}{\milli\metre} to \SI{40}{\milli\metre}. These measurements are presented in Table~\ref{tab:caseStudySSECooled}. The size-of-source effect measurements in Table~\ref{tab:caseStudySSECooled} are the mean of three independent measurement sets and the size-of-source effect value is calculated as the ratio of indicated radiance measurement compared to the radiance measurement at the largest aperture.

\begin{table}[H]
\centering
\setlength\extrarowheight{-3pt}
\caption{Results from the size-of-source effect measurements for each integration time from the cooled thermal imager.}
\begin{tabular}{ M{2.00cm}M{3.50cm}M{2.50cm}M{3.50cm}M{2.50cm}}
\toprule
& \multicolumn{2}{c}{\SI{50}{\micro\second}} & \multicolumn{2}{c}{\SI{75}{\micro\second}} \\
Aperture / \SI{}{\milli\metre} &  DUT Measurement / \SI{}{\celsius} & Size-of-Source Effect & DUT Measurement / \SI{}{\celsius} & Size-of-Source Effect \\
\cmidrule(lr){1-5}
9	&	168.5	&	0.9805	&	168.4	&	0.9798	\\
12	&	168.7	&	0.9858	&	168.7	&	0.9850	\\
15	&	168.9	&	0.9881	&	168.9	&	0.9885	\\
20	&	169.0	&	0.9908	&	169.0	&	0.9917	\\
25	&	169.2	&	0.9943	&	169.2	&	0.9945	\\
30	&	169.3	&	0.9964	&	169.3	&	0.9955	\\
40	&	169.6	&	1.0015	&	169.5	&	1.0000	\\
\bottomrule
\end{tabular}
\label{tab:caseStudySSECooled}
\end{table}

\subsubsection{Uncertainty assessment}\label{subsubsec:caseStudyUncertainty}
{\em Uncooled}

The calculated measurement uncertainty for each measurement setpoint is shown in Table~\ref{tab:caseStudyUncertaintyUncooled}. The reported expanded uncertainties are based on standard uncertainties multiplied by the coverage factor, \(k\), given in the table, providing a coverage probability of approximately \SI{95}{\percent}. 

The uncertainty values in the table include components for the: calibration of the reference source, stability of the reference source, stability of the DUT, size-of-source effect reproducibility, uniformity across the central \(\num{262}\times\SI{262}{\px}\), the residual of the calibration fit, stability over a \SI{30}{\minute} period, the effect of distance variation, the alignment to the mirror and the manufacturer stated noise equivalent temperature difference.

\begin{table}[H]
\centering
\setlength\extrarowheight{-3pt}
\caption{The measurement uncertainties of the uncooled DUT from the data measured.}
\begin{tabular}{ M{4.00cm}M{4.00cm}M{4.00cm}}
\toprule
{\em t}\textsubscript{90} / \SI{}{\celsius} & Uncertainty / \SI{}{\celsius} & \(k\) \\
\cmidrule(lr){1-3}
10 & 2.50 & 2.0 \\
20 & 2.15 & 2.0 \\
30 & 2.20 & 2.0 \\
40 & 1.85 & 2.1 \\
50 & 1.95 & 2.1 \\
60 & 2.10 & 2.1 \\
70 & 2.30 & 2.0 \\
80 & 2.40 & 2.0 \\
90 & 2.75 & 2.0 \\
100 & 3.20 & 2.0 \\
\bottomrule
\end{tabular}
\label{tab:caseStudyUncertaintyUncooled}
\end{table}

{\em Cooled}

The calculated measurement uncertainty for each measurement setpoint is shown in Table~\ref{tab:caseStudyUncertaintyCooled}. The reported expanded uncertainties are based on standard uncertainties multiplied by the coverage factor, \(k\), given in the table, providing a coverage probability of approximately \SI{95}{\percent}. 

The uncertainty values in the table include components for the: calibration of the reference source, stability of the reference source, stability of the DUT, resolution of the DUT, size-of-source effect reproducibility and the repeatability of the calibration. Effects from distance variation or detector uniformity have not been accounted for.

\begin{table}[H]
\centering
\setlength\extrarowheight{-3pt}
\caption{The measurement uncertainties of the cooled DUT from the data measured.}
\begin{tabular}{ M{3.00cm}M{2.50cm}M{2.50cm}M{2.50cm}M{2.50cm}}
\toprule
Range / \SI{}{\celsius} & 20\ldots100 & 20\ldots150 & 60\ldots200 & 80\ldots225 \\
IT / \SI{}{\micro\second} & 500 & 175 & 75 & 50 \\
\cmidrule(lr){2-5}
Uncertainty / \SI{}{\celsius} & 0.50 & 0.35 & 0.30 & 0.30 \\
\(k\) & 2.3 & 2.2 & 2.2 & 2.2 \\
\bottomrule
\end{tabular}
\label{tab:caseStudyUncertaintyCooled}
\end{table}

\newpage
\section{Calibration summary}\label{sec:calibration_summary}
\begin{figure}[H]
  \centering
  \input{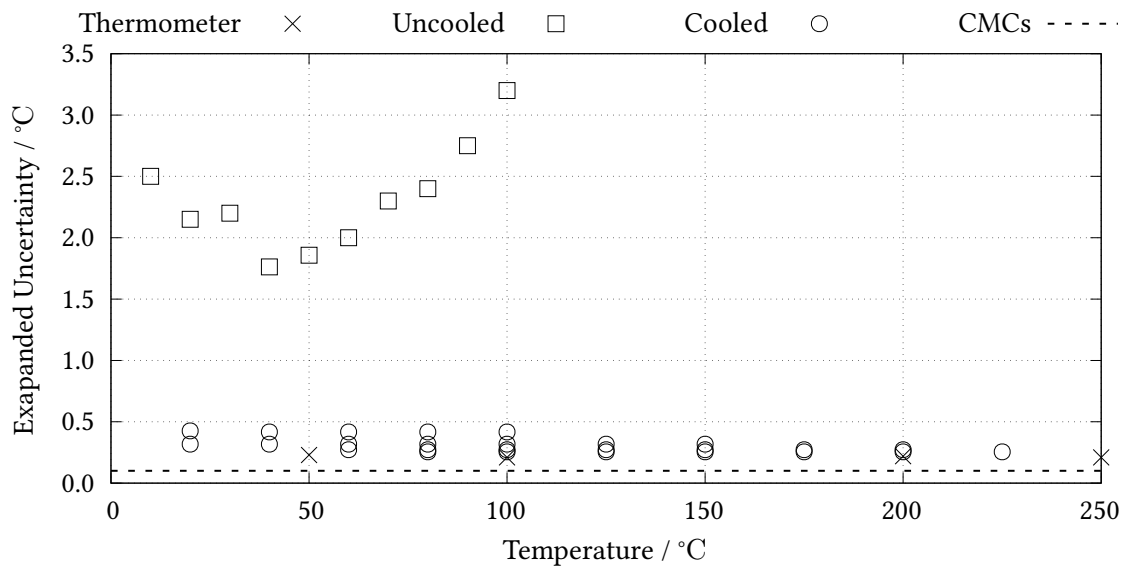}
  \caption{The uncertainties detailed in the corresponding case study tables for the radiation thermometer, uncooled thermal imager and cooled thermal imager respectively. These have been normalised to a coverage factor of \num{2.0} and are shown against the CMC capability.}
  \label{fig:thermalImagerUncertainties}
\end{figure}

In comparison between the radiation thermometer and the thermal imagers evaluated the radiation thermometer is able to achieve the lowest uncertainties that are the closest to the best achievable CMCs; and between the two imagers the cooled thermal imager achieved lower uncertainties. A comparison between the uncertainty components considered for the three instruments are presented in Table~\ref{tab:uncertaintyComponentComparison}. These results are displayed in Figure~\ref{fig:thermalImagerUncertainties}.

\begin{table}[H]
  \centering
  \caption{A comparison between the uncertainty components considered for each of the instruments: radiation thermometer, uncooled thermal imager and a cooled thermal imager.}
  \vspace*{\floatsep}
  \begin{tabular}{ M{5.0cm}M{3.0cm}M{3.0cm}M{3.0cm} }
    \toprule
    Uncertainty Component	&	Radiation Thermometer	&	Uncooled Thermal Imager	& Cooled Thermal Imager \\
    \cmidrule(lr){1-4}
	DUT stability				&	\(\checkmark\)	&	\(\checkmark\)	&	\(\checkmark\)	\\
	NPL reference sources		&	\(\checkmark\)	&	\(\checkmark\)	&	\(\checkmark\)	\\
	SSE reproducibility			&	\(\checkmark\)	&	\(\checkmark\)	&	\(\checkmark\)	\\
	Transfer function residual	&	\(\checkmark\)	&	\(\checkmark\)	&	\(\checkmark\)	\\
    \cmidrule(lr){2-4}
	DUT resolution				&	\(\checkmark\)	&					&	\(\checkmark\)	\\
	Alignment					&	\(\checkmark\)	&					&					\\
    \cmidrule(lr){2-4}
	Distance effect				&					&	\(\checkmark\)	&					\\
	Drift						&					&	\(\checkmark\)	&					\\
	Mirror alignment			&					&	\(\checkmark\)	&					\\
	NETD						&					&	\(\checkmark\)	&					\\
	ROI uniformity				&					&	\(\checkmark\)	&					\\
    \bottomrule
  \end{tabular}
  \label{tab:uncertaintyComponentComparison}
\end{table}

The uncooled thermal imager uncertainty budget is the most comprehensive and considers the greatest number of sources. The radiation thermometer and cooled thermal imager consider similar sources of uncertainty, where the thermal imager omits an alignment component.

A large temperature dependence for the uncooled thermal imager measurement uncertainty is observed and the source of this is the non-uniformity component. The largest contributors to the larger uncertainties for the thermal imagers in these case studies compared to the radiation thermometer are due to: non-uniformity, temporal stability and repeatability. The latter two can be addressed by improving the thermal stability of the environment local to the thermal imager, the former can also be addressed through greater uniformity across the detector but is largely subject to the required field of view size of the application.

Results from the thermometer and cooled thermal imager are similar and appear dwarfed by the uncooled imager, but it should be noted that a non-uniformity was not considered within the cooled thermal imager case study. Specific consideration to the instrument in use should be made because it is expected this behaviour varies from instrument to instrument.

\clearpage
\newpage

\section{Conclusion}\label{sec:thermal_imaging_metrology_conclusion}

Chapter~\ref{chap:radiance} introduced the core concepts supporting radiance temperature measurement including thermal imager apparent radiance temperature calibration. For the calibration of radiation thermometers, a reference source of a known temperature and high emissivity should be used; this source may achieve this temperature stability and appropriate emissivity through: zoned furnace design, specialised coatings, narrow cylindrical cavities and heat-pipe liners. These furnaces are typically scoped to accommodate the temperature range from \SIrange{-60}{1000}{\celsius} and can enable calibration uncertainties from \SIrange{0.05}{0.3}{\celsius} (\(k = 2\)). The calibration of a radiation thermometer -- a Heitronics TRT IV.82 -- against the available reference sources was completed; the results from an ITS-90 comparison were presented, alongside the size-of-source effect assessment and the resultant instrument uncertainties of less than \SI{1.0}{\celsius} across this range.

Deploying the reference sources for radiation thermometers to the calibration of thermal imagers introduces limitations to characterisation. The aperture diameter to the sources are typically close to \SI{50}{\milli\metre} and so subtend a small fraction of the field of view for thermal imagers under test. This constrains the scope of the assessment to a local region of the image array and does not enable a full characterisation of the array uniformity. Beholden to these constraints, a calibration of both a cooled and uncooled thermal imager against the respective sources was carried out and best attempts to measure the uniformity was made. Each imager was successfully calibrated and each of the assessments made, the uncooled and cooled thermal imagers demonstrated uncertainties from \SIrange{1.85}{3.20}{\celsius} (\(k=2\)) and from \SIrange{0.35}{0.50}{\celsius} (\(k=2\)) respectively in the temperature range from \SIrange{10}{100}{\celsius}.

To apply a calibrated apparent radiance temperature from either a radiation thermometer or thermal imager to an application, to compare to other surface thermometry methods, a correction for the non-unity emissivity and thermal surroundings must be made. Introduced in Chapter~\ref{chap:laboratoryMeasurement} a comparison between different radiance correction methods will highlight the situations in which each approximation demonstrates an acceptable deviation in accuracy.

The techniques described within this chapter have provided a framework for approaching and evaluating radiance temperature measurements using a thermal imager and have introduced the concepts pertinent to low uncertainty measurement. In Chapter~\ref{chap:laboratoryMeasurement}, the methods presented will be used to characterise a plate of stainless steel and in Chapter~\ref{chap:inSituMeasurement} they will be used to assess a special nuclear material container in an inactive store. Calibration of the measured data is pertinent in these applications because the enable measurements of a single container to be reliably monitored from year to year to ensure appropriate management decisions can be made.

\clearpage
\newpage
\markedchapter{Laboratory Characterisation}{Laboratory Surface Temperature Characterisation}\label{chap:laboratoryMeasurement}
\section{Introduction}\label{sec:inSituIntroduction}
Throughout Chapter~\ref{chap:radiance} calibration methods for thermal imagers were explored and the use of emissivity to determine radiance temperature introduced. Prior to deploying a thermal imager to measure the surface temperature of inactive nuclear material containers the measurement capability within a laboratory environment was explored. Within the controlled laboratory scenario a representative surface to the storage containers was designed, manufactured and characterised. With the assistance of an uncertainty budget analysis the surface temperature measured by a set of thermocouples and a thermal imager was made.

The two containers evaluated were the Magnox and THermal Oxide Reprocessing Plant (THORP) Special Nuclear Material (SNM) containers used by Sellafield Ltd. Accessibility to SNM containers in stores is limited due to required control measures but ongoing monitoring and inspection is a necessity to retain confidence in, plan and manage storage. The non-contact characterisation of container integrity and activity using instrumentation translated along an inspection rail to view the container surfaces would provide Sellafield with critical information to support nuclear material management decisions. External surface temperature has the ability to provide insight to the internal temperature that is analogous to the radioactivity of the contents.

Non-destructive testing methods applied to SNM evaluation include: radiographic, ultrasound, eddy current, magnetic particle and penetrant testing. These each have their merits and drawbacks that make them suitable for their respective inspection activities \cite{ref:NDTCapabilities}. In particular eddy current inspection of nuclear material containers has been demonstrated alongside an automated maintenance inspection facility \cite{ref:eddyInspection}. Additional testing methods include phosphor thermometry, this has been applied to a range of nuclear material containers \cite{ref:novelThermometryNuclearContainers,ref:phosphorThermometryWasteStorage,ref:ilw_paper_2018}.

Thermal imaging is an optical instrumentation technique that measures the apparent surface radiance temperature from objects within its field of view. The determination of surface temperature can be inferred as well as an evaluation of surface properties through geometrical characterisation of the thermal imager.

Deployment of a thermal imager for the maintenance of SNM containers could enable both a non-destructive technique for surface inspection (either through passive or active thermal imaging \cite{ref:activeThermography}) and the surface temperature measurement of containers to infer internal radioactivity characteristics \cite{ref:ilw_paper_2018,ref:ilw_paper_2020}. Thermal imaging is beneficial over visual imaging \cite{ref:opticalImaging} in this application due to the absence of illumination within the stores that does not inhibit thermal imaging; to enable optical inspection methods a local illumination source would be required, introducing further sources of error.

The thermal imager was calibrated up to \SI{225}{\celsius} to ensure complete calibration temperature coverage, however given that the surface emissivity of the plate was much lower than \num{1.0}, the highest apparent radiance temperature measured was \SI{130}{\celsius}. It was anticipated that the container temperature would have ranged from \SIrange{10}{200}{\celsius} and its environmental temperature from \SIrange{10}{60}{\celsius}.

A thermal management assembly was designed and manufactured. To support an adjacent defect detection capability study, a series of surface artefacts were engineered and dimensionally characterised. Whilst the detection capability for these artefacts will not be discussed, the comparison between radiance temperature and thermocouple measured surface temperature will be explored.

A set of radiance correction methods will be cross-examined within the scope of thermal conditions relevant to this application. Here their suitability in a variety of thermal environments will be profiled.

\clearpage
\newpage

\section{Experimental setup}\label{sec:laboratoryExperimentalSetup}

In order to validate the measurement capabilities on a controlled system prior to nuclear material container assessment, a plate was designed and characterised; this plate is one component within a larger thermal management assembly. The plate was designed to simulate the material type used in the application for storage of special nuclear materials with a range of considerations to minimise sources of uncertainty. The measurement campaign comprises a measurement of each face of the designed plate, one of these faces was coated with higher emissivity targets and the reverse side omitted these. These faces were defined as coated and uncoated respectively.

Once assembled, this plate was observed with a cooled Medium-Wave InfraRed (MWIR) thermal imager (the InfraTec ImageIR 8300 described in the Section~\ref{subsec:case_study_thermal_imager} case study), the radiance temperature determined was compared with that temperature measured by sub-surface thermocouples.

\subsection{Defect definition}\label{subsec:defectDefinition}
An independent thread of research within this project was the detection of surface defects using a thermal imager. The four types of artefact that were investigated both in the coated and uncoated configurations were: scratches, dents, thinning and pitting. It was expected that scratches, dents and pitting would introduce geometrical emissivity enhancement, that would be observable as an apparent radiance temperature step-change. Internal thinning would introduce measurable surface temperature perturbations; corrosion and surface contamination would also lead to a change in the apparent radiance temperature. The specifications for these defects are described below.

\begin{figure}[H]
\centering
	\subcaptionbox{Scratch defect. \label{fig:simulantScratch}}
{\includegraphics[width=0.46\textwidth,keepaspectratio]{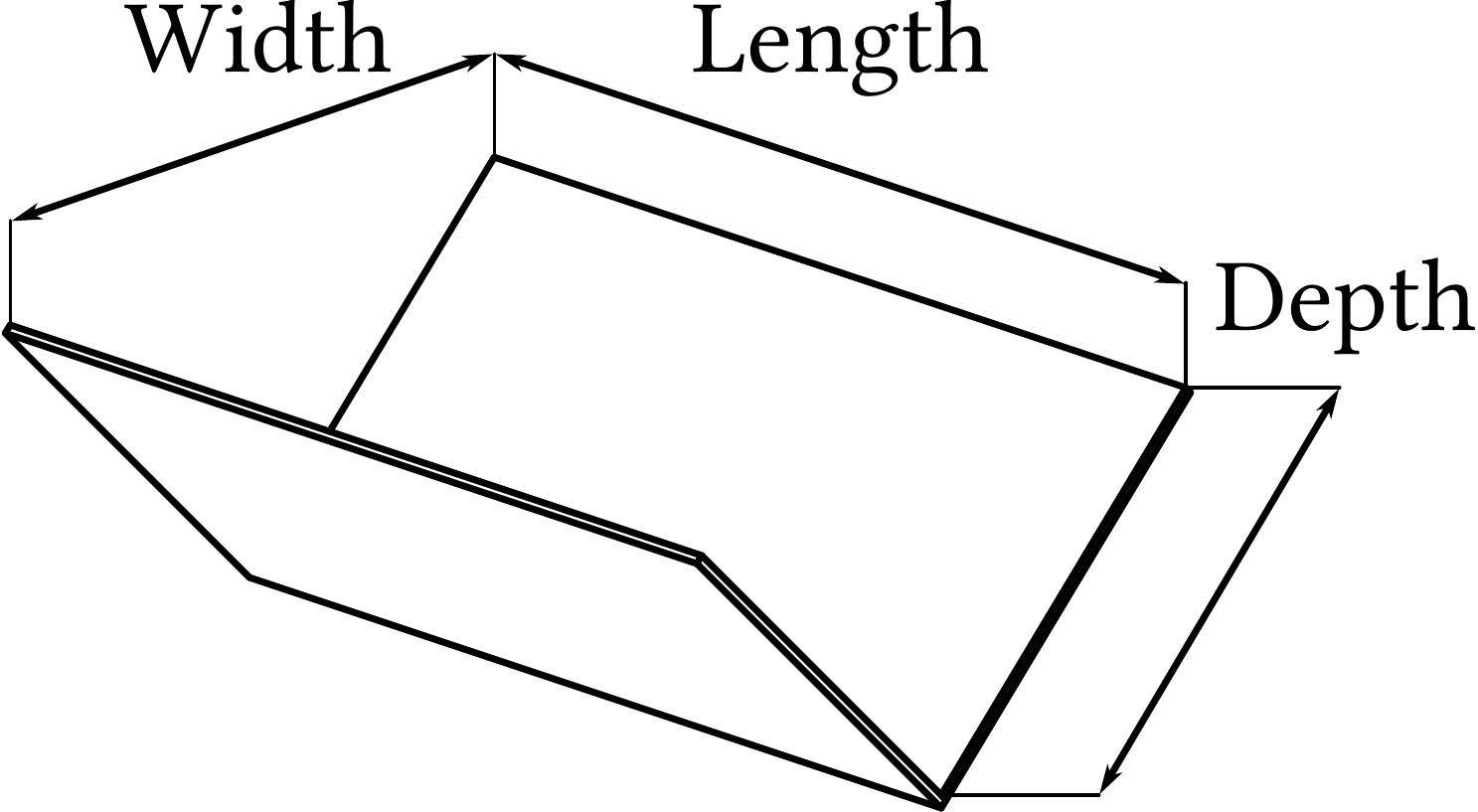}}
	\subcaptionbox{Dent defect. \label{fig:simulantDent}}
		{\includegraphics[width=0.46\textwidth,keepaspectratio]{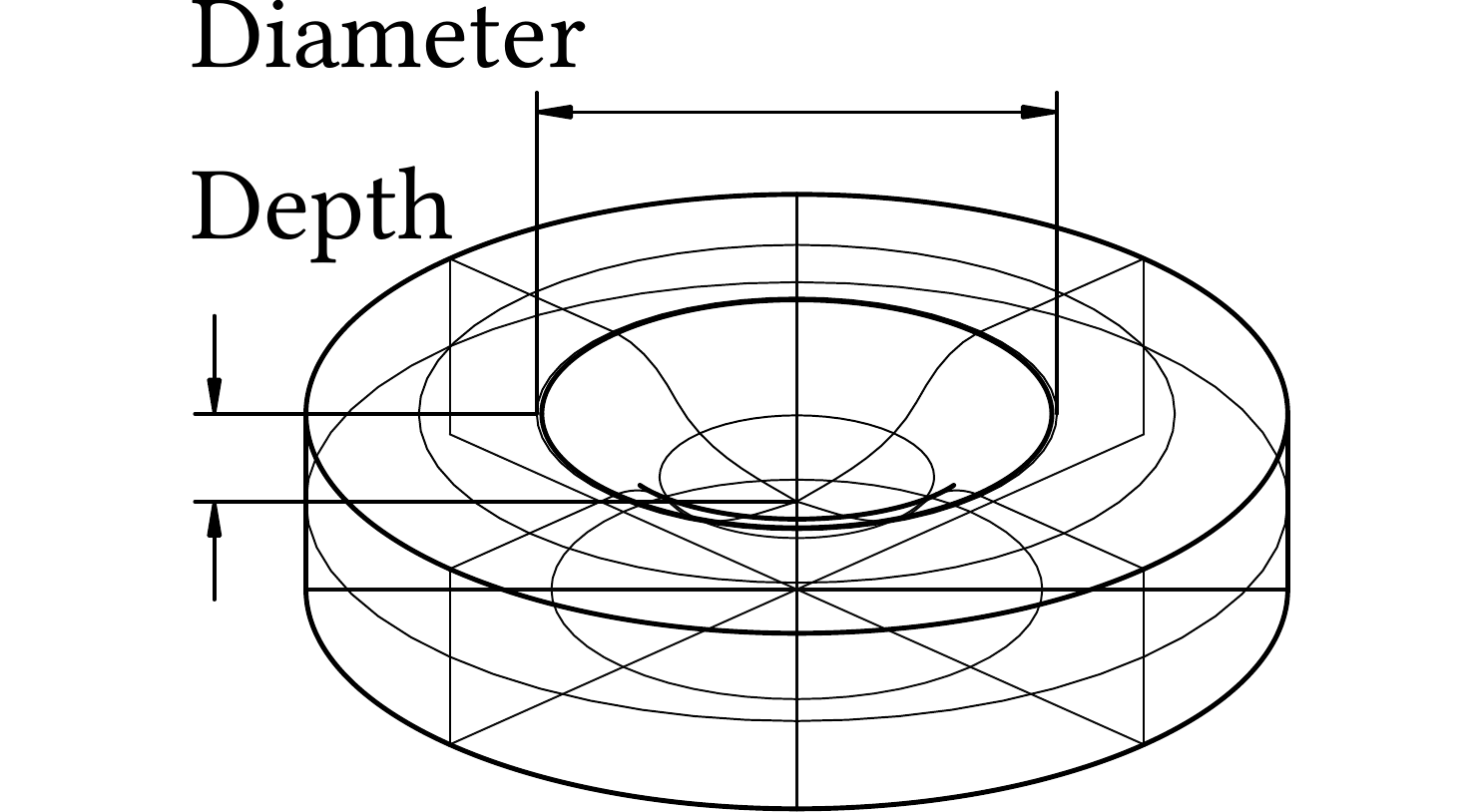}}
	\vskip\baselineskip
	\subcaptionbox{Thinning defect. \label{fig:simulantThinning}}
		{\includegraphics[width=0.46\textwidth,keepaspectratio]{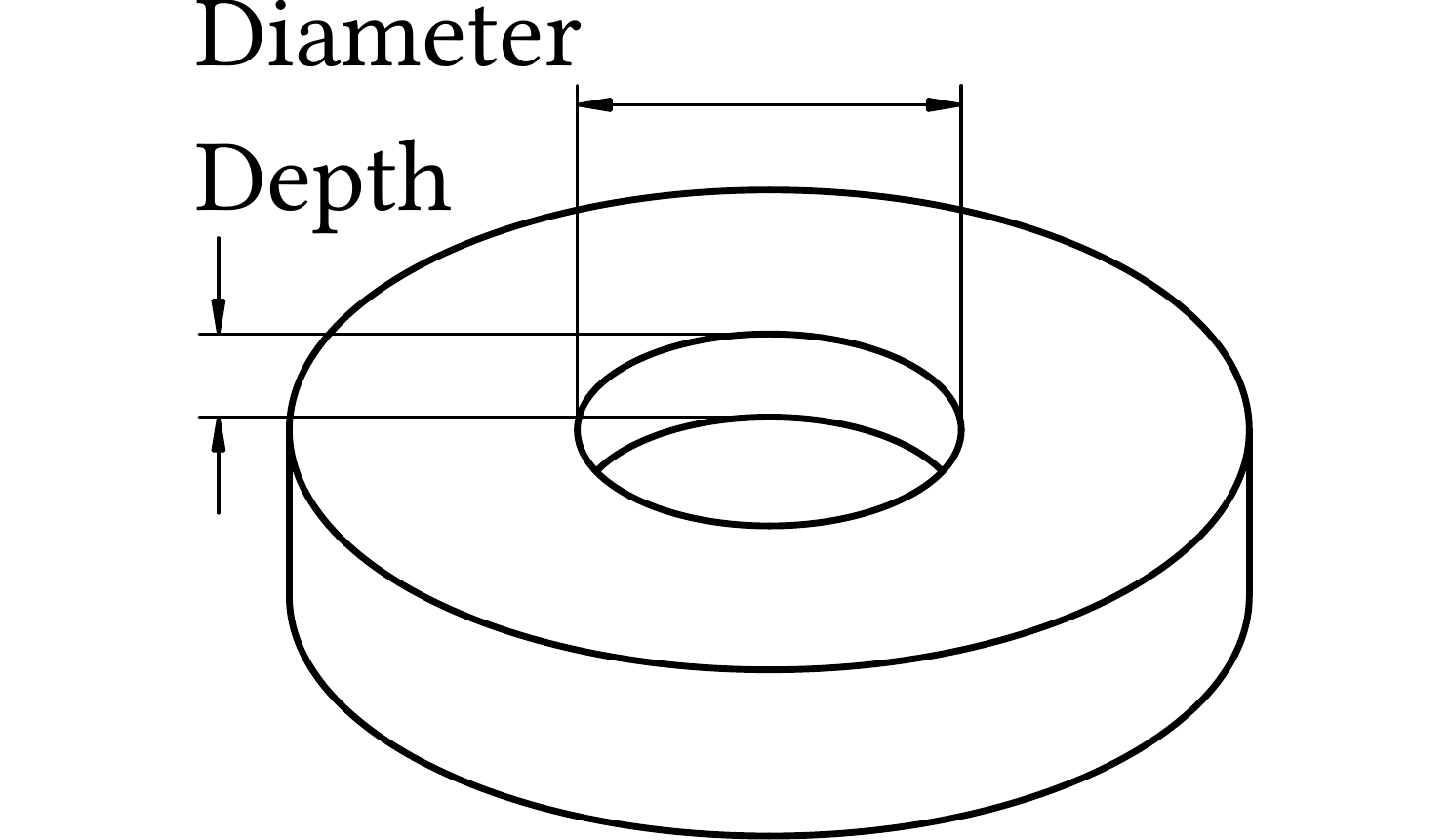}}
	\subcaptionbox{Pitting defect. \label{fig:simulantPitting}}
		{\includegraphics[width=0.46\textwidth,keepaspectratio]{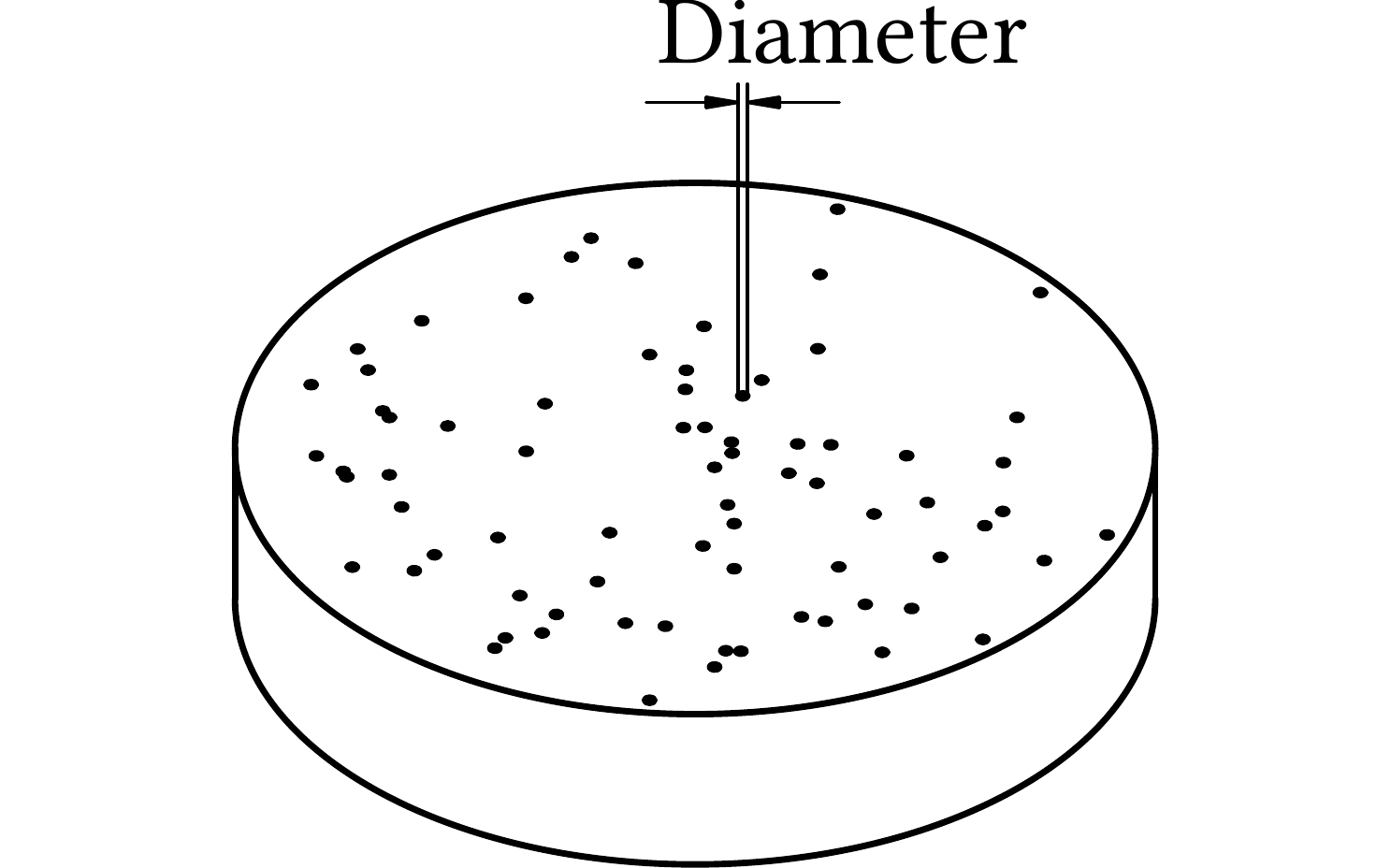}}
	\caption{The array of surface defects engineered and their geometrical definition.}
	\label{fig:defectDefinition}
\end{figure}

A scratch will be defined as a vee-groove with nominally a \SI{45}{\degree} slope gradient. Scratch depth is the distance between the apex of the groove and the height of the neighbouring surface. This is shown in \ref{fig:simulantScratch}. The width is the distance between the two top edges of the vee. The length of the scratch is the end-to-end distance of the groove. The depth is the vertical distance between the apex and the top edge of the groove.

Dents are ellipsoidal impressions in a surface, this are shown in \ref{fig:simulantDent}. The depth of the impression is the distance between the top edge and the centre of the base. The diameter is the distance between two top edges where the impression begins.

Surface thinning is a recess into a surface, this is shown in \ref{fig:simulantThinning}. The depth of the recess is the distance between the top edge and the base. The diameter is the distance between the two top edges where the recess begins.

Surface pitting is a random array of cylindrical voids of a specified diameter and depth, this is shown in \ref{fig:simulantPitting}. The diameter of the void distance between the two top edges where the void begins. The depth of the voids is defined and identical. The arrangement of the voids is arbitrary but is localised to the artefact region.

Each of these defects will introduce a change to the surface emissivity due to geometrical enhancement, this will be measured by the apparent radiance temperature images of the surface.

\clearpage
\subsection{Assembly manufacture}\label{subsec:assemblyManufacture}

\begin{figure}[H]
\centering
	\begin{subfigure}[t]{0.7\textwidth}
	\centering
	\includegraphics[width=\textwidth,keepaspectratio]{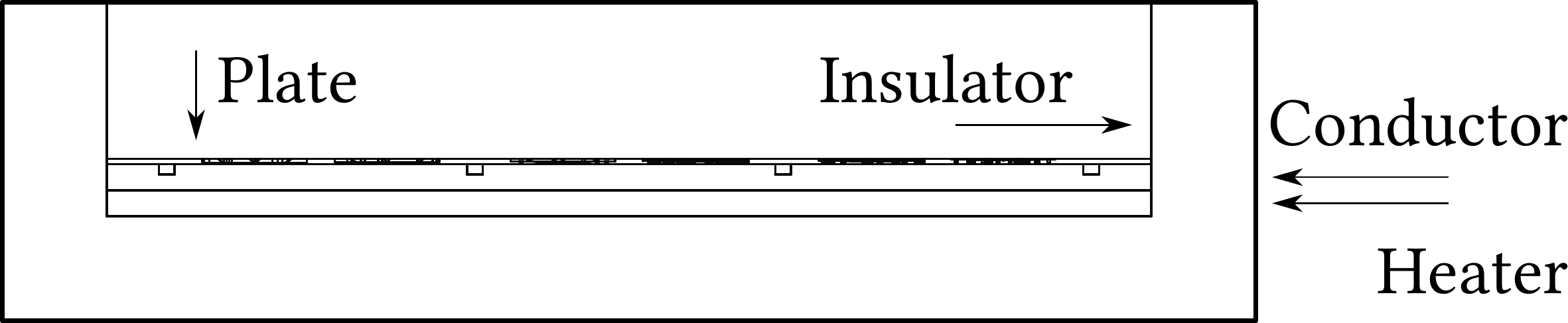}
	\caption{Cross-section of the assembly.}
	\label{fig:simulantAssemblyCrossSection}
	\end{subfigure}
	\vspace{\floatsep}
	\begin{subfigure}[t]{0.7\textwidth}
	\centering
	\includegraphics[width=\textwidth,keepaspectratio]{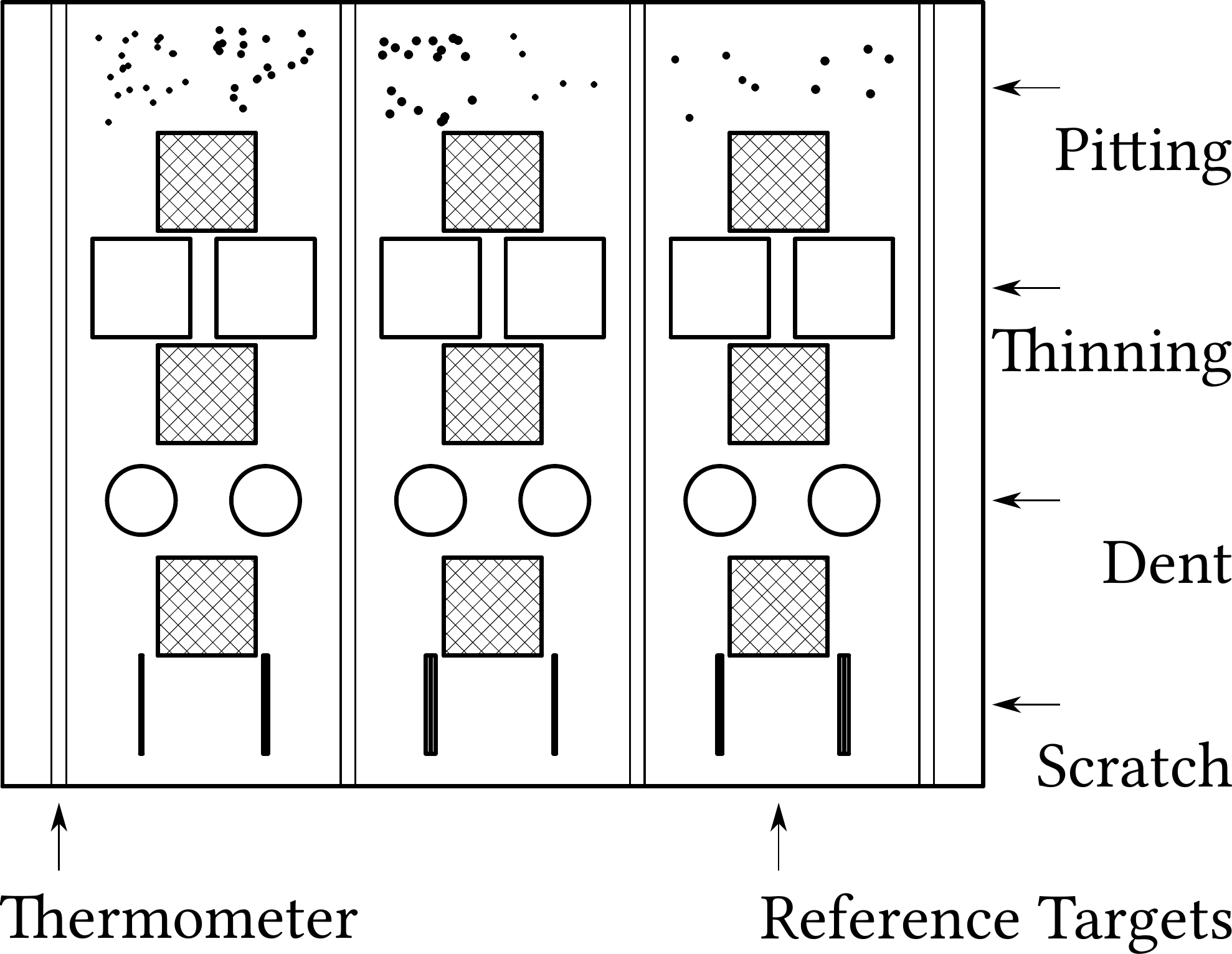}
	\caption{Lateral view of the plate.}
	\label{fig:simulantAssemblyTargetPlate}
	\end{subfigure}
\caption{The assembly designed for this investigation. (a) Cross-section of the assembly, from inside to out: radiating plate, thermal conductor plate, heater and housing. (b) Lateral view comprising the defect zones and coated reference targets. From left to right, the thermocouples located in each channel were 0, 1, 2 and 3 respectively.}
\label{fig:simulantAssembly}
\end{figure}

The primary considerations for low uncertainty surface temperature determination are: high rate of heat transfer from the heating element to radiating surface, low and known heat loss mechanisms, sufficient contact thermometer coverage, and known and low uncertainty photo-thermal properties for the radiating surface. The assembly components are shown in Figure~\ref{fig:simulantAssembly}.

The housing, insulator and conductor components have been manufactured by ProTec Ltd (Leamington Spa) and the heater was manufactured by Hawco Limited (Godalming). The insulator was machined from a cement based high-temperature insulation board Sindanyo H91, which has a thermal conductivity of \SI{0.5}{\watt\per\metre\per\kelvin} and a wall thickness of \SI{20}{\milli\metre}. The conductor plate is \SI{5}{\milli\metre} thick aluminium with \SI{4}{\milli\metre} by \SI{2}{\milli\metre} profile thermometer channels. The radiating plate is a \SI{1}{\milli\metre} thick piece of stainless steel 316L with a cross-section size of \SI{200}{\milli\metre} by \SI{160}{\milli\metre}.

\begin{figure}[h]
  \centering
  \includegraphics[width=0.5\textwidth,keepaspectratio]{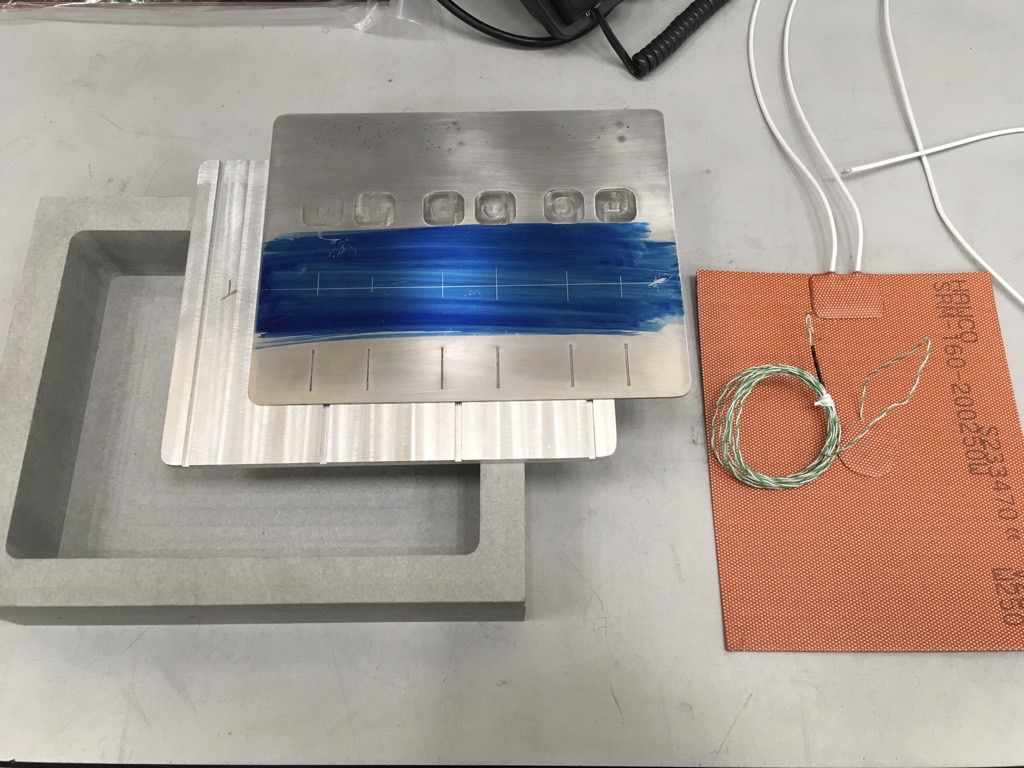}
  \caption{The manufactured assembly comprising from left to right the: insulator, conductor, plate and heater element. The blue marking indicates the regions prior to dent manufacturing.}
  \label{fig:simulantAssemblyComplete}
\end{figure}

The plate defect zones were manufactured by both ProTec Ltd and the Mass Metrology group within NPL. ProTec used a three-axis Computer Numerical Control (CNC) machine with a \SI{0.1}{\milli\metre} tolerance. The pitting was manufactured using centre drills, the scratches and thinning were manufactured using a \SI{12}{\milli\metre} carbide slot drill. The dents were manufactured using a range of ball bearings mounted within an Avery 7110 compression machine that were loaded up to \SI{30}{\kilo\newton}. From left to right (in \ref{fig:simulantAssemblyTargetPlate}) the dents were manufactured using a \SI{10}{\milli\metre} ball bearing under \SI{10}{\kilo\newton}, \SI{20}{\kilo\newton} and \SI{30}{\kilo\newton} load, and a \SI{40}{\milli\metre} ball bearing under \SI{10}{\kilo\newton}, \SI{20}{\kilo\newton} and \SI{30}{\kilo\newton} load respectively. Following the defect manufacturing the \SI{20}{\milli\metre} square reference targets were coated with Senotherm Ofen-spray schwarz 17-1644-702338 in order to ensure a higher emissivity region for surface radiance temperature measurement. From previous experience and similar coatings, it is anticipated that the emissivity of this material to be close to 0.85 \cite{ref:measurement_spectral_emissivity_ftir}, this was not measured due to project resource constraints. The completed assembly is shown in Figure~\ref{fig:simulantAssemblyComplete}.

\subsection{Assembly setup}\label{subsec:assemblySetup}

\begin{figure}[h]
  \centering
  \includegraphics[width=0.7\textwidth,keepaspectratio]{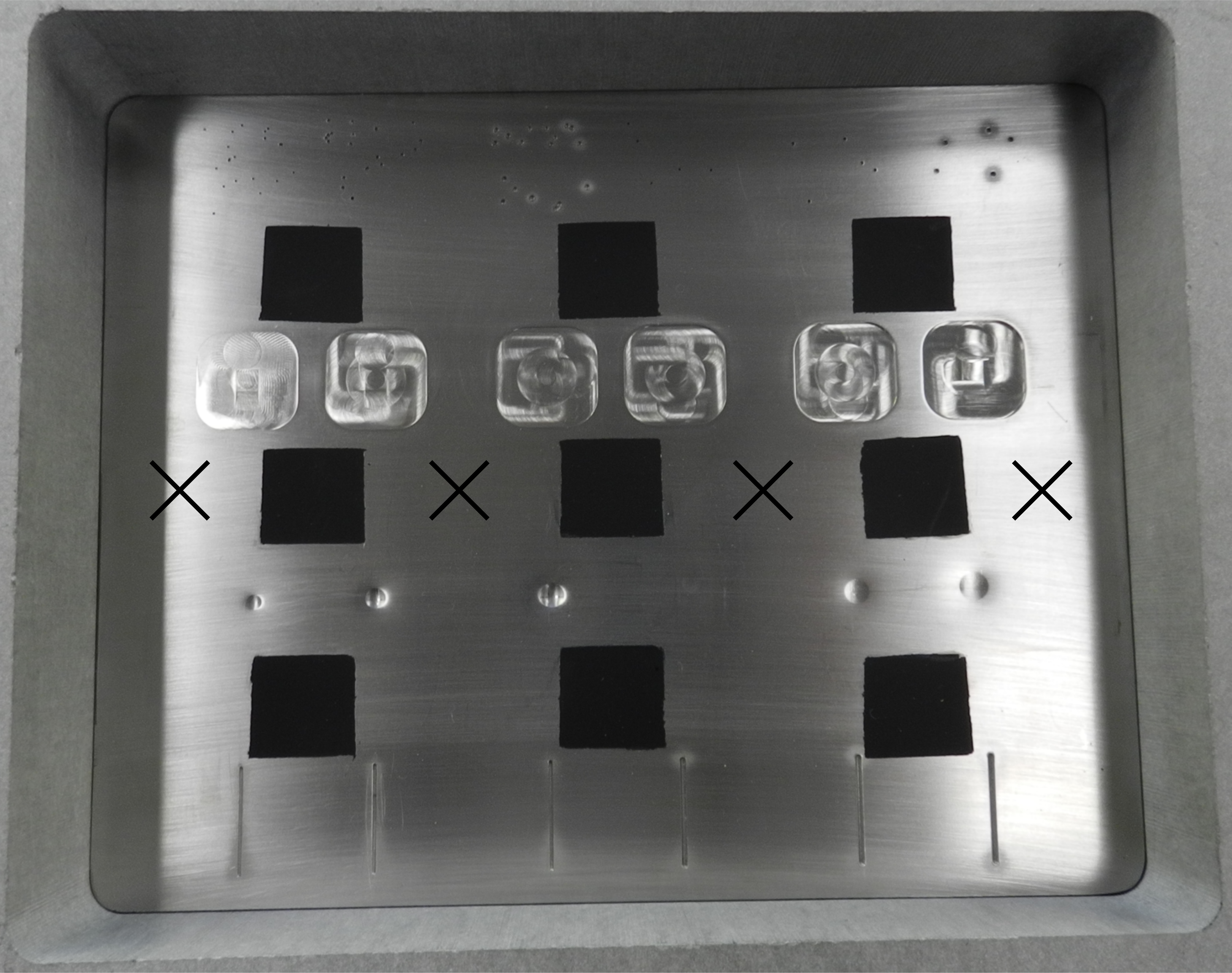}
  \caption{The finished plate is shown, the thermocouple locations are indicated by crosses.}
  \label{fig:simulantPlateImage}
\end{figure}

\begin{figure}[h]
  \centering
  \includegraphics[width=0.5\textwidth,keepaspectratio]{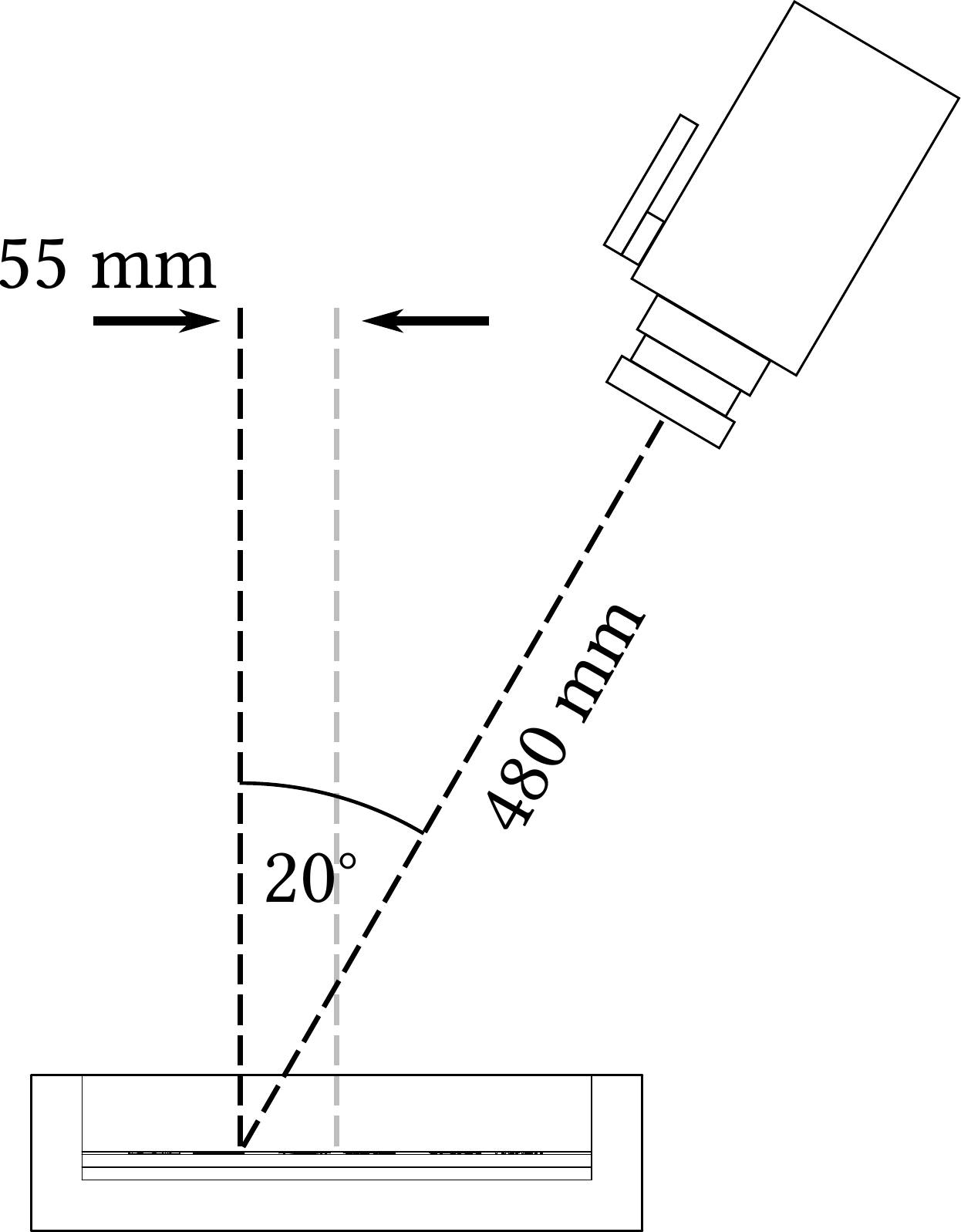}
  \caption{Schematic of the MWIR thermal imager observing the plate. The imager is \SI{20}{\degree} from the surface normal and \SI{480}{\milli\metre} from nominally the plate to the front of the lens and at its closest point the imager was \SI{450}{\milli\metre} from the plate.}
  \label{fig:simulantImagerConfiguration}
\end{figure}

The plate was setup according to the schematic in Figure~\ref{fig:simulantAssembly}, one control thermocouple was mounted to the heater surface (refer to Figure~\ref{fig:simulantAssemblyComplete}). Four monitoring thermocouples were located within the thermometer channels, one per channel, from left to right the thermocouples in each channel were 0, 1, 2 and 3 respectively. The monitoring thermocouples were connected to a Fluke 1586A Super-DAQ Precision Temperature Scanner (S/N: 44000054). Each channel was configured as a type K thermocouple and using the same internal reference point. The plate and thermocouple locations are shown in \ref{fig:simulantPlateImage}. The thermal imager was mounted \SI{20}{\degree} from the surface normal of the plate and at a distance of \SI{480}{\milli\metre} from nominally the plate to the front of the lens, this is presented in Figure~\ref{fig:simulantImagerConfiguration}. This angle of observation ensures a reflection component from the plate is located at the shroud and not from the cooled detector of the thermal imager itself. At this angle, the distance from centre of the lens to both the far and near edges on the plate varied from \SIrange{490}{450}{\milli\metre}.

To improve conductive heat transfer from the heater through to the plate, a pair of clamps were used with anodised aluminium blocks to apply pressure to the top of the plate. During the measurements a shroud was placed over the imager and assembly to reduce background reflection variations and natural convection.

For the uncoated surface measurements, the same setup was used, albeit the plate was reversed such that the pitting defects remained the farthest defect from the thermal imager, then the blocks were replaced.

\clearpage
\newpage

\section{Assembly assessment}\label{sec:assemblyAssessment}
The assembly was configured as described in Section~\ref{sec:laboratoryExperimentalSetup}, a description of the system temperature characteristics will be presented in the characterisation section followed by the uncertainty budget and a discussion. During the campaign the measurements were captured over \SI{2}{\minute} periods for each integration time at each heater setpoint using the cooled thermal imager calibration detailed in Section~\ref{subsec:case_study_thermal_imager}. The apparent radiance temperature measurements will be corrected using the method detailed in Section~\ref{subsec:surfaceTemperatureDetermination} and subsequently explored in detail in Section~\ref{subsec:multipleReflectionSingleWavelength}.

\subsection{Surface temperature determination}\label{subsec:surfaceTemperatureDetermination}
For this apparent radiance temperature correction method, first the apparent spectral radiance must be considered with respect to the local environment. Radiance from a surface \(L_{\lambda, b}\) is defined by the Planck distribution law as detailed in Section~\ref{subsec:radiation_plancks_law} as

\begin{equation}
  L_{\lambda , b}(\lambda, T_b) = \frac{c_1}{\lambda^5} \frac{1}{e^{\nicefrac{c_2}{\lambda T_b}}-1} \mathrm{.}
		\label{eq:reduced_planck_distribution_law_repeat}
\end{equation}

Here \(c_1\) and \(c_2\) are the first and second radiation constant, \(\lambda\) is the wavelength of radiation and \(T_b\) is the surface temperature.

Using the Kirchhoff law, an equivalence between the apparent radiance \(L_{app}\), emissivity and shroud radiance \(L_{\lambda, b}^{shroud}\) (anticipated reflective component detailed in Section~\ref{subsec:assemblySetup} and corresponding temperature measured by a nearby hygrometer) is given by

\begin{equation}
  L_{app} = \varepsilon L_{\lambda,b} + \left( 1 - \varepsilon \right) L_{\lambda,b}^{shroud} \mathrm{.}
		\label{eq:apparentRadianceEquivalence}
\end{equation}

This approximation assumes that the reflections are all in the specular direction and there are no diffuse contributions. Solving Eq.~\ref{eq:apparentRadianceEquivalence} for the surface temperature under this approximation gives

\begin{equation}
  T_b = \frac{c_2}{\lambda } \ln\left[ \varepsilon \left[ \frac{1}{e^{\nicefrac{c_2}{\lambda T_{app}}}-1} - \frac{\left( 1 - \varepsilon \right) }{e^{\nicefrac{c_2}{\lambda T_{shroud}}}-1} \right]^{-1} + 1 \right]^{-1} \mathrm{.}
		\label{eq:radianceCorrectionPlate}
\end{equation}

This correction can be used to determine the surface temperature using the measured temperatures, the estimated spectral mid-point and estimated emissivity of surfaces under inspection. It should be noted that usage of this approximation does not consider: potential impact from further reflections, how to assess the spectral dependency (integrate over the specific spectral range of the system and optics used), the type of emissivity value used or a transmission greater than 0. In particular, a single hemispherical total emissivity value for a surface may be used, however the effect from directional and spectral emissivity dependence is not accounted for.

\subsection{Thermal characterisation}\label{subsec:plateThermalCharacterisation}

The plate and its temperature characteristics have been profiled during the measurements at increasing heater temperature setpoints. The results from this thermal validation are detailed in the following sections.

\subsubsection{Coated plate characterisation}\label{subsubsec:coatedPlateCharacterisation}

\begin{figure}[H]
\centering
	\begin{subfigure}[t]{0.475\textwidth}
	\centering
	\includegraphics[width=\textwidth,keepaspectratio]{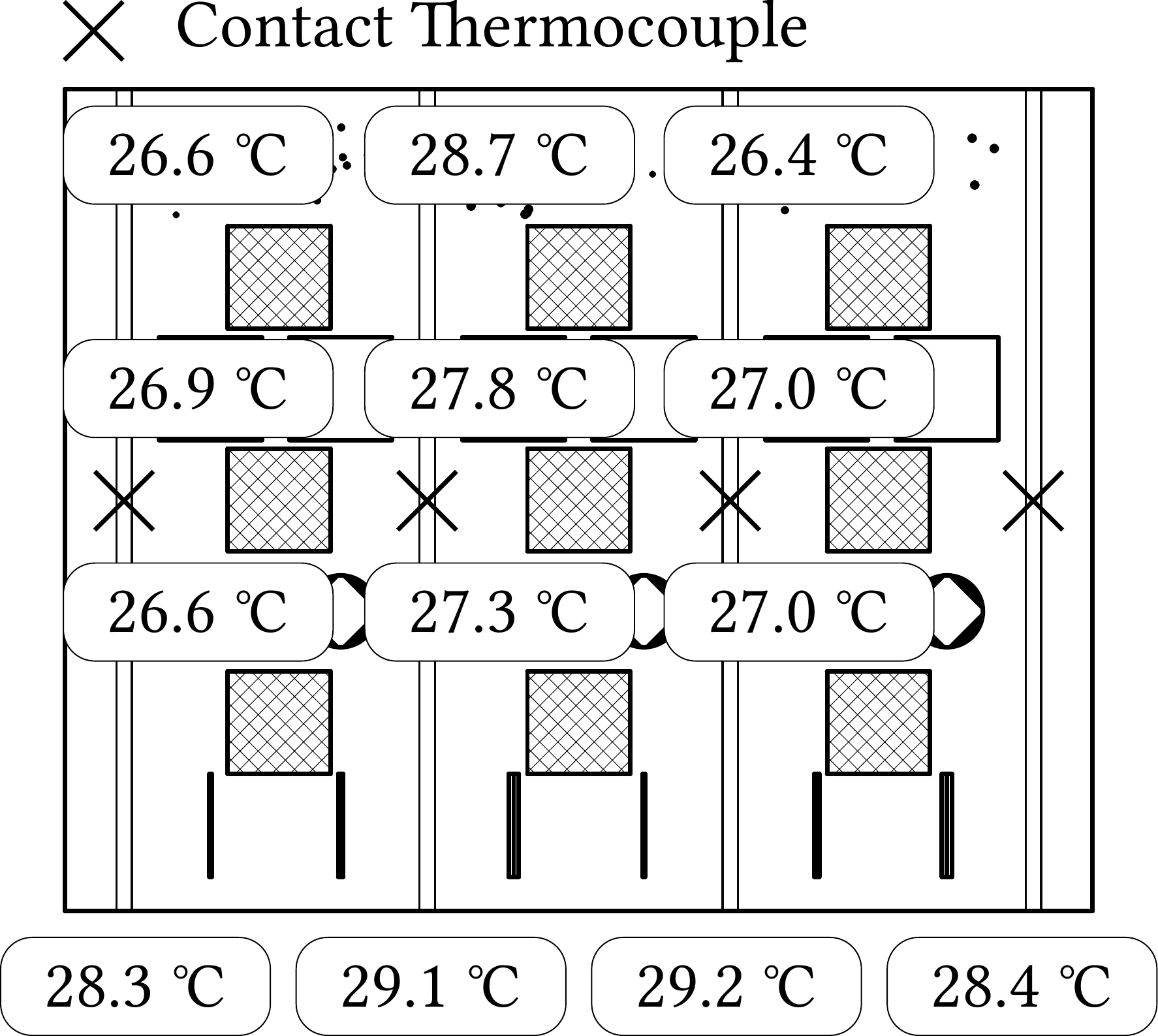}
	\caption{Temperature distribution at the \SI{30}{\celsius} temperature setpoint.}
	\label{fig:simulantCoatedDistribution30degC}
	\end{subfigure}
\hfill
	\begin{subfigure}[t]{0.475\textwidth}
	\centering
	\includegraphics[width=\textwidth,keepaspectratio]{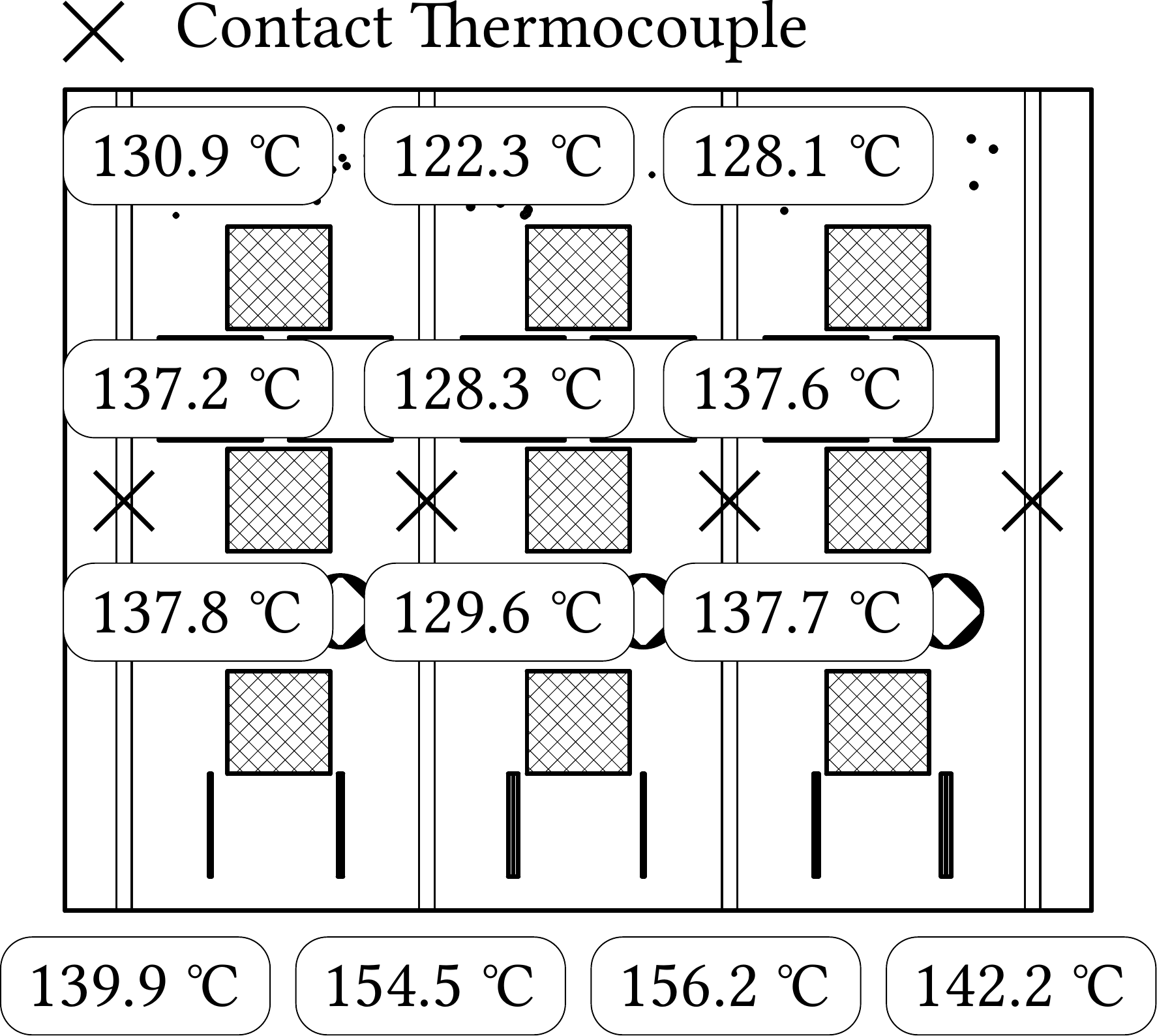}
	\caption{Temperature distribution at the \SI{170}{\celsius} temperature setpoint.}
	\label{fig:simulantCoatedDistribution170degC}
	\end{subfigure}
\caption{The measured temperature distribution of the plate during coated artefact measurements at (a) \SI{30}{\celsius} and (b) \SI{170}{\celsius}. The four measurements at the bottom correspond to the respective thermocouples indicated by the crosses. The nine central values denote the radiance temperatures of the adjacent coated regions.}
\label{fig:simulantCoatedDistribution}
\end{figure}

Figure~\ref{fig:simulantCoatedDistribution} shows the location and corresponding temperature from the thermal imager and thermocouples during the \SI{30}{\celsius} and \SI{170}{\celsius} coated plate measurements. The four measurements at the bottom correspond to the respective thermocouples. The nine central values denote the emissivity corrected radiance temperatures of the adjacent coated regions, assuming an emissivity of 0.85.

These measurements show that agreement between the radiance temperature and the mean thermocouple temperature above the \SI{30}{\celsius} measurement is not demonstrated when considering the respective measurement uncertainties (refer to Section~\ref{subsec:surfaceTemperatureUncertaintyAnalysis}).

The temperature distribution measured by the thermocouples indicates cooler temperatures towards the edges of the plate which is anticipated due to the higher heat transfer from the edges of the heater to the surrounding insulator. However, this behaviour was not observed from the radiance temperature measurements, the regions closer to the edge were hotter than at the centre by up to \SI{8}{\celsius}.

\subsubsection{Uncoated plate characterisation}\label{subsubsec:uncoatedPlateCharacterisation}

For the uncoated plate measurements, the surface emissivity was estimated using the apparent radiance temperature of the uncoated regions during the coated plate measurements and comparing to the coated regions. The emissivity of the stainless steel was estimated to be nominally 0.20. This value is supported by existing literature reporting the hemispherical total emissivity of polished stainless steel 316L to be nominally 0.26 at \SI{200}{\celsius} \cite{ref:steelEmissivity}.

\begin{figure}[H]
\centering
	\begin{subfigure}[t]{0.475\textwidth}
	\centering
	\includegraphics[width=\textwidth,keepaspectratio]{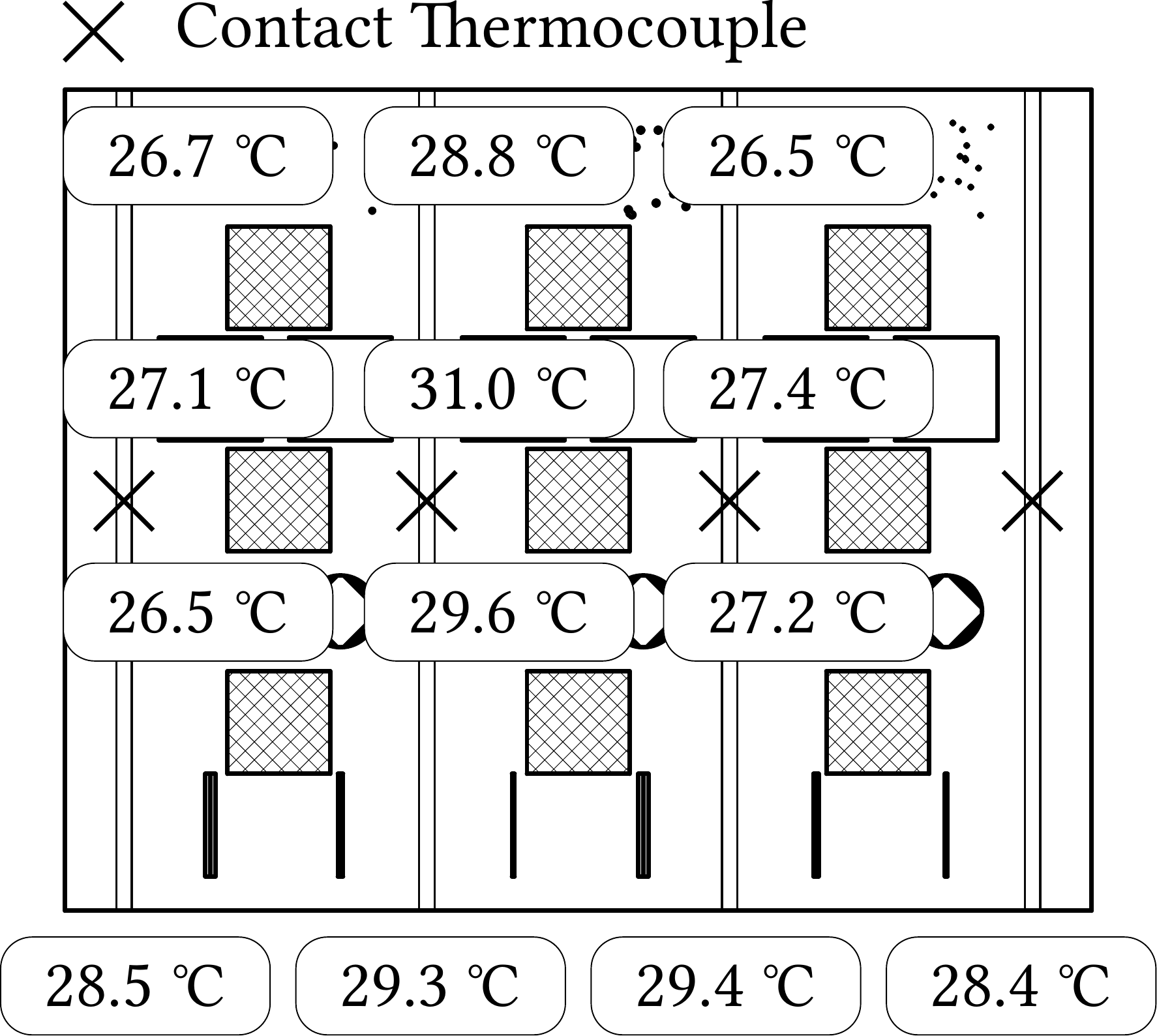}
	\caption{Temperature distribution at the \SI{30}{\celsius} temperature setpoint.}
	\label{fig:simulantUncoatedDistribution30degC}
	\end{subfigure}
\hfill
	\begin{subfigure}[t]{0.475\textwidth}
	\centering
	\includegraphics[width=\textwidth,keepaspectratio]{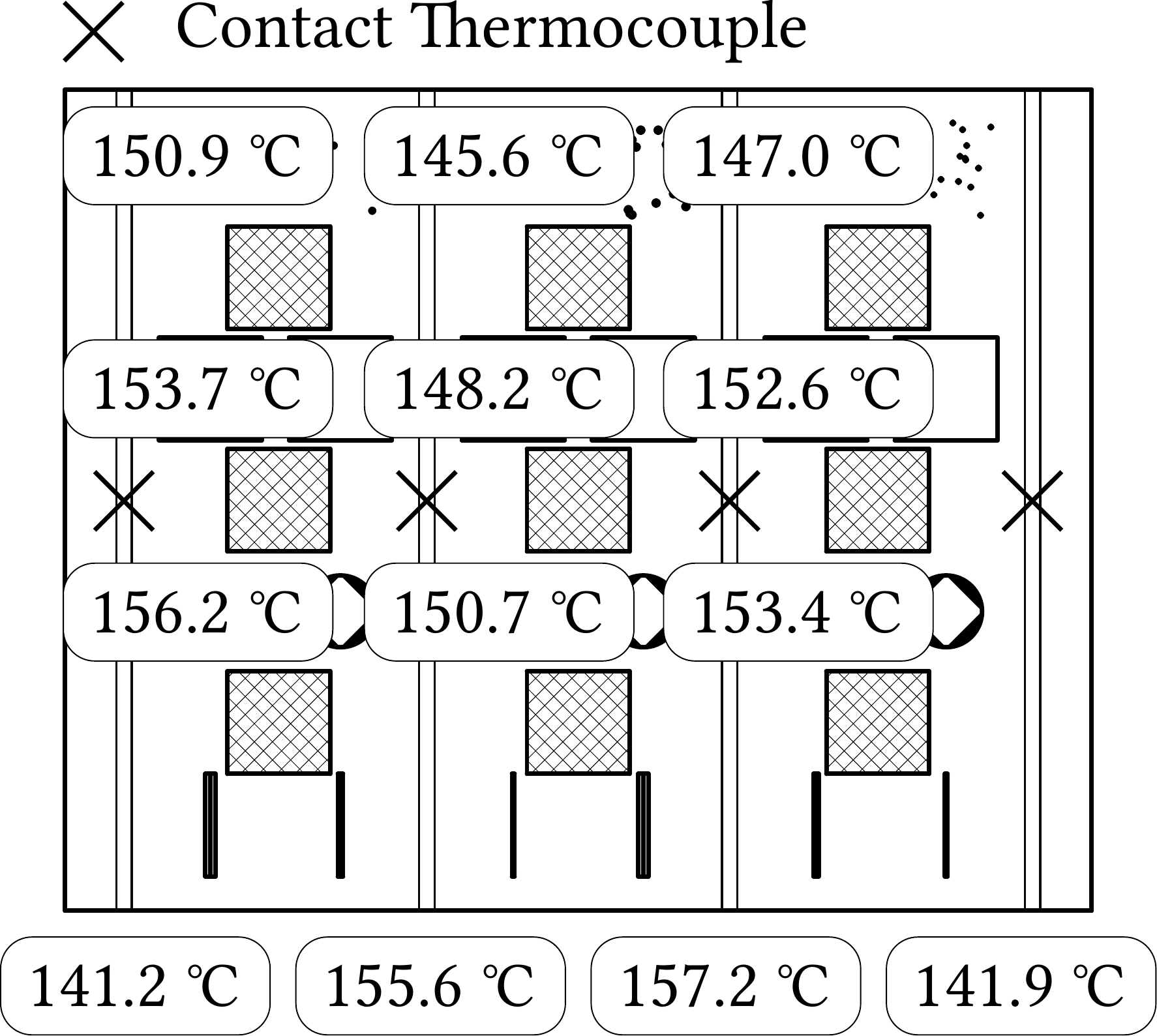}
	\caption{Temperature distribution at the \SI{170}{\celsius} temperature setpoint.}
	\label{fig:simulantUncoatedDistribution170degC}
	\end{subfigure}
\caption{The measured temperature distribution of the plate during uncoated measurements at (a) \SI{30}{\celsius} and (b) \SI{170}{\celsius}. The four measurements at the bottom correspond to the respective thermocouples indicated by the crosses. The nine central values denote the radiance temperatures of the adjacent coated regions. Note that the defects are shown for context but both coated regions and artefacts were present on the rear surface from view.}
\label{fig:simulantUncoatedDistribution}
\end{figure}

Compared to the coated plate measurements, there is much closer agreement between the mean thermocouple temperature and radiance temperature. Similarly, for the \SI{30}{\celsius} and \SI{170}{\celsius} measurement setpoint, the respective temperature distributions between the thermocouple and radiance temperatures are depicted in Figure~\ref{fig:simulantUncoatedDistribution}. The surface artefacts are presented in this image for context, the same regions of interest interrogated from the surface radiance temperature images were applied to the uniform and smooth surface at the rear of the target plate. The plate was flipped along its short edge such that from the imager perspective the defects remain in the same vertical order.

These measurements demonstrate identical thermocouple distribution to the coated plate measurements at the conductor plane. The radiance temperature range is also comparable at \SI{10}{\celsius} to the coated region measurements. However there is much closer agreement between the mean thermocouple temperature and mean radiance temperature; the difference between the mean temperatures during the coated measurement was \SI{16}{\celsius} and here the mean difference is \SI{2}{\celsius}. This may be due to the estimated emissivity of 0.20 used. But this may also be a result of a variation in the thermal conduction between the conductor and target plate due to a different force applied from the clamped aluminium blocks.

\subsection{Surface temperature uncertainty analysis}\label{subsec:surfaceTemperatureUncertaintyAnalysis}

Each individual thermocouple and thermal imager uncertainty components have been evaluated with respect to their uncertainty value, probability distribution and sensitivity to the measurand. The standard uncertainty is then combined in quadrature to determine a combined uncertainty. The uncertainty budgets have been considered according to the Guide to Uncertainty in Measurement \cite{ref:gum}. 

Each component has been attributed to the respective instrumentation source and further discretised into an intrinsic instrumentation uncertainty and an extrinsic application uncertainty.

\subsubsection{Thermocouple uncertainty}\label{subsubsec:laboratoryThermocoupleUncertainty}
Thermocouple uncertainty components are detailed here. 

\paragraph{Instrumentation}\label{para:laboratoryThermocoupleInstrumentation}
{\em Standard tolerance}

BS EN 60584-1:2013 Thermocouples describes the standard tolerance of a type K thermocouple to be \SI{1.5}{\celsius} or \(0.004 \cdot T\) (where \(T\) is the measured temperature in Celsius), whichever is larger \cite{ref:bsiThermocouples}. Given that the measurements in this application consider temperatures below \SI{375}{\celsius} this component is consistently a rectangular distribution throughout each budget.

{\em Thermometer bridge accuracy}

The manufacturer stated accuracy for temperature measurement using thermocouples is \SI{0.2}{\celsius}. This is applied as a normal distribution component in each budget.

\paragraph{Application}\label{para:thermocoupleApplication}
{\em Measurement stability}

The mean standard deviation of the four thermocouples over the measurement at each temperature setpoint was calculated. This corresponds to the temporal stability of the complete thermocouple assembly but does not consider the spatial uniformity measured by the thermocouple arrays. The respective values were applied as a normal distribution component in each budget.

\subsubsection{Thermal imager uncertainty}\label{subsubsec:laboratoryThermalImagerUncertainty}
Thermal imager uncertainty components are detailed here. The application in this case is the set of thermal response components responsible for temperature determination.

\paragraph{Instrumentation}\label{para:laboratoryThermalImagerInstrumentation}
{\em Calibration}

The calibration of the thermal imager is detailed in Section~\ref{subsec:case_study_thermal_imager}. This includes components for: the calibration of the reference source, stability of the reference source, stability of the imager, resolution of the imager, drift of the reference source and the residual of the calibration fit. This varied between the four integration times but was below \SI{0.5}{\celsius} (\(k = 2.3\)) and the appropriate values were applied as a normal distribution to each budget.

{\em Housing temperature stability}

During each measurement, the housing temperature of the thermal imager was recorded and the standard deviation during each respective dataset was applied to the budgets as a normal distribution. The relationship between measured temperature and the housing temperature is not one-to-one and is assumed to have less of an effect than this and so this an over estimation of this component. This effect however does not account for the variation of the housing temperature in the laboratory from that during calibration.

{\em Size-of-source effect}

The size-of-source effect was evaluated as described in Section~\ref{subsubsec:caseStudySSE}, the repeatability of \SI{0.18}{\celsius} was scaled from \SI{170}{\celsius} to the respective temperatures measured throughout each measurement. These values were applied as a rectangular distribution component in each budget.

{\em Distance effect}

The distance effect was evaluated as described in Section~\ref{subsubsec:caseStudyDistance} , the repeatability of \SI{0.29}{\celsius} was scaled from \SI{170}{\celsius} to the respective temperatures measured throughout each measurement. These values were applied as a rectangular distribution component in each budget.

\paragraph{Application}\label{para:laboratoryThermalImagerApplication}
{\em Region of interest uniformity}

During each measurement set, the mean spatial standard deviation of apparent radiance temperature measured by the thermal imager at the Regions Of Interest (ROIs) was evaluated. This was considered as a normal distribution component in each budget. The sensitivity function for this component is described by Eq.~\ref{eq:sensitivityApparentTemperature}. Where \(T_b\) is the surface temperature, \(T_{app}\) is the apparent radiance temperature, \(T_{amb}\) the ambient room temperature, \(\varepsilon\) the hemispherical total emissivity, \(\lambda\) the measurement wavelength and \(c_2\) the second radiation constant \cite{ref:radiation_thermometry}. This sensitivity was derived from Eq.~\ref{eq:radianceCorrectionPlate},

\begin{align}
		\begin{aligned}
				\frac{\partial T_b}{\partial T_{app}} =& \left[ \frac{c_2}{\lambda T_{app}} \right]^2 \left[ \varepsilon e^{\nicefrac{c_2}{\lambda T_{app}}} \left( e^{\nicefrac{c_2}{\lambda T_{amb}}} - 1 \right)^2 \right] \cdot \\
				& \left[ \varepsilon e^{\frac{c_2}{\lambda} \left( \frac{1}{T_{app}} + \frac{1}{T_{amb}} \right)} + \left( \varepsilon - 1 \right) e^{\nicefrac{c_2}{\lambda T_{amb}}} + e^{\nicefrac{c_2}{\lambda T_{app}}} \right]^{-1} \cdot \\
				& \left[ \varepsilon \left( e^{\nicefrac{c_2}{\lambda T_{app}}} - 1 \right) - e^{\nicefrac{c_2}{\lambda T_{app}}} + e^{\nicefrac{c_2}{\lambda T_{amb}}} \right]^{-1} \cdot \\
				\ln	& \left[ 1 + \frac{\varepsilon}{ \frac{1}{ e^{\nicefrac{c_2}{\lambda T_{app}}} - 1 } + \frac{\varepsilon - 1}{ e^{\nicefrac{c_2}{\lambda T_{amb}}} - 1 } } \right]^{-2} \mathrm{.}
		\label{eq:sensitivityApparentTemperature}
		\end{aligned}
\end{align}

{\em Emissivity measurement}

For the coated regions on the surface, the emissivity value of 0.85 was approximated from the reference measurement \cite{ref:measurement_spectral_emissivity_ftir}. This value was measured using instrumentation with an uncertainty reported to be \num{0.04} (no confidence factor was reported, so was assumed to be 1). This value was applied as a normal distribution to each measurement set. The sensitivity function for this component is described by Eq.~\ref{eq:sensitivityEmissivity}. The same uncertainty was used for both coated and uncoated regions. This sensitivity was derived from Eq.~\ref{eq:radianceCorrectionPlate},

\begin{align}
		\begin{aligned}
		  \frac{\partial T_b}{\partial \varepsilon} = & \left[ \frac{c_2}{\lambda} \right] \left[ \frac{1}{e^{\nicefrac{c_2}{\lambda T_{amb}}} - 1} - \frac{1}{e^{\nicefrac{c_2}{\lambda T_{app}}} - 1} \right] \cdot \\
				& \left[ \frac{1}{ e^{\nicefrac{c_2}{\lambda T_{app}}} - 1 } + \frac{\varepsilon - 1}{ e^{\nicefrac{c_2}{\lambda T_{amb}}} - 1 } \right]^{-1} \cdot \\
				& \left[ \varepsilon + \frac{1}{ e^{\nicefrac{c_2}{\lambda T_{app}}} - 1 } + \frac{\varepsilon - 1}{ e^{\nicefrac{c_2}{\lambda T_{amb}}} - 1 } \right]^{-1} \cdot \\
				\ln	& \left[ 1 + \frac{\varepsilon}{ \frac{1}{ e^{\nicefrac{c_2}{\lambda T_{app}}} - 1 } + \frac{\varepsilon - 1}{ e^{\nicefrac{c_2}{\lambda T_{amb}}} - 1 } } \right]^{-2} \mathrm{.}
		\label{eq:sensitivityEmissivity}
		\end{aligned}
\end{align}

{\em Emissivity interpretation}

For the coated regions on the surface, the emissivity value of 0.85 was approximated from the reference measurement \cite{ref:measurement_spectral_emissivity_ftir}. This approximation was evaluated based on emissivity data at three different temperatures and over the spectral range from \SIrange{2.7}{5.0}{\micro\metre}. The uncertainty incurred from estimating from this data is \num{0.05}. This value was applied as a normal distribution to each measurement set. The sensitivity function for this component is described by Eq.~\ref{eq:sensitivityEmissivity}. The same uncertainty was used for both coated and uncoated regions.

{\em Ambient temperature measurement}

Due to the non-unity emissivity of the surfaces measured, the uncertainty of the ambient temperature has an influence on the calculated surface temperature. The measurement uncertainty of the calibrated hygrometer is \SI{0.1}{\celsius} (\(k=1\)) across the appropriate temperature range and the variation of room temperature during each measurement dataset remained below this value. This value was applied as a normal distribution throughout each dataset using the sensitivity function described by Eq.~\ref{eq:sensitivityAmbientTemperature}. This sensitivity was derived from Eq.~\ref{eq:radianceCorrectionPlate},

\begin{align}
		\begin{aligned}
				\frac{\partial T_b}{\partial T_{amb}} =& \left[ \frac{c_2}{\lambda T_{amb}} \right]^2 \left[ \varepsilon \left( \varepsilon - 1 \right) e^{\nicefrac{c_2}{\lambda T_{amb}}} \left( e^{\nicefrac{c_2}{\lambda T_{app}}} - 1 \right)^2 \right] \cdot \\
				& \left[ \varepsilon e^{\frac{c_2}{\lambda} \left( \frac{1}{T_{app}} + \frac{1}{T_{amb}} \right)} + \left( \varepsilon - 1 \right) e^{\nicefrac{c_2}{\lambda T_{amb}}} + e^{\nicefrac{c_2}{\lambda T_{app}}} \right]^{-1} \cdot \\
				& \left[ \varepsilon \left( e^{\nicefrac{c_2}{\lambda T_{app}}} - 1 \right) - e^{\nicefrac{c_2}{\lambda T_{app}}} + e^{\nicefrac{c_2}{\lambda T_{amb}}} \right]^{-1} \cdot \\
				\ln	& \left[ 1 + \frac{\varepsilon}{ \frac{1}{ e^{\nicefrac{c_2}{\lambda T_{app}}} - 1 } + \frac{\varepsilon - 1}{ e^{\nicefrac{c_2}{\lambda T_{amb}}} - 1 } } \right]^{-2} \mathrm{.}
		\label{eq:sensitivityAmbientTemperature}
		\end{aligned}
\end{align}

\subsubsection{Combined uncertainty budget}\label{subsubsec:laboratoryCombinedUncertainty}

Considering the components introduced in Section~\ref{subsubsec:laboratoryThermocoupleUncertainty} and Section~\ref{subsubsec:laboratoryThermalImagerUncertainty}, an example complete budget is detailed for the temperature determination methods used: thermocouple and thermal imager.

\begin{table}[ht]
		\centering 
		\caption{An example complete uncertainty budget for temperature determination at the \SI{170}{\celsius} setpoint with the coated plate and \SI{50}{\micro\second} integration time. Each component is attributed to the respective instrument (thermocouple or thermal imager). The uncertainty, \(u\), is reported in degrees Celsius except for the emissivity components, the necessary divisor, sensitivity, and the standard uncertainty, \(U\), are presented. These were then combined in quadrature and multiplied by the coverage factor.}
		\begin{tabular}{ L{13.0em} M{3.0em} M{3.0em} M{5.0em} M{3.0em} }
		\toprule
		Source & \(u\) & Divisor & Sensitivity & \(U\) / \SI{}{\celsius} \\
		\cmidrule(lr){1-5}
		\multicolumn{5}{c}{Thermocouple} \\
		\cmidrule(lr){1-5}
		Tolerance & 1.50 & 1.73 & 1.00 & 0.87 \\
		Thermometry bridge accuracy & 0.20 & 1.00 & 1.00 & 0.20 \\
		Stability & 0.03 & 1.00 & 1.00 & 0.03 \\
		\cmidrule(lr){2-5}
		\multicolumn{4}{r}{Expanded uncertainty (\(k = 2\))} & {\bf 1.8} \\
		\cmidrule(lr){1-5}
		\multicolumn{5}{c}{Thermal Imager} \\
		\cmidrule(lr){1-5}
		Calibration & 0.30 & 2.20 & 1.00 & 0.14 \\
		Housing temperature & 0.03 & 1.00 & 1.00 & 0.03 \\
		Size-of-source & 0.18 & 1.73 & 1.00 & 0.10 \\
		Distance & 0.29 & 1.73 & 1.00 & 0.17 \\
		ROI non-uniformity & 1.06 & 1.00 & 1.04 & 1.10 \\
		Emissivity measurement & 0.04 & 1.00 & 50.03 & 2.00 \\
		Emissivity interpretation & 0.05 & 1.00 & 50.03 & 2.50 \\
		Ambient temperature & 0.10 & 1.00 & 0.01 & 0.00 \\
		\cmidrule(lr){2-5}
		\multicolumn{4}{r}{Expanded uncertainty (\(k = 2\))} & {\bf 6.8} \\
		\bottomrule
		\end{tabular} 
		\label{tab:laboratoryCombinedUncertainty}
\end{table}

Table~\ref{tab:laboratoryCombinedUncertainty} describes the complete budget describing the measurements at the \SI{170}{\celsius} temperature setpoint for the coated plate using the \SI{50}{\micro\second} integration time. Each component is attributed to the instrument it originates from and either an intrinsic instrumentation uncertainty or extrinsic application uncertainty. The uncertainty value \(u\) is reported as degrees Celsius unless otherwise stated (for the emissivity components). The divisor describes the probability distribution of the component, either a normal or rectangular distribution. The sensitivity is 1.00 for surface temperature components but requires conversion for other units to a standard uncertainty in degrees Celsius. For each instrumentation and application, the respective component standard uncertainties were combined in quadrature. The combined uncertainty includes both instrumentation and application uncertainty components and multiplies by \num{2} for a \SI{95}{\percent} confidence interval.

A reduced summary of the measurement uncertainty at each coated region of interest temperature setpoint is presented in Table~\ref{tab:laboratoryUncertaintyCoated}, the equivalent values for the uncoated measurements are presented in Table~\ref{tab:laboratoryUncertaintyUncoated}. These describe the greatest uncertainty between the relevant integration times for each temperature setpoint.

\begin{table}[H]
  \centering
  \caption{A summary of the instrumentation, application and combined measurement uncertainties for each temperature setpoint during the coated plate measurements. These values correspond to the respective uncertainty budgets (refer to Table~\ref{tab:laboratoryCombinedUncertainty}) constructed for each measurement. The greatest of the appropriate integration time uncertainties at each setpoint is reported.}
  \vspace*{\floatsep}
  \begin{tabular}{ M{2.5cm}M{1.5cm}M{1.5cm}M{2.5cm}M{1.5cm}M{1.5cm}M{2.5cm} }
    \toprule
	&	\multicolumn{3}{c}{Thermocouple Uncertainty / \SI{}{\celsius}} & \multicolumn{3}{c}{Thermal Imager Uncertainty / \SI{}{\celsius}} \\
		Temperature Setpoint / \SI{}{\celsius} & Instrument & Application & Expanded (\( k= 2\)) & Instrument & Application & Expanded (\(k = 2\)) \\
    \cmidrule(lr){1-7}
	30	&	0.9	&	0.0	&	1.8	&	0.2	&	0.5	&	1.0	\\
	70	&	0.9	&	0.0	&	1.8	&	0.2	&	1.7	&	3.4	\\
	110	&	0.9	&	0.1	&	1.8	&	0.3	&	2.4	&	4.8	\\
	130	&	0.9	&	0.1	&	1.8	&	0.3	&	2.8	&	5.5	\\
	150	&	0.9	&	0.0	&	1.8	&	0.2	&	3.1	&	6.2	\\
	170	&	0.9	&	0.0	&	1.8	&	0.2	&	3.4	&	6.8	\\
    \bottomrule
  \end{tabular}
  \label{tab:laboratoryUncertaintyCoated}
\end{table}

\begin{table}[H]
  \centering
  \caption{A summary of the instrumentation, application and combined measurement uncertainties for each temperature setpoint during the uncoated plate measurements. These values correspond to the respective uncertainty budgets (refer to Table~\ref{tab:laboratoryCombinedUncertainty}) constructed for each measurement. The greatest of the appropriate integration time uncertainties at each setpoint is reported.}
  \vspace*{\floatsep}
  \begin{tabular}{ M{2.5cm}M{1.5cm}M{1.5cm}M{2.5cm}M{1.5cm}M{1.5cm}M{2.5cm} }
    \toprule
	&	\multicolumn{3}{c}{Thermocouple Uncertainty / \SI{}{\celsius}} & \multicolumn{3}{c}{Thermal Imager Uncertainty / \SI{}{\celsius}} \\
		Temperature Setpoint / \SI{}{\celsius} & Instrument & Application & Expanded (\( k= 2\)) & Instrument & Application & Expanded (\(k = 2\)) \\
    \cmidrule(lr){1-7}
	30	&	0.9	&	0.0	&	1.8	&	0.2	&	2.1 	&	4.2		\\
	70	&	0.9	&	0.0	&	1.8	&	0.2	&	8.1 	&	16.2	\\
	110	&	0.9	&	0.0	&	1.8	&	0.3	&	11.1	&	22.1	\\
	130	&	0.9	&	0.0	&	1.8	&	0.3	&	12.4	&	24.7	\\
	150	&	0.9	&	0.0	&	1.8	&	0.3	&	13.6	&	27.3	\\
	170	&	0.9	&	0.1	&	1.8	&	0.3	&	15.1	&	30.1	\\
    \bottomrule
  \end{tabular}
  \label{tab:laboratoryUncertaintyUncoated}
\end{table}

These show that thermocouple measurement is dominated by the standard manufacturer tolerance component; addressing this would require calibrated thermometers with a traceable calibration. The application components contribute the most to the thermal imager temperature measurement, in particular the largest of which are the emissivity and measured spatial non-uniformity. It is unclear whether the ROI uniformity is a result of the thermal imager detector non-uniformity, variation in coating emissivity (due to application method) or a surface temperature variation.

The uncertainty of the emissivity measurement is an assumed value and so does not correspond to a value reported through a calibration certificate. Despite this, the sensitivity function (Eq.~\ref{eq:sensitivityEmissivity}) indicates that the magnitude of this component is primarily a result of the apparent and ambient temperature difference, and non-unity emissivity.

\subsection{Discussion}\label{subsec:artefactDiscussion}

Evaluation of the plate measurements in both coated and uncoated configurations was presented in the preceding sections. The dimensional characterisation results are discussed in detail in the published manuscript \cite{ref:evaluationTestPlate}. Section~\ref{subsec:plateThermalCharacterisation} broadly summarised the contact thermocouple and surface radiance temperatures; an excerpt of the surface temperature distributions is presented. The primary route of heat loss from the system was conduction through the base of the insulator and from convection at the surface was identified. The uncertainty analysis is detailed in Section~\ref{subsec:surfaceTemperatureUncertaintyAnalysis}, the individual components for each system are introduced, an example budget is stated and the combined uncertainties are presented for the different measurement conditions.

Determination of radiance temperature from apparent radiance temperature requires a good knowledge of the surface emissivity for accurate measurements. In a simple case that assumes only one instance of reflection, and that both the object surface and ambient environment possess spectrally independent and Lambertian emissivity, the surface temperature can be solved through Planck's law \cite{ref:radiation_thermometry}.

\begin{figure}[H]
  \centering
  \input{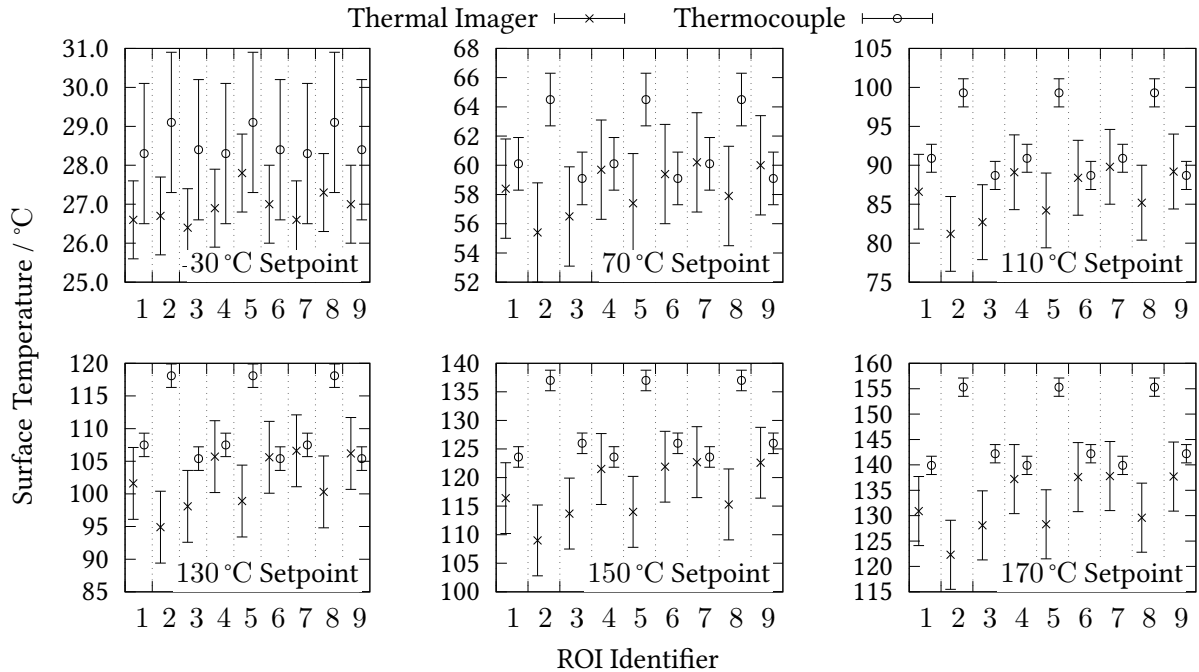}
  \caption{A comparison between the surface temperature determinations from the thermal imager and thermocouples with their equivalent uncertainties (\(k = 2\)) at each temperature setpoint. The ROI identifier corresponds to the nine coated regions as shown in Figure~\ref{fig:simulantCoatedDistribution} numbered in reading order.}
  \label{fig:simulantROICoatedComparison}
\end{figure}

When comparing the temperatures for the coated plate: thermocouple and surface radiance temperatures shown in Figure~\ref{fig:simulantCoatedDistribution} there is poor broad agreement, up to \SI{1.8}{\celsius} at the \SI{30}{\celsius} setpoint and \SI{16}{\celsius} at the \SI{170}{\celsius} setpoint. For these measurements a comparison between the nine ROIs shown in Figure~\ref{fig:simulantUncoatedDistribution} (numbered in reading order), the radiance temperature and the surface temperature evaluated from the thermocouples are presented in Figure~\ref{fig:simulantROICoatedComparison}.

Through the \SI{70}{\celsius}, \SI{110}{\celsius} and \SI{130}{\celsius} temperature setpoints there is a stratification of thermocouple values where each of the three central ROIs measured greater temperatures than radiance temperature measurements. This is caused by the large non-uniformity between thermometer channels (refer to Figure~\ref{fig:simulantPlateImage}) and is not observed by the thermal imager data. This may be a result of an air gap between the plate and conductor plate growing in size due to thermal expansion and heat transfer reducing. Visual inspection of images captured during testing indicate the flexibility of the target plate has resulted in some bowing for the central region of the plate. At \SI{170}{\celsius}, the ROIs with the poorest agreement to contact thermocouples are those along the top row and the central column.

Additional sources of disparity between thermocouple and surface radiance temperature measurement may be from variation in contact pressure between the plate and conductor plate due to inconsistent clamp force. A comparison between the mean thermocouple temperature measured at identical heater temperature setpoints during the coated (Figure~\ref{fig:simulantCoatedDistribution}) and uncoated (Figure~\ref{fig:simulantUncoatedDistribution}) measurements demonstrate less than \SI{1}{\celsius} difference (\SI{2}{\celsius} at the \SI{110}{\celsius} setpoint) and so this is considered to be negligible in this case with respect to the thermocouple measurements.

\begin{figure}[H]
  \centering
  \input{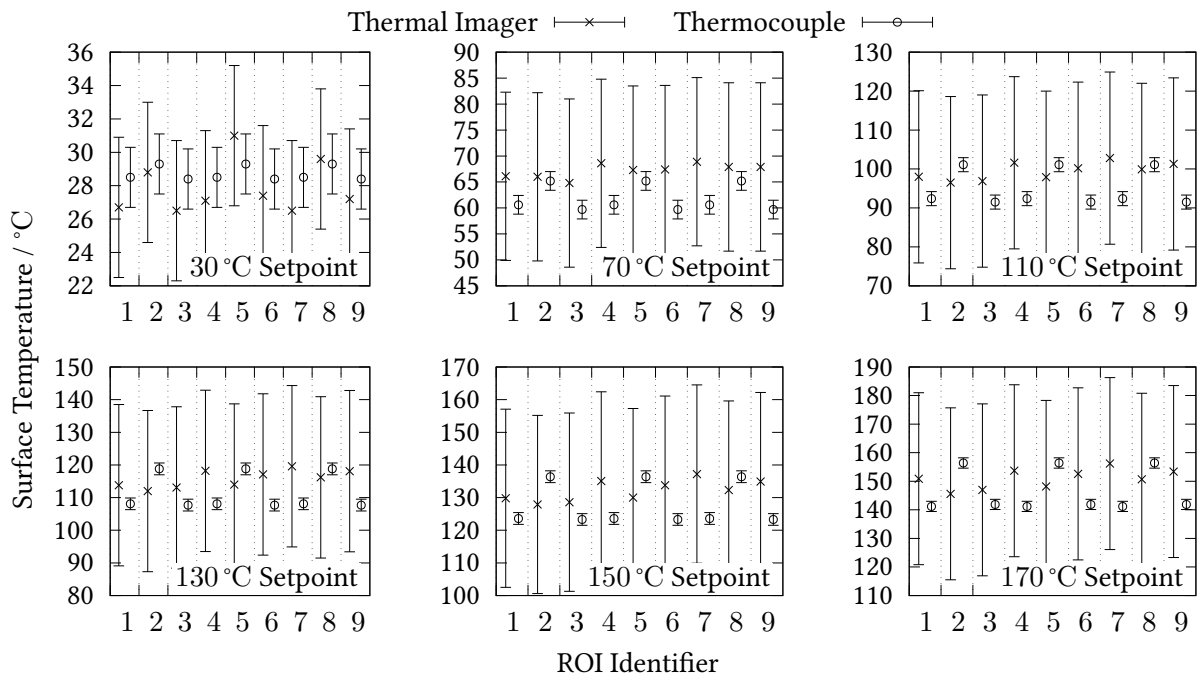}
  \caption{A comparison between the surface temperature determinations from the thermal imager and thermocouples with their equivalent uncertainties (\(k = 2\)) at each temperature setpoint. The ROIs were observing the uncoated regions at identical positions to the coated regions.}
  \label{fig:simulantROIUncoatedComparison}
\end{figure}

For the uncoated plate measurements a comparison between the nine ROIs shown in Figure~\ref{fig:simulantCoatedDistribution}, the radiance temperature and an approximation of the surface temperature evaluated from the thermocouples is presented in Figure~\ref{fig:simulantROIUncoatedComparison}.

Due to the increased sensitivity for the emissivity components for the thermal imager temperature determination, the error bars are the dominant feature in this comparison. Compared to the coated measurements, there is better agreement between the methods up to \SI{110}{\celsius}. This suggests that the assumed emissivity value for stainless steel is more appropriate than that used for the Senotherm coating. It is clear from these results the importance for low uncertainty emissivity measurement when using radiance temperature techniques, which is especially critical when measuring reflective surfaces.

\begin{figure}[h]
  \centering
  \input{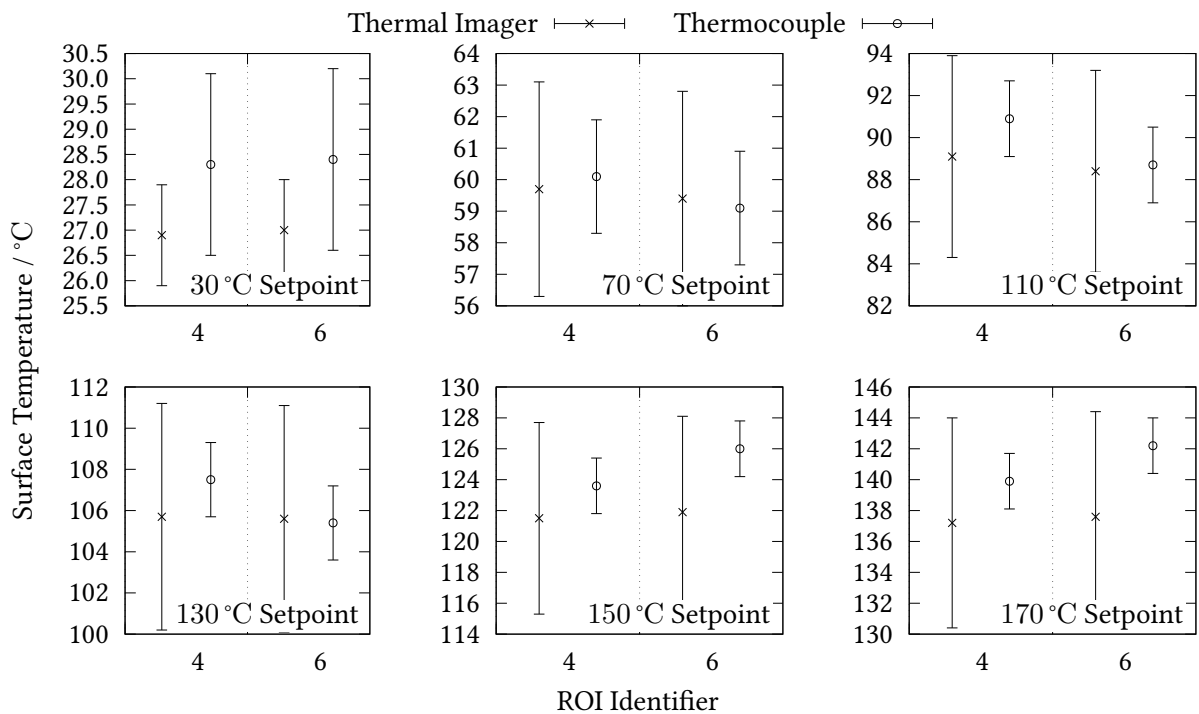}
  \caption{A reduced comparison between the surface temperature determinations from the thermal imager and thermocouples at each temperature setpoint. This data is from the coated region plate measurements shown in Figure~\ref{fig:simulantROICoatedComparison}.}
  \label{fig:simulantROICoatedComparisonReduced}
\end{figure}

\begin{figure}[H]
  \centering
  \input{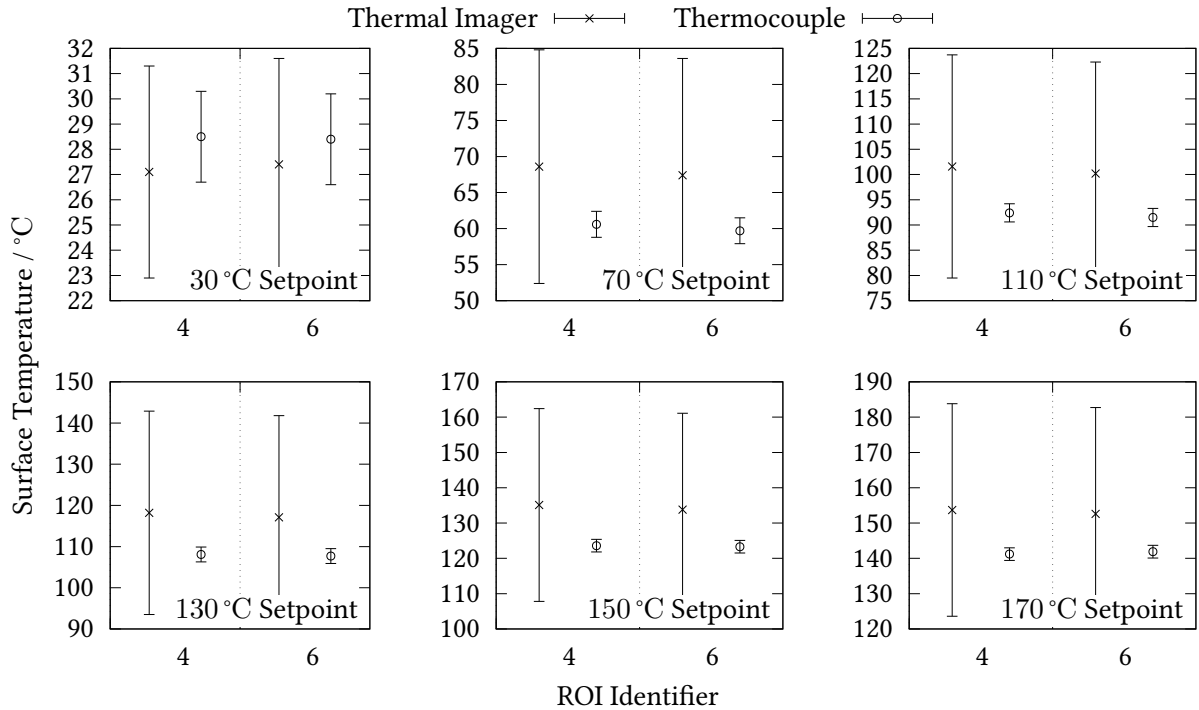}
  \caption{A reduced comparison between the surface temperature determinations from the thermal imager and thermocouples at each temperature setpoint. This data is from the uncoated region plate measurements shown in Figure~\ref{fig:simulantROIUncoatedComparison}.}
  \label{fig:simulantROIUncoatedComparisonReduced}
\end{figure}

\clearpage
A deeper analysis into the correspondence between thermocouple and radiance temperature can demonstrate a clear conclusion to the plate measurements. As discussed previously, it is understood that as the plate distorts and due to the fixed anchor points, the plate may have flexed in the centre. Investigation into this distortion suggests this may have been up to \SI{0.1}{\milli\metre} at the highest temperatures. A revised design was not implemented due to the project time constraints. For a more robust comparison between thermometry techniques, this leads to an omission of the central column of regions of interest (ROI 2, ROI 5 and ROI 8) for the analysis. Additionally the thermocouples were spatially located (as shown in Figure~\ref{fig:simulantPlateImage}) along the centre of the channels and so only those radiance temperature ROIs were considered: i.e. ROI 4 and ROI 6. A reduced comparison of the data in Figure~\ref{fig:simulantROICoatedComparison} and Figure~\ref{fig:simulantROIUncoatedComparison} is shown in Figure~\ref{fig:simulantROICoatedComparisonReduced} and Figure~\ref{fig:simulantROIUncoatedComparisonReduced}.

Here it is shown for the coated measurements that there is good agreement between the thermocouple and thermal imager at all temperatures within the prescribed uncertainty. Aside from the \SI{70}{\celsius} setpoint, the data suggests the radiance temperature is too low; therefore the emissivity correction applied is not sufficient and a lower emissivity of 0.80 may be more suitable. For the uncoated regions the thermocouple and radiance measurements are in agreement at all temperature setpoints. In this case the radiance temperature is typically greater than the contact measurement; therefore a greater emissivity of 0.25 may be more appropriate. Both of these reduced dataset comparisons support a moderate reduction in the measurement uncertainty of each instrument, assuming the agreement with a thermal model can be demonstrated.

A tabular representation of the difference between radiance and thermocouple temperature is presented in Table~\ref{tab:amendedEmissivityCoatedResults} and Table~\ref{tab:amendedEmissivityUncoatedResults} for the coated and uncoated measurements respectively. The corresponding differences for respective emissivity values of 0.80 and 0.25 are also presented.

\begin{table}[H]
	\centering 
	\caption{Comparison between the temperature difference between the thermocouple and radiance temperature during the coated measurements. The original emissivity corresponds to 0.85 and this data is shown in Figure~\ref{fig:simulantROICoatedComparisonReduced}, and the revised emissivity is 0.80.}
	\begin{tabular}{ M{3.0cm} M{2.5cm} M{2.5cm} M{2.5cm} M{2.5cm} }
	\toprule
	\multirow{2}{3cm}{Temperature Setpoint / \SI{}{\celsius}}	&	\multicolumn{2}{c}{Original Temperature Error / \SI{}{\celsius}}	&	\multicolumn{2}{c}{Revised Temperature Error / \SI{}{\celsius}} \\
	&	ROI 4	&	ROI 6	&	ROI 4 &		ROI 6	\\
	\cmidrule(lr){1-5}
	30	&	1.5	&	 1.4	&	 1.2	&	 1.0	\\
	70	&	0.4	&	-0.3	&	-1.0	&	-1.7	\\
	110	&	1.8	&	 0.3	&	-0.1	&	-1.6	\\
	130	&	1.8	&	-0.2	&	-0.4	&	-2.4	\\
	150	&	2.2	&	 4.1	&	-0.3	&	 1.7	\\
	170	&	2.9	&	 4.7	&	 0.2	&	 2.0	\\
	\cmidrule(lr){1-5}
	Mean Error	&	 1.8	&	 1.7	&	-0.1	&	-0.2	\\
	\bottomrule
	\end{tabular} 
	\label{tab:amendedEmissivityCoatedResults}
\end{table}

\begin{table}[H]
	\centering 
	\caption{Comparison between the temperature difference between the thermocouple and radiance temperature during the uncoated measurements. The original emissivity corresponds to 0.20 and this data is shown in Figure~\ref{fig:simulantROIUncoatedComparisonReduced}, and the revised emissivity is 0.25.}
	\begin{tabular}{ M{3.0cm} M{2.5cm} M{2.5cm} M{2.5cm} M{2.5cm} }
	\toprule
	\multirow{2}{3cm}{Temperature Setpoint / \SI{}{\celsius}}	&	\multicolumn{2}{c}{Original Temperature Error / \SI{}{\celsius}}	&	\multicolumn{2}{c}{Revised Temperature Error / \SI{}{\celsius}} \\
	&	ROI 4	&	ROI 6	&	ROI 4 &		ROI 6	\\
	\cmidrule(lr){1-5}
	30	&	 1.4	&	 1.1	&	 2.4	&	 2.1	\\
	70	&	-8.0	&	-7.7	&	-2.5	&	-2.2	\\
	110	&	-9.2	&	-8.7	&	-1.6	&	-1.1	\\
	130	&	-10.1	&	-9.4	&	-1.6	&	-1.0	\\
	150	&	-11.5	&	-10.5	&	-2.1	&	-1.1	\\
	170	&	-12.5	&	-10.8	&	-2.2	&	-0.4	\\
	\cmidrule(lr){1-5}
	Mean Error	&	-8.3	&	-7.7	&	-1.2	&	-0.6	\\
	\bottomrule
	\end{tabular} 
	\label{tab:amendedEmissivityUncoatedResults}
\end{table}

\begin{figure}[H]
  \centering
  \input{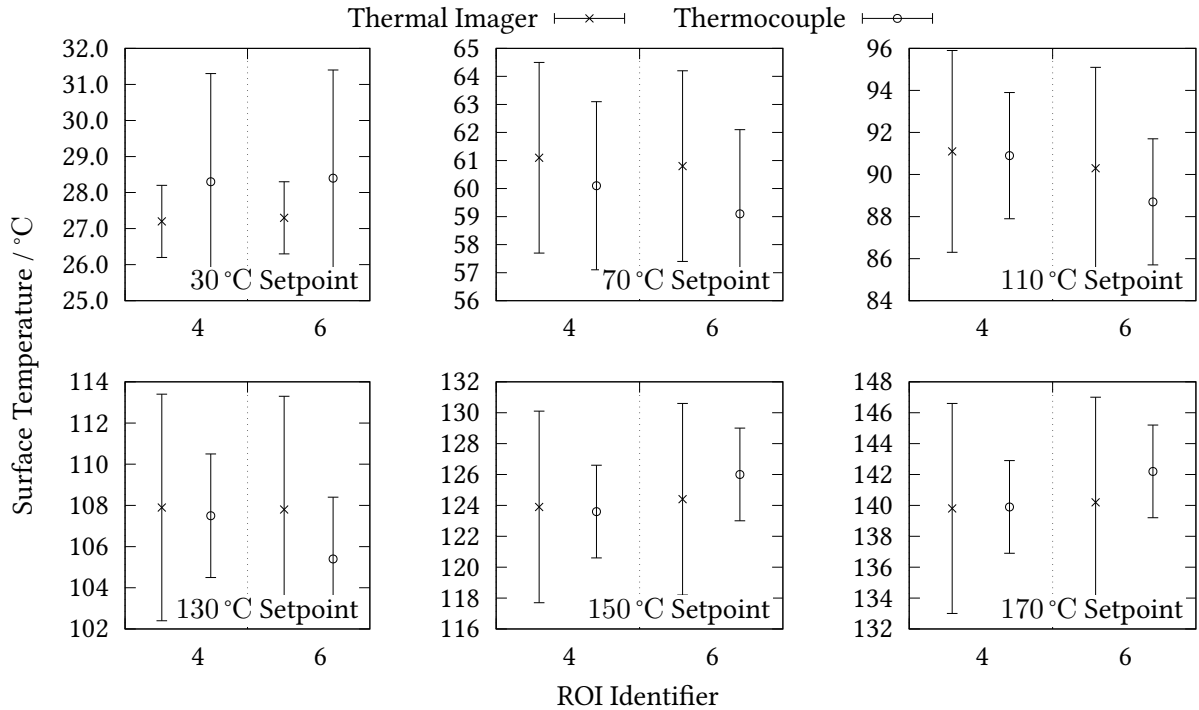}
  \caption{Following the revised emissivity for the coated region measurements in Table~\ref{tab:amendedEmissivityCoatedResults}, a comparison in the same form as Figure~\ref{fig:simulantROICoatedComparisonReduced} is presented.}
  \label{fig:simulantROICoatedComparisonReducedOptimised}
\end{figure}

These tables show that currently the largest offset for the surface temperature measurement is for the \SI{-12.5}{\celsius} at \SI{170}{\celsius} during the uncoated measurements. Following an evaluation of the applied emissivity correction, the greatest offset is reduced to \SI{-2.5}{\celsius} at \SI{70}{\celsius} during the uncoated measurements. Furthermore the mean error is greatly reduced for both regions of interest during both coated and uncoated measurements. To determine this error, the greatest magnitude of error at either setpoint was recorded in order to identify an appropriate emissivity value. A visualisation of this emissivity revision is shown in Figure~\ref{fig:simulantROICoatedComparisonReducedOptimised} where the same uncertainties are presented alongside revised surface temperature calculations for the thermal imager.

\begin{figure}[H]
  \centering
  \input{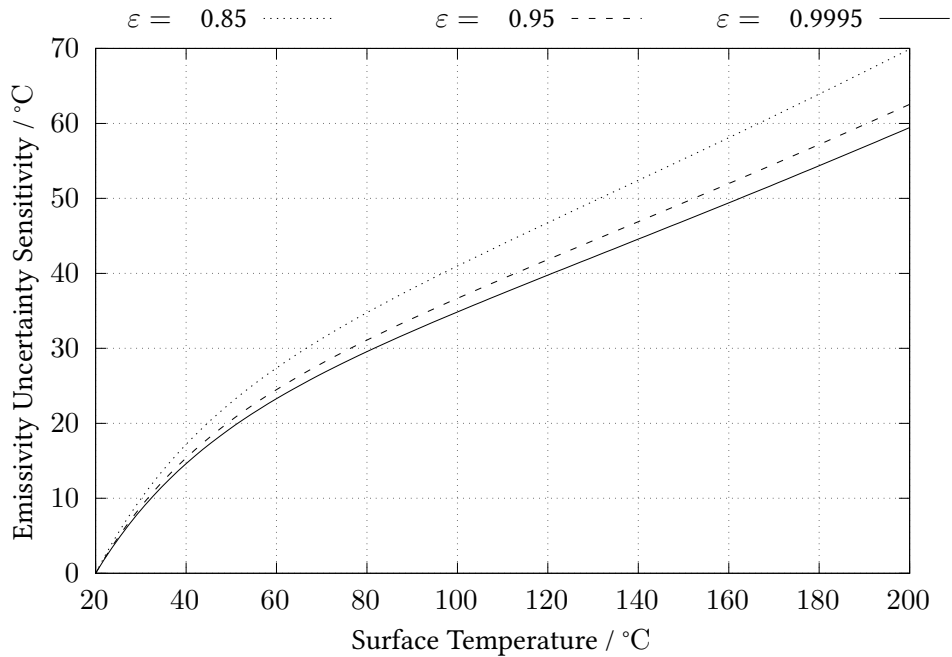}
  \caption{Functional form of the emissivity sensitivity (Eq.~\ref{eq:sensitivityEmissivity}) for increasing temperature at a number of emissivities. Note that this sensitivity is largely dominated by the difference between apparent and ambient temperatures and that the emissivity itself has a smaller effect.}
  \label{fig:sensitivityEmissivity}
\end{figure}

The uncertainty budgets for the surface temperature determination of high emissivity regions detail the measurement uncertainty reaching \SI{6.8}{\celsius} (\(k = 2\)) at \SI{170}{\celsius}. The intrinsic thermal imager uncertainty is nominally consistent at \SI{0.3}{\celsius} (\(k = 2\)) across this temperature range and as presented in Table~\ref{tab:laboratoryCombinedUncertainty}, the greatest source of uncertainty for temperature determination is from the emissivity measurement. The estimation for emissivity uncertainty of approximately 0.05 is close to national measurement institute grade measurement (as good as 0.01 (\(k = 1\)) depending on material) and so this particular component would reduce to a standard uncertainty of \SI{0.5}{\celsius}. Reduction of this component further requires reducing the sensitivity component (Eq.~\ref{eq:sensitivityEmissivity}) which is dominated by the disparity between apparent and ambient temperatures as opposed to the emissivity itself (refer to Figure~\ref{fig:sensitivityEmissivity}).

It should be noted that these equations are dependent on the spectral response of the instrument in use. Throughout these calculations a value of \SI{3.85}{\micro\metre} has been used, but if a long-wave infrared instrument would be considered then that would result in greater sensitivity coefficients throughout.

From these observations it is shown that the measurement of surface temperature is a challenging task. It is affected by both differing materials and varying surface properties, applying this to an application will require careful consideration to each of these aspects.

\subsection{Assessment summary}\label{subsec:assessmentSummary}

An aspect of surface (and by inference, internal) temperature measurement using an external thermal imager was evaluated. The challenges introduced by emissivity on radiance temperature measurement were explained in detail, with the largest contributor to uncertainty being the absence of traceable emissivity measurement. Verification of the surface radiance measurements against the contact thermocouple temperatures indicated that there exist discrepancies above \SI{70}{\celsius} during the coated region measurements and above \SI{110}{\celsius} during the uncoated region measurements. These may be a result of the value of emissivity used throughout the calculations or varying contact pressure across the plate. The systematic offset observed during the coated measurements suggest the emissivity value may be inappropriate, and this was supported by the different emissivity results presented in Table~\ref{tab:amendedEmissivityCoatedResults} and Table~\ref{tab:amendedEmissivityUncoatedResults}. However the poor agreement between two experimental temperature measurements would be supported if the presence of a thermal gradient from thermometer channel to radiating surface was present. An estimation of the surface emissivity was demonstrated and the recalculation of temperature error between thermocouple and radiance temperature, using this emissivity, enabled a reduction of error to \SI{1.2}{\celsius}.

\clearpage
\newpage

\section{Radiance correction methods}\label{sec:radianceCorrection}
For radiance temperature measurement methods (e.g.\ radiation thermometers or thermal imagers) to be deployed outside of a controlled laboratory environment with approximated blackbody sources, the effect from emissivity on radiance temperature must be considered. A selection of radiance correction methods will be introduced and explored; specifically they will be compared with respect to an existing thermal environment for the SNM containers. This study should be considered independent from the laboratory assessment in Section~\ref{sec:assemblyAssessment}, although there are conclusions to be made with implications there and in the following chapter. Following the correction method comparison, environments where the respective correction methods incur sufficiently low uncertainty contributions will be presented.

Within the literature, the principles of radiance summation are detailed in \cite{ref:radiation_thermometry} and from this any series of radiance contributions to an observation could be described. Summation of radiance contributions that are truncated to that emitted from the surface and that reflected from the surface is typical and often encountered \cite{ref:influenceBlackbodyEmissivityCorrection,ref:radiometricTechniquesEmissivityTemperature}. This presupposes that the radiance observed from the surroundings originates from a blackbody source and not another object with a non-unity emissivity; whilst this assumption can be made for many environments this should be stated when used. Considering multiple reflections as proposed in both \cite{ref:overviewRadiationThermometryChapter2,ref:accuracyMethodsEmissivity} provide robust methodologies for considering all radiance contribution sources.

Consideration of the geometry within a thermal environment is considered in \cite{ref:reflectionErrorReformer,ref:radiationThermometryPetrochemical} where the view factor from an object to a target surface is considered to ascertain its radiance contribution component in the emissivity correction.

Alternative emissivity compensation techniques include: dual-wavelength pyrometers, Fourier-transform infrared spectrometry or polarisation pyrometry \cite{ref:overviewRadiationThermometryChapter1}.

To evaluate the radiance correction methods the corrected radiance temperature (representative of surface temperature) will be compared to the surface temperature. For each of the correction methods under investigation ambient environment and apparent radiance temperatures will be provided. A representative spectral range for instrument response will be defined and a set of traceable directional spectral emissivity data provided. To calculate this apparent radiance temperature a known representative value of a complex scenario must be known; this value may either be measured in a well defined environment or simulated computationally. Under this comparison the baseline that each correction method will be compared against will be the ray tracing method introduced in Section~\ref{subsec:rayTracing}, this method will be used to generate the apparent radiance temperature interpreted by each correction method when calculating surface temperature.

\subsection{Environment setup}\label{subsec:environmentSetup}
As introduced in Section~\ref{sec:inSituIntroduction}, the thermal system being explored in this chapter comprises the properties outlined in Table~\ref{tab:environmentProperties}. The case study will extend beyond the shrouded blackbody background by introducing a container reflection.

\begin{table}[ht]
		\centering 
		\caption{Constraints to the thermal system under investigation here are detailed, this includes the: instrument spectral range; surface and ambient temperature ranges; and the surface emissivity. The reflection container temperature is the initial reflection component in position one.}
		\begin{tabular}{ M{5.0cm} M{5.0cm} }
		\toprule
		Characteristic & Description \\
		\cmidrule(lr){1-2}
		Instrument spectral range & \SIrange{7.5}{14.0}{\micro\metre} \\
		Surface temperature range & \SIrange{10}{200}{\celsius} \\
		Ambient temperature range & \SIrange{10}{60}{\celsius} \\
		Reflection container temperature & \SI{50}{\celsius} \\
		Target container hemispherical total emissivity & \num{0.06} to \num{0.94} \\
		\bottomrule
		\end{tabular} 
		\label{tab:environmentProperties}
\end{table}

Specifically the environment will comprise a pair of SNM containers within an enclosure being observed from two positions, one where the reflection component will include the second container and a second position where the reflection component observes the enclosure wall. This enclosure arrangement is shown in Figure~\ref{fig:radianceCorrectionEnvironment}, these images are rendered within \texttt{pbrt-v3} (unmodified) \cite{ref:pbrt_v3} and represent the relative radiance observed from each of the two observation positions. Each of the parameters detailed in Table~\ref{tab:environmentProperties} will be explored through a parametric study.

\begin{figure}[H]
\centering
	\begin{subfigure}[t]{0.63\textwidth}
	\centering
	\includegraphics[width=\textwidth,keepaspectratio]{./Images/radianceComparisonSchematicHalfSpherePositionOne.pdf}
	\caption{Position one.}
	\label{fig:radianceCorrectionEnvironmentPositionOne}
    \end{subfigure}
	\vspace{\floatsep}
	\begin{subfigure}[t]{0.63\textwidth}
	\centering
	\includegraphics[width=\textwidth,keepaspectratio]{./Images/radianceComparisonSchematicHalfSpherePositionTwo.pdf}
	\caption{Position two.}
	\label{fig:radianceCorrectionEnvironmentPositionTwo}
    \end{subfigure}
	\caption{Construction of the radiance correction comparison environment. The two containers are displayed and the two observation positions are either side of the lower container. At position one the reflection container is in the specular direction, at position two the walls are in the specular direction.}
  \label{fig:radianceCorrectionEnvironment}
\end{figure}

Figure~\ref{fig:radianceCorrectionEnvironment} describes the geometry of the test environment to be considered. A pair of containers were positioned within a spherical environment at the ambient temperature, the surface temperature of the observation container (the lower container in \ref{fig:radianceCorrectionEnvironment}) was set to the surface temperature variable. Two positions were evaluated, position one will arrange the reflection container (the higher container in \ref{fig:radianceCorrectionEnvironment}) in the specular direction from the observation surface position two arranges the walls in the specular direction from the observation surface. Between test cases, the emissivity of the entire observation container was described by the surface emissivity parameter. Throughout all tests the reflection container was maintained at a temperature of \SI{50}{\celsius} and demonstrate the medium emissivity material. The walls were maintained at the ambient temperature and will demonstrate an emissivity of a blackbody.

\begin{figure}[H]
  \centering
  \input{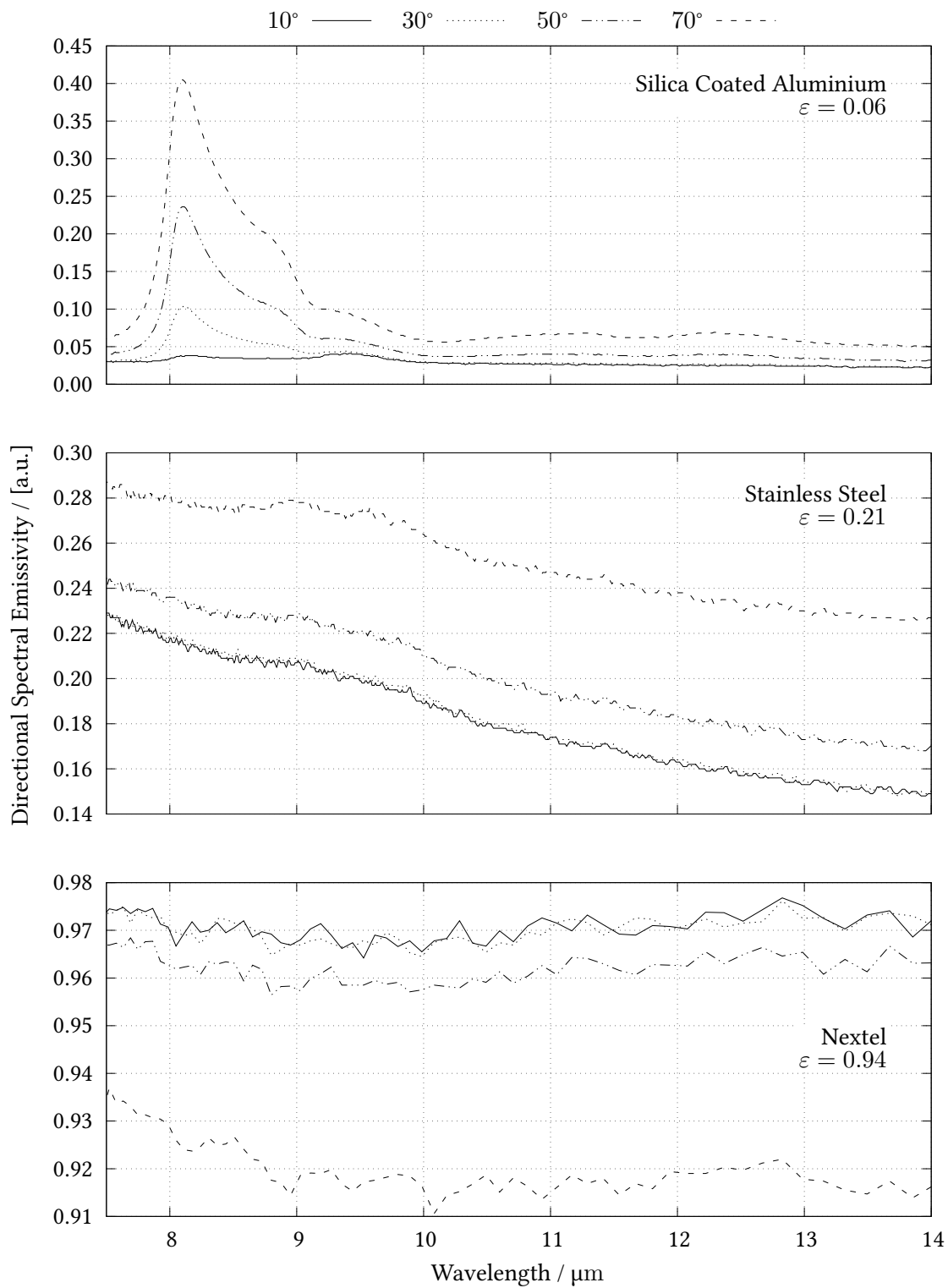}
  \caption{Measured directional spectral emissivity data used to describe the surfaces in the radiance correction comparison environment. Each angle corresponds to the zenith angle the emissivity was measured at, where \SI{0}{\degree} denotes the surface normal. The indicated value in each plot is the hemispherical total emissivity used during one of the correction methods.}
  \label{fig:emissivityMeasurementData}
\end{figure}

The three emissivity materials used -- low, medium and high -- correspond to silica coated aluminium, stainless steel and a Nextel coating respectively. Measurements for these materials are shown in Figure~\ref{fig:emissivityMeasurementData}, low and medium materials were acquired through commercial projects (and permitted for dissemination) \cite{ref:3d_thermal_imaging,ref:esaTTITN6,ref:nplPBRT} and the high emissivity was reported by PTB \cite{ref:emissivity_data_coatings}. The temperatures these were evaluated at were \SI{100}{\celsius}, \SI{15}{\celsius} and \SI{120}{\celsius}, respectively.

\clearpage

Silica coated aluminium is largely spectrally and angularly independent aside from the region at \SI{8}{\micro\metre}. The reflective region is principally a result of the aluminium interaction with radiation as detailed in literature \cite{ref:theory_off_specular_reflection_roughened_surfaces}. Reflectivity of steel decreases as wavelength increases and becomes more reflective at greater angles of incidence; the Nextel is spectrally and angularly independent within its measurement uncertainty.

\subsection{Uncorrected radiance temperature}\label{subsec:uncorrectedRadiance}
The scale of the realised error from omitting a correction for emissivity depends entirely on the disparity between surface and ambient temperature and the emissivity itself. In the event of surface temperature and ambient temperature reaching equilibrium then regardless of the emissivity, apparent radiance temperature will represent the surface temperature. Should there be a difference between surface temperature and ambient temperature then the emissivity will have a direct impact on the magnitude of incurred error. For this scenario the radiance temperature \(T_{b}^{unc}\) is given by Eq.~\ref{eq:radianceCorrectionUncorrected},

\begin{equation}
  T_b^{unc} = T_{app} \mathrm{.}
  \label{eq:radianceCorrectionUncorrected}
\end{equation}

\subsection{Multiple reflection single wavelength}\label{subsec:multipleReflectionSingleWavelength}
For the Multiple Reflection Single Wavelength (MRSW) correction method, in order to correct radiance temperature first the apparent spectral radiance must be considered with respect to the local environment. This derivation follows on from that in Section~\ref{subsec:surfaceTemperatureDetermination} and extends it to an additional reflection instance. Radiance from a surface \(L_{\lambda, b}\) is defined by the Planck distribution law as detailed in Section~\ref{subsec:radiation_plancks_law} as

\begin{equation}
  L_{\lambda , b}(\lambda, T_b) = \frac{c_1}{\lambda^5} \frac{1}{e^{\nicefrac{c_2}{\lambda T_b}}-1} \mathrm{.}
		\label{eq:reduced_planck_distribution_law_repeat}
\end{equation}

Using the Kirchhoff law, an equivalence between the apparent radiance \(L_{app}\), target container emissivity, target container radiance \(L_{\lambda, b}^{target}\), reflection container emissivity, reflection container radiance \(L_{\lambda, b}^{reflection}\) and ambient radiance \(L_{\lambda, b}^{wall}\) under the assumption of two surface reflections is given by

\begin{equation}
  L_{app} = \varepsilon^{target} L_{\lambda,b}^{target} + \left[ 1 - \varepsilon^{target} \right] \left[ \varepsilon^{reflection} L_{\lambda,b}^{reflection} + \left( 1 - \varepsilon^{reflection} \right) L_{\lambda, b}^{wall} \right] \mathrm{.}
		\label{eq:apparentRadianceSR}
\end{equation}

This approximation assumes that the reflections are all in the specular direction and there are no diffuse contributions. Solving Eq.~\ref{eq:apparentRadianceSR} for the surface temperature under this approximation gives

\begin{equation}
  T_b^{mrsw} = \frac{c_2}{\lambda } \ln\left[ \varepsilon^{target} \left[ \frac{1}{e^{\nicefrac{c_2}{\lambda T_{app}}}-1} - \frac{\left( 1 - \varepsilon^{target} \right) }{e^{\nicefrac{c_2}{\lambda T_{reflection}}}-1} \right]^{-1} + 1 \right]^{-1} \mathrm{.}
		\label{eq:radianceCorrectionMRSW}
\end{equation}

Equation~\ref{eq:radianceCorrectionMRSW} is valid for a single reflection instance where the reflection container emissivity equals one. When an additional reflection is considered such as the case in this radiance comparison the reflection container temperature can be described by

\begin{equation}
  T_{reflection} = \frac{c_2}{\lambda} \left[ \ln \left( 1 + \left( \frac{\varepsilon^{reflection}}{e^{\nicefrac{c_2}{\lambda T_{\lambda, b}^{\prime reflection}}} - 1} + \frac{1- \varepsilon^{reflection}}{e^{\nicefrac{c_2}{\lambda T_{\lambda, b}^{wall}}} - 1} \right)^{-1} \right) \right]^{-1} \mathrm{.}
  \label{eq:radianceCorrectionMRSWAdditionalReflection}
\end{equation}

Where \(T_{\lambda, b}^{\prime reflection}\) is the reflection container temperature under this additional reflection condition. This correction can be used to determine the surface temperature using the measured temperatures, the estimated spectral mid-point and estimated emissivity of surfaces under inspection. It should be noted that usage of this approximation does not consider: potential impact from further reflections, how to assess the spectral dependency (integrate over the specific spectral range of the system and optics used), the type of emissivity value used or a transmission greater than 0. In particular, a single hemispherical total emissivity value for a surface may be used, however the effect from directional and spectral emissivity dependence is not accounted for.

\subsection{Multiple reflection all wavelengths}\label{subsec:multipleReflectionAllWavelengths}
To develop from Section~\ref{subsec:multipleReflectionSingleWavelength} for the Multiple Reflection All Wavelengths (MRAW) method, the spectral response of the system can be introduced in the approximation by integrating the spectral radiance over \(\mathrm{d}\lambda\) to determine the radiance. This will in turn require an extension beyond the hemispherical total emissivity to the hemispherical spectral emissivity. Under this approximation the spectral response of the imager, \(r(\lambda)\), can be introduced if known,

\begin{equation}
  L_b(T) = \int\limits_{\lambda} L_{\lambda,b}(\lambda,T) \cdot r(\lambda) \mathrm{d}\lambda \mathrm{.}
  \label{eq:radianceCorrectionMRAW}
\end{equation}

Using the same reflection assumption from Eq.~\ref{eq:apparentRadianceSR}, the surface temperature using this approximation, \(T_b^{mraw}\), can be computed. Contrary to Eq.~\ref{eq:radianceCorrectionMRSW}, an analytical solution cannot be solved and instead this must be approached numerically. Further to this the sensitivity coefficients can only be understood through computation of their magnitude under different environments.

The calculations for this method were made using software provided by Michael Hayes, National Physical Laboratory. The software approximates the calculations shown in Eq.~\ref{eq:radianceCorrectionMRAW} using the stated input parameters \cite{ref:esaTTITN6}.

\subsection{Ray tracing}\label{subsec:rayTracing}
An alternative to discrete descriptions of the system under the approximations given above is to evaluate the complete system using ray tracing. One of the core principles of ray tracing methods is to be physically based, which means that the units and algorithms used are based on and constrained by the theory of electromagnetism. Ray tracing is based on geomtrical optics rather than wave optics and so does not consider interference or polarisation effects. For the correction methods presented above they will each be compared using the measurements simulated with this technique.

Reflectivity measurements have been used by graphical rendering groups within the fields of video games and cinematography; however it has broader applications for metrology in land and sea surface temperature measurements as well as in the visualisation of painted materials, advertisements and product descriptions. The lack of attention to physically-based models has limited its use within research areas where robust physical descriptions are critical, such as in architecture or the lighting industry.

A measured reflectivity databases (such as the Mitsubishi Electric Research Laboratories database \cite{ref:data_driven_reflectance_model} can be processed within Monte Carlo rendering engines such as: Hyperion by Disney \cite{ref:sorted_deferred_shading}, Arnold by Sony \cite{ref:sony_pictures_arnold}, the RTX platform by NVIDIA, Filament by Google, \texttt{pbrt} originating from Stanford university \cite{ref:pbrt_v3} and the Mitsuba renderer by Jakob \cite{ref:comprehensive_framework_layered_materials}. It is through the use of these rendering systems that the complete architecture from radiation emission, to surface behaviour and radiation detection, can be evaluated and an image generated.

The radiation transport equation employed by these rendering systems, which describes the exitant radiance \(L_{\lambda,o}(p;\theta_r,\varphi_r)\) incident on an image pixel from point \(p\) on a surface in the scene is shown in Eq.~\ref{eq:pbrtLightTransport} \cite{ref:pbrt_v3},

\begin{equation}
  L_{\lambda,o}(p;\theta_r,\varphi_r) = L_{\lambda,em}(p;\theta_r,\varphi_r) + \int\limits_{0}^{2\pi} \int\limits_{0}^{\nicefrac{\pi}{2}} \rho(p;\lambda;\theta_i,\varphi_i;\theta_r,\varphi_r;T) L_{\lambda,i}(p;\theta_i,\varphi_i) \lvert \cos\theta_i \rvert \mathrm{d}\theta_i \mathrm{d}\varphi_i \mathrm{.}
\label{eq:pbrtLightTransport}
\end{equation}

Where \(L_{\lambda,em}(p;\theta_r,\varphi_r)\) is the emitted radiance; \(\rho(p;\lambda;\theta_i,\varphi_i;\theta_r,\varphi_r;T)\) is the Bi-directional Reflectivity Distribution Function (BRDF) at point \(p\); and \(L_{\lambda,i}(p;\theta_i,\varphi_i)\) is the incident radiance on the surface. This BRDF is a principle component in the complete description of a synthesised scene.

Using this ray tracing method alongside both known geometry of the system including all sources of infrared radiation and their respective directional spectral emissivities, an accurate estimation of the apparent radiance temperature can be realised. The silica coated aluminium and stainless steel materials were defined to be entirely specular, whilst the Nextel coating was diffuse. It should be noted that whilst this method accounts for the behaviour of multiple reflections, it is not a fully realised thermal model that simulates the three heat transfer modes (i.e. increased radiative heat transfer from container to container at higher surface emissivities).

Apparent radiance temperatures calculated using this method was parsed by the two respective correction methods in Section~\ref{subsec:multipleReflectionSingleWavelength} and Section~\ref{subsec:multipleReflectionAllWavelengths}. The calculations for this method were made using software provided by Michael Hayes, National Physical Laboratory. This software is based on a heavily modified version of \texttt{pbrt-v2} that appropriately handles the Planck distribution radiometrically, input from directional spectral emissivity data and atmospheric absorption \cite{ref:esaTTITN6}.

\subsection{Environment correction comparison}\label{subsec:environmentComparison}

As described in Section~\ref{subsec:environmentSetup}, during this radiance correction comparison, the parameter space to be explored is defined by Table~\ref{tab:environmentProperties}. Within this space, for the two observation positions described in Figure~\ref{fig:radianceCorrectionEnvironment} and for the three observation container surface emissivities, the temperature difference between corrected and defined surface temperature will be presented through an intensity plot; this difference is described by Eq.~\ref{eq:radianceCorrectionComparison} where \(T_{b}^\prime\) is the corrected surface temperature for the respective methods (e.g. \(T_b^{unc}\), \(T_b^{mrsw}\) or \(T_b^{mraw}\)),

\begin{equation}
  \Delta T = T_{b}^\prime - T_b \mathrm{.}
  \label{eq:radianceCorrectionComparison}
\end{equation}

\begin{figure}[H]
  \centering
  \input{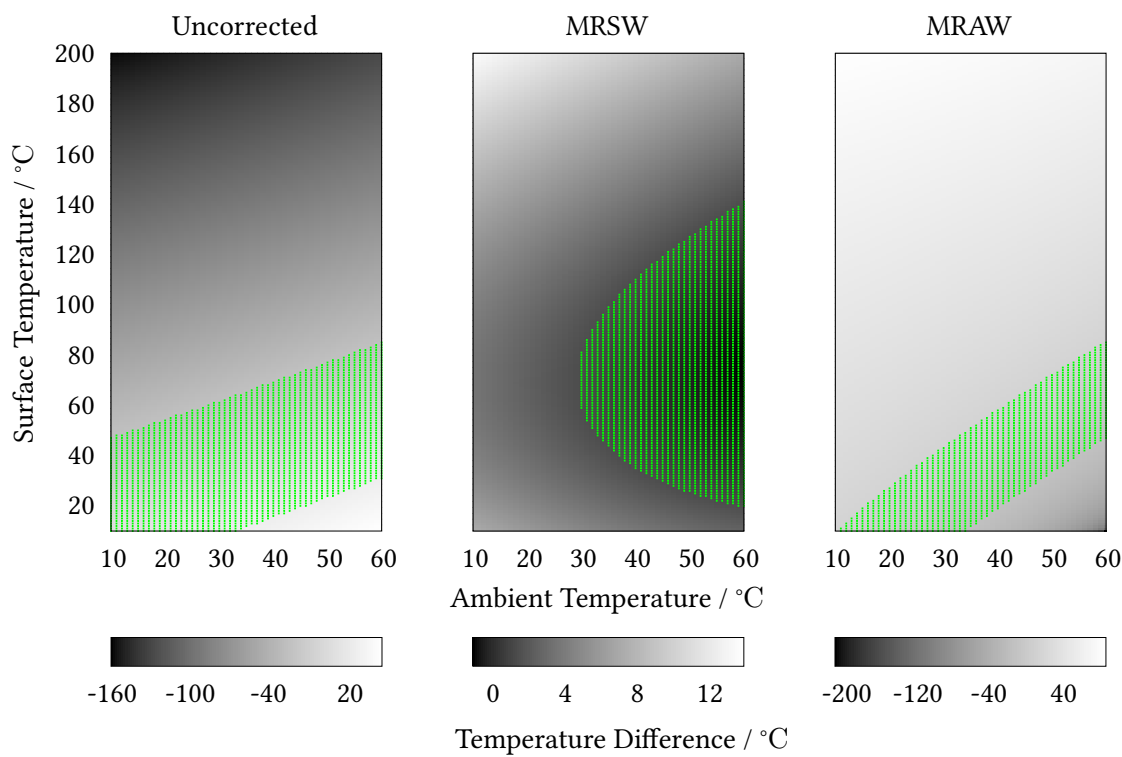}
  \caption{Position one, emissivity 0.06. The standard deviation of the datasets indicating the highlighted ranges are: \SI{51.4}{\celsius}, \SI{3.1}{\celsius} and \SI{30.1}{\celsius}.}
  \label{fig:radianceCorrectionComparisonPositionOneEmissivityLow}
\end{figure}

\begin{figure}[H]
  \centering
  \input{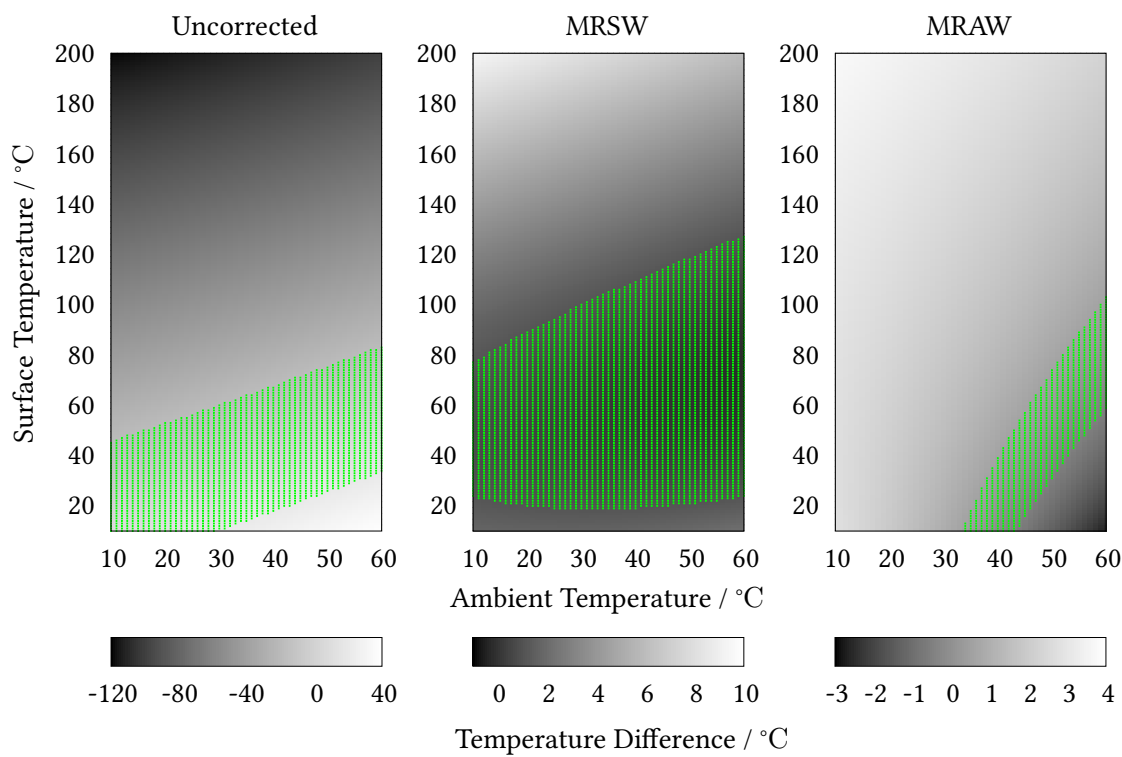}
  \caption{Position one, emissivity 0.21. The standard deviation of the datasets indicating the highlighted ranges are: \SI{39.9}{\celsius}, \SI{2.3}{\celsius} and \SI{1.2}{\celsius}.}
  \label{fig:radianceCorrectionComparisonPositionOneEmissivityMedium}
\end{figure}

\begin{figure}[H]
  \centering
  \input{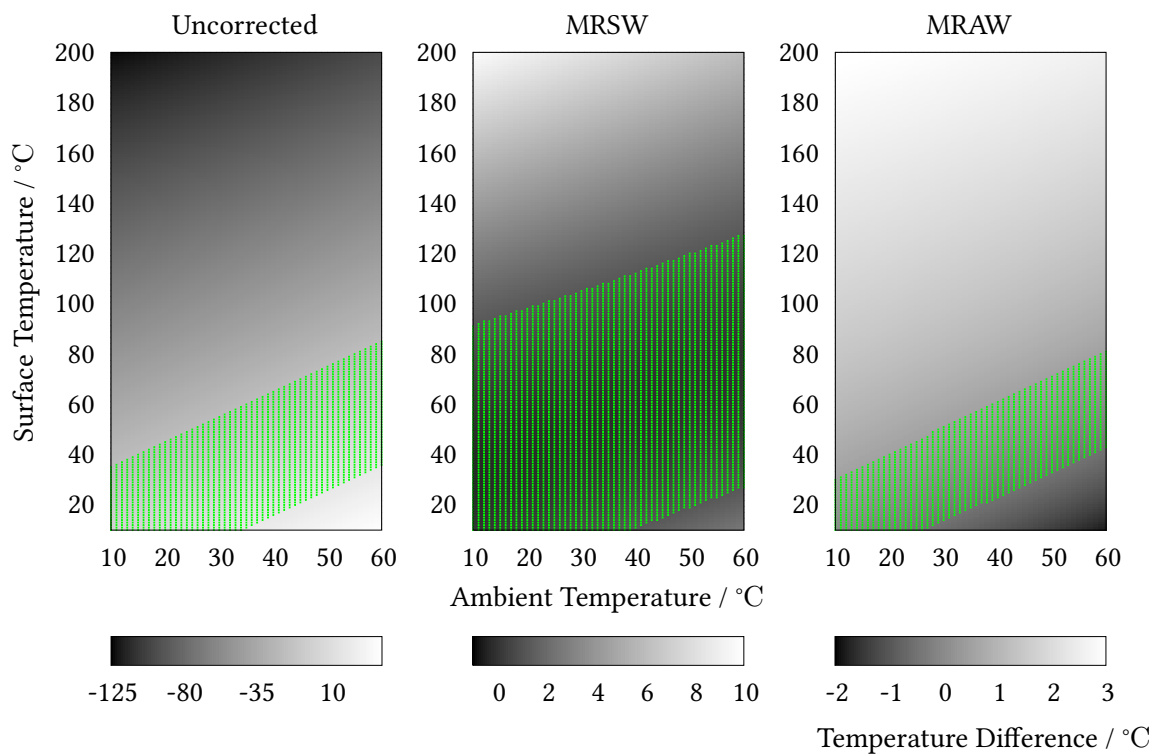}
  \caption{Position two, emissivity 0.21. The standard deviation of the datasets indicating the highlighted ranges are: \SI{40.1}{\celsius}, \SI{2.5}{\celsius} and \SI{1.1}{\celsius}.}
  \label{fig:radianceCorrectionComparisonPositionTwoEmissivityMedium}
\end{figure}

Any unique approaches taken by either of the correction methods are described in the corresponding sections above. These results are shown in Figure~\ref{fig:radianceCorrectionComparisonPositionOneEmissivityLow} through Figure~\ref{fig:radianceCorrectionComparisonPositionTwoEmissivityHigh}; where the value is close to zero, the method correction error is minimal. To facilitate interpretation of the plots and their varied colour bar scales, a green highlighted region is overlaid; this denotes the region about zero (corresponding to zero correction error) and spans a single standard deviation from the respective results array. The magnitude of these low correction error regions are detailed in the figure captions; this region can be used to identify comparable low uncertainty regions between the plots.

\clearpage

\begin{figure}[H]
  \centering
  \input{./Images/radianceCorrectionComparisonPositionTwoEmissivityLow}
  \caption{Position two, emissivity 0.06. The standard deviation of the datasets indicating the highlighted ranges are: \SI{52.0}{\celsius}, \SI{3.1}{\celsius} and \SI{30.0}{\celsius}.}
  \label{fig:radianceCorrectionComparisonPositionTwoEmissivityLow}
\end{figure}

\begin{figure}[H]
  \centering
  \input{./Images/radianceCorrectionComparisonPositionOneEmissivityHigh}
  \caption{Position one, emissivity 0.94. The standard deviation of the datasets indicating the highlighted ranges are: \SI{1.5}{\celsius}, \SI{1.3}{\celsius} and \SI{0.2}{\celsius}.}
  \label{fig:radianceCorrectionComparisonPositionOneEmissivityHigh}
\end{figure}

\begin{figure}[H]
  \centering
  \input{./Images/radianceCorrectionComparisonPositionTwoEmissivityHigh}
  \caption{Position two, emissivity 0.94. The standard deviation of the datasets indicating the highlighted ranges are: \SI{1.4}{\celsius}, \SI{1.3}{\celsius} and \SI{0.2}{\celsius}.}
  \label{fig:radianceCorrectionComparisonPositionTwoEmissivityHigh}
\end{figure}

\clearpage

Throughout these comparisons it can be observed that for all methods, as the emissivity increases, the magnitude of error decreases; nominally the error decreases from \SIrange{217}{1}{\celsius}.

For an uncorrected radiance temperature the largest underestimation occurs in the upper-left region and the low error region is consistently where \(T_b \approx T_{amb}\). Under the high emissivity container the range in error across the parameter space is reduced to \SI{7}{\celsius}.

For the MRSW correction the range of error decreases from \SI{14.8}{\celsius}, through \SI{10.1}{\celsius} to \SI{5.8}{\celsius} as the emissivity increases. For the wall reflection environment (position two) the parameter space is linear from the upper-left to lower-right region, where the largest correction error is when the surface temperature is high whilst ambient temperature is low. For large regions of the parameter space, the MRSW correction is close to zero and does not incur a large error (demonstrated by the wide highlighted bands).

At lower observation container emissivities where the radiance contribution from the reflection container (position one environment) becomes greater, a local minima of error is shown about the reflection container temperature. For the medium emissivity environment a minor deformation of the highlighted region from position two to position one is shown that increases the correction error at regions further from the reflection container temperature (in both planes). For the low emissivity environment this deformation is greater due to the larger contribution to apparent radiance temperature from the walls.

During the plate measurements in Section~\ref{sec:assemblyAssessment} the MRSW correction was applied. Across this narrow parameter space (surface temperatures from \SIrange{20}{80}{\celsius} and ambient temperatures from \SIrange{21}{23}{\celsius}) that could be represented by position two with the wall in the specular direction from the uncoated stainless steel, the incurred temperature error from this correction method was from \SIrange{-0.3}{0.4}{\celsius}. Therefore the correction used was suitable in this application.

For the MRAW correction the range of error decreases from \SI{303}{\celsius}, through \SI{6.5}{\celsius} to \SI{1.0}{\celsius} as the emissivity increases. This large reduction of error suggests the MRAW method is too sensitive to uncertainty in the input parameters as a result from the low emissivity. For the medium and high emissivity comparisons the MRAW correction demonstrates the lowest correction error across each of the parameter spaces. The region of lowest error occurs when \(T_b \approx T_{amb}\), this is demonstrated in Figure~\ref{fig:emissivityError} when the apparent radiance temperature passes through the stated ambient temperature and is because the contributions from all sources for the apparent radiance are comparable. A local region of large correction error occurs when the surface temperature is low and ambient temperature is high (lower-right region).

Between positions one and two, the difference in correction error is more apparent at lower emissivities because the radiance contribution from the reflection on the observation container increases. This shows that the approximations made by each method are applicable for most background environments in the specular direction, but care must be taken when the difference between the reflection container and wall surface temperatures is great (for example when outside of the shaded green region). This may be likely in a case study of the SNM store where a hot container is in the specular direction but was suitable for the laboratory investigation detailed in this chapter.

\subsection{Emissivity uncertainty}\label{subsec:emissivityUncertainty}

When accounting for the effect on surface temperature measurement from emissivity uncertainty, this can be evaluated depending on the correction method used. The multiple reflection single wavelength method from Section~\ref{subsec:multipleReflectionSingleWavelength} was used in the representative test case in this chapter (refer to Section~\ref{sec:assemblyAssessment}), this analytical solution enables a direct sensitivity analysis for each parameter to be calculated. By evaluating the partial differential of surface temperature with respect to emissivity, this sensitivity can be identified (refer to Eq.~\ref{eq:sensitivityEmissivity}).

\begin{figure}[h]
  \centering
  {\footnotesize \input{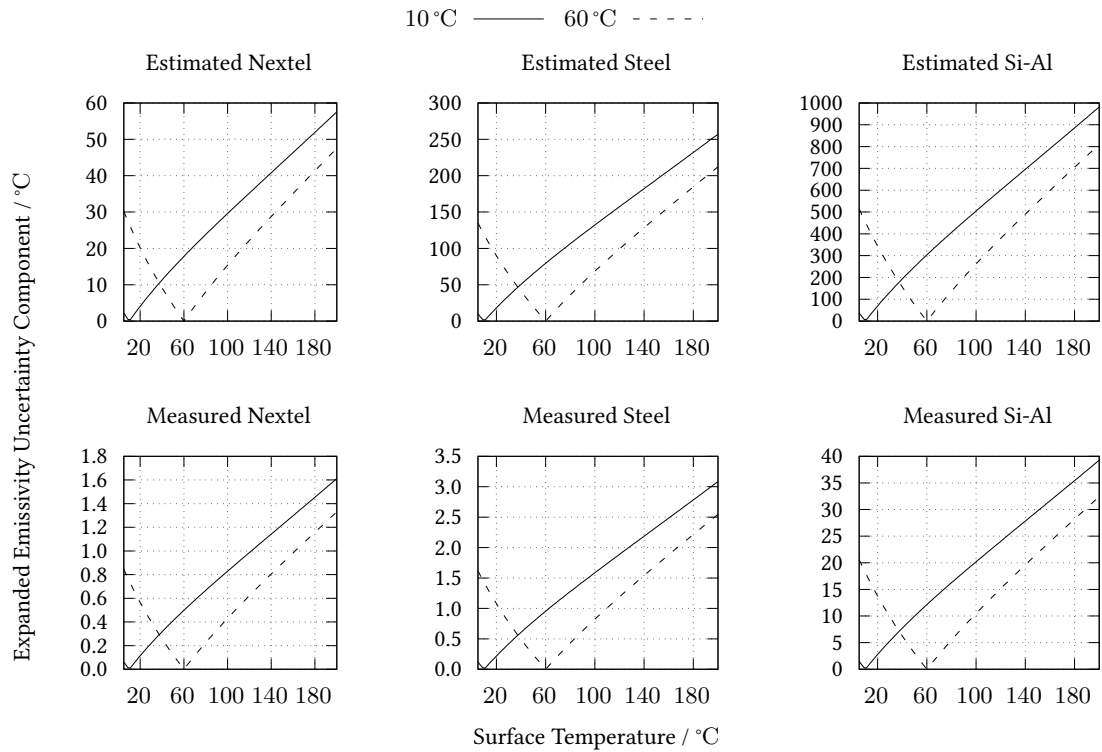}}
  \caption{The emissivity uncertainty component for a potential surface temperature measurement budget in both a \SI{10}{\celsius} and \SI{60}{\celsius} environment. Estimated materials have a \num{0.2} \(\left( k = 1 \right)\) uncertainty and measured materials are \num{0.006} \(\left( k = 1 \right)\), \num{0.024} \(\left( k = 1 \right)\) and \num{0.008} \(\left( k = 1 \right)\) respectively. The emissivity values are \num{0.94}, \num{0.21} and \num{0.06} for Nextel, steel and silica coated aluminium respectively.}
  \label{fig:emissivity_uncertainty_component}
\end{figure}

A comparison between the magnitude of this emissivity uncertainty component with respect to the temperature parameters defined in Table~\ref{tab:environmentProperties} is shown in Figure~\ref{fig:emissivity_uncertainty_component}. For the estimated material properties, the emissivity uncertainty is \num{0.2} (\(k=1\)); for the three measured emissivity values the estimated emissivity uncertainty was \num{0.008} (\(k=1\)), \num{0.024} (\(k=1\)) and \num{0.006} (\(k=1\)) respectively.

From these results it is clear that for most materials when the ambient temperature is greater, the corresponding uncertainty component is minimised due to the reduced difference between ambient and surface temperatures. For the lower emissivity measured material, this reaches close to \SI{50}{\celsius} in the most extreme case, but this remains much lower than the estimated material impact that approaches \SI{1000}{\celsius}.

\begin{figure}[h]
  \centering
  \input{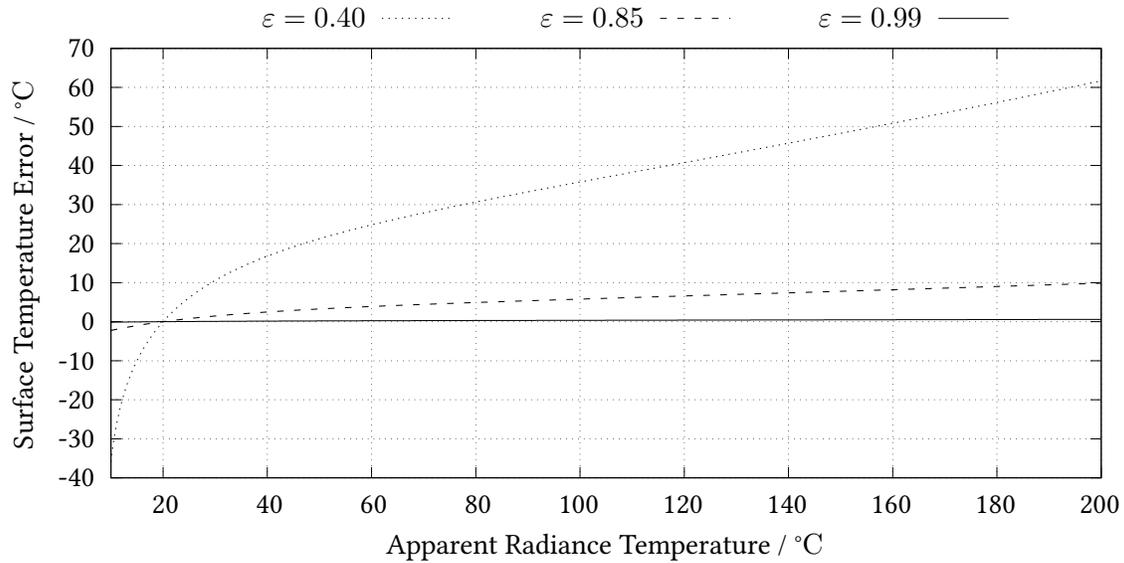}
  \caption{The difference between the apparent radiance and surface temperature for three materials with a \SI{20}{\celsius} ambient temperature and using an instrument with the same spectral response as used in the measurements from this chapter.}
  \label{fig:emissivityError}
\end{figure}

The sensitivity of surface temperature determination is shown in Figure~\ref{fig:sensitivityEmissivity} with a \SI{20}{\celsius} ambient background temperature, for three different emissivity values, the difference between apparent radiance temperature and calculated surface temperature at increasing surface temperatures is shown in Figure~\ref{fig:emissivityError}. For a surface of 0.85 emissivity, a measured apparent radiance temperature of \SI{200}{\celsius} will be nominally \SI{10}{\celsius} below the surface temperature; additionally the difference in temperature error at this temperature between an emissivity of 0.85 and 0.99 is \SI{9}{\celsius}. Further, the equivalent difference in error between emissivity values of 0.26 and 0.40 is \SI{35}{\celsius}. Therefore the importance of low uncertainty emissivity measurement (traceable and metrologically determined) for surface temperature determination using a thermal imager is clear. The gradient of the error increases for materials of lower emissivities and so the scope for higher uncertainties increases when measuring surfaces with lower emissivity. The effect of this is presented in Table~\ref{tab:laboratoryUncertaintyUncoated} where the contribution from emissivity -- with similar emissivity uncertainty -- has a greater effect on the standard uncertainty.

\subsection{Summary}\label{subsec:correctionSummary}

Throughout the subsequent laboratory and in-situ measurements, a simplification of the MRSW radiance correction method has been implemented. This could be described by using Eq.~\ref{eq:radianceCorrectionMRSW} and assuming the reflection surface is a blackbody, this was equivalent to the scenario in position two. To ascertain the suitability of this method an investigation into its behaviour across a range of surface and ambient temperatures, a set of measured emissivities and a pair of background reflection profiles. This was baselined against an uncorrected measurement and another correction method that considered both the spectral response of emissivity and thermal imager response.

It was shown that the magnitude of correction error increases for lower emissivity materials due to the disparity between surface and apparent radiance temperatures. The MRSW correction method was largely appropriate across each of the six environments but incurred the largest errors when the surface temperature was much greater than the ambient temperature. The region of lowest error typically occurred when the surface temperature was approximately equal to the ambient temperature; this was deviated from during the MRSW measurements with the reflection container in the specular direction (position one). Contrary to expectations, the MRAW method was not able to successfully correct the radiance temperature under all conditions and demonstrated large errors during the low emissivity computations.
\clearpage
\newpage

\section{Conclusions}\label{sec:artefactConclusions}
The primary objectives of this study were to investigate and evaluate the feasibility of using a thermal imager to measure container surface temperature, in order to monitor container activity for ongoing container monitoring and management.

The primary activities undertaken in this investigation were: 

\begin{enumerate}
  \item Design and manufacture of a thermal management assembly
  \item Thermal evaluation of the target plate
  \item Assessment of measurement capability through an uncertainty budget review
  \item Definition of radiance correction methods
  \item Quantitative comparison of correction methods
\end{enumerate}

Both surface and internal temperatures were evaluated using the plate and there was reasonable agreement between the experimental measurements in most instances. Utilisation of a cooled thermal imager in laboratory conditions demonstrated the state-of-the-art with regard to instrumentation and controlled environment measurement; deployment to an active store will necessitate a reduction in hardware form factor and specifications to uncooled microbolometers (due to the instrument size) to accommodate the store assessment port size. Employing a microbolometer will marginally increase instrumentation uncertainty but this component is not anticipated to significantly affect the overall uncertainty budget, with the largest uncertainty still remaining as the surface emissivity uncertainty. The reduced operational stand-off poses a more challenging optical configuration using wide field of view lenses where image distortion is greater, this will be further investigated in the following in-situ deployment experiment in Chapter~\ref{chap:inSituMeasurement}.

During the plate measurements, unanticipated assembly spatial non-uniformity increased the measurement uncertainty achieved by the thermal imager and reduced the confidence in surface temperature determination for a number of regions of interests. Additional measures to improve this uniformity have been identified to be deployed for future studies using the plate. This would also be supplemented with further reduced uncertainty contact thermometers. Following a detailed investigation into the experienced non-uniformity -- principally plate distortion -- the agreement between the thermal imager radiance and thermocouple temperature was within the measurement uncertainties and close to anticipated levels. There was moderate agreement between the experimental surface temperature measurements during the plate campaign, for the high emissivity coated regions the mean temperature difference across all temperature setpoints was \SI{1.8}{\celsius}; for the lower emissivity uncoated regions this difference increased to \SI{8.3}{\celsius}. Outstanding discrepancies between these methods are believed to be caused by an incomplete description of the thermal pathways present in the laboratory system. In all cases the lack of robust emissivity data was the dominant component in temperature determination uncertainty budgets, leading to an expanded temperature measurement uncertainty up to \SI{6.8}{\celsius} (\(k=2\)). The absence of this data becomes more prevalent when measuring surface of high reflectivity because it scales non-linearly with temperature and the uncertainty reached \SI{30.1}{\celsius} (\(k=2\)).

During the laboratory measurements the insight and progress made towards successful deployment of a thermal imager to an active container store was significant. The design and manufacture of the plate, as well as the supporting measurement data and future improvements have produced an effective tool in the characterisation of a thermal imager for artefact detection through the appropriate range of temperatures. Instrumentation of the plate enabled good thermal management and robust thermometry, as well as providing insight into the thermal distribution that may be present in active containers.

During the radiance correction comparison the magnitude of correction error increased at lower emissivities from \SIrange{5.8}{14.8}{\celsius} (for the MRSW method) as a result of the increased difference between surface and apparent temperatures. The correction method used through the plate measurements was demonstrated to be suitable across the majority of the parameter space and had a low incurred error spanning \SI{0.7}{\celsius} within the region of these measurements.

The laboratory assessment observations here are specific to these materials: stainless steel 316L and a Senotherm coating on this steel; when investigating real world surfaces, the effect from changing surface properties (e.g. surface wear, corrosion or powder deposition) should be evaluated. As discussed in Section~\ref{subsec:surfaceTemperatureUncertaintyAnalysis} the sensitivity of uncertainty components evaluated were dependent on each of the input parameters and radiance correction method employed. As a rule of thumb the sensitivity -- and by inference -- temperature uncertainty increases both as the surface emissivity decreases and when the apparent radiance temperature differs greatly from the ambient temperature.

The temperature measurement uncertainty budgets support the quality of the thermal imager calibration carried out and the hardware used; across the temperature range investigated the intrinsic components of the thermal imager remained well below the manufacturer stated accuracy and within a reasonable bounds of surface temperature determination applications. Given traceable emissivity measurement, the confidence in surface temperature measurement will improve greatly.

The principle conclusions to be drawn from this study are:

\begin{enumerate}
  \item Surface temperature measurement of containers with thermal imaging is feasible 
  \item Temperature determination agreed to contact thermometers within the estimated measurement uncertainties 
  \item The proposed radiance correction method is appropriate within the application parameter space
\end{enumerate}

\clearpage
\newpage
\markedchapter{Container Characterisation}{Characterisation of Nuclear Storage Containers}\label{chap:inSituMeasurement}
\section{Introduction}\label{sec:introduction}
Chapter~\ref{chap:laboratoryMeasurement} explored a set of radiance correction methods and deployed one of these to a laboratory environment and manufactured test plate. This deployment was investigated further within this chapter through the use of instrumented containers used in both a laboratory and a store environment. The two containers evaluated are the THermal Oxide Reprocessing Plant (THORP) and Magnox containers.

This chapter discusses the development of thermal mapping technology for in-situ, periodic container monitoring. This thermal mapping was evaluated through direct surface temperature comparisons between surface mounted thermocouples and emissivity corrected radiance temperature measurements from a thermal imaging system.

For each experimental setup, the suitability of a thermal imager for the determination of surface temperature was evaluated. This comparison was drawn against surface mounted thermocouples on the containers and considered beside the respective uncertainty budgets and radiance correction methods. By doing so, the viability of deploying a thermal imager to an active container store was determined.

During the laboratory measurements a medium wave cooled thermal imager was used to observe the instrumented containers. These containers had internal thermocouples and heaters positioned, the external surface remained unchanged.

For the in-situ measurements a pair of long wave uncooled thermal imagers were positioned perpendicularly to the surface, whilst a pair of off-axis mirrors were used to direct the surface radiation from the container to the imagers. For these measurements the internal instrumentation of the container was nominally identical to the laboratory measurements; however the external surface was instrumented with thermocouples, a high emissivity coating and a series of both scratch and pitting artefact regions. This setup was then placed within a test store at the James Fisher Nuclear Limited (JFNL) facility to replicate an active warehouse setup.

This work used test Magnox and THORP containers instrumented to provide equivalent temperatures to those containers held in store (i.e. \SI{110}{\celsius} to \SI{170}{\celsius}), located in a replica test store with an inspection system mounted on a rail below to provide equivalent set up and access challenges to those posed by active stores.

Prior to deployment within the test store, the instrumented containers were setup in a laboratory environment using the same instrumentation as the plate in Chapter~\ref{chap:laboratoryMeasurement}. During these measurements it was not possible to make a direct comparison between temperature determination methods and instead a computational model correction was introduced to the thermocouple analysis. Despite this correction, the two methods did not compare favourably across the entire temperature range.

Changes were made when moving to the test store environment and externally mounted thermocouples were introduced. During this campaign instrumentation challenges were encountered with the thermal stabilisation of the thermal imagers. It was demonstrated that this stabilisation error could be mitigated through rigorous procedure control and when this was omitted, the two temperature determination methods compared favourably across a range of temperatures and optical configurations.

A successful measurement campaign will progress the testing, assessment, and uncertainty analysis for test containers in a replica store and provide evidence for the suitability of this technique for further deployment.

\newpage

\section{Experimental setup}\label{sec:containerExperimentalSetup}
For both the laboratory and store measurements, a thermal imaging system was positioned to observe the external surface of both THORP and Magnox containers whilst surface mounted thermocouples evaluated surface temperatures.

For the two measurement campaigns two different thermal imaging systems were used. In the laboratory a single cooled Medium-Wave InfraRed (MWIR) thermal imager was used and during store measurements a pair of uncooled Long-Wave InfraRed (LWIR) thermal imagers setup perpendicularly to the inspection surface with corresponding off-axis mirrors. The specifications for each of these three imagers are given in Table~\ref{tab:thermalImagerSpecifications} as well as their mean Focal Plane Array (FPA) temperatures during calibration. The thermal imager naming conventions correspond to Wide Field Of View (WFOV) and Narrow Field Of View (NFOV).

\begin{table}[H]
  \renewcommand{\arraystretch}{0.75}
  \centering
  \caption{The specifications of the thermal imagers used throughout the measurements. The MWIR imager was used during the laboratory measurements in Section~\ref{sec:laboratoryContainerAssessment}. The two LWIR imagers were used during the store assessments in Section~\ref{sec:storeContainerAssessment}.}
  \vspace*{\floatsep}
  \begin{tabular}{L{4.0cm}L{3.5cm}L{3.5cm}L{3.5cm}}
    \toprule
	Characteristic								&	MWIR Imager							&	WFOV LWIR Imager						&	NFOV LWIR Imager						\\
    \cmidrule(lr){1-4}
	Manufacturer								&	InfraTec							&	FLIR									&	FLIR									\\
	Model										&	ImageIR 8300hp						&	Boson									&	Boson									\\
	Product code								&	-									&	20320H092-9PAAX							&	20320A050-9CAAX							\\
	Serial number								&	8312592								&	S0055574								&	S0032888								\\
	Detector									&	Indium antimonide					&	Vanadium oxide							&	Vanadium oxide							\\
	Pitch										&	\SI{15}{\micro\metre}				&	\SI{12}{\micro\metre}					&	\SI{12}{\micro\metre}					\\
	Resolution									&	\(640 \times 512\)\SI{}{\px}		&	\(320 \times 256\)\SI{}{\px}			&	\(320 \times 256\)\SI{}{\px}			\\
	Spectral response							&	\SIrange{2.0}{5.7}{\micro\metre}	&	\SIrange{7.5}{14.0}{\micro\metre}		&	\SIrange{7.5}{14.0}{\micro\metre}		\\
	Noise equivalent temperature difference		&	\(<\)\SI{20}{\milli\kelvin}			&	\(<\)\SI{50}{\milli\kelvin}				&	\(<\)\SI{60}{\milli\kelvin}				\\
	Lens horizontal field of view				&	\SI{22}{\degree}					&	\SI{92}{\degree}						&	\SI{50}{\degree}						\\
	Lens focal length							&	\SI{25.0}{\milli\metre}				&	\SI{2.3}{\milli\metre}					&	\SI{4.3}{\milli\metre}					\\
	Software									&	IRBIS 3 (version 3.1.90)			&	FLIR Boson Application (version 1.4.4)	&	FLIR Boson Application (version 1.4.4)	\\
	Calibration FPA temperature					&	\(24.7 \pm 2.7\)\SI{}{\celsius}		&	\(34.8 \pm 0.3\)\SI{}{\celsius}			&	\(33.1 \pm 0.5\)\SI{}{\celsius}			\\

    \bottomrule
  \end{tabular}
  \label{tab:thermalImagerSpecifications}
\end{table}

\subsection{Laboratory setup}\label{subsec:laboratorySetup}
Special Nuclear Material (SNM) is defined to be: plutonium, uranium-233 or uranium-235 \cite{ref:atomicEnergyAct}, these are usually stored in stainless steel 316 containers. SNM containers will corrode over time and therefore it is important that degradation is effectively monitored. Inspection of active SNM containers requires high quality quantitative data to demonstrate that they will reach their design lifetime. The detection of defects on the external surface of containers is a technical challenge due to the harsh environment, reduced access and requirement for non-contact inspection. Special nuclear materials emit gamma rays and neutron radiation that may result in the production of heat which can be measured by surface thermometry techniques.

\begin{figure}[H]
  \centering
  \includegraphics{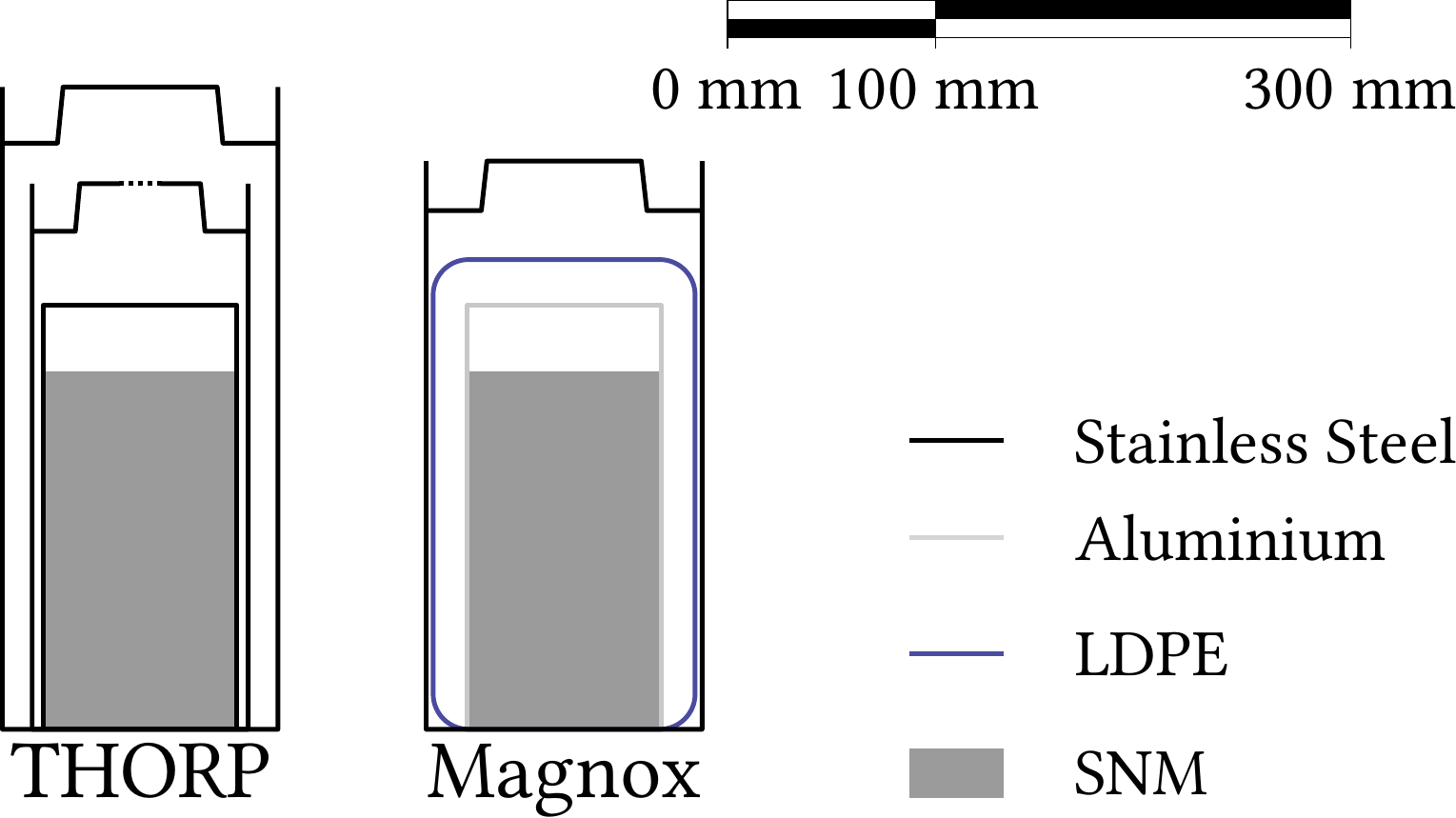}
  \caption{Two container types provided by Sellafield and their schematic construction. The containers were measured from their longer edge.}
  \label{fig:snmContainerComparison}
\end{figure}

The two container types provided by Sellafield, comprising both Magnox and THORP designs (summarised in Figure~\ref{fig:snmContainerComparison}), were inspected using a calibrated cooled MWIR thermal imager (calibration described in Section~\ref{subsec:case_study_thermal_imager} \cite{ref:thermalImagerCertificate2020030001}). The Magnox containers as provided did not have an intact Low-Density PolyEthylene (LDPE) layer due to their previous heat treatment, instead the remnants of this layer were observed on the appropriate surfaces. The two Magnox containers each comprised the anticipated outer container, but the internal container was the THORP filter-less internal container. The two THORP containers each comprised one outer, one intermediate and one internal container.  

For the Magnox measurements a typical Magnox stainless steel outer container was used with the THORP inner container. For the THORP measurements the appropriate outer, intermediate and inner containers were used together.

\begin{figure}[H]
\centering
	\begin{subfigure}[t]{0.475\textwidth}
	\centering
	\includegraphics[height=5cm,keepaspectratio]{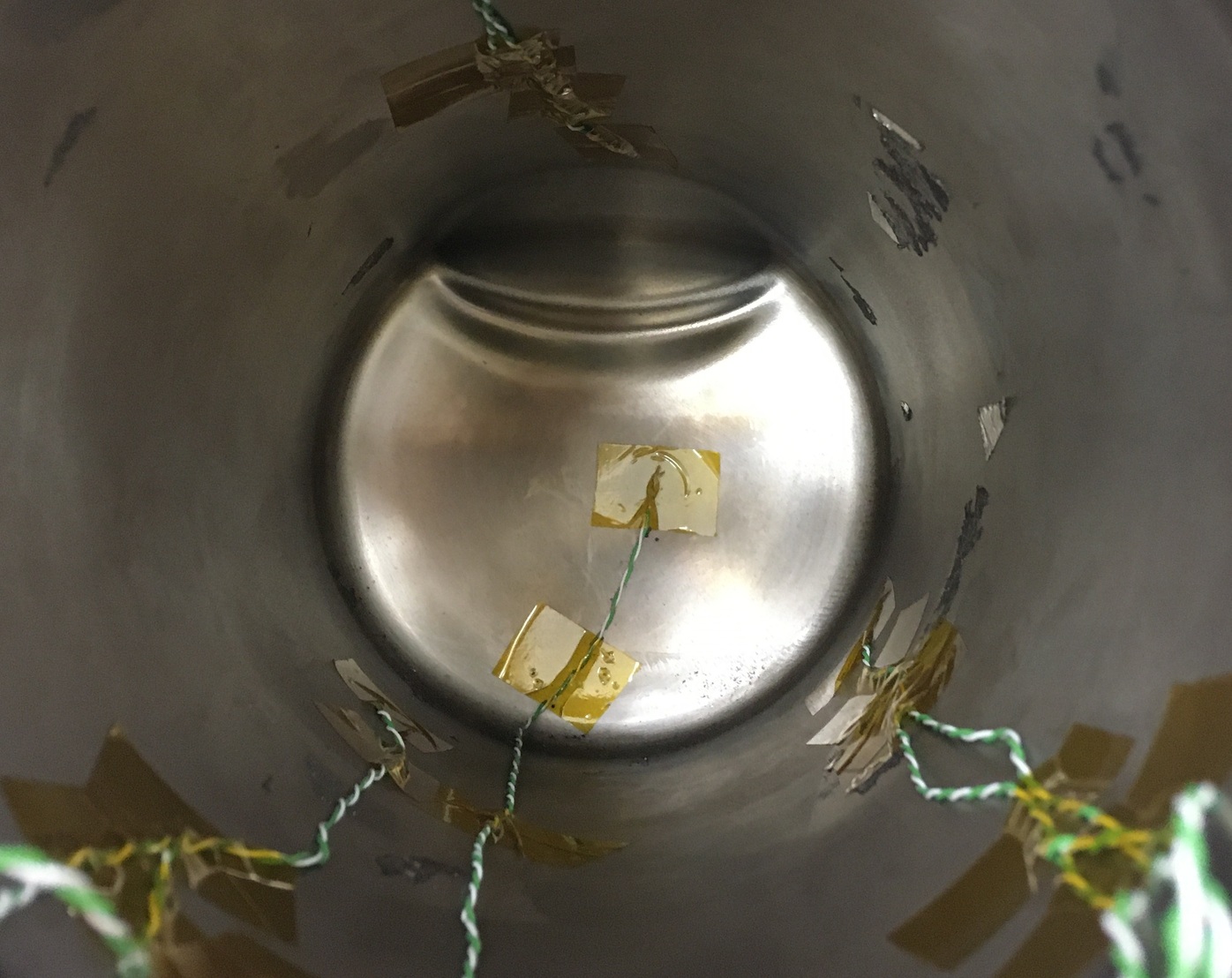}
	\caption{Internal container with monitoring thermocouples affixed.}
	\label{fig:snmLaboratoryContainerInternalThermocouples}
	\end{subfigure}
\hfill
	\begin{subfigure}[t]{0.475\textwidth}
	\centering
	\includegraphics[height=5cm,keepaspectratio]{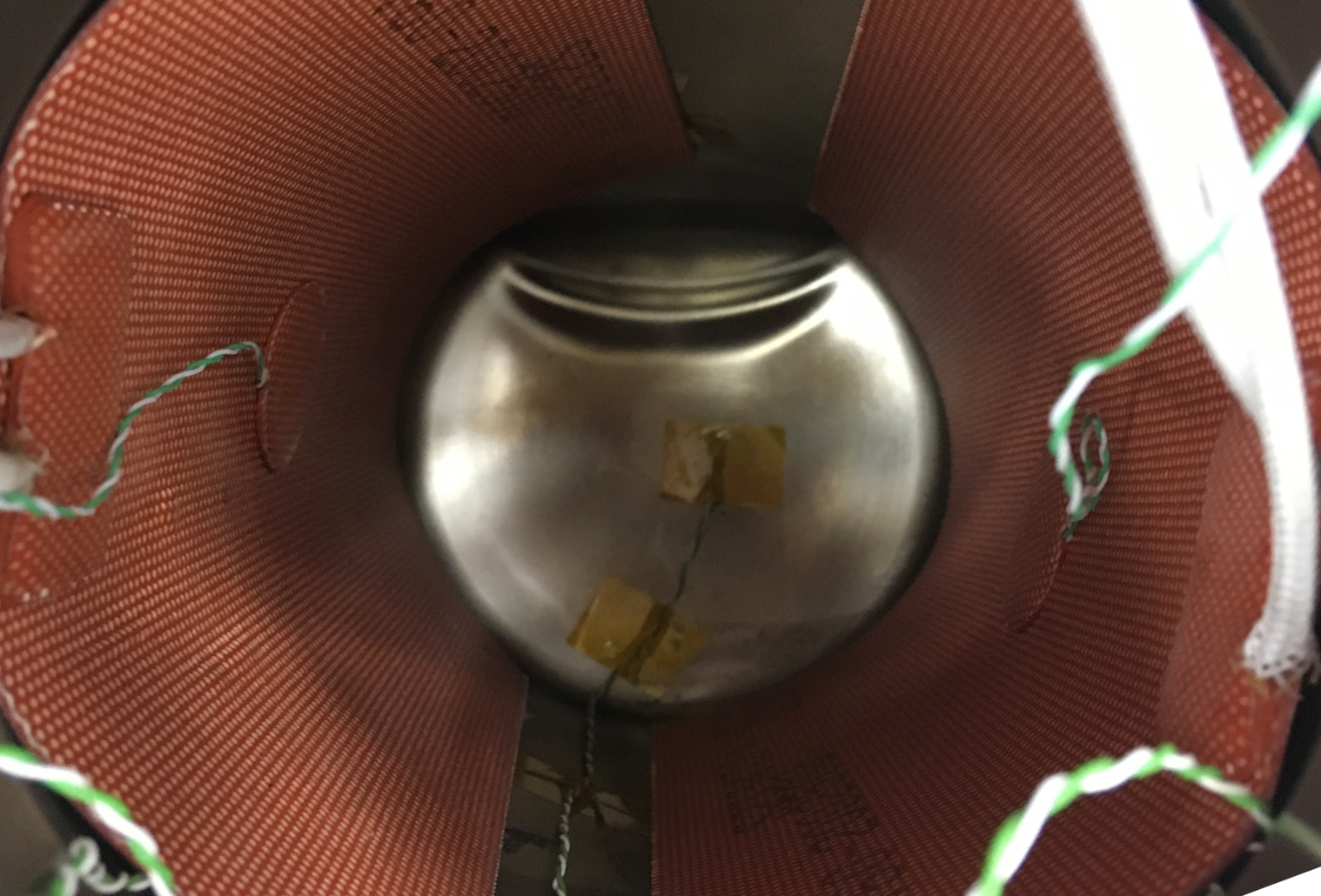}
	\caption{Internal container with heaters mounted above the monitoring thermocouples.}
	\label{fig:snmLaboratoryContainerInternalHeaters}
	\end{subfigure}
	\caption{The thermocouple and heater instrumentation of the inner container. One thermocouple is attached to the far end side and one other to the inside of the lid (not in view), three radially spaced \SI{65}{\milli\metre} from the far end, two radially spaced towards the longitudinal centre (\SI{130}{\milli\metre} from far end) and three radially spaced \SI{195}{\milli\metre} from the far end. The heater mats were pressed against the inside wall and the affixed thermocouples facing the centre.}
\label{fig:snmLaboratoryContainerInternal}
\end{figure}

\begin{figure}[H]
  \centering
  \includegraphics[width=\textwidth,keepaspectratio]{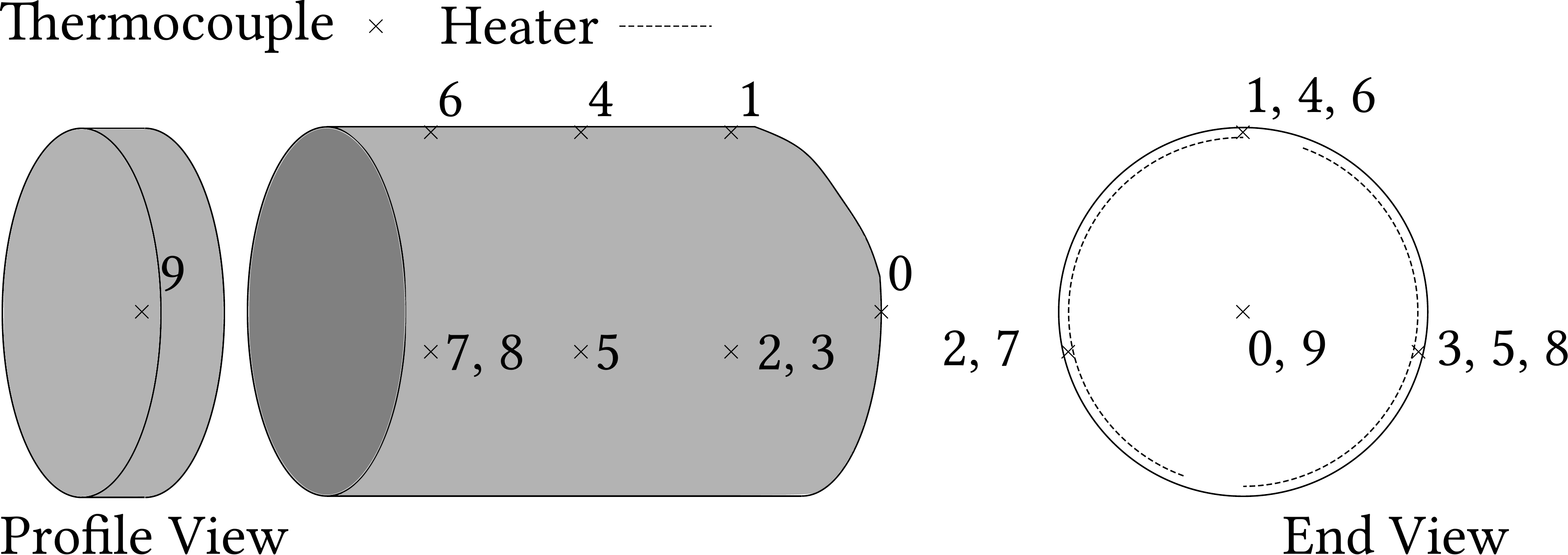}
  \caption{A schematic of the internal container instrumentation. The dashed lines represent the heaters and the crosses indicate the location of the ten thermocouples and their respective identification numbers. The container is presented in its profile perspective and the end view corresponds to the images in Figure~\ref{fig:snmLaboratoryContainerInternal}.}
  \label{fig:snmLaboratoryContainerSchematic}
\end{figure}

The internal container was instrumented with ten type K monitoring thermocouples and two heater mats, these are shown in Figure~\ref{fig:snmLaboratoryContainerInternal}. A schematic of these thermocouple locations are shown in Figure~\ref{fig:snmLaboratoryContainerSchematic}.

The two heater mats were slightly mis-aligned with the central axis of the container such that thermocouples 1, 4 and 6 were in closer contact with the heater. The thermocouples were arranged to be equally distributed across the length of the container and radially distributed around its circumference. A pair of thermocouples were affixed to the bottom face and the lid at the radial origin. The internal container was filled with the steel shot originally packaged within the container (as a nuclear material proxy) and the instrumentation cables directed through the appropriate lids.  

\begin{figure}[H]
  \centering
  \includegraphics[width=0.6\textwidth,keepaspectratio]{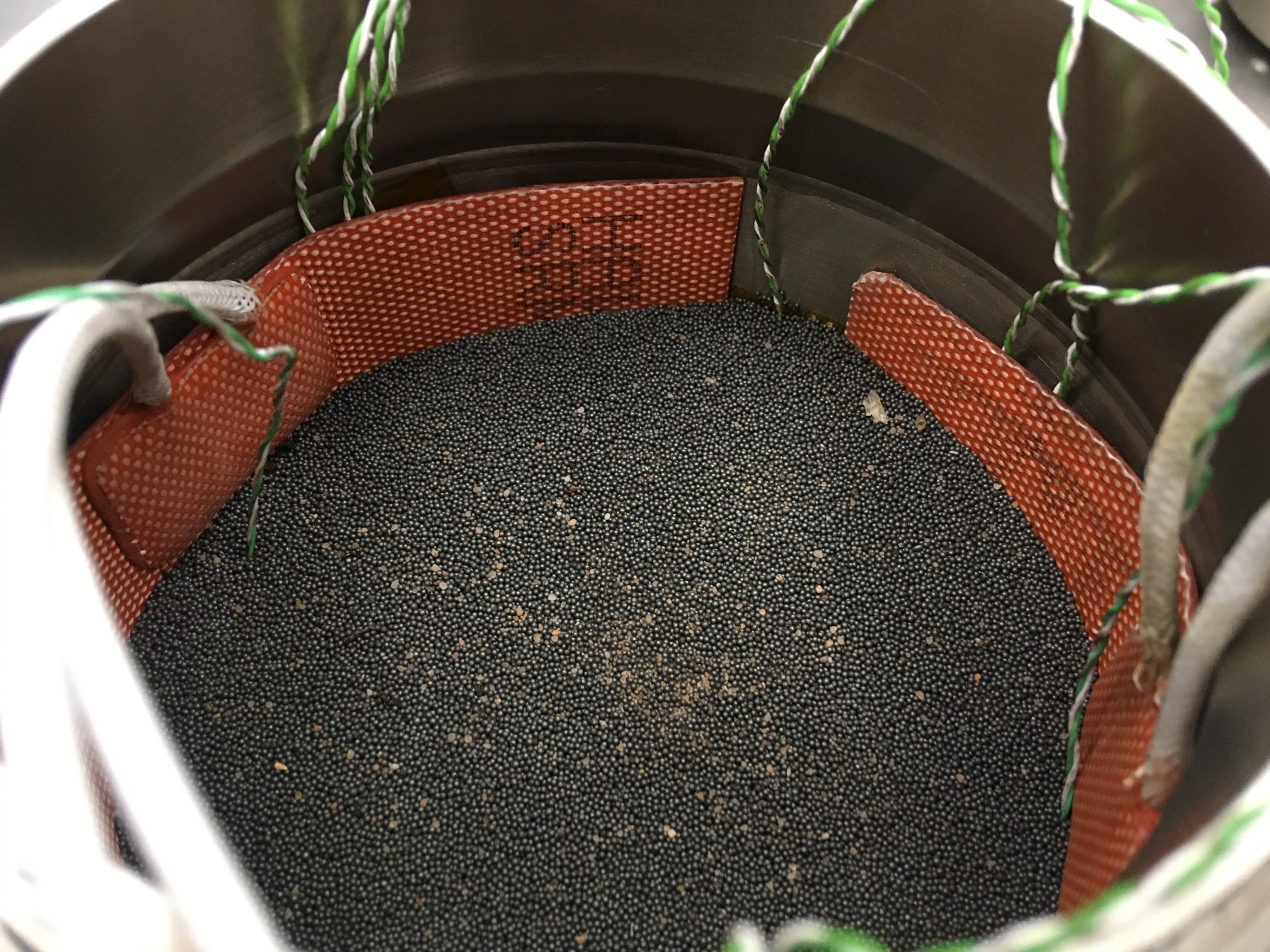}
  \caption{The loaded internal container with the provided steel shot.}
  \label{fig:snmLaboratoryContainerShot}
\end{figure}

For the Magnox container measurements, the internal container was sealed shut with the threaded lid. This was then placed within the outer container D33-5, the outer container lid was sealed using high-temperature adhesive tape.  

\begin{figure}[H]
  \centering
  \includegraphics[width=0.6\textwidth,keepaspectratio]{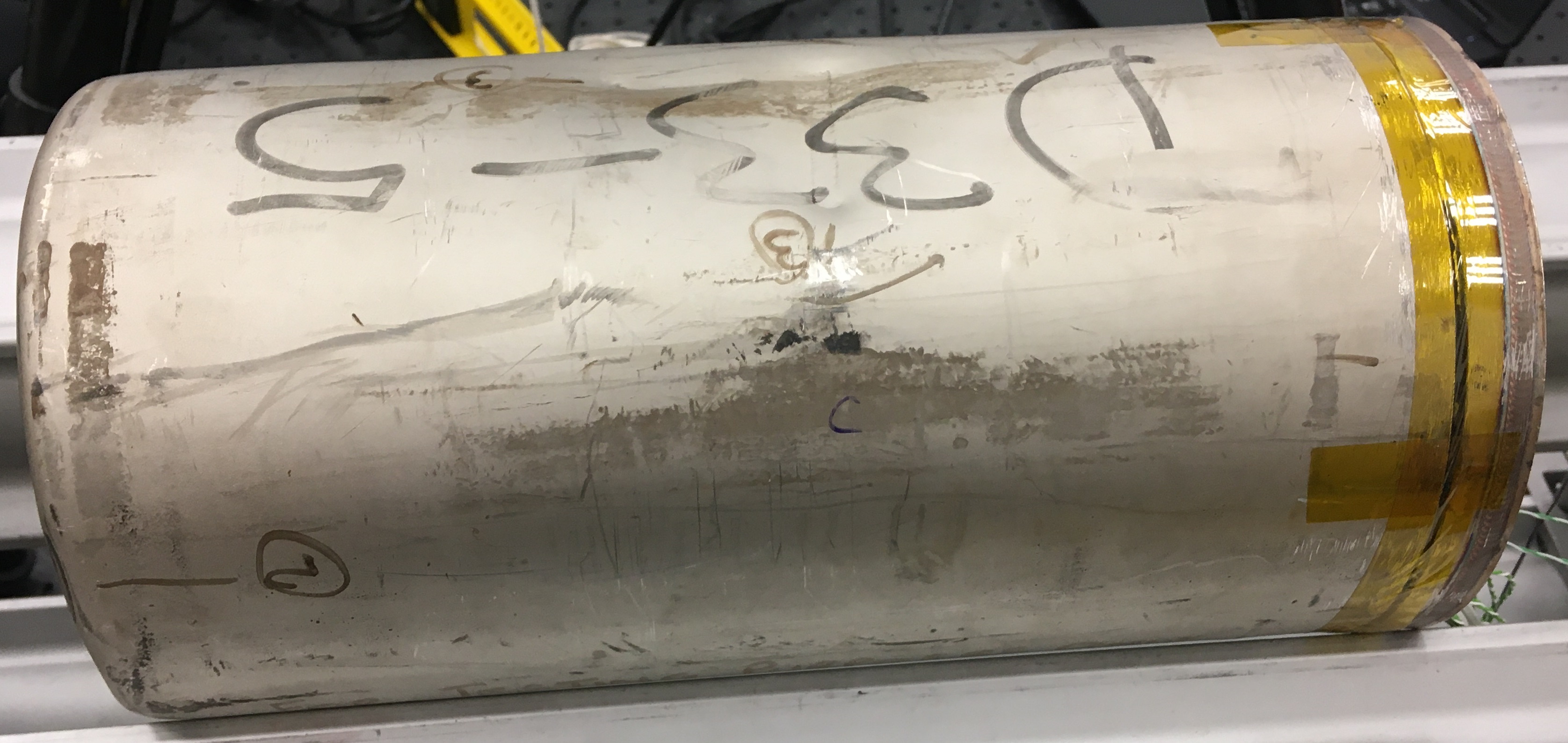}
  \caption{The internal container within the Magnox outer. The instrumentation cables were passed through openings on both container lids and the outer lid was held in place using high-temperature adhesive tape.}
  \label{fig:snmLaboratoryContainerExternalMagnox}
\end{figure}

Following the Magnox container measurements, the internal container was removed and placed within the intermediate THORP container. This was sealed with high-temperature adhesive tape, placed within the THORP outer container which was also sealed.

\begin{figure}[H]
  \centering
  \includegraphics[width=0.6\textwidth,keepaspectratio,trim={0cm 4cm 0cm 4cm},clip]{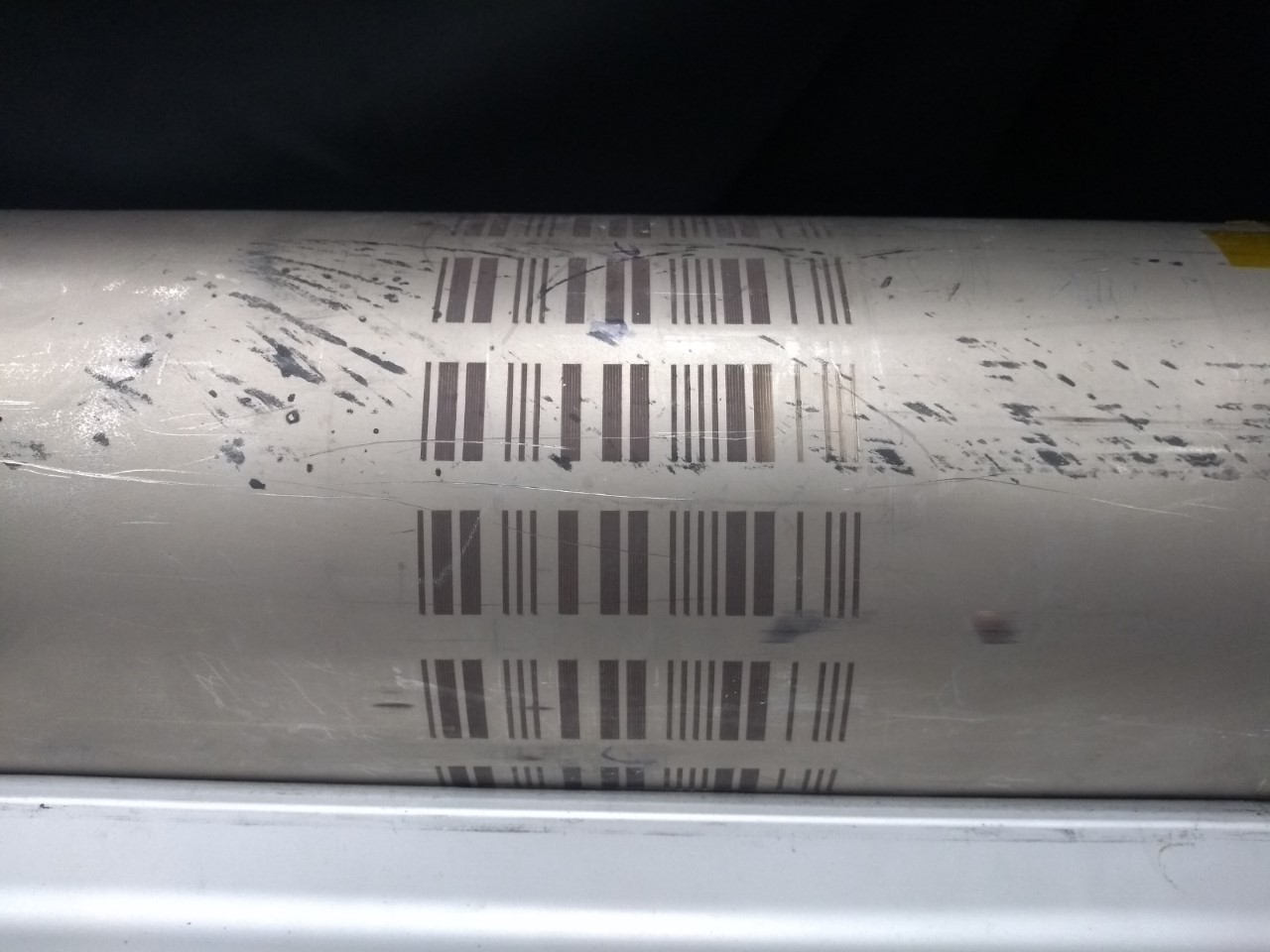}
  \caption{The internal container within the THORP intermediate, sealed and then placed within the THORP outer container. The instrumentation cables were passed through openings on each container lid, the intermediate and outer lids were held in place using high-temperature adhesive tape. The bar code serves as identification for the container whilst in storage.}
  \label{fig:snmLaboratoryContainerExternalTHORP}
\end{figure}

\begin{figure}[H]
\centering
	\begin{subfigure}[t]{0.490\textwidth}
	\centering
	\includegraphics[height=8.9cm,keepaspectratio]{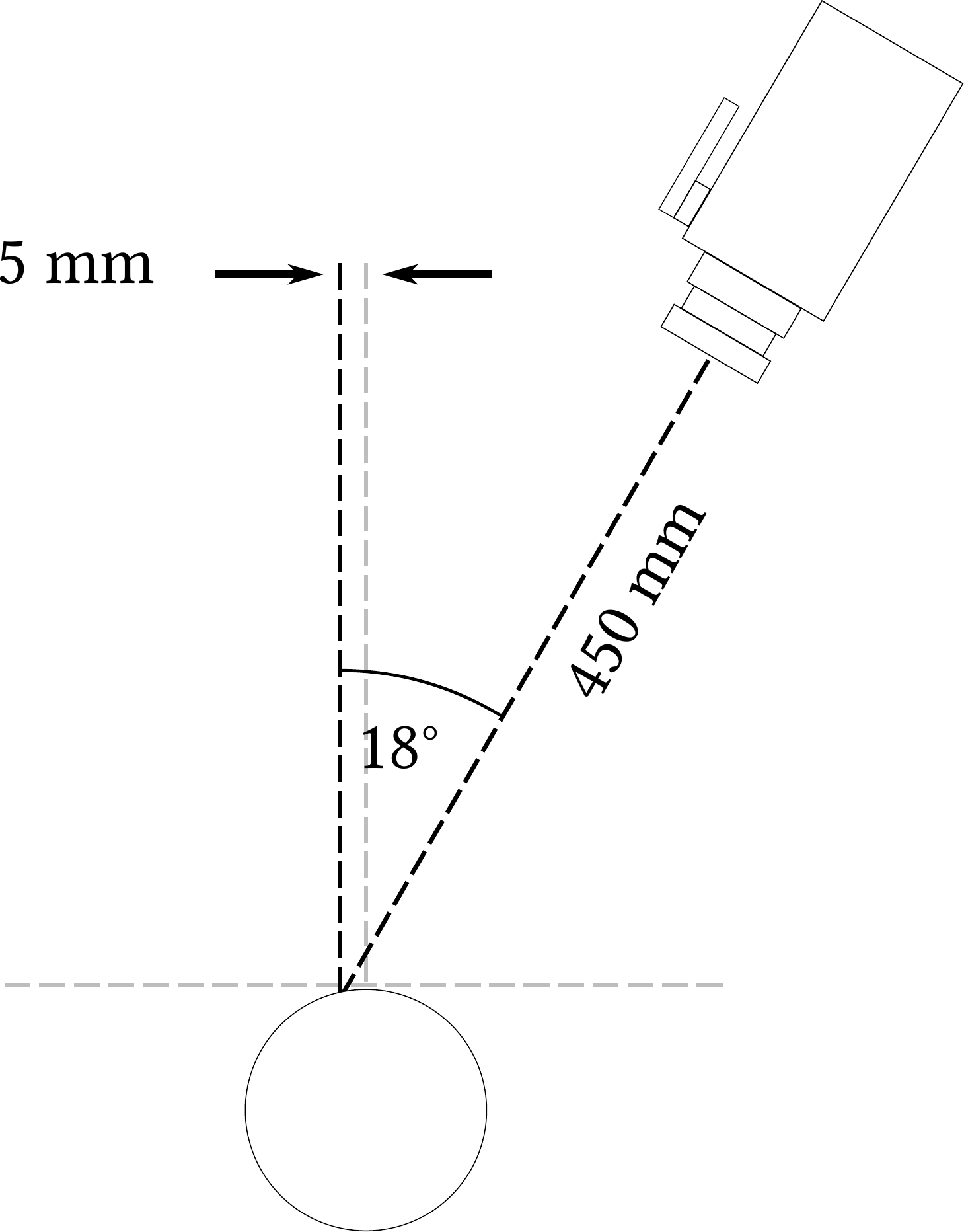}
	\caption{Magnox configuration.}
	\label{fig:snmLaboratoryContainerImagerConfigurationMagnox}
	\end{subfigure}
\hfill
	\begin{subfigure}[t]{0.490\textwidth}
	\centering
	\includegraphics[height=8.9cm,keepaspectratio]{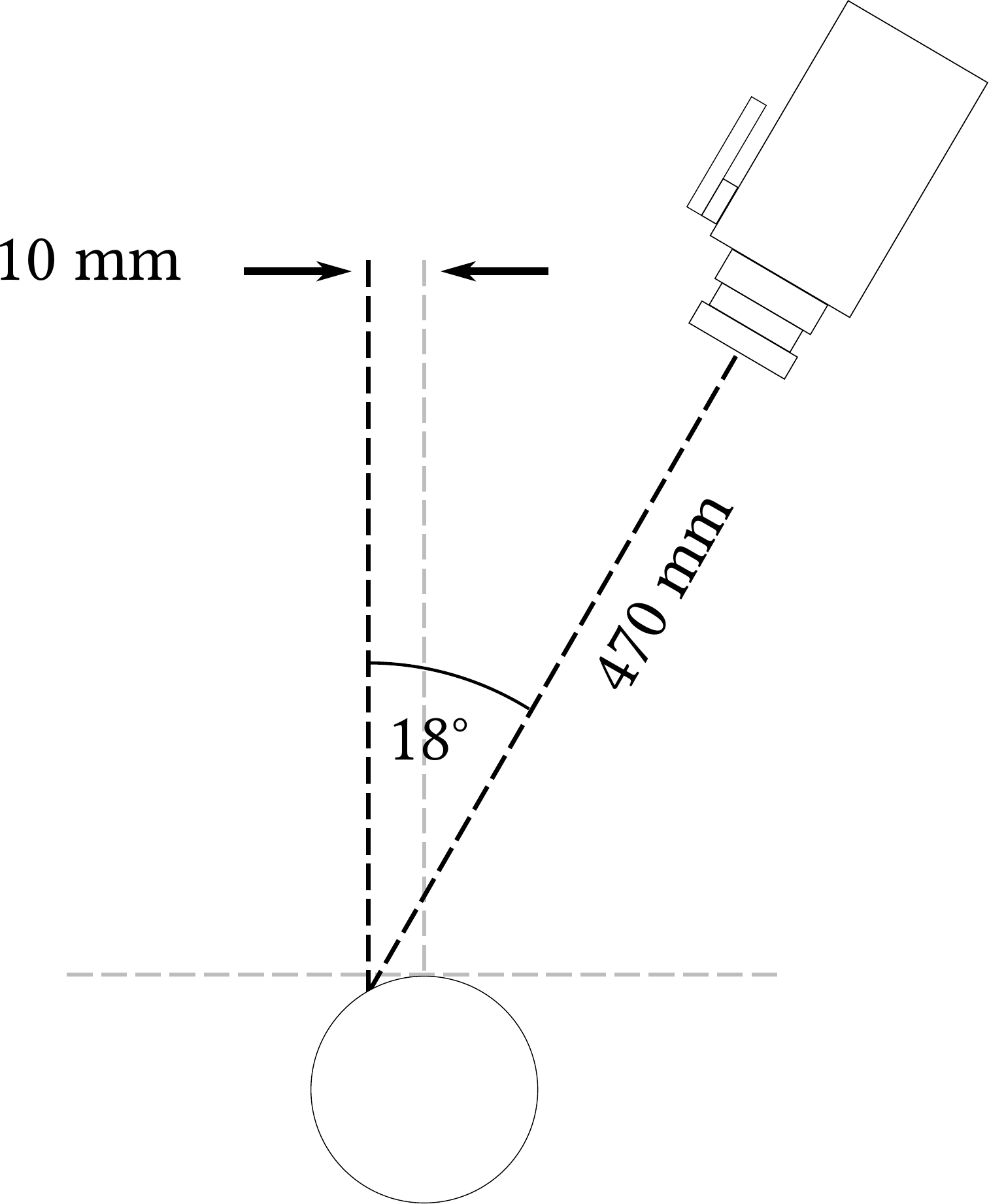}
	\caption{THORP configuration.}
	\label{fig:snmLaboratoryContainerImagerConfigurationTHORP}
	\end{subfigure}
	\vspace{\floatsep}
	\begin{subfigure}[t]{\textwidth}
	\centering
	\includegraphics[width=0.65\textwidth,keepaspectratio,trim={0cm 15cm 0cm 0cm},clip]{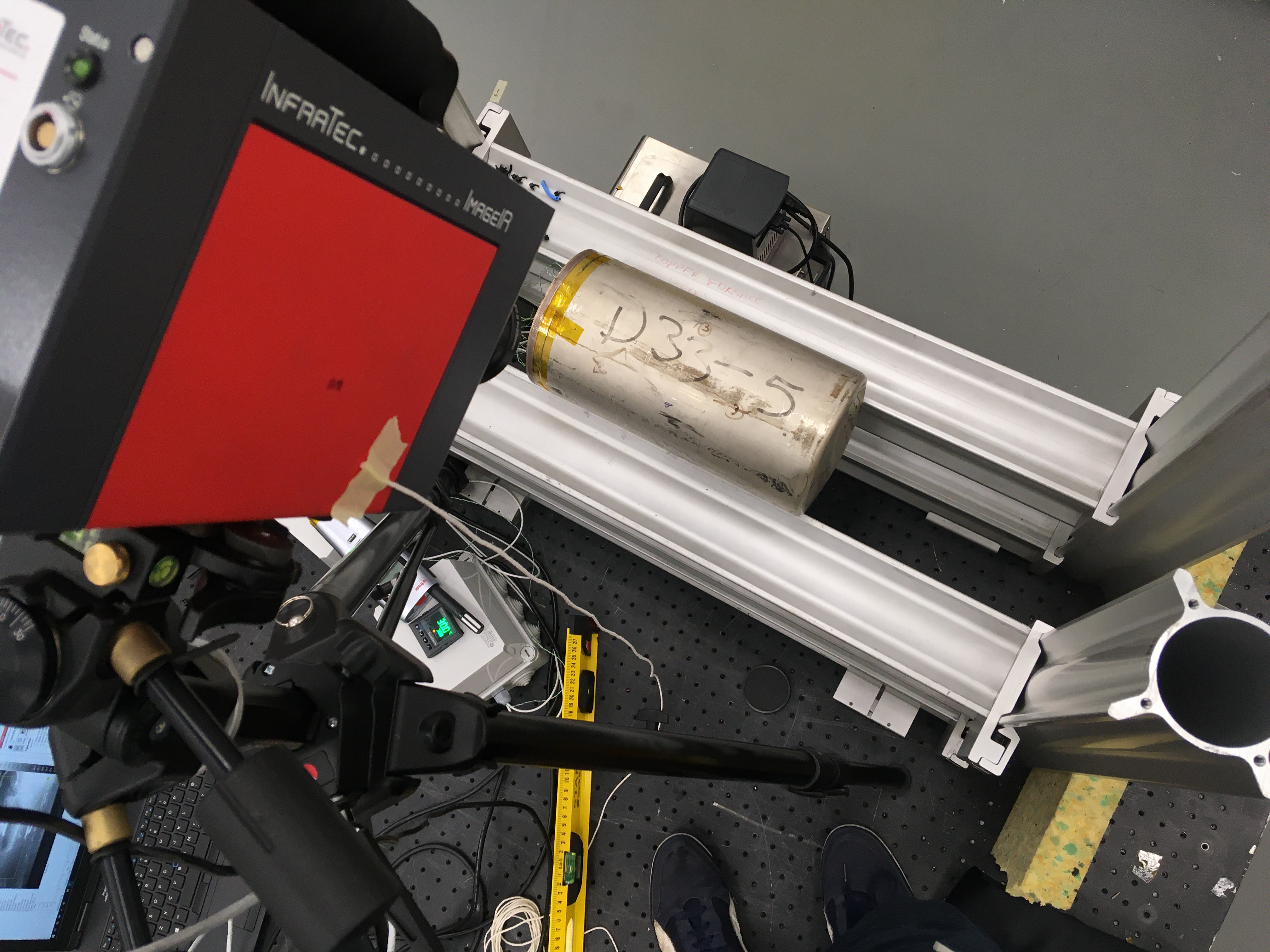}
	\caption{Perspective of configuration.}
	\label{fig:snmLaboratoryContainerImagerConfigurationImage}
	\end{subfigure}
	\caption{Schematic of the thermal imager observing the both the Magnox and THORP container. The imager is \SI{18}{\degree} from the plane normal in both cases. There is a \SI{450}{\milli\metre} separation to the Magnox container and \SI{470}{\milli\metre} between the imager and THORP container.  The image shows the Magnox container prior to inspection.}
    \label{fig:snmLaboratoryContainerImagerConfiguration}
\end{figure}

During the measurements, the heater mats and two control thermocouples affixed to each heater surface were controlled using bespoke Eurotherm controllers. The ten monitoring thermocouples were located as shown in Figure~\ref{fig:snmLaboratoryContainerSchematic}, each was monitored by the Fluke 1586A Super-DAQ Precision Temperature Scanner. The thermal imager was mounted \SI{18}{\degree} to the normal from the base plane and at a distance of \SI{450}{\milli\metre} and \SI{470}{\milli\metre} from the front of the lens, to the Magnox and THORP containers respectively. This is sown in Figure~\ref{fig:snmLaboratoryContainerImagerConfiguration}.

During the measurements a shroud was placed over the imager and container to reduce background reflection variations and natural convection.
\clearpage

\subsection{Store setup} \label{subsec:storeSetup}
The following section details the necessary preparatory activities and experimental set up for the store measurements. Namely: thermal imager calibration; instrumented container set up; and set up for store measurements at the JFNL test store. The thermal imager calibration is described in Section~\ref{sec:thermal_imager_calibration}, this calibration included the mirror setup so the calibrated apparent radiance temperature accounted for the non-unity reflectivity of the mirrors.

\subsubsection{SNM container instrumentation}\label{subsubsec:containerInstrumentation}
The two special nuclear material container types provided by Sellafield comprising of both Magnox and THORP designs (summarised in Figure~\ref{fig:snmContainerComparison}) were each inspected using the pair of thermal imagers (one of which was detailed in the uncooled thermal imager case study in Section~\ref{subsec:case_study_thermal_imager}). 

For the THORP measurements, the appropriate inner, intermediate and outer containers were used together. For the Magnox measurements, the Magnox outer, THORP intermediate and THORP inner containers were used together; note that this configuration differs from the laboratory configuration described in Section~\ref{subsec:laboratorySetup}.

\begin{figure}[H]
\centering
	\begin{subfigure}[t]{0.49\textwidth}
	\centering
	\includegraphics[height=6cm,keepaspectratio]{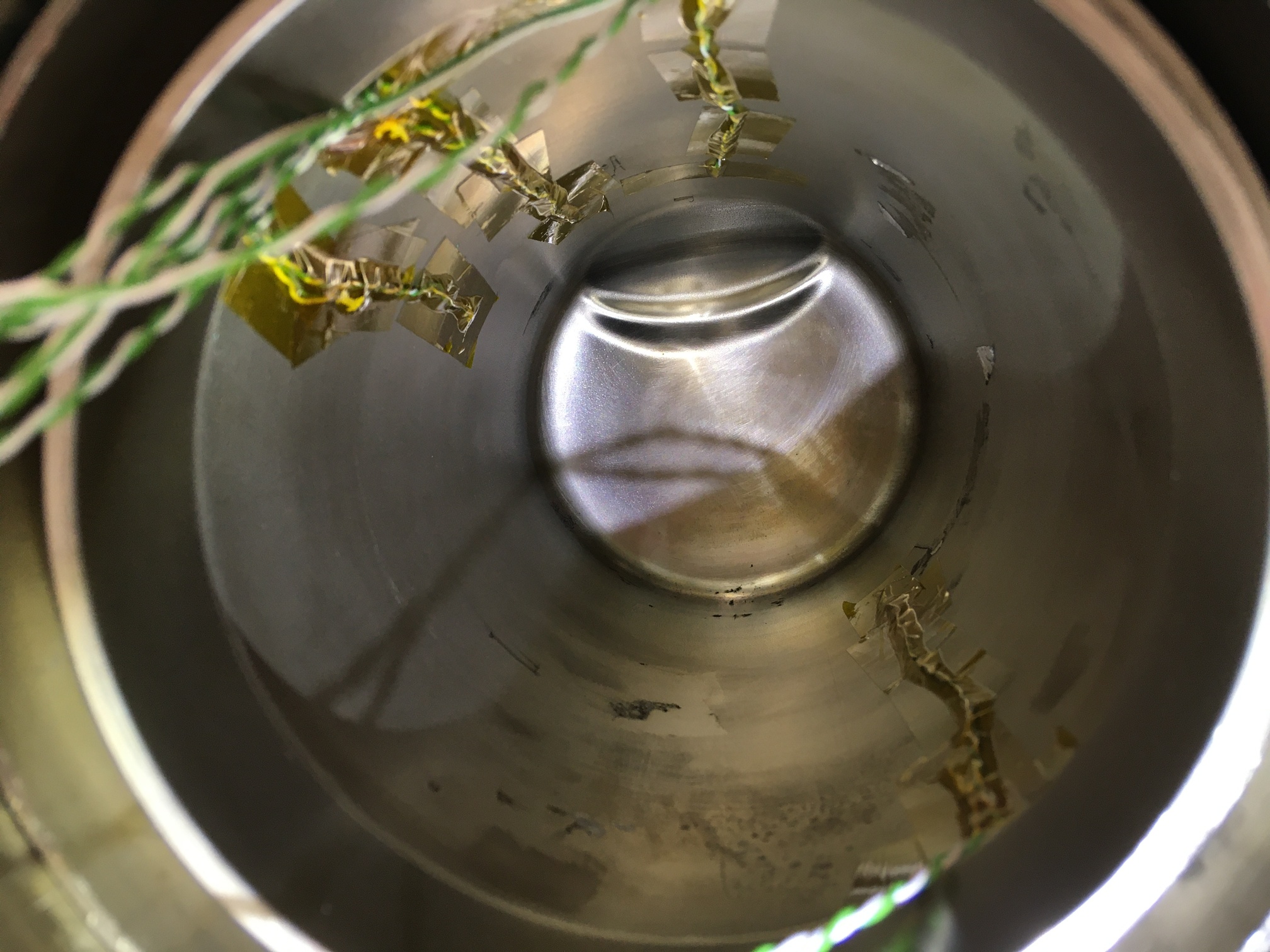}
	\caption{Internal container with monitoring thermocouples affixed.}
	\label{fig:snmStoreContainerInternalThermocouples}
	\end{subfigure}
\hfill
	\begin{subfigure}[t]{0.49\textwidth}
	\centering
	\includegraphics[height=6cm,keepaspectratio]{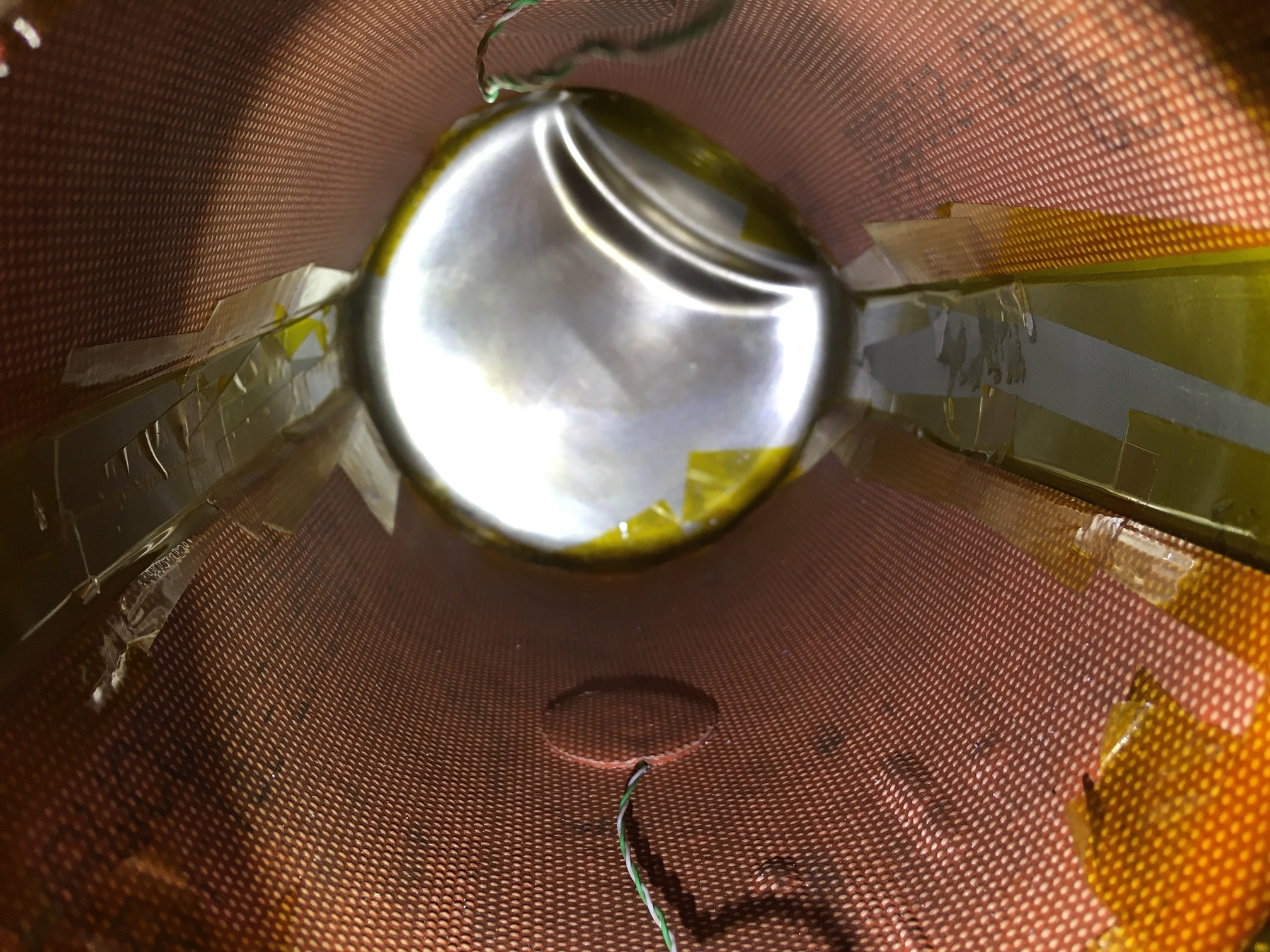}
	\caption{Internal container with heaters mounted above the monitoring thermocouples.}
	\label{fig:snmStoreContainerInternalHeaters}
	\end{subfigure}
	\caption{The thermocouple and heater instrumentation of the inner container. Seven thermocouples were located close together both radially and longitudinally distributed to correlate with the outer container configuration. A further two thermocouples were located opposite to the prior thermocouples to accommodate thermal model comparisons. The heater mats were pressed against the inside wall and the monitoring thermocouples facing the centre.}
\label{fig:snmStoreContainerInternal}
\end{figure}

\begin{figure}[H]
  \centering
  \includegraphics[width=0.9\textwidth,keepaspectratio]{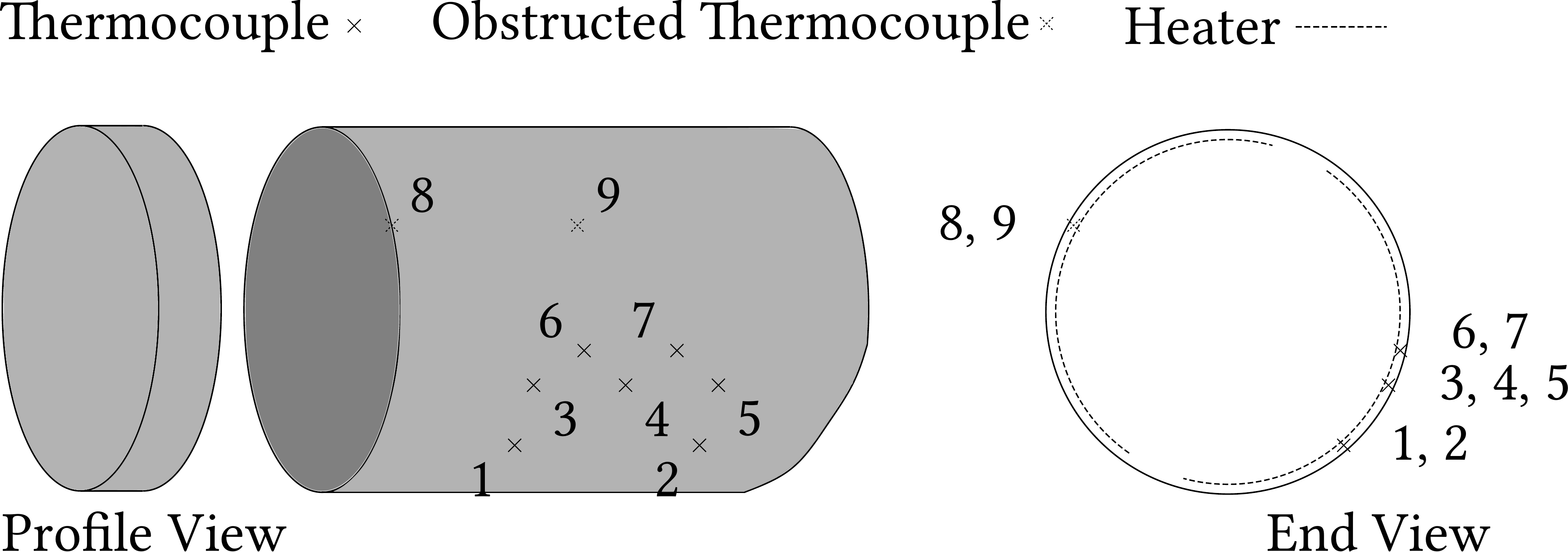}
  \caption{A schematic of the internal container instrumentation. The dashed lines represent the heaters and the crosses indicate the location of the nine thermocouples and their respective identification numbers.  The container is shown in its profile perspective and the end view corresponds to the images in Figure~\ref{fig:snmStoreContainerInternal}.}
  \label{fig:snmStoreContainerInternalSchematic}
\end{figure}

The internal container was instrumented with nine monitoring thermocouples (type K) and two heater mats, these are shown in Figure~\ref{fig:snmStoreContainerInternal}. A schematic of these thermocouple locations is shown in Figure~\ref{fig:snmStoreContainerInternalSchematic}. The thermocouples were directly fixed to the surface using adhesive tape.

\clearpage

\begin{figure}[H]
  \centering
  \includegraphics[width=\textwidth,keepaspectratio,trim={0cm 13cm 0cm 16cm}, clip]{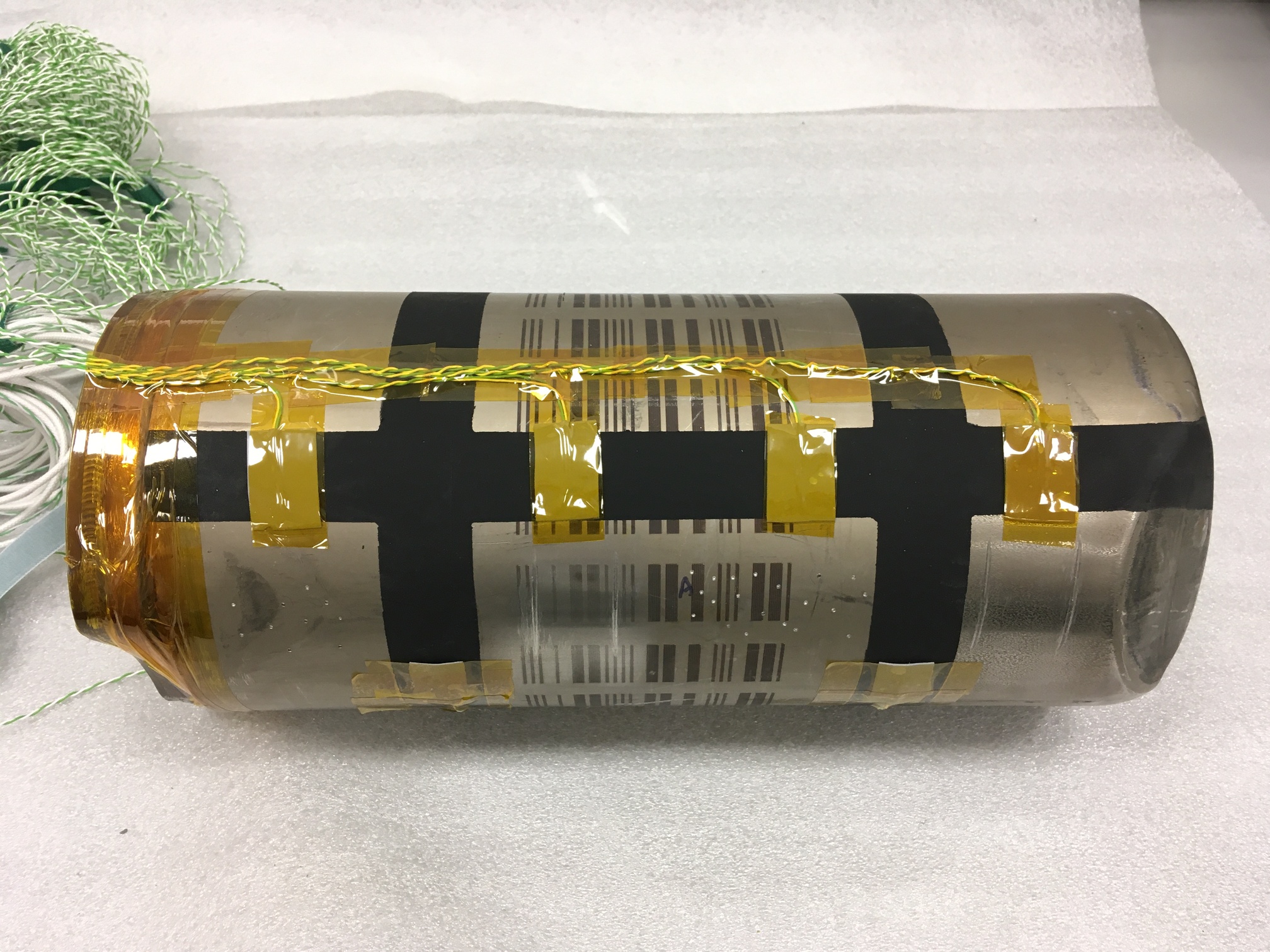}
  \caption{The internal container within the THORP intermediate and outer. The instrumentation cables were passed through openings on both container lids and the outer lid was held in place using high-temperature adhesive tape. The external surface was coated with high emissivity paint and thermocouples positioned across the surface.}
  \label{fig:snmStoreContainerExternal}
\end{figure}

The two heater mats were slightly mis-aligned with the central axis of the container such that thermocouples 1 through 7 were in closer contact with the heater. A Laird Technologies Tflex HR620 thermal gap pad was used (\SI{0.5}{\milli\metre} thick), this was cut to the size of the heater mat and placed between the heater mat and internal container surface. To minimise steel shot moving between the heater mat and container surface, the edges of the heater mat were sealed with Kapton adhesive tape. The internal container was then filled with the steel shot and sand mixture originally packaged within the unit and the instrumentation cables directed through the appropriate lids (completed shown in Figure~\ref{fig:snmStoreContainerExternal}).

Following the THORP container measurements, the intermediate container was removed and placed within the outer Magnox container. This was sealed with high-temperature adhesive tape, then placed on the inspection rail.

\begin{figure}[H]
  \centering
  \includegraphics[width=0.8\textwidth,keepaspectratio]{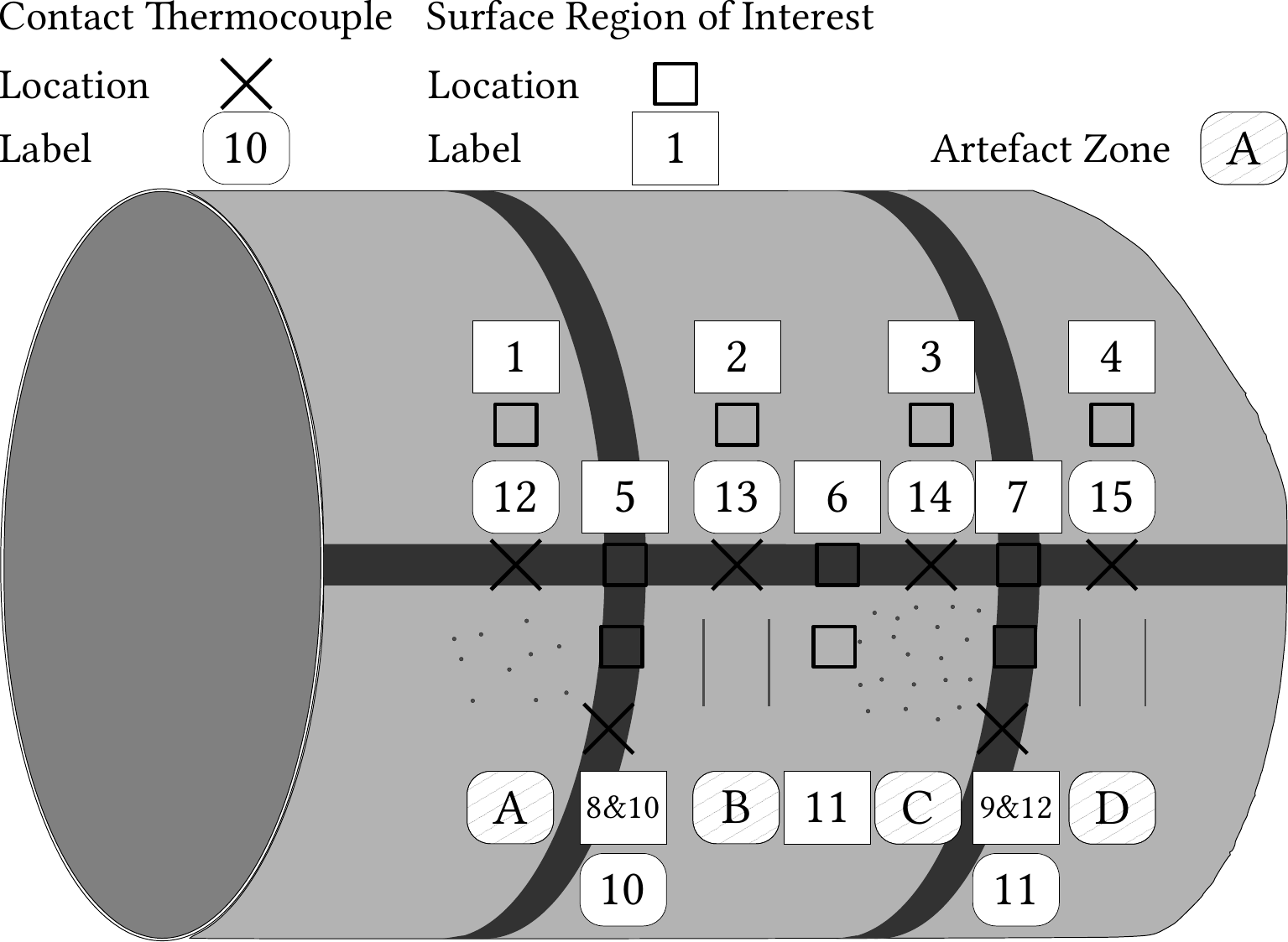}
  \caption{A schematic of the external container instrumentation. The dark regions represent the coating, the crosses indicate the location of the thermocouples and squares represent the thermal imager regions of interest, both with their respective identification numbers.}
  \label{fig:snmStoreContainerExternalSchematic}
\end{figure}

A schematic of the external surface and the respective thermocouple and region of interest identifiers is shown in Figure~\ref{fig:snmStoreContainerExternalSchematic}. A set of artefacts are denoted by the zones A through D as shown in Figure~\ref{fig:snmStoreContainerExternalSchematic} and labelled by the rounded-hatched boxes. Zone A consist of ten pits, whilst zone C comprises twenty pits. Two sets of scratches were manufactured on the surface in zones B and D, each containing two scratches. 

Prior to assembly the two outer containers were sent to Goldburn Finishers Ltd (Lindford, Hampshire) and a coating of Nextel velvet 811-21 was applied to the containers according to the diagram in Figure~\ref{fig:snmStoreContainerExternalSchematic} \cite{ref:emissivity_data_coatings}. The coating was stated to be \SI{80}{\micro\metre} thick and the widths of the strips were \SI{20}{\milli\metre}, the finished coating is shown in Figure~\ref{fig:snmStoreContainerExternal}.

\subsubsection{Store configuration}\label{subsubsec:storeConfiguration}
The James Fisher Nuclear Limited test store is a to scale mock up (rails, access ports) of the active store facility that allows inactive facility testing of new condition monitoring and inspection technologies. Whilst a number of different container channels were available, a single channel was used for testing the viability and feasibility of thermal imaging.

On the inspection channel a number of non-instrumented containers were placed, these containers operated as additional observation containers with no thermal management or content; the inspection container under test was placed toward the front of the channel. During the measurement capture the facility was used in a closed state, the surrounding housing was closed up similar to the laboratory work with an equivalent -- albeit larger -- shroud to that in the laboratory work.

\begin{figure}[H]
  \centering
  \begin{subfigure}[t]{\textwidth}
	\centering
  	\includegraphics[width=\textwidth,keepaspectratio]{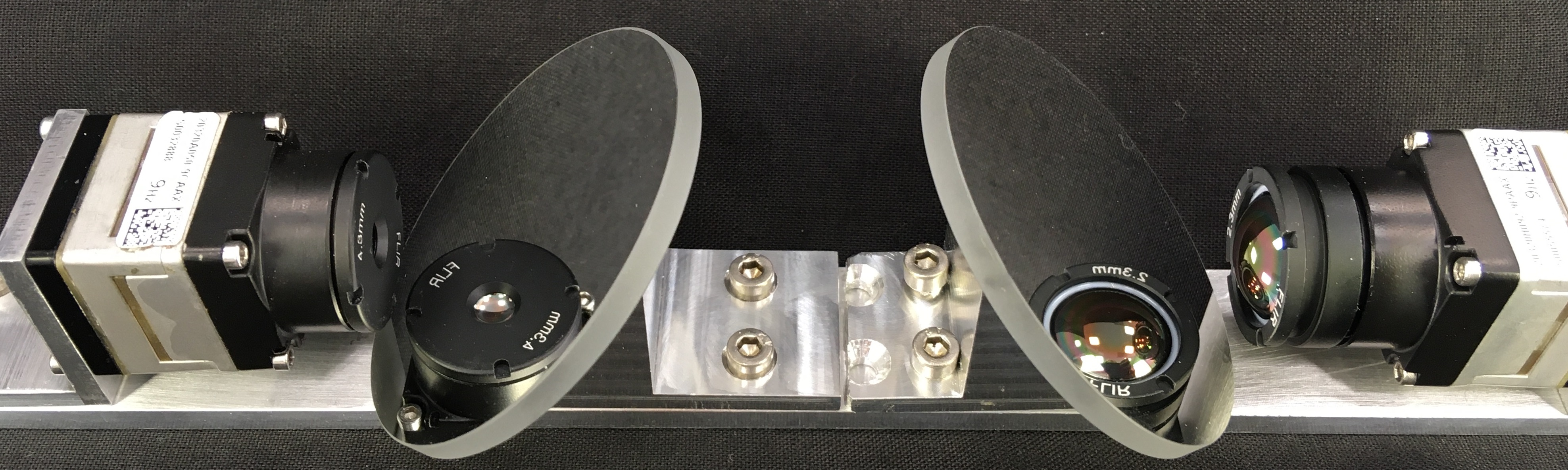}
	\caption{The optical carriage to mount a pair of uncooled thermal imagers during the store measurements.}
  	\label{fig:snmStoreContainerInspectionCarriage}
  \end{subfigure}
  \vspace{\floatsep}
  \begin{subfigure}[t]{\textwidth}
	\centering
  	\includegraphics[width=\textwidth,keepaspectratio]{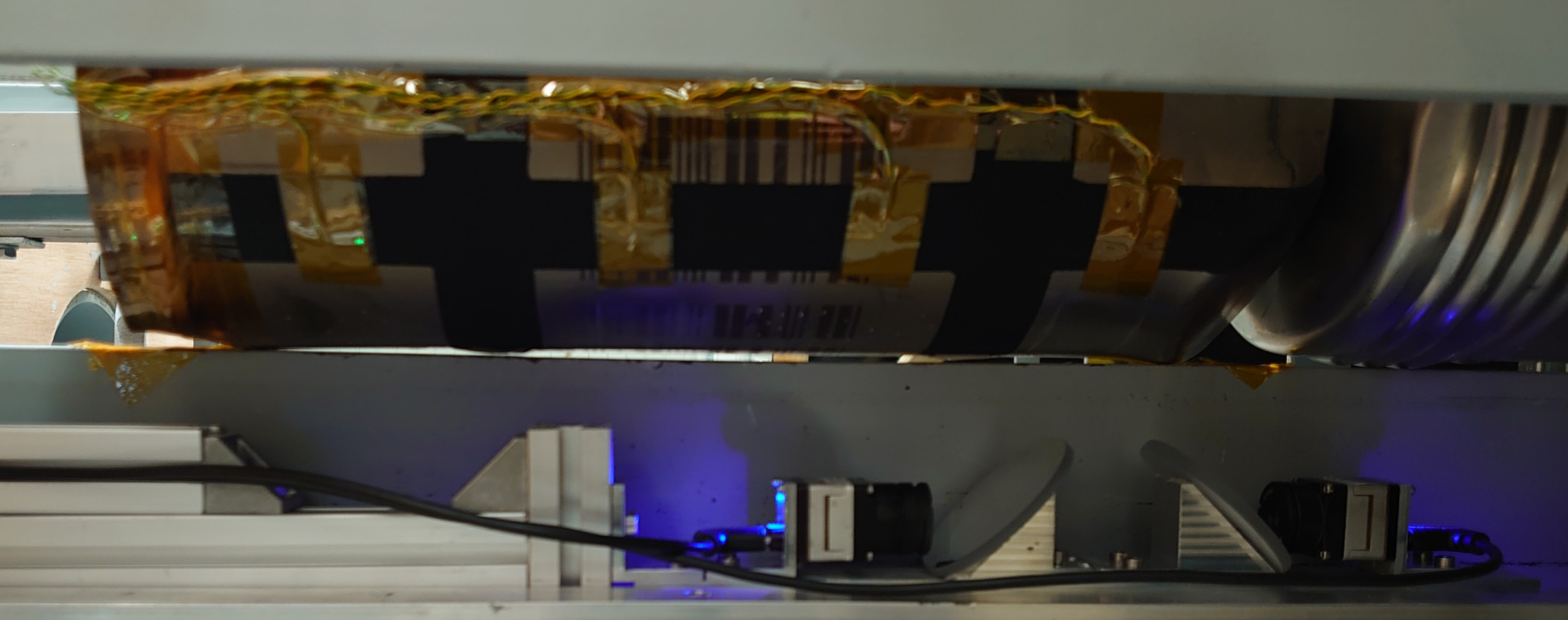}
  	\caption{Image of the THORP container installed on the inspection channel, viewed from the inspection side with the thermal imaging inspection trolley underneath. Imager on the left is the WFOV (S/N: S0055574) and on the right is the NFOV (S/N: S0032888).}
  	\label{fig:snmStoreContainerInspectionPosition}
  \end{subfigure}
\caption{Both a) the inspection carriage and b) the carriage in-situ below a container.}
  \label{fig:snmStoreContainerInspection}
\end{figure}

The instrumented container for inspection was orientated on the rail to provide a view of the regions of interest. The thermal imaging inspection trolley (refer to \ref{fig:snmStoreContainerInspectionCarriage}) was mounted to view from directly under the instrumented container and able to translate along the inspection rail underneath the container (refer to \ref{fig:snmStoreContainerInspectionPosition}).

For each measurement temperature setpoint the thermal imaging inspection trolley was manually translated along the rail underneath the instrumented container from the front to the back at fixed capture positions along the length of the container. For each capture position along the \SI{30}{\milli\metre} step, images were recorded at \SI{10}{\second} intervals for \SI{1}{\minute} from each imager using the same software used in the thermal imager calibration.

\clearpage
\newpage
\section{Laboratory container assessment}\label{sec:laboratoryContainerAssessment}

The containers were assembled as described in Section~\ref{subsec:laboratorySetup}, within this section a description of the system temperature characteristics will be presented; the results from the thermal measurements will be summarised; followed by the uncertainty budget and a discussion.  

The objectives of this activity were:

\begin{itemize}
  \item To set up and validate the thermal operation of each container
  \item To determine the radiance temperature of each container as a function of heater power
  \item To calculate the estimated surface temperature measurement uncertainty for a given thermal imager
\end{itemize}

The thermocouples used during this assessment were not calibrated and were used with manufacturer tolerances, this was noted accordingly in the uncertainty budget.

\subsection{Thermal characterisation}\label{subsec:laboratoryContainerThermalCharacterisation}

\begin{figure}[H]
  \centering
  \includegraphics[width=0.75\textwidth,keepaspectratio]{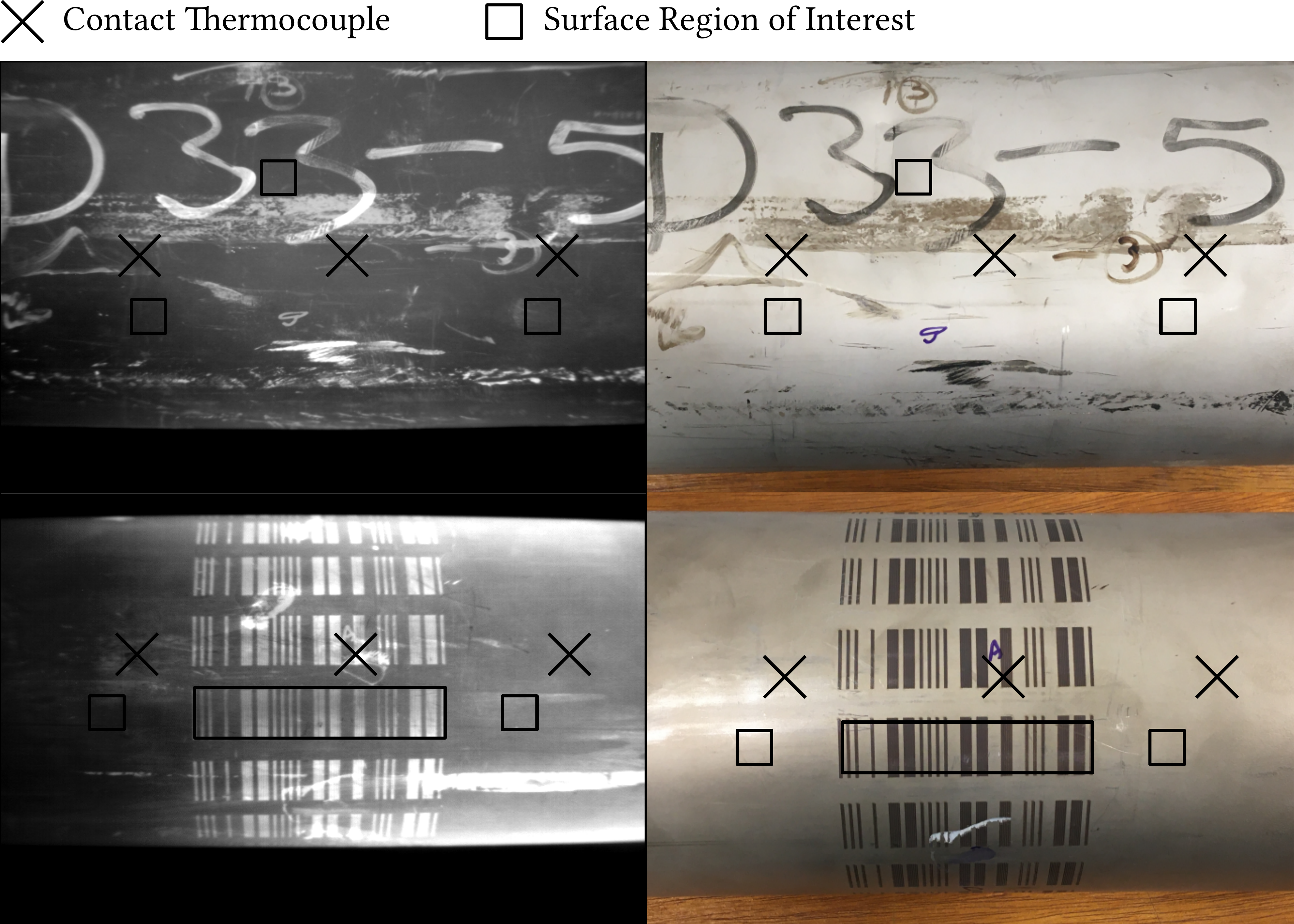}
  \caption{A comparison between thermal (left) and visual (right) images for the Magnox (top) and THORP (bottom) containers. Locations of the contact thermocouples below the surface in the internal container are denoted by crosses. The square regions of interest are shown beside these thermocouple locations.}
  \label{fig:snmLaboratoryContainerROIComparison}
\end{figure}

The containers and their temperature characteristics have been profiled during the container measurements at increasing heater temperature setpoints; the results from this validation are presented in the following section. The Region Of Interest (ROI) placement was arbitrary but attempted to provide a distributed coverage across the surface within the field of view of the thermal imaging system whilst minimising surface emissivity perturbations. Radiance temperature is the apparent radiance temperature with an emissivity correction applied for the respective room conditions and assumed emissivity value (as introduced in Section~\ref{sec:thermometer_calibration}). The container surface emissivity value was evaluated from the data presented in \cite{ref:steelEmissivity}. It is clear that this is likely an inadequate estimate due to the large emissivity variation caused by different surface finishes. Despite the clear emissivity difference between the barcode region of the THORP container compared to the surrounding surface, the same emissivity correction will be applied to the apparent radiance temperature. It is not simple to estimate the increase in emissivity introduced by this barcode, however it will be demonstrably clear from the results the effect from the emissivity increase. The ROI and estimated thermocouple location for both containers are shown in Figure~\ref{fig:snmLaboratoryContainerROIComparison}.

\clearpage

\begin{figure}[H]
\centering
	\begin{subfigure}[t]{0.475\textwidth}
	\centering
	\includegraphics[height=7cm,keepaspectratio]{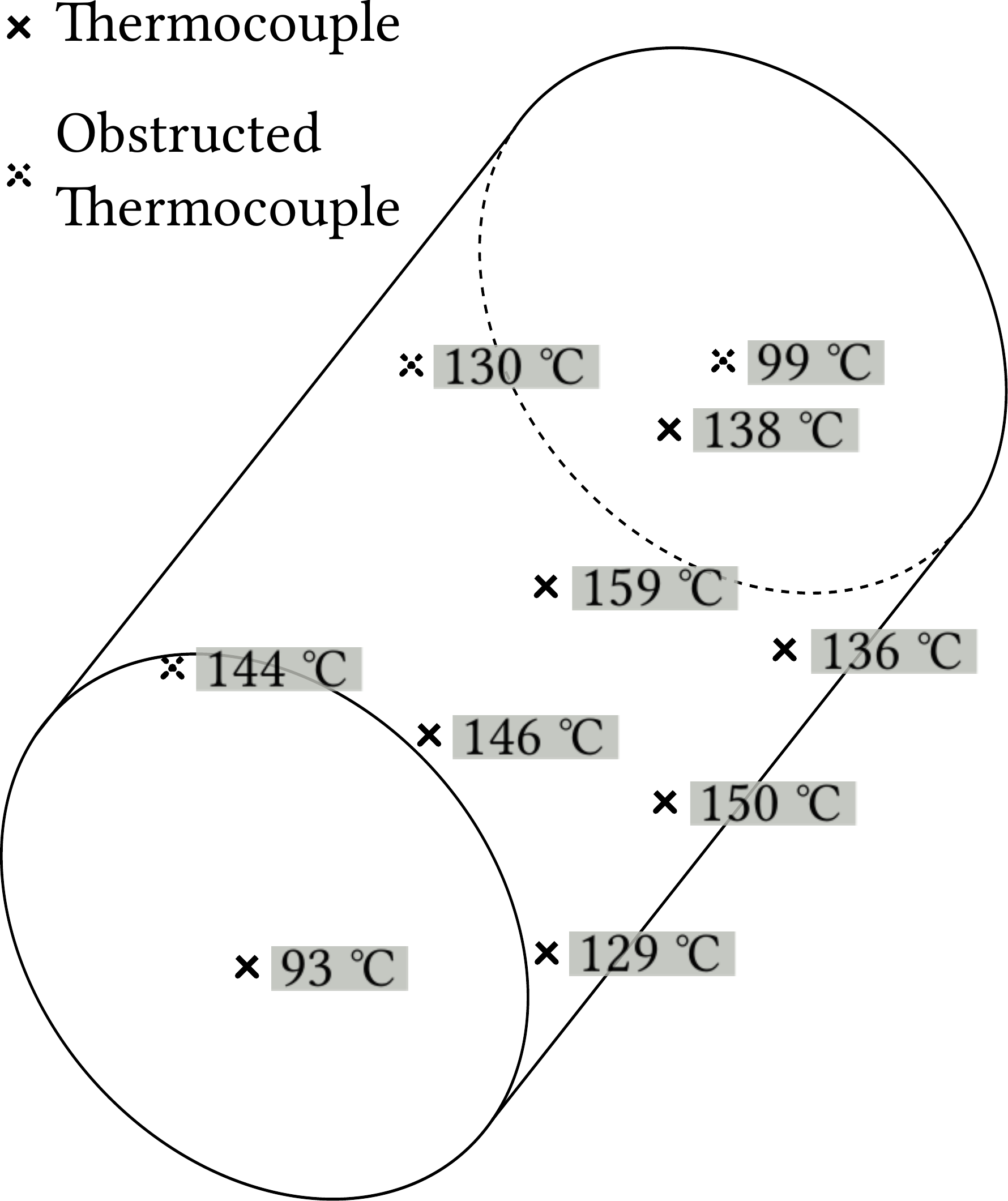}
	\caption{Magnox container.}
	\label{fig:snmLaboratoryContainerThermocoupleComparisonMagnox}
	\end{subfigure}
\hfill
	\begin{subfigure}[t]{0.475\textwidth}
	\centering
	\includegraphics[height=7cm,keepaspectratio]{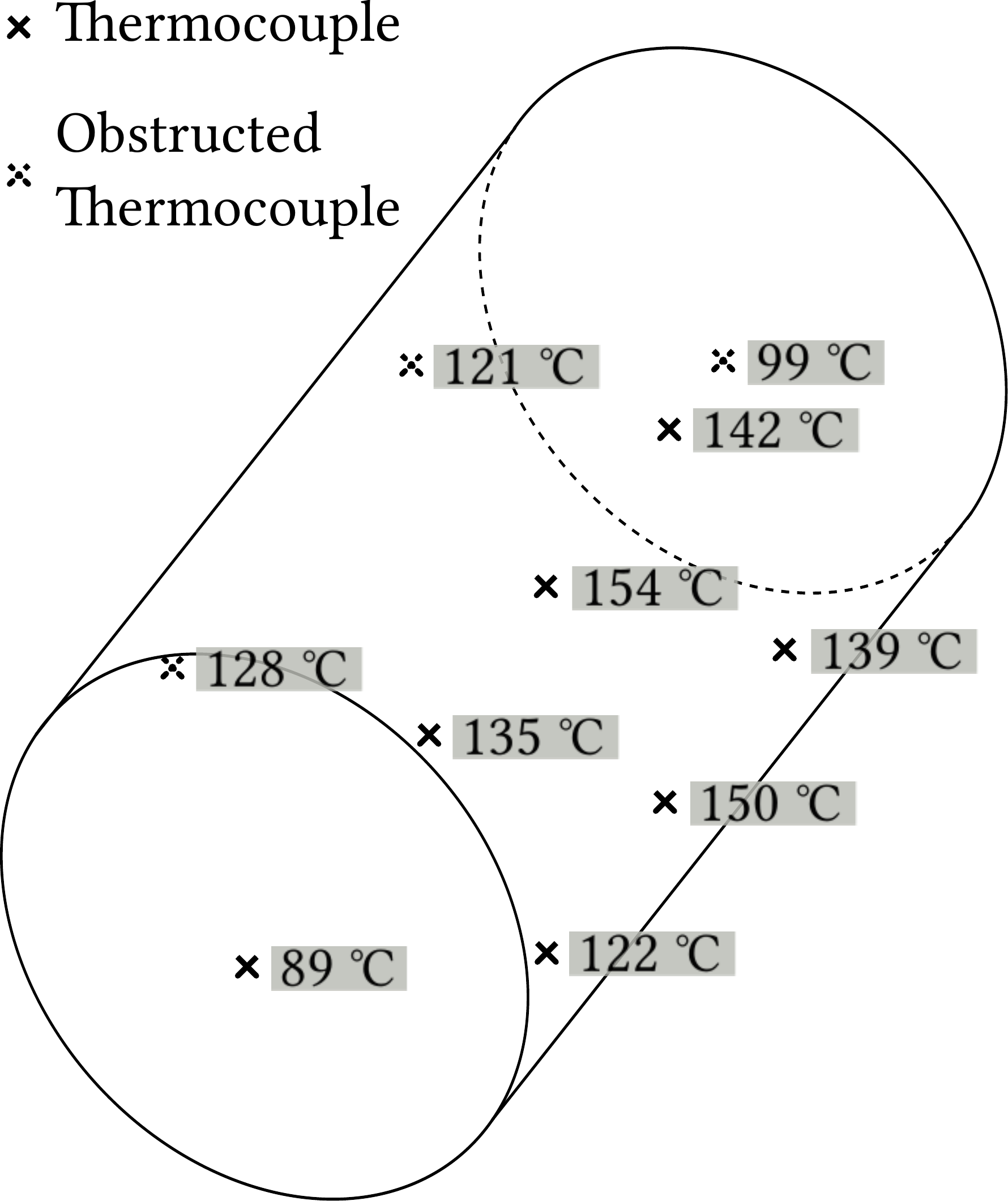}
	\caption{THORP container.}
	\label{fig:snmLaboratoryContainerThermocoupleComparisonTHORP}
	\end{subfigure}
\caption{The thermocouple measurements at the \SI{170}{\celsius} heater setpoint for the a) Magnox, and b) THORP containers. The open face of the containers is positioned to the left.}
\label{fig:snmLaboratoryContainerThermocoupleComparison}
\end{figure}

A visualisation of the thermocouple measurements at the \SI{170}{\celsius} heater setpoint is shown in Figure~\ref{fig:snmLaboratoryContainerThermocoupleComparison} for both containers. Here the thermocouples measure greater temperatures in the Magnox container, it was observed that the open end thermocouples measure relatively cooler than the corresponding thermocouples for the THORP container.  

\begin{figure}[H]
\centering
\includegraphics[width=0.90\textwidth,keepaspectratio]{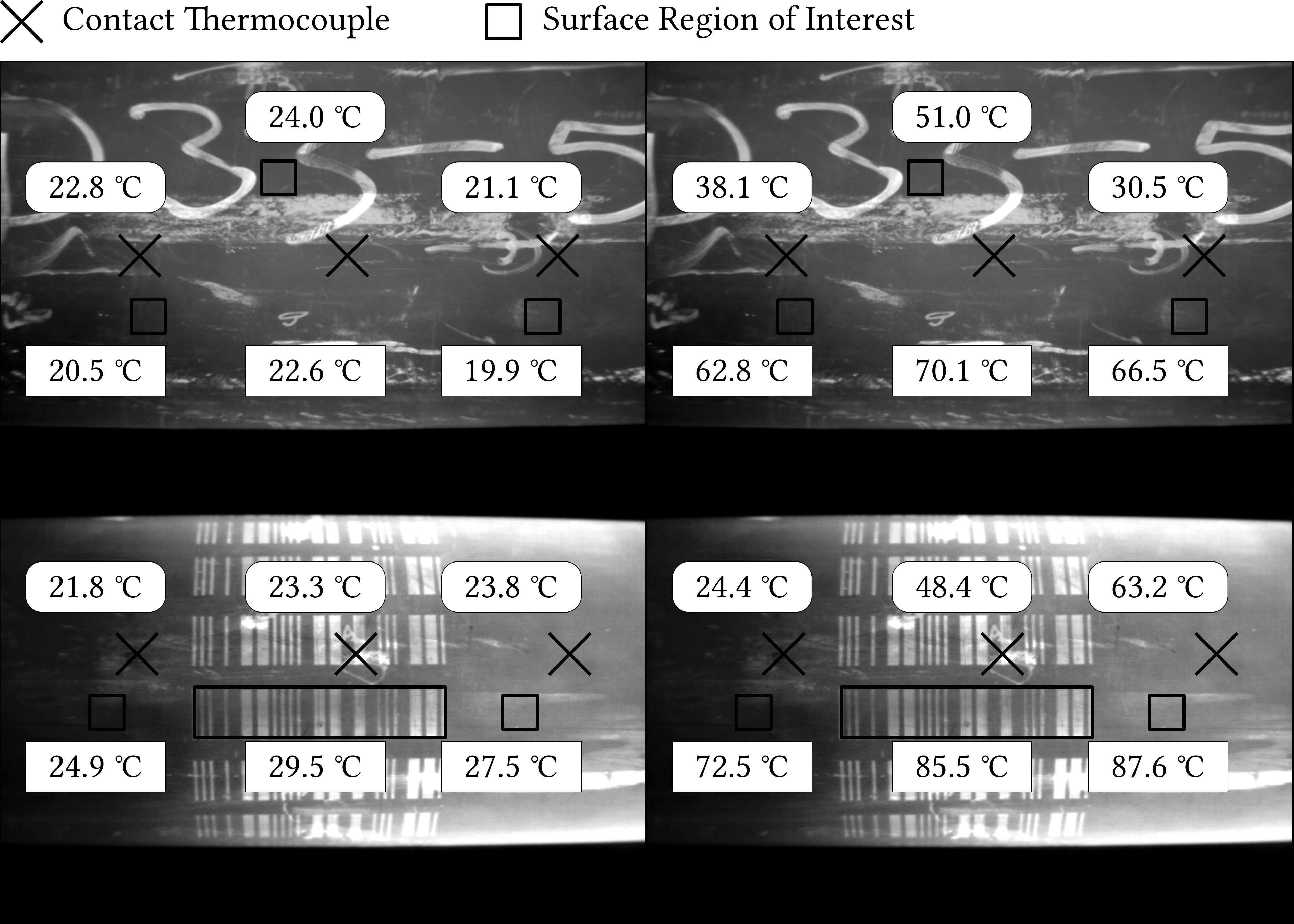}
	\caption{Comparison between the Magnox (top) and THORP (bottom) container measurements at \SI{30}{\celsius} (left) and \SI{170}{\celsius} (right). The radiance temperature measurements assume an emissivity of 0.30 and are denoted by the square boxes. The thermocouple temperature is the model-corrected thermocouple measurement introduced in Section~\ref{subsec:laboratoryContainerThermalModel} and is indicative of the surface temperature, these are denoted by the rounded boxes.}
\label{fig:snmLaboratoryContainerRadianceComparison}
\end{figure}

A visual comparison between the Magnox and THORP container, radiance temperature and thermocouple measurement at both \SI{30}{\celsius} and \SI{170}{\celsius} is shown in Figure~\ref{fig:snmLaboratoryContainerRadianceComparison}. The radiance temperature measurements (square boxes) assume a surface emissivity of 0.30 and were measured from the indicated regions of interest. Thermocouple measurements (rounded boxes) from within the internal container and are corrected for by the offset calculated from the thermal model (detailed in Section~\ref{subsec:laboratoryContainerThermalModel}).  

The ThermoCouple (TC) measurements from both container measurements indicated the temperature non-uniformity across both the radial and longitudinal axes of the containers.

Both containers demonstrated cooler edges (TC 0 and 9) at each temperature setpoint, so for both cases there were greater thermal losses (due to conductive and convective heat transfer) at the end faces. At \SI{30}{\celsius}, the end faces were \SI{1}{\celsius} warmer for the THORP container. These observations can be attributed to each setpoint measurement (albeit with increasing differences) and suggest that there was more uniform thermal losses from the THORP container as opposed to the Magnox container that is more non-uniform in temperature distribution.  

For both containers at each temperature, the mean temperature of TC 1, 4 and 6 that were affixed to the top edge were greater than that of the remaining circumference-mounted thermocouples. This indicates that the effects from internal convection heating the higher edges and the thermal conductive path through the rails each container rests on were non-negligible effects, resulting in nominally \SI{10}{\celsius} differences.  

\subsection{Thermal model}\label{subsec:laboratoryContainerThermalModel}

A thermal model was developed to support the assessment of the measurement data using COMSOL Multiphysics. It was not possible to compare temperatures measured internally with the thermocouples and externally with the thermal imager due to the thermal gradients throughout the containers. Therefore to compare the contact and non-contact temperatures measured during the laboratory container measurements, the contact thermometer measurements will introduce a correction based on the model calculations. 

The model used the geometry of the containers and their respective environments, known material properties and computationally solved conductive, convective and radiative heat transfer interactions at each temperature setpoint.

Using the developed thermal model, the surface temperature at both the external region of interest location (refer to Figure~\ref{fig:snmLaboratoryContainerROIComparison}) and equivalent thermocouple in close proximity was evaluated and difference calculated. For each container and region of interest, this correction was calculated and applied to the thermocouple measurements in the laboratory container measurements.

This computational model was conceptualised and developed by Sofia Korniliou, National Physical Laboratory \cite{ref:SNMTN3TN4,ref:SNMTN5}.

\subsection{Uncertainty analysis}\label{subsec:laboratoryContainerUncertaintyAnalysis}

As carried out with the plate measurements in Section~\ref{subsec:surfaceTemperatureUncertaintyAnalysis}, the same components have been considered for the surface temperature measurement uncertainty budgets during this assessment. An additional component was included to account for the {\em Model correction error} introduced to the thermocouple measurement. This was calculated from the difference between the model surface temperature and thermocouple model-corrected surface temperature, it was considered a rectangular distribution so a \(2\sqrt{3}\) divisor was used and the largest difference from the three ROIs was selected. An example complete budget is detailed in Table~\ref{tab:snmLaboratoryUncertaintyCompleteMagnox} and Table~\ref{tab:snmLaboratoryUncertaintyCompleteTHORP} for each container and the two temperature determination methods used: thermocouple and thermal imager.

\clearpage
\begin{table}[H]
  \centering
  \caption{An example complete uncertainty budget for temperature determination at the \SI{170}{\celsius} setpoint with the Magnox container and \SI{175}{\micro\second} integration time.}
  \vspace*{\floatsep}
  \begin{tabular}{L{5.0cm}M{1.5cm}M{1.5cm}M{1.5cm}M{1.5cm}}
    \toprule
	Source	&	\(u\)	&	Divisor	&	Sensitivity	&	\(U\) / \SI{}{\celsius} \\
    \cmidrule(lr){1-5}
	\multicolumn{5}{c}{Thermocouple} \\
    \cmidrule(lr){1-5}
	Tolerance	&	1.50	&	1.73	&	1.00	&	0.87	\\
	Thermometry bridge accuracy	&	0.20	&	1.00	&	1.00	&	0.20	\\
	Model correction error	&	33.03	&	3.46	&	1.00	&	9.54	\\
	Stability	&	0.03	&	1.00	&	1.00	&	0.03	\\
	\cmidrule(lr){2-5}
	\multicolumn{4}{r}{Expanded uncertainty (\(k=2\))}				&	{\bf 19.3} 	\\
	\cmidrule(lr){1-5}
	\multicolumn{5}{c}{Thermal Imager} \\
	\cmidrule(lr){1-5}
	Calibration	&	0.30	&	2.20	&	1.00	&	0.14	\\
	Housing temperature	&	0.00	&	1.00	&	1.00	&	0.00	\\
	Size-of-source effect	&	0.18	&	1.73	&	1.00	&	0.10	\\
	Distance effect	&	0.29	&	1.73	&	1.00	&	0.17	\\
	ROI non-uniformity	&	0.50	&	1.00	&	1.69	&	0.85	\\
	Emissivity interpretation	&	0.08	&	1.00	&	82.68	&	6.61	\\
	Ambient temperature	&	0.10	&	1.00	&	0.60	&	0.06	\\
	\cmidrule(lr){2-5}
	\multicolumn{4}{r}{Expanded uncertainty (\(k=2\))}				&	{\bf 13.3} 	\\
    \bottomrule
  \end{tabular}
  \label{tab:snmLaboratoryUncertaintyCompleteMagnox}
\end{table}

\begin{table}[H]
  \centering
  \caption{An example complete uncertainty budget for temperature determination at the \SI{170}{\celsius} setpoint with the THORP container and \SI{500}{\micro\second} integration time.}
  \vspace*{\floatsep}
  \begin{tabular}{L{5.0cm}M{1.5cm}M{1.5cm}M{1.5cm}M{1.5cm}}
    \toprule
	Source	&	\(u\)	&	Divisor	&	Sensitivity	&	\(U\) / \SI{}{\celsius} \\
    \cmidrule(lr){1-5}
	Tolerance	&	1.50	&	1.73	&	1.00	&	0.87	\\
	Thermometry bridge accuracy	&	0.20	&	1.00	&	1.00	&	0.20	\\
	Model correction error	&	35.38	&	3.46	&	1.00	&	10.21	\\
	Stability	&	0.03	&	1.00	&	1.00	&	0.03	\\
    \cmidrule(lr){2-5}
	\multicolumn{4}{r}{Expanded uncertainty (\(k=2\))}				&	{\bf 20.5} 	\\
    \cmidrule(lr){1-5}
	\multicolumn{5}{c}{Thermal Imager} \\
    \cmidrule(lr){1-5}
	Calibration	&	0.50	&	2.30	&	1.00	&	0.22	\\
	Housing temperature	&	0.04	&	1.00	&	1.00	&	0.04	\\
	Size-of-source effect	&	0.18	&	1.73	&	1.00	&	0.10	\\
	Distance effect	&	0.29	&	1.73	&	1.00	&	0.17	\\
	ROI non-uniformity	&	1.46	&	1.00	&	1.53	&	2.24	\\
	Emissivity interpretation	&	0.08	&	1.00	&	98.72	&	7.90	\\
	Ambient temperature	&	0.10	&	1.00	&	0.41	&	0.04	\\
    \cmidrule(lr){2-5}
	\multicolumn{4}{r}{Expanded uncertainty (\(k=2\))}				&	{\bf 16.4} 	\\
    \bottomrule
  \end{tabular}
  \label{tab:snmLaboratoryUncertaintyCompleteTHORP}
\end{table}

A summary of the measurement uncertainty at each Magnox temperature setpoint is presented in Table~\ref{tab:snmLaboratoryUncertaintySummaryMagnox}, the equivalent values for the THORP measurements are presented in Table~\ref{tab:snmLaboratoryUncertaintySummaryTHORP}. These describe the greatest uncertainty between the relevant integration times for each temperature setpoint.  

\begin{table}[H]
  \centering
  \caption{A summary of the instrumentation, application and combined measurement uncertainties for each temperature setpoint during the Magnox container measurements.}
  \vspace*{\floatsep}
  \begin{tabular}{ M{2.5cm}M{1.5cm}M{1.5cm}M{2.5cm}M{1.5cm}M{1.5cm}M{2.5cm} }
    \toprule
	&	\multicolumn{3}{c}{Thermocouple Uncertainty / \SI{}{\celsius}} & \multicolumn{3}{c}{Thermal Imager Uncertainty / \SI{}{\celsius}} \\
	Temperature Setpoint / \SI{}{\celsius} & Instrument & Application & Expanded (\( k= 2\)) & Instrument & Application & Expanded (\(k = 2\)) \\
    \cmidrule(lr){1-7}
	30	&	0.9	&	0.8	&	2.3		&	0.2	&	1.1	&	2.2			\\
	70	&	0.9	&	4.1	&	8.3		&	0.2	&	3.0	&	6.1			\\
	110	&	0.9	&	6.8	&	13.7	&	0.3	&	4.7	&	9.4			\\
	130	&	0.9	&	7.6	&	15.3	&	0.3	&	5.4	&	10.9		\\
	150	&	0.9	&	8.6	&	17.3	&	0.3	&	6.1	&	12.1		\\
	170	&	0.9	&	9.5	&	19.2	&	0.2	&	6.7	&	13.3		\\
    \bottomrule
  \end{tabular}
  \label{tab:snmLaboratoryUncertaintySummaryMagnox}
\end{table}

\begin{table}[H]
  \centering
  \caption{A summary of the instrumentation, application and combined measurement uncertainties for each temperature setpoint during the THORP container measurements.}
  \vspace*{\floatsep}
  \begin{tabular}{ M{2.5cm}M{1.5cm}M{1.5cm}M{2.5cm}M{1.5cm}M{1.5cm}M{2.5cm} }
    \toprule
	&	\multicolumn{3}{c}{Thermocouple Uncertainty / \SI{}{\celsius}} & \multicolumn{3}{c}{Thermal Imager Uncertainty / \SI{}{\celsius}} \\
	Temperature Setpoint / \SI{}{\celsius} & Instrument & Application & Expanded (\( k= 2\)) & Instrument & Application & Expanded (\(k = 2\)) \\
    \cmidrule(lr){1-7}
	30	&	0.9	&	0.5 	&	2.1		&	0.2	&	1.8	&	3.7		\\
	70	&	0.9	&	3.3 	&	6.8		&	0.2	&	4.4	&	8.8		\\
	110	&	0.9	&	5.7 	&	11.5	&	0.2	&	6.3	&	12.5	\\	
	130	&	0.9	&	7.2 	&	14.6	&	0.3	&	6.9	&	13.9	\\	
	150	&	0.9	&	8.7 	&	17.6	&	0.2	&	7.6	&	15.3	\\	
	170	&	0.9	&	10.2	&	20.5	&	0.3	&	8.2	&	16.4	\\	
    \bottomrule
  \end{tabular}
  \label{tab:snmLaboratoryUncertaintySummaryTHORP}
\end{table}

In comparison to the plate measurements (refer to Section~\ref{sec:assemblyAssessment}), the thermocouple uncertainties were nominally equivalent, but there is a clear variation in thermal imager application uncertainty. The principle component contributing to this variation is the sensitivity of the emissivity uncertainty.  Because this sensitivity function (refer to Eq.~\ref{eq:sensitivityEmissivity}) is sensitive to emissivity it is shown that for the lowest emissivity measurements (polished steel from uncoated plate measurements) this component is the greatest, whilst being the least for the Senotherm coated regions on the plate, and in between for the containers. The variation between the containers themselves is due to the both the spatial uniformity and emissivity sensitivity being greater for the THORP container.  

\subsection{Discussion}\label{subsec:laboratoryContainerDiscussion}
Evaluation of the laboratory container measurements was presented in the preceding sections. Section~\ref{subsec:laboratoryContainerThermalCharacterisation} broadly summarised the contact thermocouple and surface radiance temperatures; an excerpt of the surface temperature distributions is presented. The uncertainty analysis is detailed in Section~\ref{subsec:laboratoryContainerUncertaintyAnalysis}, the individual components for each system were identical to those from the plate budgets; an example budget is stated and the combined temperature measurement uncertainties are presented for the two container types.  

\begin{figure}[H]
  \centering
  \input{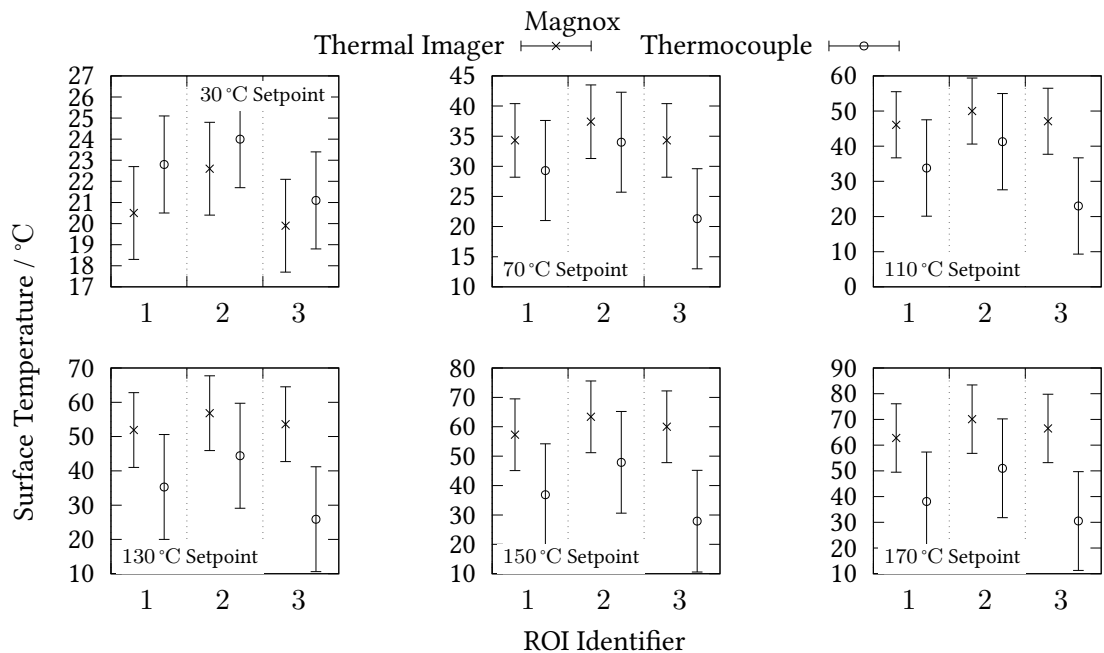}
  \caption{A comparison between the surface temperature determinations from the thermal imager and thermocouples with their equivalent uncertainties (\(k = 2\)) at each temperature setpoint. The ROIs correspond to those in Figure~\ref{fig:snmLaboratoryContainerROIComparison} during the Magnox container measurements. Note that the thermocouple measurements include a correction based on the temperature difference described by the thermal model in Section~\ref{subsec:laboratoryContainerThermalModel}.}
  \label{fig:snmLaboratoryContainerTemperatureComparisonMagnox}
\end{figure}

\begin{figure}[H]
  \centering
  \input{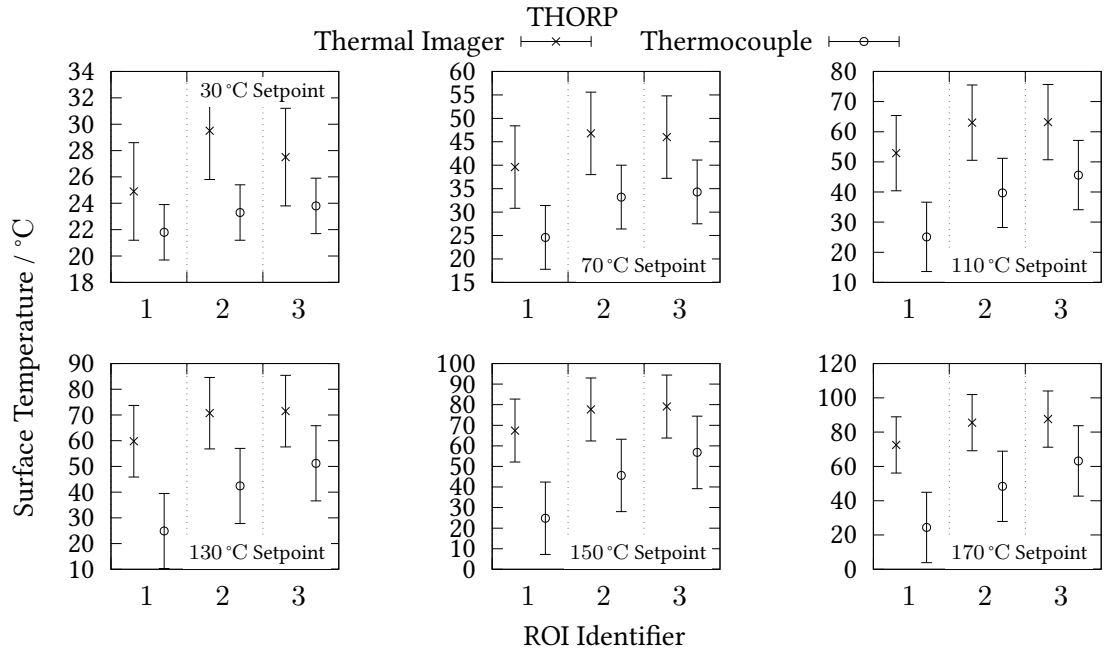}
  \caption{A comparison between the surface temperature determinations from the thermal imager and thermocouples with their equivalent uncertainties (\(k = 2\)) at each temperature setpoint. The ROIs correspond to those in Figure~\ref{fig:snmLaboratoryContainerROIComparison} during the THORP container measurements. Note that the thermocouple measurements include a correction based on the temperature difference described by the thermal model in Section~\ref{subsec:laboratoryContainerThermalModel}.}
  \label{fig:snmLaboratoryContainerTemperatureComparisonTHORP}
\end{figure}

As with the plate a direct comparison between the two temperature determination methods at three regions of interest in Figure~\ref{fig:snmLaboratoryContainerROIComparison} has been carried out. As shown through both the measurement temperatures in Section~\ref{subsec:laboratoryContainerThermalCharacterisation} there is a large variation from the thermocouple locations to the radiating surface temperature. To facilitate this comparison between methods, a correction to the thermocouple measurements based on the calculated temperature differences using a thermal model was applied. The thermocouples used for ROI 1 through ROI 3 were TC 6, TC 4 and TC 1 respectively. This correction incurs a component of uncertainty from any discrepancy between the computational and experimental thermal descriptions and was appropriately considered. The comparison plots for the Magnox and THORP containers are shown in Figure~\ref{fig:snmLaboratoryContainerTemperatureComparisonMagnox} and Figure~\ref{fig:snmLaboratoryContainerTemperatureComparisonTHORP} respectively.  

\newpage

The three Magnox ROIs used were distributed both longitudinally and radially; one through three were positioned from the lid side to the base end. At \SI{30}{\celsius} both radiance and thermocouple values were in good agreement. However the thermocouple measurements became increasingly disparate from the remaining measurements. It was observed from the excerpt of thermocouple measurements in Figure~\ref{fig:snmLaboratoryContainerThermocoupleComparison} that the distribution shown is consistent with Figure~\ref{fig:snmLaboratoryContainerTemperatureComparisonMagnox}: that the container is cooler at the edges but with greater coolth towards the base end. The model correction applied is nominally longitudinally invariant, but this data suggests that the correction applied is too great at ROIs 1 and 3.

As hypothesised in Section~\ref{subsec:laboratoryContainerThermalCharacterisation}, during the THORP container measurements the emissivity correction applied to ROI 2 was too great; this is shown by the radiance temperature being consistently higher than the other ROIs. And so the emissivity of the barcode region is measurably greater than the surrounding surface. As with the Magnox container the thermocouple measurements were not suitably comparable to the radiance temperatures. In this configuration the correction applied to the thermocouple measurements does vary along the longitudinal axis. Therefore, there is likely greater disparity, compared to the Magnox container, between the experimental and model thermal descriptions, which is not sufficiently described by this model.

The difference in configuration from the Magnox to THORP container, where there was increased heat transfer to the radiating surface due to larger diameter, s shown through the temperature measurements. Compared to the plate measurements where there is nominally \SI{3}{\milli\metre} between the base of the thermometer channel and radiating surface, for the Magnox container there is at most a \SI{30}{\milli\metre} separation through varying sizes of air spaces. The mean thermocouple temperature is lower due to more heat transfer to the external environment, the increased temperature range shows greater variation across the internal container.  Additionally, the surface radiance temperature is higher because the thermal conduction is greater in this configuration.  

\subsection{Summary}\label{subsec:laboratoryContainerSummary}

From the temperature measurements of the container, it is apparent that there is a clear distinction between the thermal properties measured for the two container configurations. It should be noted that the Magnox configuration is not representative of typical in store Magnox containers and would typically use a low-density polyethylene container surrounding an aluminium inner container of a different geometry. This arose from the containers available. Despite this the laboratory measurements indicated a cooler temperature at the edges of the Magnox container as opposed to a linear increase towards the base of the THORP containers. In the experimental setup used, the inner container was hotter during the Magnox measurements due to reduced thermal losses, whilst the THORP outer container surface was hotter than the Magnox container. The laser etched material in the barcode region demonstrated a measurably greater temperature than the surrounding surface. 

In general the radiance and thermocouple temperatures could not be sufficiently compared due to the thermal model correction implemented.

\clearpage
\newpage

\section{Store container assessment}\label{sec:storeContainerAssessment}

During the previous section, the thermal characteristics of two special nuclear material containers were evaluated using a thermal imager under laboratory conditions. The next phase of work was to carry out a modified test campaign using the James Fisher Nuclear Ltd replica test store to assess containers in an environment equivalent to an active store. The two containers evaluated within this trial were the Magnox and THORP SNM test containers as described during the setup in Section~\ref{subsec:storeSetup}.

The objectives of this test campaign were:

\begin{itemize}
  \item Demonstrate suitability of an uncooled microbolometer in an inactive store
  \item Implement surface temperature determination in a mock store environment
  \item To assess whether the difference between radiance temperature and externally mounted thermocouples is less than the estimated measurement uncertainty at a confidence level of \SI{95}{\percent} at each setpoint temperature
\end{itemize}

The thermal imagers were successfully calibrated, deployed within JFNL test store, and an evaluation of surface temperatures and comparison to contact thermometers has been completed. The results are given in this section. The following section details the JFNL test store measurement campaign results for the thermal characterisation of the Magnox and THORP instrumented containers.

The measurement data capture during this trial was carried out by Sofia Korniliou and Rob Simpson, National Physical Laboratory.

\subsection{Thermal characterisation}\label{subsec:storeContainerThermalCharacterisation}
The following section details the measurement campaign results for the thermal characterisation of the Magnox and THORP instrumented containers.

\subsubsection{Thermocouple evaluation}\label{subsubsec:storeContainerThermocoupleEvaluation}
Prior to use the thermocouples had been calibrated over the range from \SIrange{5}{200}{\celsius} and the deviation from the International Temperature Scale of 1990 reported. For each thermocouple a calibration correction to the measured temperature was applied either through a polynomial fit or fixed offset, depending on the function of deviation with temperature \cite{ref:thermocoupleCertificate2020120016,ref:thermocoupleCertificate2021030425}.

\begin{figure}[H]
  \centering
  \includegraphics[width=0.95\textwidth,keepaspectratio]{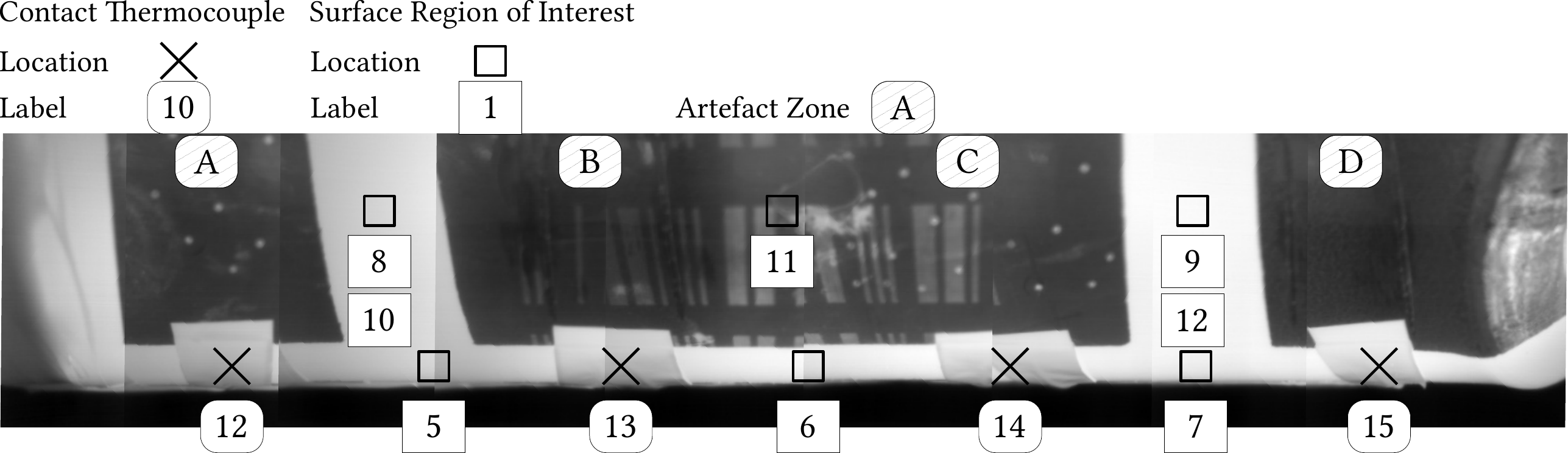}
  \caption{Collated qualitative representation from the THORP \SI{170}{\celsius} NFOV distortion-corrected measurements. The indicated thermocouples and regions of interest are those denoted in Figure~\ref{fig:snmStoreContainerExternalSchematic}.}
  \label{fig:snmStoreContainerROIDefinition}
\end{figure}

For each surface temperature region of interest (shown in Figure~\ref{fig:snmStoreContainerExternalSchematic}) the adjacent thermocouples used as the comparator were sampled. The regions of interest and respective thermocouples are presented in Table~\ref{tab:storeROIDefinition} and visually overlaid on the thermal image data in Figure~\ref{fig:snmStoreContainerROIDefinition}.

\begin{table}[H]
  \renewcommand{\arraystretch}{0.75}
  \centering
  \caption{The regions of interest (refer to Figure~\ref{fig:snmStoreContainerExternalSchematic}) and the respective thermocouples used to assess the reference surface temperature measurement. The size of the ROI is shown for each imager.}
  \vspace*{\floatsep}
  \begin{tabular}{M{2.4cm}M{2.4cm}M{2.4cm}M{2.4cm}}
    \toprule
	ROI	&	Thermocouple Comparators	&	NFOV ROI Size / \SI{}{\px}	&	WFOV ROI Size / \SI{}{\px}	\\
    \cmidrule(lr){1-4}
	1	&	12				&	15	&	30	\\
	2	&	13				&	15	&	30	\\
	3	&	14				&	15	&	30	\\
	4	&	15				&	15	&	30	\\
	5	&	12, 13			&	15	&	30	\\
	6	&	13, 14			&	15	&	30	\\
	7	&	14, 15			&	15	&	30	\\
	8	&	10				&	15	&	30	\\
	9	&	11				&	15	&	30	\\
	10	&	10				&	50	&	120	\\
	11	&	10, 11, 13, 14	&	15	&	30	\\
	12	&	11				&	50	&	120	\\
    \bottomrule
  \end{tabular}
  \label{tab:storeROIDefinition}
\end{table}

\subsubsection{Radiance temperature evaluation}\label{subsubsec:storeContainerRadianceEvaluation}
Using the calibration data from the respective thermal imager calibrations, a calibrated apparent radiance temperature was calculated from the Digital Level (DL) measurement and mean FPA temperature during each measurement set.

\paragraph{Region of interest evaluation}\label{para:regionOfInterestEvaluation}
For each measurement set, the surface ROIs were identified for each respective stopping position with respect to the regions shown in Figure~\ref{fig:snmStoreContainerExternalSchematic}. Due to the geometry of the storage rail and distance between imager and container, it was not possible to observe the full ROI set proposed in the campaign preparations. Notably ROI one through to four were not observable and so ROI eleven was added to represent an uncoated region.  Due to the large area of the FOV at ROI eight and nine, an additional pair of ROIs were added (ten and twelve) with a larger ROI size. This enabled the evaluation of the effect from ROI size on temperature determination, especially due to larger spatial temperature distribution.

The raw \texttt{tiff} thermal imaging files were evaluated using GNU Octave (version 5.2.0) and the digital level values from each ROI were exported to individual \texttt{csv} files. The respective ROI files were assessed for each data set, reporting the mean, temporal standard deviation (standard deviation of each mean from each extracted ROI) and spatial standard deviation (mean of the standard deviations from each extracted ROI). This exercise was repeated for each thermal imager and temperature setpoint for all observable regions of interest.

\paragraph{Calibration interpretation of measurements}\label{para:calibrationInterpretation}
A challenge encountered at this stage was the disparity between FPA temperature during the calibration and during each measurement set within the test store facility. Due to the sensitivity of the signal transfer function shown in Eq.~\ref{eq:transfer_function}, changes to the FPA temperature beyond that experienced during the calibration process would have a large impact on temperature measurement accuracy. To illustrate this, the mean FPA temperatures at each set are detailed in Table~\ref{tab:storeContainerFPATemperatures} alongside the mean value the calibration fit was based on.

\begin{table}[h]
  \renewcommand{\arraystretch}{0.75}
  \centering
  \caption{FPA temperatures during each of the measurements sets compared to those used to build the calibration transfer function (Eq.~\ref{eq:transfer_function}). Note measurements were not carried out at \SI{30}{\celsius} for the Magnox container.}
  \vspace*{\floatsep}
  \begin{tabular}{lllM{5cm}}
    \toprule
	Container	&	Setpoint Temperature / \SI{}{\celsius}	&	Imager	&	Mean FPA Temperature / \SI{}{\celsius}	\\
    \cmidrule(lr){1-4}
								&	\multirow{2}{3.0cm}{Calibration}	&	NFOV	&	32.9	\\
								&										&	WFOV	&	34.9	\\
	\cmidrule(lr){1-4}
	\multirow{6}{2.0cm}{THORP}	&	\multirow{2}{3.0cm}{30}				&	NFOV	&	26.0	\\
								&										&	WFOV	&	27.8	\\
								&	\multirow{2}{3.0cm}{110}			&	NFOV	&	29.5	\\
								&										&	WFOV	&	31.2	\\
								&	\multirow{2}{3.0cm}{170}			&	NFOV	&	31.7	\\
								&										&	WFOV	&	33.5	\\
								\cmidrule(lr){2-4}
	\multirow{6}{2.0cm}{Magnox}	&	\multirow{2}{3.0cm}{30}				&	NFOV	&	-		\\
								&										&	WFOV	&	-		\\
								&	\multirow{2}{3.0cm}{110}			&	NFOV	&	30.2	\\
								&										&	WFOV	&	32.4	\\
								&	\multirow{2}{3.0cm}{170}			&	NFOV	&	32.0	\\
								&										&	WFOV	&	33.1	\\
    \bottomrule
  \end{tabular}
  \label{tab:storeContainerFPATemperatures}
\end{table}

What can be noted from this data is that there was a discrepancy between the imager temperature during calibration (when the transfer function was defined in Eq.~\ref{eq:transfer_function}) and when deployed to the store. The entries in the table are presented chronologically and so what is shown is each imager gradually warming from set to set.

\begin{figure}[H]
  \centering
  \input{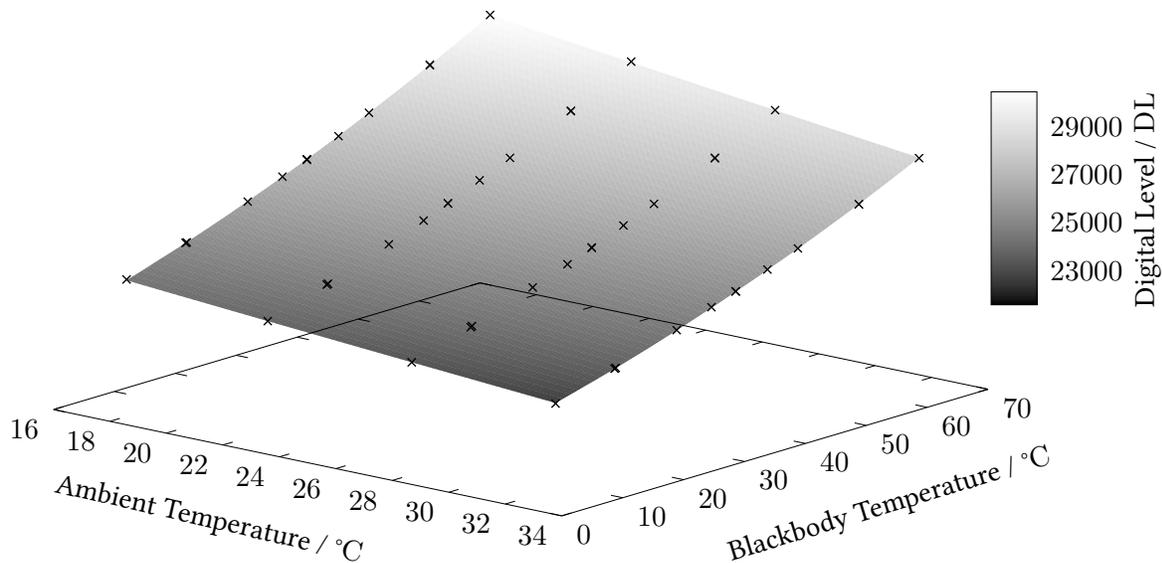}
  \caption{A separate calibration of a different FLIR Boson (model: 20640A050-9PAAX) that was calibrated within an environmental chamber. The plane describes the function relating digital level, blackbody temperature and FPA temperature. This was used to correct the store measurements of the thermal imagers.}
  \label{fig:thermalImagerChamberCalibration}
\end{figure}

The effect this discrepancy had on apparent temperature determination varied between the two imagers, but the WFOV system was more sensitive to this variation. As a result, a correction from this FPA temperature offset was applied; the magnitude of this correction was calculated by applying the results from an independent calibration of a similar instrument (FLIR Boson 20640A050-9PAAX). The data from that calibration was used to define a plane (two-dimensional function) of the same form used to build the deployed thermal imager calibration function (shown in Figure~\ref{fig:thermalImagerChamberCalibration}). The digital level correction was calculated using this plane at the respective FPA temperature and corrected thermocouple measurement. An exemption to this correction is for the THORP measurements using the WFOV imager at \SI{30}{\celsius} and \SI{110}{\celsius}; here this correction was not sufficient to account for the FPA temperature offset and so an additional correction of \SI{10150}{\digitallevel} and \SI{1900}{\digitallevel} was added to the functional correction.

Using the measured digital level, calculated FPA correction and FPA temperature, the apparent radiance temperature was calculated. This function is detailed in \cite{ref:thermalImagerCertificate2021040168_1,ref:thermalImagerCertificate2021040168_2}.

\paragraph{Temperature evaluation}\label{para:temperatureEvaluation}
An emissivity corrected surface radiance temperature for each measurement set was determined using the multiple reflection single wavelength method described in Section~\ref{subsec:multipleReflectionSingleWavelength}. Requirements for this calculation are: apparent radiance temperature (from the calibrated thermal imager), reflected background radiance temperature (from the calibrated hygrometer remotely positioned by the workstation), wavelength of the instrument and hemispherical total emissivity of the surface. It was assumed that the reflected component evaluated by the hygrometer originated from a blackbody surrounding the measured surface.

Due to the complexity in determining the particular directional emissivity in each thermal image, instead the hemispherical total emissivity was used; this assumption should be considered an under-estimation of the emissivity due to the angle of measurement (greater than \SI{10}{\degree} from the surface plane). The absence of spectrally considered radiance calculation (e.g. multiple reflection all wavelength radiance correction) should also be considered carefully if there are large emissivity variations within the spectral range being considered. The hemispherical total emissivity for the Nextel coating used was assumed to be \(0.9381 \pm 0.0056\) (\(k=1\)) \cite{ref:emissivity_data_coatings} and the uncoated region was estimated to have an emissivity of \(0.45 \pm 0.08\) (\(k=1\)) \cite{ref:steelEmissivity,ref:endToEndSNM}. The spectral response of the thermal imagers is from \SIrange{7.5}{14.0}{\micro\metre} and the value used for lambda was \SI{10.75}{\micro\metre} in these calculations.

\begin{figure}[H]
  \centering
  \input{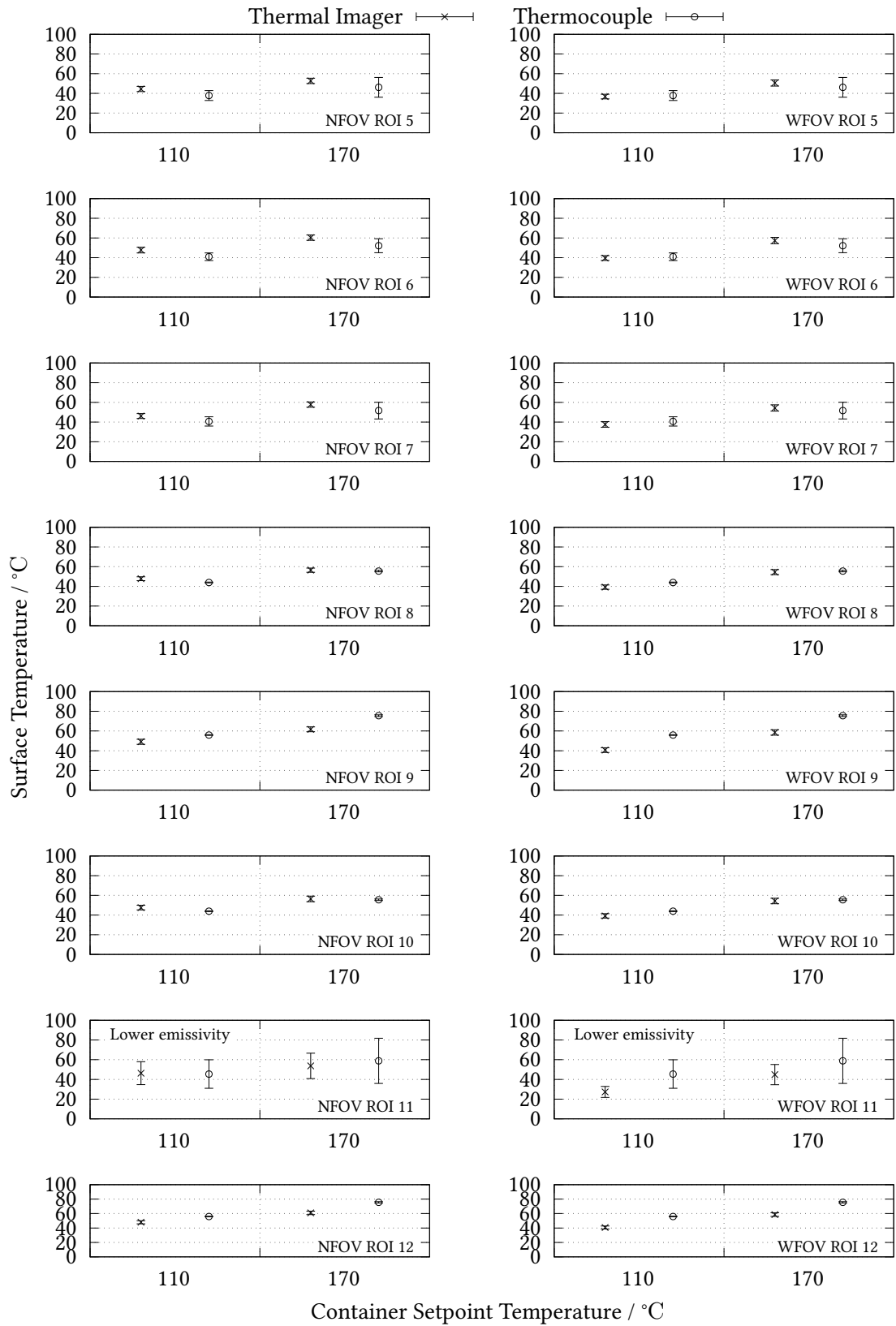}
  \caption{Magnox surface temperature determined using the thermal imager and thermocouple. ROIs are at high emissivity coating locations unless specified.}
  \label{fig:snmStoreContainerTemperatureComparisonMagnox}
\end{figure}

\begin{figure}[H]
  \centering
  \input{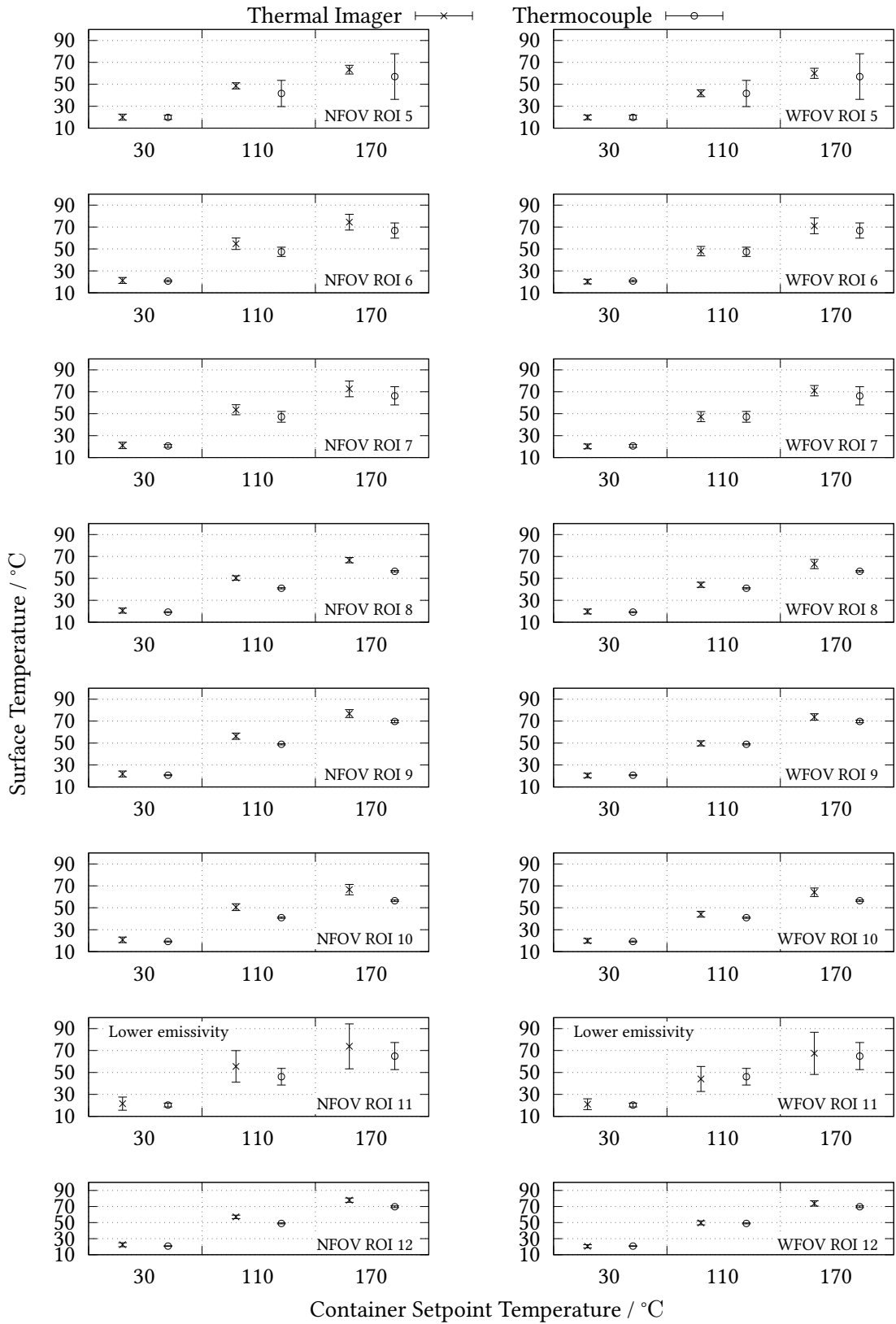}
  \caption{THORP surface temperature determined using the thermal imager and thermocouple. ROIs are at high emissivity coating locations unless specified.}
  \label{fig:snmStoreContainerTemperatureComparisonTHORP}
\end{figure}

The surface temperature determinations from thermal imaging and thermocouples for each respective region of interest, temperature setpoint and imager is shown for the Magnox container in Figure~\ref{fig:snmStoreContainerTemperatureComparisonMagnox}. The corresponding THORP measurements are shown in Figure~\ref{fig:snmStoreContainerTemperatureComparisonTHORP}. The error bars shown represent the \(k=2\) uncertainty of each measurement, where the uncertainty budgets are detailed in Section~\ref{subsec:storeContainerUncertainty}. These error bars do not include the FPA components as discussed in Section~\ref{para:calibrationInterpretation}, these could be mitigated with a carefully deployed test campaign and would have effected a large influence on the error bars that diminishes the behaviour of other uncertainty components.

\subsection{Uncertainty analysis}\label{subsec:storeContainerUncertainty}
The following section details the uncertainties and estimated combined uncertainties for the thermal characterisation of the Magnox and THORP instrumented containers.

\subsubsection{Thermocouple uncertainties}\label{subsubsec:storeContainerThermocoupleUncertainty}
Thermocouple uncertainty components are detailed here. The thermocouples associated with each region of interest are detailed in Table~\ref{tab:storeROIDefinition}.

\paragraph{Instrumentation}\label{para:storeThermocoupleInstrumentation}
{\em Calibration uncertainty}

This is reported in the respective calibration certificates as a normal distribution and a coverage factor of 2.0 \cite{ref:thermocoupleCertificate2020120016,ref:thermocoupleCertificate2021030425}. From \SIrange{5}{65}{\celsius} this is \SI{0.05}{\celsius} (\(k=2\)) and from \SIrange{65}{200}{\celsius} this is \SI{0.75}{\celsius} (\(k=2\)).

{\em Calibration fit error}

Using the correction data from the certificates for each thermocouple either a polynomial fit or mean offset was used to describe the respective corrections. This was dependent on the characteristic of the correction across the full temperature range (i.e. did the data suitably fit a second-order polynomial). For the polynomial based corrections, this component is the root mean squared from the fit; for the mean offset corrections this is the standard deviation of the corrections reported at each temperature, both have a normal distribution.

\paragraph{Application}\label{para:storeThermocoupleApplication}
{\em Measurement stability error}

The standard deviation of the thermometer over the measurement period was recorded and in the instances of multiple thermometers defining the ROI, the mean of these respective deviations was reported. This was evaluated as a normal distribution.

{\em Measurement sampling error}

For single thermometer ROIs, this component was zero, otherwise this value was evaluated as the mean of the respective thermometer standard deviation measurements. This represents the spatial temperature differences measured by the thermometers and is a normal distribution.

\subsubsection{Radiance temperature uncertainties}\label{subsubsec:storeContainerRadianceUncertainty}
These values are each converted from their respective units (e.g. apparent radiance temperature, FPA temperature, emissivity and room temperature) to radiance temperature using sensitivity coefficients. These are described in Section~\ref{subsubsec:laboratoryThermalImagerUncertainty} and the FPA component sensitivities are described in their respective sections.

\paragraph{Instrumentation}\label{para:storeThermalImagerInstrumentation}
{\em Calibration}

This component collates all of the reported sub-components assessed within the laboratory characterisation of the instruments. This includes calibration of the reference source, stability of the reference source, stability of the instrument, size-of-source effect reproducibility, uniformity across the centrally assessed pixels, the residual of the calibration fit, stability over a \SI{30}{\minute} period, the effect of distance variation and the alignment to the mirror \cite{ref:thermalImagerCertificate2021040168_1,ref:thermalImagerCertificate2021040168_2}. This was evaluated as a normal distribution with a coverage factor as indicated by the calibration.

\paragraph{Application}\label{para:storeThermalImagerApplication}
{\em Measurement stability error}

From each respective set of images for an ROI in a measurement set, the standard deviation of the mean was evaluated. This typically represented the temporal variation of the measurement data. However in this instance where the ROI was manually positioned across the field of view this also represents the repeatability of ROI placement and temperature variation across the detector field of view. This is assessed as a normal distribution.

{\em Measurement spatial error}

Conversely to the above, the standard deviation across each ROI during the measurement was recorded, describing the temperature variation across the surface. The mean standard deviation across the appropriate ROI assessments was evaluated as a normal distribution.

{\em FPA stability}

During the measurements there was a variation of FPA temperature and this standard deviation was evaluated.  The sensitivity of radiance temperature, \(t_{90}\), to FPA temperature, \(T_{FPA}\), was derived from Eq.~\ref{eq:transfer_function} and assessed using,

\begin{equation}
  \frac{\partial t_{90}}{\partial T_{FPA}} = c \cdot \left( S_{DUT} + S_{Correction} \right) + 2d \cdot T_{FPA} + e \mathrm{.}
  \label{eq:sensitivityFPAStability}
\end{equation}

Where \(S_{DUT}\) is the digital level measured, \(S_{Correction}\) is the applied digital level correction and the coefficients \(c, d\) and \(e\) are from those described in the respective calibration certificates. This was evaluated as a normal distribution.

{\em FPA correction applied}

The plane correction and additional surplus correction applied to any measurement sets was accounted for by this component. The magnitude of the correction was considered as a rectangular distribution and so a \(2\sqrt{3}\) divisor was applied. The sensitivity of this component was derived from Eq.~\ref{eq:transfer_function} and evaluated using,

\begin{equation}
  \frac{\partial t_{90}}{\partial S} = 2a \cdot \left( S_{DUT} + S_{Correction} \right) + b + c \cdot T_{FPA} \mathrm{.}
  \label{eq:sensitivityFPACorrection}
\end{equation}

Where the coefficients \(a, b\) and \(c\) are from those described in the respective calibration certificates. This was evaluated as a normal distribution.

{\em Emissivity}

The coated and uncoated regions had values and uncertainties defined in Section~\ref{subsubsec:storeContainerRadianceEvaluation}. The corresponding sensitivity for each region of interest was calculated and this was assessed as a normal distribution.

{\em Room temperature uncertainty}

Corresponds to the instrumentation uncertainty of the hygrometer used. From the calibration certificate, this is \SI{0.2}{\celsius} (\(k=2\)). This was assessed as a normal distribution and the appropriate sensitivity calculated.

{\em Variation in room temperature}

Where the standard deviation in room temperature exceeded the instrumentation uncertainty, this value was also considered for its effect on radiance temperature determination. This was evaluated as a normal distribution with the appropriate sensitivity calculated.

\subsubsection{Compiled uncertainty budgets}\label{subsubsec:storeContainerCompiledUncertainty}
An excerpt of the uncertainty budgets (total of 80) are presented in Table~\ref{tab:snmStoreUncertaintyCompleteMagnox} and Table~\ref{tab:snmStoreUncertaintyCompleteTHORP}. Due to the nature of the measurement, there were too many fine details to demonstrate in this subset of data.  Below are the budgets for a coated ROI from the \SI{30}{\celsius} THORP measurements and an uncoated ROI from the \SI{170}{\celsius} Magnox measurements respectively. These budgets are the largest uncertainty budgets for each container.

\begin{table}[H]
  \centering
  \caption{Uncertainty budget from the \SI{30}{\celsius} THORP measurements using the WFOV imager at the coated ROI 5.}
  \vspace*{\floatsep}
  \begin{tabular}{L{5.0cm}M{1.5cm}M{1.5cm}M{1.5cm}M{1.5cm}}
    \toprule
	Source	&	\(u\)	&	Divisor	&	Sensitivity	&	\(U\) / \SI{}{\celsius} \\
    \cmidrule(lr){1-5}
	\multicolumn{5}{c}{Thermocouple} \\
    \cmidrule(lr){1-5}
	Calibration uncertainty		&	0.05	&	2.00	&	1.00	&	0.03	\\
	Calibration fit error		&	0.08	&	1.00	&	1.00	&	0.08	\\
	Measurement stability error	&	0.12	&	1.00	&	1.00	&	0.12	\\
	Measurement sampling error	&	0.94	&	1.00	&	1.00	&	0.94	\\
    \cmidrule(lr){2-5}
	\multicolumn{4}{r}{Expanded uncertainty (\(k=2\))}				&	{\bf 1.9} 	\\
    \cmidrule(lr){1-5}
	\multicolumn{5}{c}{Thermal Imager} \\
    \cmidrule(lr){1-5}
	Calibration										&	2.10	&	2.10	&	1.04	&	1.04	\\
	Measurement stability error						&	0.18	&	1.00	&	1.04	&	0.19	\\
	Measurement spatial error						&	0.19	&	1.00	&	1.04	&	0.19	\\
	FPA stability									&	0.57	&	1.00	&	22.65	&	12.82	\\
	FPA correction applied							&	5466	&	1.73	&	0.01	&	17		\\
	Emissivity										&	0.01	&	1.00	&	4.42	&	0.03	\\
	Room temperature uncertainty					&	0.10	&	1.00	&	0.06	&	0.01	\\
	Variation in room temperature					&	0.42	&	1.00	&	0.06	&	0.03	\\
    \cmidrule(lr){2-5}
	\multicolumn{4}{r}{Expanded uncertainty (\(k=2\))}									&	{\bf 42.7} 	\\
    \bottomrule
  \end{tabular}
  \label{tab:snmStoreUncertaintyCompleteTHORP}
\end{table}

\begin{table}[H]
  \centering
  \caption{Uncertainty budget from the \SI{170}{\celsius} Magnox measurements using the NFOV imager at the uncoated ROI 11.}
  \vspace*{\floatsep}
  \begin{tabular}{L{5.0cm}M{1.5cm}M{1.5cm}M{1.5cm}M{1.5cm}}
    \toprule
	Source	&	\(u\)	&	Divisor	&	Sensitivity	&	\(U\) / \SI{}{\celsius} \\
    \cmidrule(lr){1-5}
	\multicolumn{5}{c}{Thermocouple} \\
    \cmidrule(lr){1-5}
	Calibration uncertainty		&	0.05	&	2.00	&	1.00	&	0.03	\\
	Calibration fit error		&	0.28	&	1.00	&	1.00	&	0.27	\\
	Measurement stability error	&	0.69	&	1.00	&	1.00	&	0.69	\\
	Measurement sampling error	&	11.43	&	1.00	&	1.00	&	11.43	\\
    \cmidrule(lr){2-5}
	\multicolumn{4}{r}{Expanded uncertainty (\(k=2\))}				&	{\bf 22.9} 	\\
    \cmidrule(lr){1-5}
	\multicolumn{5}{c}{Thermal Imager} \\
    \cmidrule(lr){1-5}
	Calibration										&	2.25	&	2.00	&	1.92	&	2.16	\\
	Measurement stability error						&	0.57	&	1.00	&	1.92	&	1.10	\\
	Measurement spatial error						&	1.17	&	1.00	&	1.92	&	2.24	\\
	FPA stability									&	0.42	&	1.00	&	0.61	&	0.26	\\
	FPA correction applied							&	64.1	&	1.73	&	0.01	&	0.30	\\
	Emissivity										&	0.08	&	1.00	&	69.16	&	5.53	\\
	Room temperature uncertainty					&	0.10	&	1.00	&	0.91	&	0.09	\\
	Variation in room temperature					&	0.28	&	1.00	&	0.91	&	0.26	\\
    \cmidrule(lr){2-5}
	\multicolumn{4}{r}{Expanded uncertainty (\(k=2\))}									&	{\bf 12.9} 	\\
    \bottomrule
  \end{tabular}
  \label{tab:snmStoreUncertaintyCompleteMagnox}
\end{table}

The uncertainty bars presented in Figure~\ref{fig:snmStoreContainerTemperatureComparisonMagnox} and Figure~\ref{fig:snmStoreContainerTemperatureComparisonTHORP} omit the two FPA thermal imager components. The two instrument calibration components are detailed in the respective certificates. Thermocouple fit error does not vary much between ROI and measurement set. Thermocouple measurement stability is at an acceptably low level.

The first comparison to draw is the difference between the two thermocouple measurement sampling errors. ROI 5 is the mean between two thermocouples and this component is small. For the ROI 11 measurement the large temperature variation between thermocouple locations leads to a large incurred uncertainty.

For these budgets the thermal imager measurement error components do not demonstrate anything unexpected, they are of an acceptable magnitude and the spatial component is greater than the stability component. There are many budgets where the stability component is greater than the spatial component (e.g. THORP \SI{170}{\celsius} NFOV ROI 9), this is an example of either the imager uniformity or ROI placement incurring an error beyond the expected magnitude of the stability component.

A comparison between the two FPA stability components highlights the variation in sensitivity between the two imager calibration functions and their respective sensitivity to this variation. The magnitude of the uncertainty is comparable, however the effect this had on one imager is greater than that on the other. The FPA correction accounts for how far the FPA temperature was in-situ from the laboratory evaluation. Where this component is greater in the first set (first dataset captured on the campaign) compared to the last set of the campaign, the impact on the budget is considerable. If this was to be better managed, through strict protocol deployment in-situ, these components would have only a minor impact on the surface temperature determination.

As anticipated, the emissivity component had a greater effect for the uncoated region, and this is exacerbated by the increased sensitivity when observing lower emissivity surfaces. For comparison, if the uncoated region (with the estimated uncertainty) was to have a comparable uncertainty to the coated region, the corresponding standard uncertainty in this example would be \SI{0.4}{\celsius} and not a dominant component of the budget.

There is not much variation in the estimated uncertainty for the two room temperature components throughout the measurement campaign. What can be observed is the difference in sensitivity between the two regimes (coated and uncoated); this difference is caused by the emissivity values used for the two surfaces.

In comparison to the uncertainties presented in Section~\ref{sec:laboratoryContainerAssessment} where the instrument was a medium wave infrared thermal imager, the greater wavelength sensitivity of these long wavelength instruments had a detrimental effect on the uncertainty analysis. Note the wavelength component in the sensitivity coefficients (e.g. Eq.~\ref{eq:sensitivityEmissivity}) causes a greater value in this instance.

Throughout these measurements, the thermal imager determined surface temperature was measured with an uncertainty from \SI{1.9}{\celsius} (\(k=2\)) to \SI{7.2}{\celsius} (\(k=2\)) for known emissivity regions (excluding the FPA temperature components). Similarly, for regions of an estimated emissivity, surface temperature was determined to be within an uncertainty from \SI{4.9}{\celsius} (\(k=2\)) to \SI{20.5}{\celsius} (\(k=2\)). These demonstrate that the uncertainty can be managed with confidence if the predominant components are clearly understood.

\subsection{Discussion}\label{subsec:storeContainerDiscussion}
The observations from the store container measurements are discussed here. The comparison between surface temperature determination methods, the uncertainty analysis and key observations are detailed.

\subsubsection{Temperature comparisons}\label{subsubsec:storeContainerTemperatureComparison}
Each of these comparisons detail the surface temperature determined independently by the thermal imagers and thermocouples respectively. A broad observation that the external surface temperature increases as the internal heater temperature increase can be made. The rate of increase at the setpoints is spatially dependent as shown by the ROI variation.

From detailed comparison between the two methods at each of the respective regions of interest and using each imager, a finer set of conclusions can be observed. The corrected NFOV thermal imager measurements are significantly higher than the thermocouple measurements at \SI{110}{\celsius} and above outside of the combined estimated uncertainties. Whilst higher, the thermal imager measurements were not outside of the combined estimated uncertainties for ROIs 5, 6, 7 and 11 because the corresponding thermocouple uncertainties were large at those points. This observed offset between thermal imager measurements and thermocouple measurements for the NFOV is understood to be from the difference between calibration and in-situ FPA temperatures.  Additional uncertainty may also be present due to the actual surface location difference between the defined ROI and thermocouple location, leading to a thermal gradient not accounted for in the uncertainties.  The region closer to the rail should be cooler due to higher conductive heat transfer from the container through to the support rail and this holds given their respective locations. This thermal gradient effect (due to heat transfer to the support rail) can be observed in each of the ROI 8, 9, 10 and 12 measurements. These ROIs are at either end and away from the central strip. This overestimation from the thermal imager may be due to that ROI location having a greater temperature than the associated thermocouple location.

The thermocouple uncertainty at ROIs 5 and 11 are discernibly greater than at other measurement locations. ROI 5 is impacted by the large (up to \SI{15}{\celsius}) difference between thermocouples 12 and 13. This follows to ROI 11, the differences are up to \SI{13}{\celsius} but principally due to thermocouple 10 being much cooler than the other three thermocouples.

Given the additional FPA temperature correction applied to the THORP WFOV measurements at \SI{30}{\celsius} and \SI{110}{\celsius} there does not appear to be a clear systematic offset at these measurements that had not been accounted for. Note that this correction is a result of the calibration function being defined from the calibration data demonstrating a large gradient in the plane of FPA temperature beyond the laboratory conditions. This could be better managed through either greater stabilisation time of both instrument and environment ahead of the measurement, as shown in the subsequent THORP and Magnox measurements, or by a complete sensitivity calibration of the thermal imager (shown in Figure~\ref{fig:thermalImagerChamberCalibration}).

Results from the Magnox measurements, as with the THORP data ROIs 5, 6 and 7 thermal imager and thermocouple results are in agreement within the uncertainties. This is primarily due to the larger uncertainty coverage facilitating this and that the thermal imager measurement was hotter than the thermocouple during NFOV measurements, there is varying agreement for the WFOV data. There is agreement between the ROI 8 and 10 (and ROI 9 and 12) measurements suggesting the size of the ROI does not influence the measurement capability negatively.

Uncoated region measurements were in agreement between the thermal imager and thermocouple and the larger estimated uncertainty due to uncertainty in surface emissivity value is clearly observed.

Improvements to the thermal imager measurement in-situ could be achieved through a broader thermal imager calibration to incorporate FPA temperature assessment (such as that shown in Figure~\ref{fig:thermalImagerChamberCalibration}). Additionally, using a thermal model to identify regions of lower thermal gradients where ROI and thermocouple locations could have been located to minimise spatial uncertainty issues would have enabled lower uncertainty comparisons.

\subsubsection{Uncertainty component assessment}\label{subsubsec:storeContainerUncertaintyComponentAssessment}
The larger uncertainty components can be approached systematically. For ROI 5, its two largest components are thermal imager calibration (due to FPA temperature) and spatial components and so the calibration component may be addressed as discussed above. ROI 6 and 7 (centre of the strip) had an even larger spatial component but also a large temporal component that suggests the ROI locations were not well placed between imager positions.

Additional sources of uncertainty are: hygrometer location, inter-reflections and thermal imager distance variation. During these measurements the radiance correction was applied using the ambient conditions measured by the hygrometer located external to the measurement site. Ideally this ambient component would represent the surface temperature in the specular direction from observation. If this measurement underestimates the in-situ reflected radiance component, and it would actually be hotter, then the emissivity compensation will be too great and the presented radiance temperatures should be increased.

The assumption made for the radiance correction, that there is only one reflection component, is a simplification and ignores any influence multiple surface reflections would have on the correction radiance temperature. In an instance where the measured apparent radiance temperature comprises, for example, a reflection from a support rail -- as demonstrated by position two in the radiance correction case study (Section~\ref{sec:radianceCorrection}) -- then the corresponding radiance temperature would differ to that currently estimated. Approaching complex radiance scenarios such as the one presented can be challenging and prone to misinterpretation; deployment of 3D thermal imaging (ray tracing based correction) may enable improved results \cite{ref:3d_thermal_imaging,ref:nplPBRT}.

During the calibration of the thermal imagers, the effect from distance variation between the mirror and reference source was investigated from \SIrange{75}{115}{\milli\metre}. If the container surface measured was outside of this range then there would be a linear correction to be applied.  Determining this at each pixel requires the implementation of the geometrical correction and rendering of the surface point cloud. This can be expressed by comparison of the ROIs respective positions in the imager field of view. ROIs 5, 6 and 7 were further from the imager and would have measured slightly lower apparent radiance temperatures than if they were closer. Compared to ROIs 8 through 12, the apparent radiance temperature will be slightly greater than if they were positioned further from the imager.

\subsubsection{Key observations}\label{subsubsec:storeContainerKeyObservations}
Deployment of the laboratory calibrated thermal imagers to a thermally active environment given the FPA temperature sensitivity was considered a large error source. Whilst this was addressed through the deployment of additional corrective measures and the subsequent uncertainties; when highlighting the measurement sets that occurred when the system was thermally stable (i.e. Magnox and THORP at \SI{170}{\celsius}) then it can be demonsrated that these uncertainty components become negligible. When including these FPA components, the measurement data is identical to that presented in Section~\ref{subsubsec:storeContainerTemperatureComparison} however the uncertainties represented by the error bars included the two FPA temperature components. The error bars presented are considerably larger here for some sets but for those indicated above, they are equivalent to those in the presented data.

The discrepancy between the anticipated field of view (depicted by the ROI placement in Figure~\ref{fig:snmStoreContainerExternalSchematic}) and the in-situ measurement (visualised in Figure~\ref{fig:snmStoreContainerROIDefinition}) is both due to the variation in distance between laboratory estimates and the rail deployment, but primarily due to the observation angle. If the observation trolley was perpendicular to the container surface then a greater container surface coverage would be achieved.

Due to the scope of the project, the mirrors used for off-axis observation were practical but not optimised for the application.  Alignment between the imager field of view and their size led to regions of the thermal imager field of view being unusable for measurement.  Whilst the reflectivity of the mirror was not independently measured (beyond the manufacturer specification), the entire system (thermal imager and mirror) calibration accounted for this characteristic; however the effect from non-perfect reflectivity of the mirror on temperature measurement was not fully explored. This may have led to apparent radiance temperatures that were lower than that measured by a perfect reflector. Despite these concerns, the measurement data did not indicate any large systematic error that would have originated from the use of the mirrors and so any further development with bespoke off-axis thermally-regulated observation apparatus would not be adversely affected through this method.

\subsection{Summary}\label{subsec:storeContainerSummary}
Calibration of the two thermal imaging systems was carried out, this included definition of the signal to temperature function and evaluated a range of system characteristics that could have led to systematic error in application. As discussed in Section~\ref{para:calibrationInterpretation}, the coverage of the calibration in both the blackbody reference temperature and FPA temperature space was not sufficient for the application; this resulted in additional corrections to be implemented for interpretation of the results. Further deployment of these systems to a store should ensure suitable calibration ahead of the measurements to accommodate the anticipated environmental temperature range through environmental chamber calibration.

\begin{figure}[H]
  \centering
  \includegraphics[width=\textwidth,keepaspectratio]{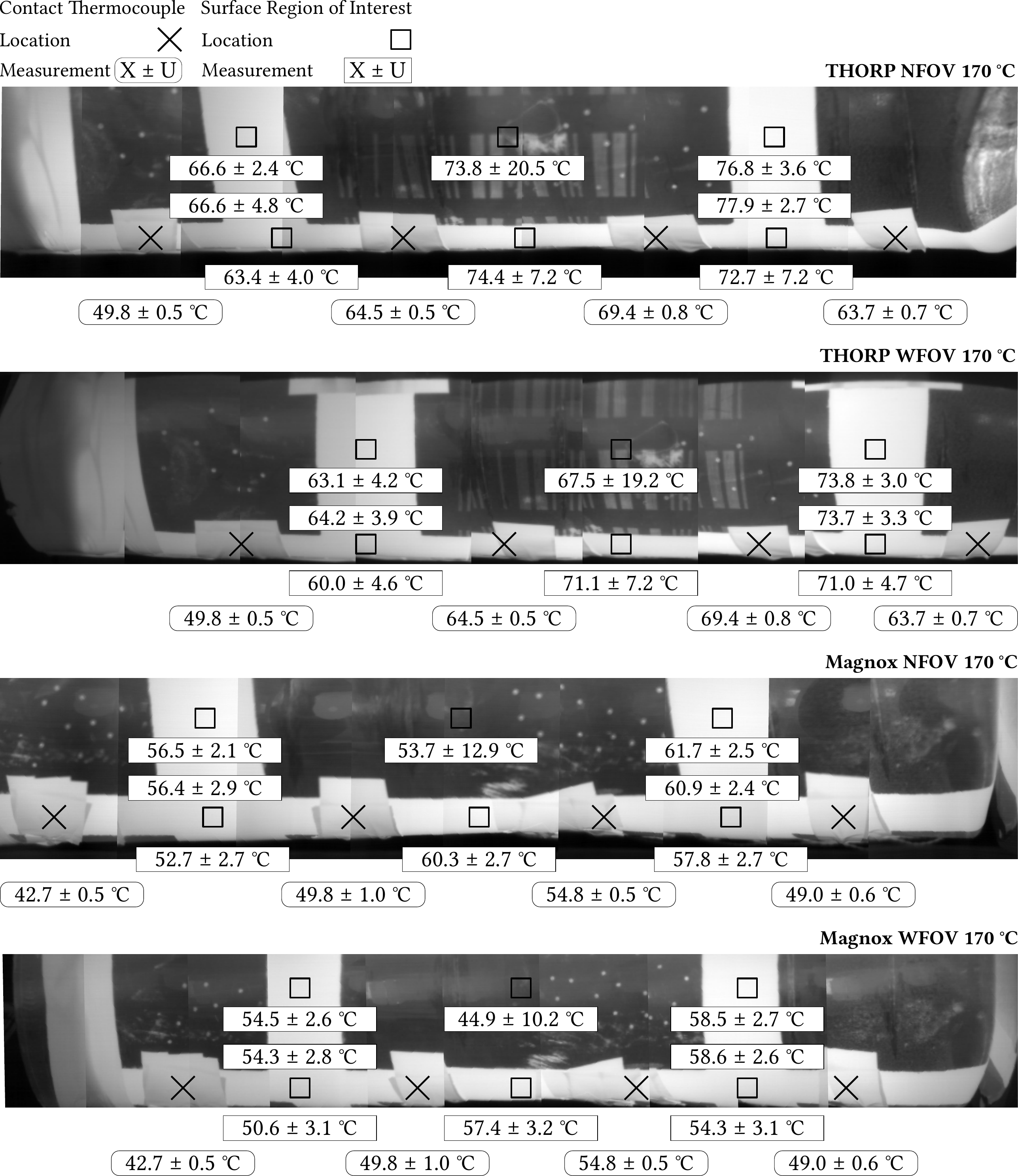}
  \caption{Summary of the surface temperature measurements overlaid on qualitative representations of the thermal imager data. The respective ROI positions correspond to those shown in Figure~\ref{fig:snmStoreContainerExternalSchematic}. Measurement uncertainty reported is the expanded combined uncertainty with a coverage factor of 2.0 and does not include the FPA components.}
  \label{fig:snmStoreContainerTemperatureSummary}
\end{figure}

Suitability of the uncooled microbolometer to the inactive store was shown to be successful when the instrument and environment thermal stability was mitigated, shown in the uncertainty budget for the uncoated measurement of the Magnox container (refer to Table~\ref{tab:snmLaboratoryUncertaintyCompleteMagnox}). Impact from the FPA temperature components here was minimal and the remaining largest uncertainty components were: emissivity effect, ROI spatial contribution and the instrument calibration. The latter two components are intrinsic to the measurement application and so cannot be actively reduced. As highlighted in Section~\ref{sec:laboratoryContainerAssessment}, the use of traceable low uncertainty emissivity measurement in radiance temperature determination activities would enable an order of magnitude improvement for the resulting expanded uncertainty. The example given in Section~\ref{subsubsec:storeContainerCompiledUncertainty} suggested that for that budget the expanded uncertainty could be reduced to \SI{6.8}{\celsius} through the employment of improved emissivity data. A summary of the surface temperature measurements of each container using the two thermal imagers are shown in Figure~\ref{fig:snmStoreContainerTemperatureSummary}. This is the undistorted data for the NFOV whilst this is the uncorrected WFOV data, due to the inability to apply a geometrical correction.

Demonstrated in Section~\ref{subsubsec:storeContainerRadianceEvaluation}, the two methods of temperature determination (thermal imager and thermocouple) were successfully employed to evaluate the behaviour of the two containers within the store environment. Broadly across the two containers, two thermal imagers and eight regions of interest, there was concurrence between the methods. The objective for this aspect of the investigation was agreement between the thermocouple and radiance temperature within the measurements uncertainty at all points. This was not achieved, however there were regions of interest that were consistently susceptible to error. These ROIs (8, 9, 10 and 12) were observed to be in error due to surface temperature gradients incurring measurable temperature variation from ROI and thermocouple location.

\clearpage
\newpage
\section{Conclusions}\label{sec:containerConclusions}
To extend beyond the laboratory assessment of the test plate in Chapter~\ref{chap:laboratoryMeasurement}, surface temperature determination methods were deployed to a pair of instrumented containers. This was explored in both an identical laboratory environment with a cooled MWIR thermal imager, and a test store representative of an active store environment using a pair of lower cost uncooled LWIR thermal imagers.

The containers during the laboratory measurements were limited in experimental design to internal thermocouples. This incurred an additional challenge to compare surface temperature methods (between thermocouples and thermal imager) because a computational model correction was required. The surface temperature comparison was then subject to the uncertainty of the model. It was shown that at cooler temperatures the methods were in agreement, but above \SI{130}{\celsius} the methods did not agree within their respective uncertainties. This comparison could have been improved through the use of either: externally mounted thermocouples to enable a direct comparison; by refining the model to reduce the incurred thermocouple correction error; or by using a traceable emissivity measurement.

Completion of the in-situ store measurements at the JFNL site enabled demonstration of two independent thermal imaging systems to observe a perpendicularly positioned instrumented container. This investigation explored multiple challenging aspects including: use of off-axis mirror, FPA temperature dependent calibration, coated and uncoated surface temperature measurement.

The use of the mirror did not inhibit the use of the instrumentation, the calibration included the reflectivity properties of the mirror and so there were no complications when deploying this to the in-situ measurements. Additionally the use of these mirrors -- as shown in \ref{fig:snmStoreContainerInspectionPosition} -- enabled reduced profile inspection trolley designs that could facilitate further applications (e.g. cross-channel observation).

Despite the limitation imposed by the calibration with reduced environmental chamber characterisation, this alongside additional data enabled the successful deployment of the imagers for apparent radiance determination. These FPA temperature components ranged from \SI{0.2}{\celsius} (\(k=1\)) to \SI{17.4}{\celsius} (\(k=1\)). Emissivity correction to the apparent radiance temperature was more successful than the laboratory measurements as demonstrated by the similarity in temperature response between the coated and uncoated regions.

As detailed in Chapter~\ref{chap:laboratoryMeasurement}, emissivity was the greatest contributor to the uncertainty budget. Shown through the two uncertainty budgets presented, in the case where emissivity uncertainty was low, the respective component was negligible compared to other larger sources. The effects from these emissivity components ranged from \SI{0.02}{\celsius} (\(k=1\)) for a coated region to \SI{8.1}{\celsius} (\(k=1\)) for an uncoated region. This further supports the necessity to deploy high confidence emissivity data in a radiance temperature measurement campaign.

A summary of the surface temperature measurements is presented in Figure~\ref{fig:snmStoreContainerTemperatureSummary}. Throughout these measurements, the thermal imager determined surface temperature was measured with an uncertainty from \SI{1.9}{\celsius} (\(k=2\)) to \SI{7.2}{\celsius} (\(k=2\)) for known emissivity regions (excluding the FPA temperature components). Similarly for regions of an estimated emissivity, surface temperature was determined to be within an uncertainty from \SI{4.9}{\celsius} (\(k=2\)) to \SI{20.5}{\celsius} (\(k=2\)). The outlined target uncertainties of \SI{10}{\celsius} (\(k=2\)) from Chapter~\ref{chap:introduction} correspond to an order of magnitude below which thermal imaging can be considered a feasible technology to be deployed for this application. Through most of the test cases and uncertainty budgets this was achieved, environments leading to uncertainties in excess of this threshold were due to emissivity. The principle cause for this emissivity uncertainty is both the emissivity uncertainty itself and the difference between ambient and surface temperatures. Under the assumption that the store environment will be warmer than that in this test environment -- and that the primary state of containers will be at a safe low temperature -- thermal imaging can be considered to be a suitable technique.

For the deployment to a future application, the following topics require addressing:

\begin{enumerate}
  \item Thermal imager environmental chamber calibration exceeding the full apparent radiance temperature and environmental temperature range
  \item Mounting instrumentation designed to accommodate sufficient stand off and rotational alignment from observation surface to maximise field of view coverage
  \item Traceable and low uncertainty emissivity measurement data representative of the observation surface, accounting for variation due to lifetime surface degradation
\end{enumerate}

Through the development of these avenues prior to active store deployment, the primary envisaged challenges will be addressed and successful simultaneous surface temperature and defect detection of SNM containers using thermal imagers will be possible.

\clearpage
\newpage
\chapter{Concluding Remarks}\label{chap:concludingRemarks}
The research detailed within this thesis spanned three technical chapters that broadly introduced surface temperature measurement using thermal imagers, then explored this within three applications. 

Two case studies detailed the calibration procedures for each a radiation thermometer and a pair of thermal imagers. The radiation thermometer performed close to the calibration and measurement capability of the calibration laboratory with an uncertainty below \SI{1}{\celsius} across the temperature range from \SIrange{-30}{1000}{\celsius}. A number of differences arose in the uncertainty budget components from the thermometer to each of the thermal imagers; over the temperature range from \SIrange{20}{100}{\celsius} the uncooled and cooled thermal imager demonstrated a measurement uncertainty below \SI{3.2}{\celsius} and \SI{0.4}{\celsius} respectively. It is expected that this discrepancy between thermal imagers would be less if the uncertainty budget components were harmonised, but each was constructed with respect to the requirements for the corresponding application cases.

Through deploying the thermal imagers to the applications in both the laboratory and store, the greatest uncertainty components identified were: estimated emissivity uncertainties in place of values from a calibration; measured spatial non-uniformity across the region of interest; and unstable thermal imager housing temperatures. In the instances of using an estimated uncertainty, the uncertainty for surface temperature ranged from \SIrange{2.5}{7.9}{\celsius}. When a Nextel coating was applied and a lower uncertainty used, the impact on surface temperature uncertainty was as low as \SI{0.03}{\celsius}. It should be noted that this known emissivity of Nextel was from literature and the ideal case would utilise an appropriate calibrated emissivity measurement and an appropriate examination of the temporal degradation of surface emissivity. Spatial non-uniformity contribution can be reduced through the application of a correction as explored in the translation-correction non-uniformity correction method. In these applications the component was solely assessed for uncertainty contribution. Shown through the chronological series of measurements in the store, the effect on surface temperature uncertainty from Focal Plane Array (FPA) components decreased as the FPA temperature approached that measured during the laboratory calibration. What is demonstrated through this measurement campaign was the impact an unstable environment would have on surface temperature determination; for improved deployment, consideration to these effects should be made.

Calculating surface temperature from an apparent radiance temperature requires an understanding of both the surface photo-thermal properties (e.g. emissivity, reflectivity) and the temperature of the environment. The quantity of information required depends on the complexity of radiance correction model employed. During the radiance correction comparison, two correction methods (Multiple Reflection Single Wavelength (MRSW) and Multiple Reflection All Wavelengths (MRAW)) were baselined against the absence of correction. The environment was setup within a sphere at the ambient temperature parameter, observing the observation container from two positions: position one arranged the reflection container in the specular direction; position two arranged the ambient sphere in the specular direction. At position one the reflection container was fixed at a single temperature throughout the computations and this arrangement necessitated additional complexity to the apparent radiance contribution sources. The parameter space explored was three discrete emissivities of silica coated aluminium, stainless steel and Nextel; the observation container temperature ranged from \SIrange{10}{200}{\celsius} and the ambient temperature  was from \SIrange{10}{60}{\celsius}.

For low emissivities, the MRAW method was too sensitive to minor emissivity variations and the error spanned from \SIrange{-215}{89}{\celsius}. Despite this, at greater emissivities this method exceeded the MRSW method by achieving a maximum range of \SI{6.5}{\celsius} and \SI{1.0}{\celsius} across the entire parameter space for the stainless steel and Nextel respectively. For the MRSW assessment at the lowest material emissivity, the error was as low as \SI{13.7}{\celsius}, the range of error for the greater emissivity materials decreased from \SI{10.1}{\celsius} through \SI{5.8}{\celsius}. When no correction is applied, naturally the apparent radiance temperature closely approximates surface temperature when the surface temperature and ambient temperature are equivalent; otherwise the maximum error for the three materials decreased from \SI{-170}{\celsius}, to \SI{-127}{\celsius} and \SI{-4.3}{\celsius}.

Specifically for the uncoated container measurements where: the emissivity was likely close to the stainless steel material; the external surface temperature of the container did not exceed \SI{80}{\celsius}; the ambient temperature was close to \SI{22}{\celsius} in most cases; and the position two arrangement best described the environment, the following values indicate the suitability of each correction method. When uncorrected, the error between apparent radiance and surface temperature reaches \SI{-44}{\celsius}. Both MRSW and MRAW corrections are approximately equal in this regime with the range of error across this reduced parameter space \SI{0.7}{\celsius} and \SI{1.4}{\celsius} respectively.

The research arc throughout this research has been developing surface temperature measurement techniques using a thermal imager from a laboratory-based proxy plate, to a pair of nuclear material containers, through to an inactive store assessment of the same containers. The primary objective through each of these campaigns was to build confidence in the use of thermal imagers in this application by demonstrating agreement between contact thermometers and the thermal imager under a range of conditions. During the initial laboratory plate measurements this agreement was clearly observed when reviewing the reduced set of inspection regions for a coated and uncoated surface. When then assessing the containers in the laboratory environment challenges were encountered from the experimental design because the comparator thermometers were below the surface. This necessitated a thermal model correction to the thermocouple measurements and despite this and the emissivity corrections, agreement was not achieved under the majority of scenarios. Finally at the inactive store campaign there were mixed successes. Due to the constraints of the environment and programme for the test schedule, insufficient time was permitted for thermal imager and environment stabilisation during the earlier measurements; this caused a large uncertainty component from the variation in FPA temperature between the calibration and store measurement. These components were shown to become negligible during the latter measurements where both environment and instrumentation achieved local thermal equilibrium. Further to this, the positioning of the container on the inspection rail led to increased thermal conduction heat loss and a larger temperature gradient between some regions of interest and their respective comparator thermometers. Given these challenges, agreement between the thermometer and thermal imager was observed for the two thermal imagers, at a range of temperatures and for both coated and uncoated surfaces.

Throughout this research the challenges encountered that affected low uncertainty surface temperature measurement with a thermal imager were: emissivity knowledge of the surface, well-considered experimental design when comparing thermometry techniques and instability of instrumentation.

Through an instrument calibration, uncertainty components appropriate for the application should be considered. If large observation distances are experienced then a component for the atmospheric absorption should be considered, whilst this would not be applicable in a close focus production line application. Critical surfaces comprise both traceable emissivity with a measurement uncertainty for the primary observation surface at a minimum and where necessary, subsequent reflection surfaces.

For each uncertainty component included and omitted, full consideration should be given to ensure a robust description of the measurement environment. Where surface temperature is necessary, a radiance correction method suitable for the application should be employed. As employed throughout this work, an analysis of the suitability of this correction method in the application parameters should be made. During a confidence building experiment any comparators should be positioned in locations of lowest thermal gradient.

Experimental setup of the thermal imager should be such to minimise erroneous reflection components. When this cannot be avoided, a controlled surface from the reflection component could be introduced. The complete environment consisting of the thermal imager, investigation surface and local environment should be permitted an appropriate time to achieve thermal equilibrium. When evaluating a region of interest, the impact from instrument non-uniformity and effect from small target sizes should be considered. Often smaller regions of interest should be used to minimise the non-uniformity effect, whilst regions of interest too small may incur an erroneous measurement due to pixel definition.

Given these challenges discussed the principle recommendations that would be given to the researcher implementing some basis of temperature metrology in their application would be:

\begin{enumerate}
  \item Employ a thermal imager calibration with appropriate uncertainty components for the application
  \item The critical surfaces (primary observation surface and any subsequent reflection component surfaces) comprise known emissivity with a measurement uncertainty
  \item Provide a robust description of uncertainty sources within the uncertainty budget
  \item A radiance correction method used is suitable for the application
  \item For a confidence building experiment, place comparators in locations of lowest thermal gradient
  \item Position the thermal imager such to minimise erroneous reflection components or introduce a known reflection surface
  \item Permit the complete environment (thermal imager, investigation surface and local environment) sufficient time to achieve thermal equilibrium
  \item For region of interest assessment consider the non-uniformity across the array and effect from small target sizes
\end{enumerate}

By giving full consideration to these recommendations in line with the appropriate research discussed in this body of work, an appropriate foundation for metrology could be achieved.

\clearpage
\newpage

\pagestyle{empty}
\clearpage
\newpage
\pagestyle{fancyStyle}

\renewcommand\chaptername{Appendix}
\renewcommand*{\thechapter}{\Alph{chapter}}
\setcounter{chapter}{0}
\chapter{Supporting Literature}\label{chap:literature}
\section{Radiation Heat Transfer}\label{sec:radiation_heat_transfer}
The principle concepts relating to radiative heat transfer will be described here. This will outline the requisite understanding for apparent radiance temperature measurement as made by radiation thermometers and thermal imagers.

\subsection{Definition of a blackbody radiator}\label{subsec:radiation_definition_blackbody}
Each material at a temperature greater than absolute zero will emit thermal radiation due to the oscillation and transition of the atoms within the system. Consider a solid object at temperature \(T_{object}\) within a vacuum and surrounded by some medium at temperature \(T_{background}\). From the three forms of heat transfer: conduction, convection and radiation; only radiative heat transfer will exist to the outside system. If \(T_{object} > T_{background}\) then the object will cool and the background will warm until the two bodies reach thermal equilibrium and \(T_{object} = T_{background}\).

This thermal radiation has two important characteristics, the radiation varies per unit wavelength interval and this is described as its spectral concentration. In addition to this, the thermal radiation emitted from a surface may possess a non-uniform directional distribution and as such the intensity of radiation may change when being observed from different directions~\cite{ref:radiation_thermometry}.

A blackbody radiator is an ideal radiating surface that has the following properties:

\begin{enumerate}
  \item	It absorbs all incident radiation with zero spectral or directional dependence
  \item	For a surface at a defined temperature and wavelength, it can not emit more thermal radiation than a blackbody radiator under identical conditions
  \item	Radiation emitted from a blackbody radiator is independent of direction and can be considered an isotropically diffuse (Lambertian) emitter
\end{enumerate}

There exist zero real surfaces that satisfy each of these criteria, however there do exist close approximations to blackbody radiators achieved through the use of cavity geometry.

\subsection{Planck distribution law}\label{subsec:radiation_plancks_law}
The spectral distribution of radiation emitted from a blackbody \(L_{\lambda,b}\) is described by the Planck distribution law~\cite{ref:radiation_thermometry},

\begin{equation}
  L_{\lambda,b}(\lambda,T) = \frac{ 2hc_0^2 }{ \lambda^5 \left[ e^{\frac{ hc_0 }{ \lambda kT }} - 1 \right]} \mathrm{.}
  \label{eq:planck_distribution_law}
\end{equation}

Where \(h\) is the Planck constant\footnote{Planck constant \SI{6.626070e-34}{\joule\second}}; \(c_0\) is the speed of light in a vacuum\footnote{Speed of light in a vacuum \SI{299792458}{\metre\per\second}}; \(\lambda\) is the wavelength of light; \(k\) is the Boltzmann constant\footnote{Boltzmann constant \SI{1.380649e-23}{\joule\per\kelvin}}~\cite{ref:bipm_constants}; and \(T\) is the surface temperature. The first\footnote{First radiation constant \SI{1.191043e-16}{\kilo\gram\metre\tothe{4}\per\second\cubed}} and second\footnote{Second radiation constant \SI{1.438777e-2}{\metre\kelvin}} radiation constants are defined as \(c_1 = 2hc_0^2\) and \(c_2 = \nicefrac{hc_0}{k}\). Equation~\ref{eq:planck_distribution_law} can be written as Eq.~\ref{eq:reduced_planck_distribution_law},

\begin{equation}
  L_{\lambda,b}(\lambda,T) = \frac{ c_1 }{ \lambda^5 \left[ e^{\frac{ c_2 }{ \lambda T }} - 1 \right]} \mathrm{.}
  \label{eq:reduced_planck_distribution_law}
\end{equation}

\begin{figure}[h!]
  \centering
  \input{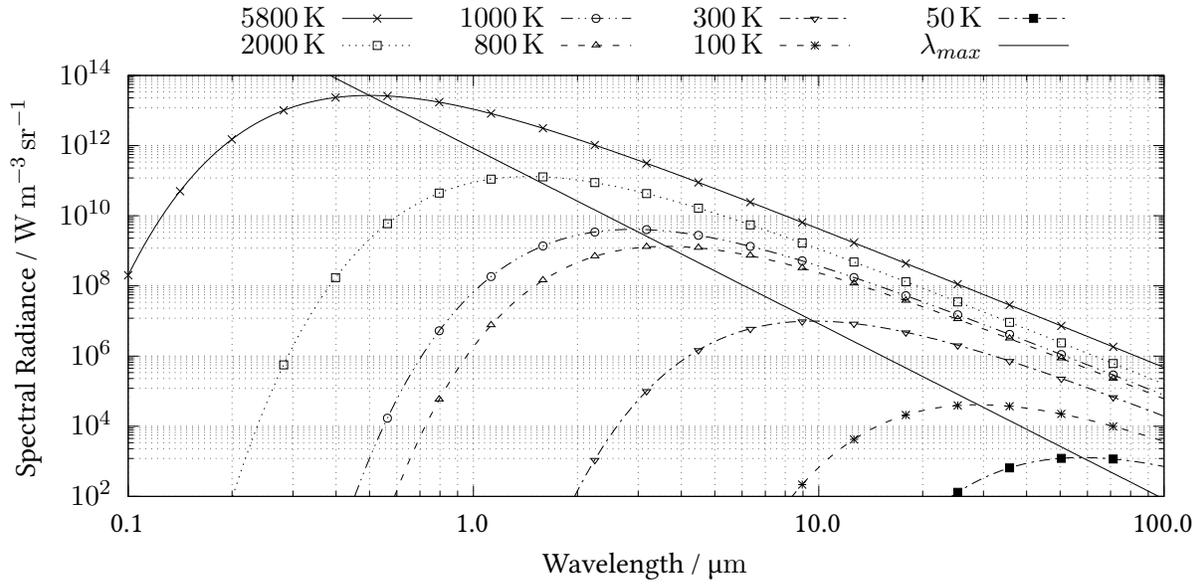}
  \caption{The Planck distribution (Eq.~\ref{eq:reduced_planck_distribution_law}) as a function of wavelength for a defined set of temperatures. The solid line corresponds to the region of peak spectral radiance as a function of wavelength, as described by the Wien displacement law.}
  \label{fig:planck_distribution_law}
\end{figure}

This distribution is shown in Figure~\ref{fig:planck_distribution_law}; it shows that the spectral radiance of thermal radiation changes as the wavelength increases, additionally at increasing temperatures the spectral radiance increases at each wavelength. The maxima of the distribution is located at shorter wavelengths as the temperature increases. This shows that the majority of thermal radiation emitted from the Sun (\SI{5800}{\kelvin}) is centred within the visible range of the electromagnetic spectrum; however, for temperatures in the range of \SIrange{300}{800}{\kelvin} this radiation is centred about the infrared region.

The relationship between the temperature of a surface and the maxima of its spectral distribution \(\lambda_{max}\) can be identified through differentiating Eq.~\ref{eq:reduced_planck_distribution_law} with respect to \(\lambda\) and solving for \(\nicefrac{dL_{\lambda,b}}{d\lambda} = 0\). The result of this is known as the Wien displacement law~\cite{ref:radiation_thermometry},

\begin{equation}
  \lambda_{max} T = c_3 \mathrm{.}
  \label{eq:wien_displacement_law}
\end{equation}

Where \(c_3\) is the third radiation constant\footnote{Third radiation constant \SI{2.8977e-3}{\metre\kelvin}}. The location of this peak spectral radiance is indicated by the solid line in Figure~\ref{fig:planck_distribution_law}.

\subsection{Stefan-Boltzmann law}\label{subsec:radiation_stefan_boltzmann}
The total exitance \(M_b\) of a blackbody can be derived from the Planck distribution law,

\begin{equation}
  M_b = \int\limits_{0}^{\infty} M_{\lambda,b} d\lambda = \int\limits_{0}^{\infty} \pi L_{\lambda,b} d\lambda = \int\limits_{0}^{\infty} \frac{ \pi c_1 }{ \lambda^5 \left[ e^{\frac{ c_2 }{ \lambda T }} - 1 \right] } d\lambda \mathrm{.}
  \label{eq:exitance_blackbody}
\end{equation}

This can be solved to show Eq.~\ref{eq:stefan_boltzmann}, the Stefan-Boltzmann law~\cite{ref:radiation_thermometry},

\begin{equation}
  M_b = \sigma T^4 \mathrm{.}
  \label{eq:stefan_boltzmann}
\end{equation}

Where \(\sigma\) is the Stefan-Boltzmann constant\footnote{Stefan-Boltzmann constant \SI{5.670374e-08}{\kilo\gram\kelvin\tothe{4}\per\second\cubed}} \(\sigma = \nicefrac{ \pi^5 c_1 }{ 15c_2^4 }\).

\clearpage
\newpage
\section{Surface Photo-thermal Properties}\label{sec:photothermal_properties}
In a laboratory, validation of measured temperature follows from International Temperature Scale of 1990 (ITS-90) through to the apparent radiance temperature determined by an instrument. To compare these non-contact surface temperature determination methods with typical instruments (e.g.\ thermocouples and resistance thermometers), the effect from the photo-thermal properties of a surface must be considered. To account for the change in radiance temperature of a surface, its emissivity must be considered.

\subsection{Definition of emissivity}\label{subsec:definition_emissivity}
Emissivity is a property of an object that describes its ability to absorb and emit electromagnetic radiation. Emissivity \(\varepsilon\) is defined as the ratio of radiance \(L_{\lambda,em}\) emitted by an object at a temperature \(T\), observation direction (zenith and azimuthal angles \((\theta_r,\varphi_r)\)) and wavelength \(\lambda\), to the radiance \(L_{\lambda,b}\) emitted by a blackbody radiator at the same temperature. Additionally the emissivity of a surface is dependent on the surface finish of the material (e.g. polished, roughened),

\begin{equation}
  \varepsilon(\lambda;\theta_r,\varphi_r;T) = \frac{ L_{\lambda,em}(\lambda;\theta_r,\varphi_r;T) }{ L_{\lambda,b}(\lambda,T) } \mathrm{.}
  \label{eq:directional_spectral_emissivity}
\end{equation}

As described in Section \ref{subsec:radiation_definition_blackbody}, it is not possible for a real object to be described as a blackbody radiator, therefore the value of emissivity is constrained to the range \num{0}\( \rightarrow \)\num{1}. An object with an emissivity of \num{0} can be considered a perfect reflector and measurement of this would describe the surrounding environment and not the inspection surface. Conversely, an object with an emissivity of \num{1} would be considered a perfect emitter and observation of this surface would describe solely this surface \cite{ref:thermal_radiation_heat_transfer}.

Four expressions of emissivity are: directional spectral emissivity, directional total emissivity, hemispherical spectral emissivity and hemispherical total emissivity.

The {\em directional spectral emissivity} has been introduced in Eq.~\ref{eq:directional_spectral_emissivity}.

{\em Directional total emissivity} is the total emissivity across the entire spectral range and defined as the ratio between the radiance from a surface integrated across all wavelengths, to the radiance from a blackbody across all wavelengths. This can be expressed by Eq.~\ref{eq:directional_total_emissivity},

\begin{equation}
  \varepsilon(\theta_r,\varphi_r;T) = \frac{ \int\limits_{0}^{\infty} L_{\lambda,em}(\lambda;\theta_r,\varphi_r;T) d\lambda }{ \int\limits_{0}^{\infty} L_{\lambda,b}(\lambda,T) d\lambda }  = \frac{ \int\limits_{0}^{\infty} \pi \varepsilon(\lambda;\theta_r,\varphi_r;T) L_{\lambda,b}(\lambda,T) d\lambda  }{ \sigma T^4 } \mathrm{.}
  \label{eq:directional_total_emissivity}
\end{equation}

{\em Hemispherical spectral emissivity} describes an average across all directions within the hemispherical space above a surface,

\begin{equation}
  \varepsilon(\lambda,T) = \frac{ \int\limits_{0}^{2\pi} \int\limits_{0}^{\nicefrac{\pi}{2}} L_{\lambda,em}(\lambda;\theta_r,\varphi_r;T) \cos\theta_r \sin\theta_r d\theta_r d\varphi_r }{ \int\limits_{0}^{2\pi} \int\limits_{0}^{\nicefrac{\pi}{2}} L_{\lambda,b}(\lambda,T) \cos\theta_r \sin\theta_r d\theta_r d\varphi_r }  = \int\limits_{0}^{\nicefrac{\pi}{2}} 2~ \varepsilon(\lambda;\theta_r,\varphi_r;T) \cos\theta_r \sin\theta_r d\theta_r \mathrm{.}
  \label{eq:hemispherical_spectral_emissivity}
\end{equation}

{\em Hemispherical total emissivity} describes the emissivity averaged over the entire spectral range and all directions in the hemisphere above a surface,

\begin{equation}
  \varepsilon(T) = \frac{ \int\limits_{0}^{\infty} \int\limits_{0}^{\nicefrac{\pi}{2}} 2\pi~ \varepsilon(\lambda;\theta_r,\varphi_r;T) L_{\lambda,b}(\lambda,T) \cos\theta_r \sin\theta_r d\theta_r d\lambda }{ \sigma T^4 } \mathrm{.}
  \label{eq:hemispherical_total_emissivity}
\end{equation}

\subsection{Kirchhoff law}\label{subsec:radiation_kirchoff}
Given an object in an isothermal enclosure and both in thermal equilibrium, then the radiation emitted and absorbed by both surfaces across each wavelength and direction must be equal due to the first law of thermodynamics. This is expressed in Eq.~\ref{eq:kirchhoff_law} \cite{ref:thermal_radiation_heat_transfer},

\begin{equation}
  \varepsilon(\lambda;\theta_r,\varphi_r;T) = \alpha(\lambda;\theta_r,\varphi_r;T) \mathrm{.}
  \label{eq:kirchhoff_law}
\end{equation}

Where \(\alpha(\lambda;\theta_r,\varphi_r;T)\) is the directional spectral absorptivity of a material.

A description of the energy conservation at a surface during the interaction between light and matter can be given as,

\begin{equation}
  \alpha(\lambda;\theta_r,\varphi_r;T) + \rho(\lambda;\theta_r,\varphi_r;T) + \tau(\lambda;\theta_r,\varphi_r;T) = 1 \mathrm{.}
  \label{eq:absorptivity_reflectivity_transmission}
\end{equation}

This states that the incident energy upon a surface must be either absorbed by the surface, \(\alpha(\lambda;\theta_r,\varphi_r;T)\), reflected, \(\rho(\lambda;\theta_r,\varphi_r;T)\), or transmitted through the medium, \(\tau(\lambda;\theta_r,\varphi_r;T)\). Following Eq.~\ref{eq:kirchhoff_law} for an opaque medium with zero transmission, Eq.~\ref{eq:absorptivity_reflectivity_transmission} can be written as,

\begin{equation}
  \varepsilon(\lambda;\theta_r,\varphi_r;T) + \rho(\lambda;\theta_r,\varphi_r;T) = 1 \mathrm{.}
  \label{eq:emissivity_reflectivity}
\end{equation}

This identity defines the relationship between emissivity and reflectivity that is employed to infer one property from another.

\subsection{Experimental determination of emissivity}\label{subsec:determination_emissivity}
\begin{figure}[hb]
  \centering
  \includegraphics[width=0.80\textwidth,keepaspectratio]{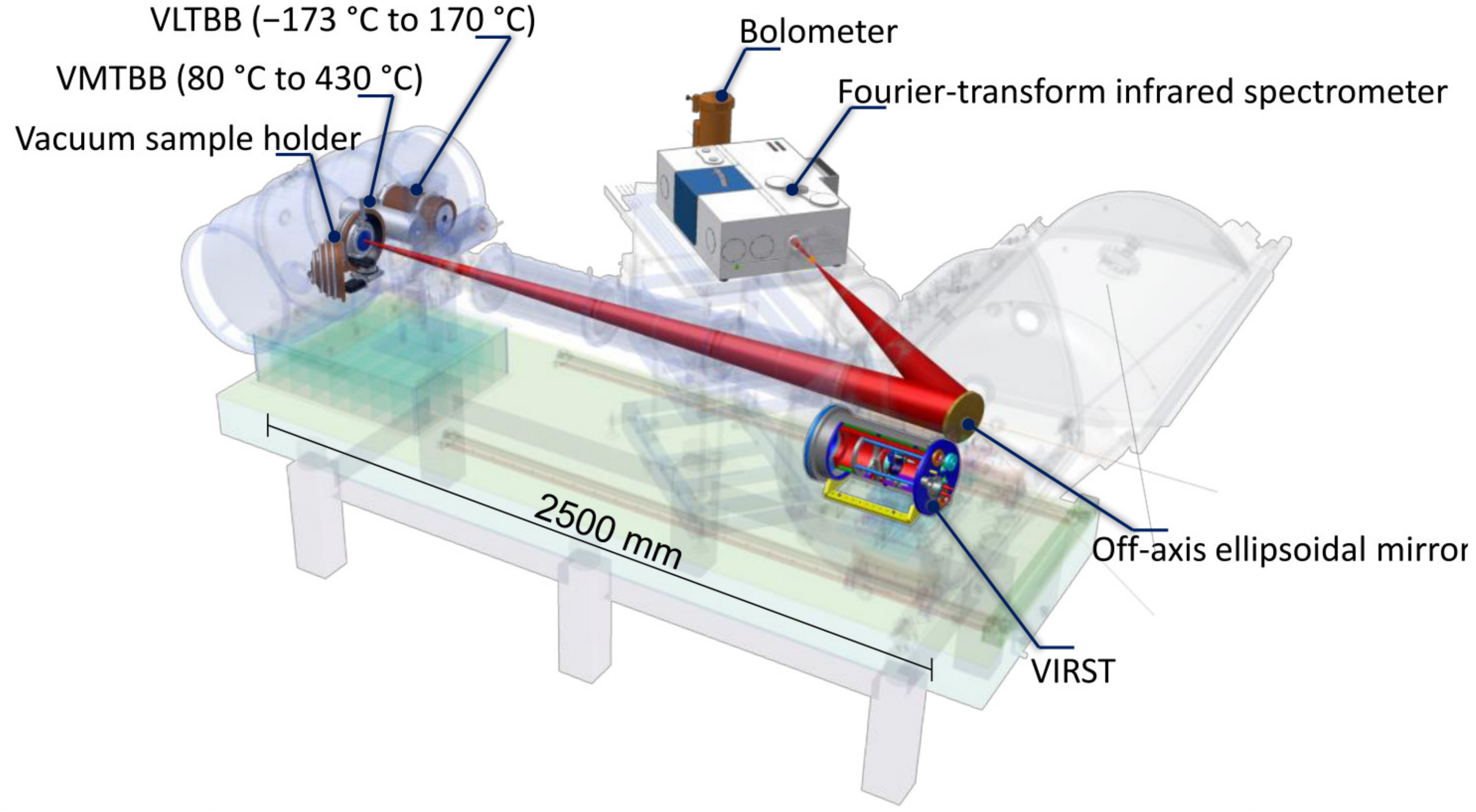}
  \caption{The reduced background calibration facility for the measurement of emissivity at the PTB. The equipment permits the comparison of radiance from either the sample, one of two blackbody sources and a liquid nitrogen cooled blackbody. The latter allows the measurement to correct for the complete radiation budget of the chamber \cite{ref:albert_adibekyan_thesis}.}
  \label{fig:ptb_emissivity}
\end{figure}

There are multiple approaches to the determination of emissivity; there exist both direct and indirect methods and for direct determination both calorimetric and radiometric methods can be used \cite{ref:albert_adibekyan_thesis}. Examples of emissivity measurement facilities include the National Measurement Institute (NMI) in Germany, Physikalisch-Technische Bundesanstalt (PTB) that employs both an in-air facility \cite{ref:uncertainties_of_emissivity_at_ptb} and a vacuum facility \cite{ref:emissivity_vacuum_ptb,ref:emissivity_data_coatings} that measures emissivity using the direct radiometric method, Figure \ref{fig:ptb_emissivity}. The in-air facility can measure the directional spectral emissivity in the temperature range of \SIrange{80}{400}{\celsius} and in the spectral range of \SIrange{4}{40}{\micro\metre}. For a sample of silicon carbide the uncertainty of measurement was determined to be less than \num{0.050} (\(k=2\)). The vacuum facility is suitable for emissivity measurements in the temperature range of \SIrange{-40}{450}{\celsius} and in the spectral range of \SIrange{4}{100}{\micro\metre}. It has been demonstrated to measure the high emissivity of materials (e.g. coatings) to less than \num{0.010} (\(k=2\)) and the low emissivity of reflective materials to less than \num{0.022} (\(k=2\)). For the vacuum facility, the directional emissivity can only be measured in the range of observation angles between \SI{0}{\degree} and \SI{75}{\degree}. The output data from the PTB calibration facility is spectrally resolved in \SI{10}{\nano\metre} steps and \SI{10}{\degree} polar angle steps.

\begin{figure}[!htbp]
  \centering
  \includegraphics[width=0.6\textwidth,keepaspectratio]{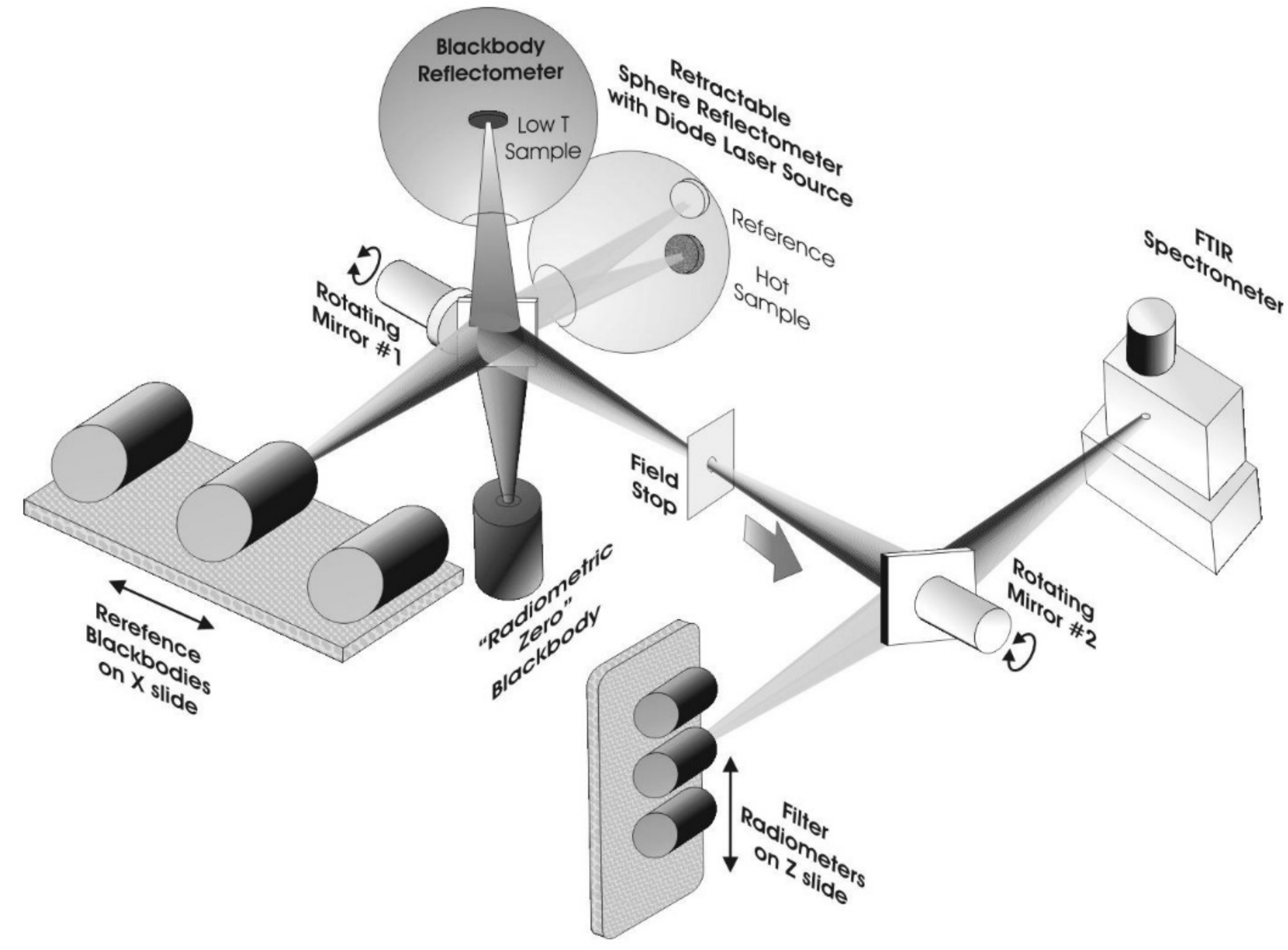}
  \caption{The arrangement of the infrared spectral emittance characterisation facility at NIST. The suite of blackbodies can be directed towards the measurement sample, this radiation can then be detected by the range of radiometers and an infrared spectrometer \cite{ref:spectral_emissivity_characterisation_nist}.}
  \label{fig:nist_emissivity}
\end{figure}

The NMI in the USA, the National Institute of Standards and Technology (NIST) employ the indirect method of an integrating sphere enclosure \cite{ref:measurement_emittance_nist} to determine directional spectral emissivity. This high reflectivity sintered polytetrafluoroethylene sphere allows the measurement of reflectivity in the infrared region at low temperatures due to the low signal necessary. The system measures emissivity in the temperature range \SIrange{327}{1127}{\celsius}, the spectral range \SIrange{1}{100}{\micro\metre} and polar angles from \SIrange{0}{75}{\degree} (shown in Figure \ref{fig:nist_emissivity}) \cite{ref:spectral_emissivity_characterisation_nist}. The uncertainty of measurement for a silicon carbide sample is less than \num{0.040} \((k=2)\). Whilst using the indirect method and relying upon the high diffuse reflectivity of the integrating sphere, its surface becomes more specular at longer wavelengths and there are technical challenges in temperature stability at higher temperatures \cite{ref:albert_adibekyan_thesis}.

The NMI in Denmark, the RIS{\O} National Laboratory use an infrared spectrometer with a spectral response from \SIrange{0.9}{20.0}{\micro\metre} measuring the collimated radiance from either a blackbody cavity reference source, or the sample under test at ambient or elevated to in temperature to \SI{500}{\celsius}. The path of radiation between the sample and detector is not contained within a vacuum, additionally the background measurement is taken at ambient temperatures as opposed to cryogenic temperatures. Due to these experimental restrictions, measurements obtained at low signal levels or within regions of atmospheric absorption require careful evaluation. The uncertainty of measurement for a diffuse spectral emissivity are less than \SI{2}{\percent}; however due to the absence of a documented uncertainty budget, the effect of stray radiation and atmospheric interference are potentially underestimated \cite{ref:emissivity_blackbody_calibrators,ref:measurement_spectral_emissivity_ftir}.

\begin{figure}[!htbp]
  \centering
  \includegraphics[width=0.6\textwidth,keepaspectratio]{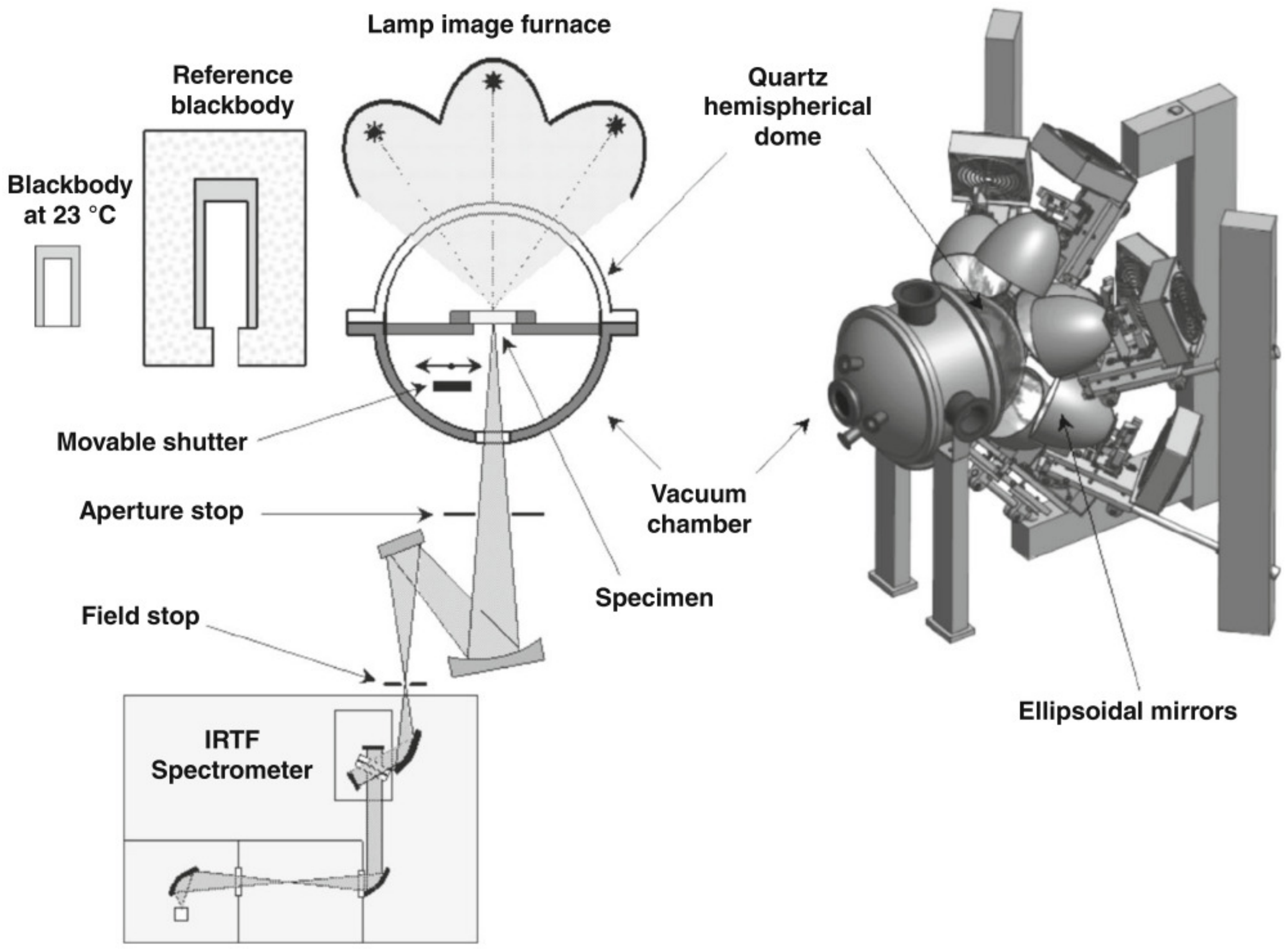}
  \caption{Measurement of directional spectral emissivity at the LNE up to surface temperatures of \SI{1500}{\celsius} using pyroreflectometry. The sample is illuminated by seven \SI{400}{\watt} halogen lamps and the radiance reflected is measured at two independent wavelengths in order to determine the directional spectral emissivity \cite{ref:new_facilities_thermophysical_lne}.}
  \label{fig:lne_emissivity}
\end{figure}

The NMI of France, Laboratoire National de M\'etrologie et d'Essais (LNE) have a range of facilities for the measurement of thermophysical properties of materials. Directional spectral and hemispherical spectral emissivity is determined using an apparatus comprising a blackbody source, a lamp source, a monochromator and set of plane mirrors to select the specified angle of incidence. This setup permits the measurement of emissivity from \SIrange{23}{120}{\celsius} and from \SIrange{1}{14}{\micro\metre} at the angles of incidence \SI{12}{\degree}, \SI{24}{\degree}, \SI{36}{\degree}, \SI{48}{\degree} and \SI{60}{\degree} \cite{ref:measurement_reflectance_emissivity_bnm_lne}. This measurement system  is capable of an uncertainty of emissivity measurement less than \num{0.045} (\(k=2\)) for a coating of Nextel Velvet Coating 811-21. Hemispherical total emissivity is measured using an evacuated chamber, situating a heated sample that permits the calculation of net heat-transfer rate of the system. The model used to describe the relationship between: electrical power supplied to heater plate, the surface temperature of the sample and the surface temperature of the surrounding environment, considers three surfaces in the system. It is built from the radiosity model that assumes that each surface has a Lambertian and spectrally independent emissivity response. The hemispherical total emissivity can be measured from \SIrange{-20}{200}{\celsius} with an uncertainty less than \num{0.030} (\(k=2\)) for materials with a thermal conductivity higher than \SI{0.2}{\watt\per\metre\per\kelvin} \cite{ref:measurement_emissivity_calorimetric_technique}. A facility that permits the measurement of directional spectral emissivity up to \SI{1500}{\celsius} is shown in in Figure \ref{fig:lne_emissivity}. This is operated over the spectral range from \SIrange{0.8}{10}{\micro\metre} through the comparison of radiance, using a Fourier transform infrared spectrometer, between the sample, a heated blackbody reference source and blackbody reference source at nominally ambient temperatures \cite{ref:new_facilities_thermophysical_lne}.

A comparison between the facilities detailed, maintained at a number of NMIs is shown in \cite{ref:cct_comparison_spectral_emissivity}. These facilities each measure emissivity upon different principles and each have independently constructed uncertainty budgets. The comparison between the normal directional spectral emissivity over the temperature range \SIrange{23}{800}{\celsius} and spectral range \SIrange{2}{14}{\micro\metre} demonstrated nominally \SI{98}{\percent} of \num{2000} measurements were in agreement.

\subsection{Traceability of emissivity measurement}\label{subsec:traceability_emissivity}
The traceability of emissivity measurement is derived through its definition of a ratio of radiances, and typically therefore demonstrates an unbroken chain of traceability through to the ITS-90. This temperature traceability originates through the use of calibrated blackbody reference sources and well-defined measurement systems. The properties of a emissivity measurement reference standard would necessitate long-term spectral and directional emission stability. Currently there are no accepted reference standards for emissivity measurement, however there are ongoing efforts from leading NMIs (NIST, PTB, LNE, Insituto Nazionale di Ricerca Metrologica (INRIM) and the National Metrology Institute of Japan (NMIJ)) to determine the suitability from a range of candidates. These reference standard candidates include: boron nitride, oxidised inconel and silicon carbide \cite{ref:cct_comparison_spectral_emissivity}.

The artefacts used in the laboratory comparison comprised cylinders with diameters varying from \SIrange{16}{45}{\milli\metre} and thickness from \SIrange{2}{10}{\milli\metre}. This artefact variation incurs an added level of complexity to the comparison between the laboratories, and so to account for this the reference artefacts were nominally manufactured within the same production run to minimise sample-to-sample variation.

\subsection{Applications of emissivity measurement}\label{subsec:application_emissivity}
Knowledge of material emissivity is a necessary for a number of applications. Primarily it is used for the characterisation of blackbody reference cavities for the fields of radiometry and radiation thermometry, but is also crucial for the correction of apparent radiance temperature measurement observed in industrial environments. The measurement of high emissivity coatings \cite{ref:common_black_coatings} including NPL super black, Halfords matte black, Nextel velvet black coating 811-21 and Pyromark series 2500 high temperature paint, which are applied to surfaces is necessary for the systems. This measurement can also be extended to the characterisation of complete blackbody cavity systems \cite{ref:blackbody_inhomogeneity_gloria,ref:large_area_blackbody_in_flight_calibration}, where the on-board blackbody cavity due to reside on a spacecraft was directionally and spectrally measured.

\subsection{Summary}\label{subsec:summary_emissivity_measurement}
From the measurement facilities presented, the common characteristics are the use of a spectrally-resolved detector to determine the ratio of radiances between the sample under test and one or more blackbody sources. Dense angular measurement in both the zenith and azimuthal planes independently is not currently possible using the existing systems.

Emissivity measurement represents a mature field of metrology; experimental determination of the property defined through the literature ensures a chain of traceability to the SI, through the kelvin and to the Boltzmann constant. Additionally the measurement apparatus utilised by each NMI has demonstrated agreement through international comparison, this provides robust confidence for the measurements and the respective uncertainties.

\chapter{Supporting Measurement Data}\label{chap:measurementData}
\section{Measurement data}\label{sec:appendixMeasurementData}
Measurement data to support the main text is presented here.

\begin{table}[h]
  \renewcommand{\arraystretch}{0.75}
  \centering
  \caption{Measurement results from the radiation thermometer case study. Reference temperature is traceable to ITS-90, instrument output is recorded from the DUT analogue output and sensitivity is determined from the analogue output}
  \vspace*{\floatsep}
  \begin{tabular}{ M{5.0cm}M{5.0cm}M{5.0cm} }
    \toprule
    Reference Temperature / \SI{}{\celsius} & Instrument Output / \SI{}{\milli\volt} & Instrument Sensitivity / \SI{}{\celsius\per\milli\volt} \\
    \cmidrule(lr){1-3}
	\num{-31.1}	&	 18.51		& 				\\
	\num{-30.1}	&	 19.50		& 	1.027		\\
	49.0		&	 94.05		& 				\\
	49.9		&	 94.94 		&				\\
	50.0		&	 94.97		& 	1.047		\\
	99.7		&	 142.12		& 				\\
	100.6		&	 142.98		& 	1.051		\\
	199.9		&	 237.11		& 				\\
	200.8		&	 238.04		& 	1.047		\\
	249.6		&	 284.26		& 				\\
	250.4		&	 285.04		& 	1.011		\\
	\cmidrule(lr){2-2}
	299.2		&	 331.28		& 				\\
	300.1		&	 332.16		& 	1.066		\\
	399.6		&	 426.40		& 				\\
	400.6		&	 427.46		& 	0.975		\\
	500.2		&	 522.05		& 				\\
	500.3		&	 522.05		& 				\\
	501.1		&	 522.83		& 	1.045		\\
	590.0		&	 607.23		& 				\\
	590.6		&	 607.65		& 				\\
	591.6		&	 608.56		& 	1.134		\\
	699.9		&	 711.10		& 				\\
	700.8		&	 712.03		& 	1.000		\\
	\cmidrule(lr){2-2}
	750.5		&	 759.19		& 				\\
	751.5		&	 760.19		& 	1.026		\\
	800.0		&	 805.98		& 				\\
	801.0		&	 806.98		& 	0.995		\\
	900.1		&	 900.96		& 				\\
	901.1		&	 901.98		& 	1.010		\\
	999.9		&	 995.39		& 				\\
	1000.8		&	 996.19		& 	1.060		\\
    \bottomrule
  \end{tabular}
  \label{tab:radiation_thermometer_calibration}
\end{table}

\begin{table}[h]
  \renewcommand{\arraystretch}{0.75}
  \centering
  \caption{Measurement uncertainties from the radiation thermometer case study.}
  \vspace*{\floatsep}
  \begin{tabular}{ M{5.0cm}M{5.0cm}M{5.0cm} }
    \toprule
    Reference Temperature / \SI{}{\celsius} & Expanded Uncertainty / \SI{}{\celsius} & Coverage Factor \\
    \cmidrule(lr){1-3}
	-30 	&	0.25	&	2.1	\\
	50  	&	0.24	&	2.1	\\
	100 	&	0.22	&	2.1	\\
	200 	&	0.23	&	2.1	\\
	250 	&	0.22	&	2.1	\\
	300 	&	0.33	&	2.0	\\
	400 	&	0.36	&	2.0	\\
	500 	&	0.60	&	2.0	\\
	590 	&	0.56	&	2.0	\\
	700 	&	0.65	&	2.0	\\
	750 	&	0.69	&	2.0	\\
	800 	&	0.74	&	2.0	\\
	900 	&	0.85	&	2.0	\\
	1000	&	0.98	&	2.0	\\
    \bottomrule
  \end{tabular}
  \label{tab:radiation_thermometer_uncertainties}
\end{table}

\clearpage

\backmatter
{\setstretch{1.0}
  \bibliographystyle{unsrt}
  \bibliography{thesis}
}

\end{document}